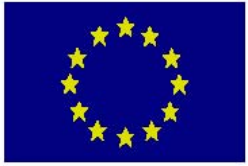
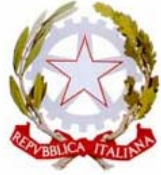
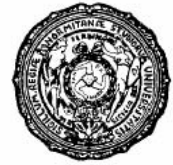

Unione Europea
Fondo Sociale Europeo

Ministero dell'Istruzione,
dell' Università e della Ricerca

Università degli studi di
Palermo

# Experimental investigation on the microscopic structure of intrinsic paramagnetic point defects in amorphous silicon dioxide

## Ph. D. Thesis

*Gianpiero Buscarino*

Supervisore: Prof. Franco Mario Gelardi
Coordinatore: Prof. Natale Robba

## DOTTORATO DI RICERCA IN FISICA

Dipartimento di Scienze fisiche ed Astronomiche

Palermo, February 2007

*To Giovanna and my family*

# Contents

# Acknowledgements

I am grateful to a number of people who helped and assisted me with my work on this Ph. D thesis.

First of all I would like to thank Prof. Franco Mario Gelardi for his guidance and advice and for being always an estimable scientist who has conveyed enthusiasm for research and science. With a busy schedule he has always been very responsive to my requests and always had great ideas about my research.

I thank Dr. Simonpietro Agnello for having shared with me a lot of time during the experiments, for his unconditioned availability and for the precious suggests both professional and human which made this work less hard.

I would like to manifest my sincere gratitude to Prof. Roberto Boscaino for his permanently renewed passion of science, his thorough insights and the time he shared with me. I also thank Prof. D. L. Griscom, Prof. M. Cannas, Prof. M. Leone, Prof. A. L. Shluger, Dr. A. Kimmel, Prof. V. A. Radzig, Prof. A. N. Trukhin, Dr. L. Skuja, Dr. P. V. Sushko, Prof. R. Passante and Prof. A. Emanuele for the interesting discussions and for having shared with me their extensive knowledge and experiences.

I would like to thank, moreover, all my friends and colleagues. Particularly, I'm grateful to Dr. Fabrizio Messina for the interesting and fructuous discussions and to Laura Nuccio for proofreading my thesis.

I am grateful to G. Napoli and G. Tricomi for the technical assistance during my experiments. I also thank E. Calderaro and A. Parlato of the Nuclear Engineering Department of the University of Palermo for taking care of γ-ray irradiation in the IGS-3 irradiator, and to Dr. B. Boizot of the Ecole Polytechnique for taking care of the β-ray irradiation in the Van de Graff accelerator, Palaiseau (France).

This thesis is dedicated to Giovanna and my family. Many thanks for your unlimited love, encouragements and sacrifices and for supporting me throughout my endeavors in academics and in life. Your faith in me and your words of encouragements will never be forgotten.

Most of all, I can not thank enough my father, Calogero. He is the source of all knowledge I deem most invaluable and will remain the greatest teacher of my life.

*Experimental investigation on the microscopic structure of intrinsic paramagnetic point defects in amorphous silicon dioxide*

# Introduction

Silicon dioxide in its crystalline (c-SiO$_2$) and amorphous (a-SiO$_2$) forms plays a fundamental role in most of the modern technologies [1, 2]. However, although the c-SiO$_2$, whose most common form is the α-quartz, is considered a key material for high precision oscillators and frequency standards, the major interest regards a-SiO$_2$ which offers a wider field of relevant applications [1, 2]. The latter is used as the core material in low-loss optical fibers, in a wide variety of refractive and trasmissive optics and it is found in the gate terminal of almost the totality of the modern metal-oxide-semiconductor (MOS) devices [1, 2]. Nevertheless, a drawback of the presence of a-SiO$_2$ in these systems is connected with the fact that, as a consequence of the manufacturing process, or exposing them to ionizing radiation, several types of point defects are generated in a-SiO$_2$, affecting the performance of the devices and, in many cases, causing their definitive failure [1, 2]. The deleterious effects of point defects are connected with their ability to give rise to optical absorption (OA) and luminescence (PL) bands, to induce refraction index variations and to act as charge traps [1, 2]. In particular, this latter property influences the electronic properties of MOS devices determining a deleterious threshold voltage shift [3-7].

In general, a point defect can be visualized [1, 2] as a local distortion of the atomic structure caused by a bond rupture, an over or undercoordinated atom, the presence of an impurity atom (homo or heterovalent substitution, interstitial,…), etc. These defects are usually indicated as *intrinsic* when they are due to irregular arrangements of the crystal atoms (Si and O for SiO$_2$), and *extrinsic* when they are related to impurities (atoms differing from Si or O). A further general classification of the point defects can be made on the basis of their electronic configuration: those having unpaired electrons constitute *paramagnetic* defects, and the others are *diamagnetic* defects. Both typologies could in principle exhibit optical activities as OA and PL of visible and ultraviolet radiation. Instead, only the paramagnetic defects have a further feature since they have a non-zero magnetic moment, due to unpaired electrons, and are responsible for the magnetic resonance absorption.

Although the definitions given above apply to both crystalline and amorphous SiO$_2$, many properties of the formed defects differ significantly depending on the nature of the hosting matrix [1, 2]. First of all, due to the larger flexibility of the disordered a-SiO$_2$ matrix with respect to that of α-quartz, a larger variety of defective structures are expected to be accommodated in a-SiO$_2$ than in α-quartz. In addition, also the properties of the same defect induced in a-SiO$_2$ or in α-quartz differ. In fact, while in the amorphous structure a point defect explores a variety of statistically distributed environments, in α-quartz the number of possible environments is limited by the number of inequivalent sites of the crystalline primitive cell in which the defect can be located [1, 2]. As a consequence of this property, the spectral features of a defect in a-SiO$_2$ usually appear more broadened than in α-quartz, so making the interpretation of the spectra more complicated [1, 2, 8, 9].

Up to date the most relevant information on the microscopic structure of the point defects in a-SiO$_2$, and more in general in insulating materials, have been obtained by continuous wave electron paramagnetic resonance (EPR) spectroscopy, which is only applicable to paramagnetic point defects. The key role played by this technique is connected with its ability to obtain relevant information on the atomic scale structure of the center under study [1, 2, 8, 10]. Once the main microscopic structure of a paramagnetic point defect has been established by EPR, then OA and PL spectroscopies are usually employed to determine the energetic levels configuration of the defect [1, 2].

In the present Ph.D. Thesis we report an experimental investigation on the effects of γ- and β-ray irradiation and of subsequent thermal treatment on many types of a-SiO$_2$ materials, differing in the production methods, OH- and Al-content, and oxygen deficiencies. Our main objective is to gain further insight on the microscopic structures of the E'$_\gamma$, E'$_\delta$, E'$_\alpha$ and triplet paramagnetic centers, which are among the most important and studied class of radiation induced intrinsic point defects in a-SiO$_2$. To pursue this objective, for the reasons reported above, we use prevalently the EPR spectroscopy. In particular, our work is focused on the properties of the unpaired electrons wave functions involved in the defects, and this aspect is mainly investigated through the study of the EPR signals originating from the interaction of the unpaired electrons with $^{29}$Si magnetic nuclei (with nuclear spin I=½ and natural abundance 4.7 %). In addition, in some cases of interest, OA measurements are also performed with the aim to further characterize the electronic properties of the defects. Furthermore, due to its relevance for electronics application, the charge state of the defects is investigated by looking at the processes responsible for the generation of the defects of interest. Once these information were gained, the possible sites that can serve as precursors for defects formation are deduced, with the definitive purpose to obtain in the future more radiation resistant a-SiO$_2$ materials in which the deleterious effects connected with the point defects are significantly reduced.

This Ph.D. thesis is organized in 8 chapters. Chapter 1 briefly summarizes the principles of the experimental techniques we have used in the present work. Chapter 2 reviews the main structural properties of α-quartz and a-SiO$_2$ together with the current understanding and open questions regarding their point defects. Chapters 3 and 4 concern experimental procedures and materials employed, respectively. In Chapters 5, 6 and 7 we report and discuss the results of our experimental investigation. Finally, in Chapter 8 we summarize our results and briefly discuss the perspectives for future works.

# Chapter 1

## *Experimental techniques*

The main interest of the present Thesis concerns the study of the point defects in amorphous silicon dioxide (a-SiO$_2$). The formation of these defects usually involves a radiation-induced removal of an electron from a chemical bond [1, 2]. As a result the bond breaks, and an unpaired electron, having a non-zero magnetic moment, remains in a non-bonding orbital. The electron paramagnetic resonance (EPR) spectroscopy is a powerful tool applied to these systems, mainly aiming to gain structural and chemical information. Once these information are gained, the possible sites that can serve as precursors for defects formation upon irradiation are investigated with the definitive purpose to obtain more radiation resistant a-SiO$_2$ materials in which the deleterious effects connected with the formation of point defects are significantly reduced [1, 2]. In addition, together with the EPR technique, the optical absorption (OA) spectroscopy is used for exploring the presence of point defects and for characterizing their electronic levels distribution.

The present chapter is devoted to introduce some theoretical backgrounds necessary to the comprehension of the reported experimental results. We mainly focus on the EPR technique, as its role is particularly relevant for the present work, but some aspects of the optical absorption technique are also treated.

## 1.1    Electron paramagnetic resonance spectroscopy

The EPR spectroscopy [11-15] can be applied to systems with non zero electronic magnetic moment, arising from both electron spin and orbital angular momenta. In a typical EPR experiment the sample under study is placed in two external magnetic fields pointing in orthogonal directions [11-15]. The first field, **H**, makes the ground state energetic levels spread as a consequence of the interaction of the electronic magnetic moment with **H**, the ensuing splitting being ~0.3 cm$^{-1}$ for magnetic fields of ~300 mT [11-15]. The second external magnetic field, **H**$_1$ $\left(\left|\mathbf{H_1}\right| \ll \left|\mathbf{H}\right|\right)$, with amplitude oscillating at a microwave frequency, is used to induce resonant transitions between the states splitted by **H** [11-15].

The relevance of the EPR spectroscopy is mainly connected with its ability to put forward important information on the chemical composition and on the microscopic structure of the paramagnetic centers under study [11-16]. Typical systems to which the EPR spectroscopy is currently applied include [11-15]:

i)   Isolated atoms and ions;
ii)  Free radicals in solid, liquid or gaseous phases;
iii) Ions of the transition metal and actinide group in solids;
iv)  Localized imperfections in solids (point defects);
v)   Systems with conduction electrons.

### 1.1.1 Spin Hamiltonian

Ions with filled electronic shells, and consequently with zero magnetic moment, are usually present in solid systems. However, this is not the only possible situation. The transition ions embedded in a solid, for example, in many cases possess partially filled electronic shells and a magnetic moment due to orbital and spin angular momenta is associated to them [12, 13, 15]. The properties of an ion in a solid are generally very different with respect to those of the isolated one, as a consequence of the interaction of the ion with the surrounding ones located at distances of about $0.2 \div 0.3$ nm. The energy of this interaction typically falls in the range from $10^2$ cm$^{-1}$ to $10^4$ cm$^{-1}$ [12].

From a general point of view, two different methods to study the magnetic properties of a paramagnetic center in a solid exist [12, 13]. The first one is the *crystalline field method* [12, 13] which consists in supposing that the electronic wave functions of the paramagnetic center are highly localized on a single ion and that the only effect of the surrounding ions on the paramagnetic center is to produce an electric field on it. As a consequence of the presence of this electric field, the electronic energy levels of the paramagnetic center undergo a *Stark shift*. A second more sophisticated method consists in considering the electronic wave function distributed over a group of atoms forming molecular orbitals of type σ and π, instead of the atomic ones [12, 13]. While the *crystalline field method* is applicable exclusively to systems with pure ionic bonds, the latter method permits to treat systems with covalent bonds.

In both physical situations, the result is the appearance of groups of levels with nearly equal energies. The details of the energy levels frame strictly depends on the symmetry properties of the paramagnetic center and they can be predicted by the *group theory method* [12]. An important theorem in this context, due to Kramers, assures that the ground state of a paramagnetic system in absence of an external magnetic field is at least doubly degenerate, provided that an odd number of electrons are involved [12].

As mentioned above, the transitions involved in the EPR measurements occur between energy levels split apart by ~ 0.3 cm$^{-1}$. Consequently, in EPR studies the only groups of energy levels of interest are those which are degenerate (or nearly degenerate) in absence of an external magnetic field. The paramagnetic system is usually studied by introducing the *effective spin Hamiltonian* (or simply *spin Hamiltonian*) which permits to investigate the effects of the external magnetic fields on the paramagnetic systems by restricting oneself to consider only the nearly degenerate group of states of interest [12, 13]. The *spin Hamiltonian* operator is written in terms of a *effective spin* operator **S** chosen in such a way that 2S+1 is equal to the number of states of the group considered [12, 13]. However, in many cases, the *spin Hamiltonian* of a paramagnetic center is deduced by empirical considerations, as it can not be obtained theoretically starting from the *true* Hamiltonian operator describing the interactions acting on the system. One of the most important cases in which the spin Hamiltonian can be obtained explicitly is that of the transition ions in a crystal host, which can be treated in the context of the crystalline field theory [12, 13]. In addition, by introducing molecular orbital wave functions, a method has been developed which permits, under opportune hypothesis, to study systems involving covalent bonds. Examples of application of these two methods are discussed in more details in the following [16].

## 1.1.2 Zeeman interaction and $\hat{g}$ tensor

In a typical EPR experiment the paramagnetic sample is placed in a static and uniform magnetic field **H**. The effect of the interaction of the microscopic magnetic moment of the system under study with the external magnetic field can be described by the Zeeman operator [11-15]:

$$\hat{\mathcal{H}}_{zeeman} = -\boldsymbol{\mu}^T \cdot \mathbf{H} \quad^{(1)} \tag{1.1}$$

where **μ** is the magnetic moment of the paramagnetic center. In the simple and common case in which the magnetic moment is due to the electronic spin angular momenta alone, the Zeeman interaction operator can be simplified as follow:

$$\hat{\mathcal{H}}_{zeeman} = g_e \mu_B \mathbf{H}^T \cdot \mathbf{S} \tag{1.2}$$

where $g_e \cong 2.00232$ is the electronic splitting factor, $\mu_B = 9.27408 \cdot 10^{-24}$ J/T is the Bohr magnetic moment and **S** is the spin operator in units of $\hbar = h/2\pi = (1/2\pi) \cdot 6.62618 \cdot 10^{-34}$ J/s. The eigenvalues of $\hat{\mathcal{H}}_{zeeman}$, which represent the energy levels of the system, are given by

$$\varepsilon_{m_s} = g_e \mu_B H m_s \tag{1.3}$$

where $m_s$ is the eigenvalue of the component of **S** along the direction of **H**. Eq. (1.3) shows that, as a consequence of the interaction of the paramagnetic system with the static magnetic field (Zeeman interaction), a splitting of the energy levels with different values of $m_s$ occurs. In EPR experiments the system is at the same time subjected to a second magnetic field $\mathbf{H_1}$ $\left(\left|\mathbf{H_1}\right| \ll \left|\mathbf{H}\right|\right)$ directed perpendicularly to **H** and with amplitude oscillating at a microwave frequency [11-15].

---

[1] The apex symbol T adjacent to an operator, as $\boldsymbol{\mu}^T$ in Eq. (1.1), indicates that the transpose of the matrix representing the operator has to be considered in the product of matrixes.

The aim of this oscillating field is to induce transitions between pairs of states energetically separated by the Zeeman interaction. These transitions occur when the energy of the microwave photons, hν, matches the energy difference between a pair of levels with $m_s$=j and $m_s$=j+1, where the magnetic dipole selection rules $\Delta m_s = \pm 1$ have been imposed [11-15]. The acquisition of an EPR spectrum consists in the measurement of the energy absorbed by the paramagnetic system from the microwave field as a function of the amplitude of **H** and at fixed amplitude and frequency of the magnetic field **H$_1$** [11-15].

For systems in which the magnetic moment μ is due to the electron spin S =½, which are of principal interest in the present Thesis, the energies of the states with $m_s$= ½ and $m_s$= -½ are simply

$$\varepsilon_{\pm\frac{1}{2}} = \pm \tfrac{1}{2} g_e \mu_B H \tag{1.4}$$

and the resonance occurs at a given field $H_r$ such that

$$h\nu = g_e \mu_B H_r \tag{1.5}$$

where ν is the frequency of the oscillating magnetic field **H$_1$**. Note that from Eq. (1.5) it follows that for a simple system of paramagnetic centers with S=½ only a transition in correspondence to a static magnetic field $H_r = \dfrac{h\nu}{g_e \mu_B}$ occurs.

Until now we have supposed that the paramagnetic centers are isolated. However, in many physical systems of interest, as for many point defects in solids, the paramagnetic centers interact with the surrounding atoms and consequently the EPR spectrum usually differs significantly with respect to that described by the Hamiltonian of Eq. (1.2) [12, 13, 15]. One of the most important consequences of these interactions is that, due to the spin-orbit interaction between the electron spin and orbital angular momenta, the electronic spectroscopic splitting factor $g_e$ has to be replaced by a matrix operator [12, 13, 15]. Consequently, the Zeeman interaction is described by the following Hamiltonian [12, 13, 15]:

$$\hat{\mathcal{H}}_{zeeman} = \mu_B \mathbf{H}^T \cdot \hat{\mathbf{g}} \cdot \mathbf{S} = \mu_B \begin{Vmatrix} H_x & H_y & H_z \end{Vmatrix} \begin{Vmatrix} g_{xx} & g_{xy} & g_{xz} \\ g_{yx} & g_{yy} & g_{yz} \\ g_{zx} & g_{zy} & g_{zz} \end{Vmatrix} \begin{Vmatrix} S_x \\ S_y \\ S_z \end{Vmatrix} \tag{1.6}$$

where $\hat{\mathbf{g}}$ is the spectroscopic splitting 3x3 matrix operator and x, y, and z are the axes of the laboratory frame of reference. The vector $\mathbf{H}^T \cdot \hat{\mathbf{g}}$ in Eq. (1.6) can be regarded as a vector resulting from a transformation of the actual field **H** to an effective one [15]

$$\mathbf{H}_{eff} = g_e^{-1} \mathbf{H}^T \cdot \hat{\mathbf{g}} \tag{1.7}$$

In general, **H$_{eff}$** and **H** point in different directions. The modulus of **H$_{eff}$** can be estimated as follow

$$H_{eff} = g_e^{-1} [(\mathbf{H^T} \cdot \mathbf{\hat{g}})^T \cdot (\mathbf{H^T} \cdot \mathbf{\hat{g}})]^{1/2} = \{ g_e^{-1} [\mathbf{n^T} \cdot (\mathbf{\hat{g}} \cdot \mathbf{\hat{g}^T}) \cdot \mathbf{n}]^{1/2} \} H = (g_{eff} / g_e) H \qquad (1.8)$$

where $\mathbf{n}=\mathbf{H}/H$ is the unit vector along $\mathbf{H}$ and $g_{eff} = [\mathbf{n^T} \cdot (\mathbf{\hat{g}} \cdot \mathbf{\hat{g}^T}) \cdot \mathbf{n}]^{1/2}$ is the effective spectroscopic splitting factor. The value of $g_{eff}$ depends on the orientation of the external magnetic field $\mathbf{H}$ with respect to the axes of symmetry of the paramagnetic center. Considering the quantization of the spin angular momentum along $\mathbf{H_{eff}}$, from Eq. (1.6) one obtains the following energy levels of the system

$$\varepsilon_{m'_s} = \mu_B g_e \mathbf{H_{eff}}^T \cdot \mathbf{S} = \mu_B g_e H_{eff} \, m'_s = g_{eff} \mu_B H \, m'_s \qquad (1.9)$$

where $m'_s$ is the eigenvalue of the component of $\mathbf{S}$ along the direction of $\mathbf{H_{eff}}$. Eq. (1.9) is similar to Eq. (1.3) but for the substitution of the isotropic $g_e$ with the directional dependent $g_{eff}$. For a system with $S=\frac{1}{2}$, in analogy with Eq. (1.4) and Eq. (1.5), the transition occurs in correspondence to a static magnetic field $H_r = \dfrac{h\nu}{g_{eff} \mu_B}$. However, in the present case the line position in the EPR spectrum corresponding to this transition depends, through $g_{eff}$, on the relative orientation of $\mathbf{H}$ with respect to the symmetry axes of the paramagnetic center.

One of the most important steps towards the identification of the relevant structural properties of a paramagnetic center consists in the determination of its $\mathbf{\hat{g}}$ matrix, as defined in Eq. (1.6) [12, 13, 15, 16]. This objective is accomplished through the knowledge of the matrix elements of the operator $\mathbf{\hat{g}} \cdot \mathbf{\hat{g}^T}$, which, in fact, is the quantity that can be obtained by EPR measurements, instead of $\mathbf{\hat{g}}$ [see Eq. (1.8)]. In particular, the matrix elements of $\mathbf{\hat{g}} \cdot \mathbf{\hat{g}^T}$, expressed with respect to the laboratory frame of reference, are obtained by evaluating the $g_{eff}$ values from the line positions of the resonances in correspondence to various relative angles between the external field $\mathbf{H}$ and the crystal axes of the sample under study [15]. In general, this procedure is concluded when all the six independent matrix elements of the symmetric operator $\mathbf{\hat{g}} \cdot \mathbf{\hat{g}^T}$ have been obtained. The successive step towards the determination of $\mathbf{\hat{g}}$ consists in the transformation of the matrix $\mathbf{\hat{g}} \cdot \mathbf{\hat{g}^T}$ into a diagonal form. This is accomplished by finding a matrix $\mathbf{C}$ such that [15]

$$\mathbf{C} \cdot (\mathbf{\hat{g}} \cdot \mathbf{\hat{g}^T}) \cdot \mathbf{C^T} = \begin{Vmatrix} C_{Xx} & C_{Xy} & C_{Xz} \\ C_{Yx} & C_{Yy} & C_{Yz} \\ C_{Zx} & C_{Zy} & C_{Zz} \end{Vmatrix} \cdot \begin{Vmatrix} (\mathbf{\hat{g}} \cdot \mathbf{\hat{g}^T})_{xx} & (\mathbf{\hat{g}} \cdot \mathbf{\hat{g}^T})_{xy} & (\mathbf{\hat{g}} \cdot \mathbf{\hat{g}^T})_{xz} \\ (\mathbf{\hat{g}} \cdot \mathbf{\hat{g}^T})_{yx} & (\mathbf{\hat{g}} \cdot \mathbf{\hat{g}^T})_{yy} & (\mathbf{\hat{g}} \cdot \mathbf{\hat{g}^T})_{yz} \\ (\mathbf{\hat{g}} \cdot \mathbf{\hat{g}^T})_{zx} & (\mathbf{\hat{g}} \cdot \mathbf{\hat{g}^T})_{zy} & (\mathbf{\hat{g}} \cdot \mathbf{\hat{g}^T})_{zz} \end{Vmatrix} \cdot \begin{Vmatrix} C_{Xx} & C_{Yx} & C_{Zx} \\ C_{Xy} & C_{Yy} & C_{Zy} \\ C_{Xz} & C_{Yz} & C_{Zz} \end{Vmatrix} =$$

$$= \begin{Vmatrix} (\mathbf{\hat{g}} \cdot \mathbf{\hat{g}^T})_X & 0 & 0 \\ 0 & (\mathbf{\hat{g}} \cdot \mathbf{\hat{g}^T})_Y & 0 \\ 0 & 0 & (\mathbf{\hat{g}} \cdot \mathbf{\hat{g}^T})_Z \end{Vmatrix} = {}^D(\mathbf{\hat{g}} \cdot \mathbf{\hat{g}^T}) \qquad (1.10)$$

where **C** is the matrix which transforms $\hat{\mathbf{g}} \cdot \hat{\mathbf{g}}^T$ into its diagonal form $^D(\hat{\mathbf{g}} \cdot \hat{\mathbf{g}}^T)$. The matrix elements of **C** are the direction cosines connecting the principal axes of the paramagnetic center, X, Y and Z, to the axes of the laboratory frame of reference, x, y and z. Once $^D(\hat{\mathbf{g}} \cdot \hat{\mathbf{g}}^T)$ is known, the $\hat{\mathbf{g}}$ matrix is obtained by taking the positive square roots of the diagonal elements of $^D(\hat{\mathbf{g}} \cdot \hat{\mathbf{g}}^T)$, which give $^D\hat{\mathbf{g}}$, and then applying the reverse transformation of Eq. (1.10): $\hat{\mathbf{g}} = \mathbf{C}^T \cdot {}^D\hat{\mathbf{g}} \cdot \mathbf{C}$ [15]. From a general point of view, although the $\hat{\mathbf{g}}$ matrix obtained with this method correctly reproduces the position and the intensity observed in the experiments, it could differ with respect to that defined in Eq. (1.6). In fact, if this latter matrix is asymmetric, then its principal-axes system is not an orthogonal one, thus differing from that obtained from $\hat{\mathbf{g}} \cdot \hat{\mathbf{g}}^T$. Furthermore, since the diagonal elements of $^D\hat{\mathbf{g}}$ are obtained as square roots of the diagonal elements of $^D(\hat{\mathbf{g}} \cdot \hat{\mathbf{g}}^T)$, their signs are, in general, undetermined. As a consequence, in principle, the "conventional" $\hat{\mathbf{g}}$ matrix obtained experimentally should be distinguished by the "true" one defined in Eq. (1.6) [15]. In the present Thesis, the principal values of the $\hat{\mathbf{g}}$ matrix estimated by EPR measurements refers to the "conventional" ones, although it will not be further indicated.

Now that the main properties of the $\hat{\mathbf{g}}$ matrix have been discussed, the way in which the spin-orbit coupling affects the energy levels and gives rise to anisotropic features of the EPR spectrum is treated in more details. The contribution to the overall spin Hamiltonian of a paramagnetic center comprising orbital magnetic moment and spin-orbit coupling is [14, 17]

$$\hat{\mathcal{H}}_{\text{spin}} = g_e \mu_B \mathbf{H}^T \cdot \mathbf{S} + \frac{g_e e}{2m^2c^2}\mathbf{S}^T \cdot \left[\mathbf{E}(\mathbf{r}) \times (\mathbf{p} + e\mathbf{A}_0)\right] + \frac{e}{2m}\left(\mathbf{p}^T \cdot \mathbf{A}_0 + \mathbf{A}_0^T \cdot \mathbf{p}\right) + \frac{e^2}{2m}\mathbf{A}_0^2 \hat{\mathbf{1}} \quad (1.11)$$

where **E**(**r**) is the electric field, **p** is the linear momentum operator, **A**$_0$ is the vector potential associated to the external magnetic field **H**, c the velocity of light in vacuum, whereas e and m are the electronic charge and mass, respectively. The first term is the usual Zeeman energy associated with the spin magnetic moment, the second term describes the spin-orbit interaction, whereas the last two terms take into account the coupling of the orbital magnetic moment with the external magnetic field [14, 17]. For a system consisting in an electron highly localized on a single atom weakly interacting with the surrounding, the relevant terms of Eq. (1.11) can be simplified as follows [14]

$$\hat{\mathcal{H}}_{\text{spin}} = \mu_B \mathbf{H}^T \cdot (\mathbf{L} + g_e \mathbf{S}) + \lambda \mathbf{L}^T \cdot \mathbf{S} \quad (1.12)$$

where λ is the spin-orbit coupling constant and **L** the angular momentum operator in units of ℏ. Among the systems which can be properly described by the spin Hamiltonian of Eq. (1.12), a subclass can be defined which deserves a dedicated treatment. These systems are characterized by the fact that the Zeeman and the spin-orbit interactions are small compared to the atomic Hamiltonian comprising electronic kinetic and potential energy terms. Furthermore, we suppose that the ground state orbital angular momentum is zero. Under these hypothesis, by using a perturbative approach up to second order corrections, it is possible to show that the spin Hamiltonian of Eq. (1.12) can be written as [15]

$$\hat{\mathcal{H}}_{spin} = \mu_B \mathbf{H}^T \cdot \hat{\mathbf{g}} \cdot \mathbf{S} + \mathbf{S}^T \cdot \hat{\mathbf{D}} \cdot \mathbf{S} \tag{1.13}$$

where

$$\hat{\mathbf{g}} = g_e \mathbf{1} + 2\lambda \hat{\mathbf{\Lambda}} \tag{1.14a}$$

$$\hat{\mathbf{D}} = \lambda^2 \hat{\mathbf{\Lambda}} \tag{1.14b}$$

with

$$\hat{\mathbf{\Lambda}} = \begin{Vmatrix} \Lambda_{xx} & \Lambda_{xy} & \Lambda_{xz} \\ \Lambda_{yx} & \Lambda_{yy} & \Lambda_{yz} \\ \Lambda_{zx} & \Lambda_{zy} & \Lambda_{zz} \end{Vmatrix} = -\sum_{n \neq G} \frac{\langle G|\mathbf{L}|n\rangle\langle n|\mathbf{L}|G\rangle}{E_n - E_G} \tag{1.15}$$

where $|G\rangle$ and $|n\rangle$ are the spatial parts of the ground state and of the excited states of the system, respectively, whereas $E_G$ and $E_n$ are their energies. The term $\mathbf{S}^T \cdot \hat{\mathbf{D}} \cdot \mathbf{S}$ in Eq. (1.13) describes the electronic quadrupole interaction and affects the energy levels of a paramagnetic system only if $S > \frac{1}{2}$. Eq. (1.14a) evidences that the deviations of $\hat{\mathbf{g}}$ from the value of the free electron, $g_e$, and its anisotropic nature are due to the term $2\lambda\hat{\mathbf{\Lambda}}$, originating from the spin-orbit interaction. In general, the calculation of the elements $\Lambda_{ij}$ of the matrix $\hat{\mathbf{\Lambda}}$ is not a simple task, mainly because realistic expressions of $|n\rangle$ and $E_n - E_G$ are not known. However, in many cases of interest a large number of properties of the paramagnetic center can be deduced from Eqs. (1.13)-(1.15) by applying simple symmetry considerations.

Another class of systems which deserves a dedicated treatment involves the paramagnetic centers consisting in an unpaired electron prevalently localized in the dangling bond orbital of an atom coordinated to other three atoms by covalent bonds [16, 17]. The overall molecular system forms the broken tetrahedron structure schematically shown in Figure 1.1 (a). Among the paramagnetic centers which fall in this class, the most important and pertinent to the arguments discussed in the present Thesis are the E' point defects in crystalline and amorphous $SiO_2$, for

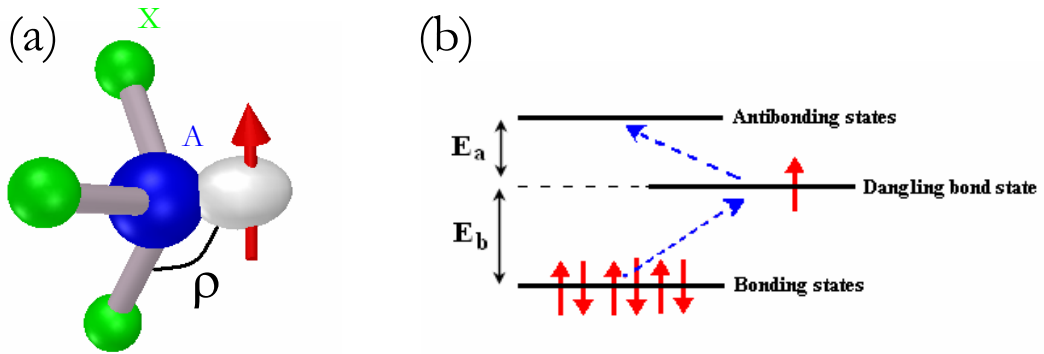

**Figure 1.1** (a) Broken tetrahedron structure. (b) Energetic levels scheme for the structure shown in (a). In (b), red arrows represent electrons, whereas blue arrows indicate the two possible transitions among the electronic levels of the molecule.

which A=Si and X=O in Figure 1.1 (a) [18]. The principal g values of the broken tetrahedron structure have been obtained starting from the general spin Hamiltonian of Eq. (1.11) and constructing the proper local molecular orbitals of the system by the LCAO (Linear Combination of Atomic Orbitals) method [16-18]. In particular, this procedure was used by Watkins and Corbett [16, 17] in the study of the E center in crystalline silicon, and by Silsbee [19] and Feigl and Anderson [18] for the E'$_1$ center in α-quartz. In its most general form, the principal g values of these defects can be written as [18]

$$g_{||} = g_e \tag{1.16}$$

$$g_\perp = g_e + |\lambda| C_{3p}^2 \left[ \frac{(1-\gamma)(1-\delta)}{E_b} - \frac{(1+\gamma)(1+\delta)}{E_a} \right] \tag{1.17}$$

where $C_{3p}^2$ is the percentage of the 3p atomic orbital of A involved in the unpaired electron wave function, whereas $E_b$ and $E_a$ [shown in Figure 1.1 (b)] are the differences between the energy of the dangling bond state and those of the valence (made by the A-X bonding orbitals) and conduction (made by the A-X antibonding orbitals) states, respectively [18]. The parameters γ and δ are small corrections (γ«1, δ«1) to the $g_\perp$ value introduced in order to make the hybrid orbitals of the outer shell of the A atom orthogonal to the core states and to take into account a partial ionic character of the A-X bonds, respectively [18]. It is worth to note that, although the rough approximations introduced in the treatment of the problem limit the possibility to obtain from Eqs. (1.16) and (1.17) principal g values in quite good quantitative agreement with those estimated experimentally, the functional dependences expressed by these equations could give important information on the system under study. In particular, on the basis of the Eq. (1.16) it is expected that the $g_{||}$ value of different paramagnetic centers, all characterized by the broken tetrahedron structure, should differ little from each other and with respect to the free electron value $g_e$, whereas from Eq. (1.17) one can obtain a *sketch* of the functional dependence of $g_\perp$ on the physical quantities $|\lambda|$, $C_{3p}^2$, $E_a$, $E_b$, γ and δ characterizing the system.

### 1.1.2.1  *The powder EPR line shape*

When a paramagnetic sample consists of powdered crystal or it is in amorphous form, the EPR spectrum changes considerably with respect to that of the single crystal [11, 15, 20, 21]. In the former case, in fact, the observed EPR spectrum arises from the superposition of a multitude of single crystal spectra, each one corresponding to a specific orientation of the principal symmetry axes of the paramagnetic center with respect to the direction of the external magnetic field [11, 15, 20, 21]. Nevertheless, as it is shown in the present paragraph, the principal g values of the paramagnetic center can be obtained from the resonance line again by estimating the positions of some specific spectral features in the EPR spectrum [11, 15, 20, 21].

To outline the procedure which permits to obtain the EPR line shape of a powdered crystal or of an amorphous sample, we consider the simple case of a $S=\frac{1}{2}$ paramagnetic center with axial symmetry. For such a center the single crystal resonance magnetic field is given by [12, 15, 21]

$$H_r = \frac{h\nu}{g_{eff}(\theta)\mu_B} = \frac{h\nu}{\mu_B}\left(g_\perp^2 \sin^2\theta + g_\parallel^2 \cos^2\theta\right)^{-\frac{1}{2}} \qquad (1.18)$$

where $\theta$ is the angle of the external magnetic field with respect to the symmetry axis of the paramagnetic center and it has been posed $g_{eff}(\theta) = \left(g_\perp^2 \sin^2\theta + g_\parallel^2 \cos^2\theta\right)^{\frac{1}{2}}$. Since, for hypothesis, all the angles $\theta$ in the range $0° \div 180°$ occur with the same probability, the fraction $\frac{dN}{N}$ of paramagnetic centers having an angle between $\theta$ and $\theta+d\theta$ is simply equal to the fraction of solid angle swept by $d\theta$ (shown in Figure 1.2):

$$\frac{dN}{N} = \frac{1}{2}\sin\theta \, d\theta \qquad (1.19)$$

where N is the total number of paramagnetic centers of the system. From Eqs. (1.18) and (1.19), after straightforward manipulation, it is found that [21]

$$\frac{dN}{N} = \frac{1}{2}\left(\frac{4H_0^2}{H^3}\right)\left\{\left(g_\parallel^2 - g_\perp^2\right)\left[\left(\frac{2H_0}{H}\right)^2 - g_\perp^2\right]\right\}^{-\frac{1}{2}} dH \qquad (1.20)$$

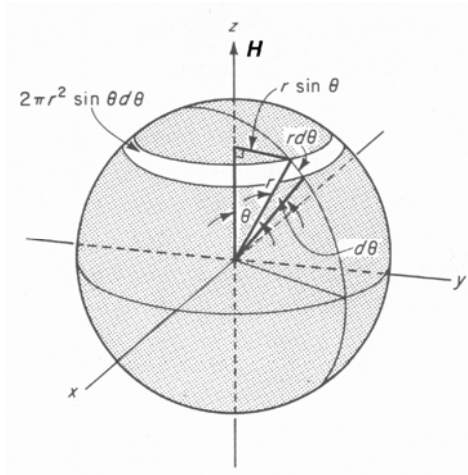

**Figure 1.2** Portion of solid angle comprised between $\theta$ and $\theta+d\theta$. After Ref. 15.

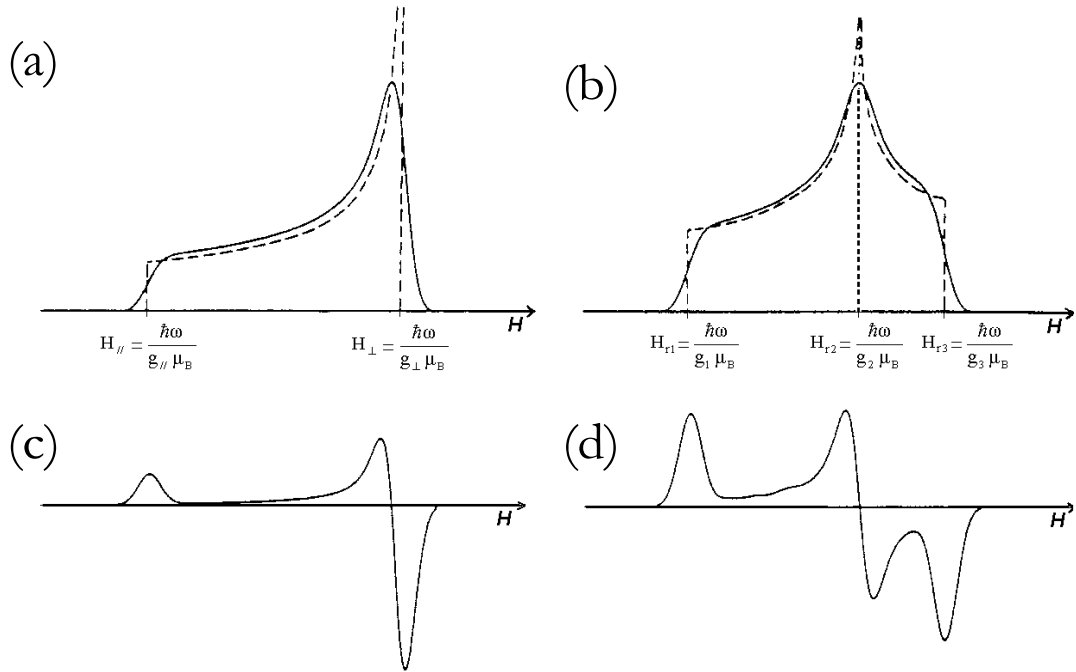

**Figure 1.3** EPR absorption line shapes (continuous lines) for paramagnetic centers with (a) axial and (b) orthorhombic symmetry obtained from the convolution of the density of states $D(H)=\frac{dN}{dH}$ (broken lines) with the single center line shape (not shown). The lines in (c) and (d) are the first derivatives of those in (a) and (b), respectively. Adapted from Ref. 8.

with $H_0 = \frac{h\nu}{2\mu_B}$. The distribution function, defined as $D(H)=\frac{dN}{dH}$, can be easily obtained from Eq. (1.20) and gives the broken line shown in Figure 1.3 (a). With a similar procedure, the distribution function for the general case of orthorhombic symmetry can be also found in terms of the principal g values $g_1$, $g_2$ and $g_3$. The result of this latter calculation is shown with a broken line in Figure 1.3 (b).

In order to obtain the true EPR line shape it is necessary to take into account the single center line shape $f(H-H_r)$ which, in general, can be represented by a Gaussian profile, a Lorentzian profile or a combination of both, depending on the specific interactions acting on the paramagnetic system [12]. The single center EPR line shape is taken into account by applying the following convolution procedure [20-22]

$$A(H) = \int f(H - H_r) D(H_r) dH_r \quad (1.21)$$

The function $A(H)$ is proportional to the EPR absorption line shape and is shown as a continuous line in Figure 1.3 for the cases of (a) axial and (b) orthorhombic symmetries. As shown, the main effect in considering the single center line shape is the elimination of the divergences of the density function $D(H)$ occurring in correspondence to $g_\perp$ [Figure 1.3 (a)] and $g_2$ [Figure 1.3 (b)]. Since the EPR spectra usually consist of the derivative of the absorption lines, in Figure 1.3 (c) and (d) the first derivative of the line shapes shown in Figure 1.3 (a) and (b), respectively, are reported. It is worth to note that for the simplest case of an ideal spherically symmetric paramagnetic center, the distribution function D(H) is given by the Dirac delta function and the absorption profile coincides with the single center line shape.

From inspection of Figure 1.3 (a)-(d), it is evident that from the EPR spectrum of a powdered crystal or of an amorphous sample the principal g values can be estimated, although the degree of precision is lower with respect to the case of a single crystal system. For example, for a paramagnetic system with orthorhombic symmetry [see Figure 1.3 (d)], the principal $g_1$, $g_2$ and $g_3$ values can be obtained by measuring in the first derivative spectra the magnetic field positions at which the first maximum, the zero crossing and the second minimum of the resonance line occur. A similar method can be used to estimate the principal g values in the cases of higher symmetry.

### 1.1.2.2 *Influence of the amorphous-state disorder on the EPR line shape*

Although in the previous paragraph the EPR line shapes of powdered crystals and those of amorphous materials have been treated in the same way, it is important to underline that an important difference exists between these two physical systems. This difference consists in the fact that in the amorphous system, in addition to the randomness of the relative orientations of the principal symmetry axes of the paramagnetic centers with respect to the external magnetic field, a statistical distribution of the spin Hamiltonian parameters has to be considered, arising from the statistical distribution of bond lengths and angles characteristic of the amorphous state.

The main consequence of this amorphous-state disorder is the smearing of the EPR powder line shapes of Figure 1.3 [2]. In some cases, this smearing effect is so large that some spectroscopic features observed for a paramagnetic center in a powdered crystal system become experimentally unobservable when the same paramagnetic center is embedded in an amorphous host. Such effect, for example, is believed to occur for the extrinsic $[AlO_4]^0$ center, as it came out by comparing the EPR spectra of this point defect in crystalline (powder) and amorphous $SiO_2$ materials [23]. When such a situation occurs, the only way to obtain a reasonable estimation of the principal g values of the paramagnetic center is to perform computer simulations of the experimental spectra including in the software the possibility to account for statistically distributed spin Hamiltonian parameters [2, 23-25]. At variance, in many cases of interest, the smearing effect on the EPR spectrum, due to the amorphous-state disorder, is not so large to invalidate the observation of all the relevant spectroscopic features of a paramagnetic center and, consequently, the principal g values can, again, be estimated by measuring the position in the EPR spectrum of proper features of the resonance line, as discussed at the end of the previous paragraph.

## 1.1.3 Hyperfine interaction

In many systems of interest in EPR studies, the unpaired electron responsible for the paramagnetic properties is located nearby to one or more atoms with non-zero nuclear spin which, consequently, possess a non-zero nuclear magnetic moment. The interaction of the magnetic moment of the unpaired electron with those of the nearby nuclei is described by the hyperfine Hamiltonian [12-15]. The spin Hamiltonian of a paramagnetic system with S=½, including the most relevant terms of the hyperfine interaction, can be written as follows [12-15]

$$\hat{\mathcal{H}}_{spin} = \mu_B \mathbf{H}^T \cdot \hat{\mathbf{g}} \cdot \mathbf{S} + \sum_{i=1}^{n}\left(a_i \mathbf{S}^T \cdot \mathbf{I}_i + \mathbf{S}^T \cdot \hat{\mathbf{T}}_i \cdot \mathbf{I}_i\right) =$$
$$= \mu_B \mathbf{H}^T \cdot \hat{\mathbf{g}} \cdot \mathbf{S} + \sum_{i=1}^{n}\left(\mathbf{S}^T \cdot \hat{\mathbf{A}}_i \cdot \mathbf{I}_i\right) \qquad (1.22)$$

where it has been posed $\hat{\mathbf{A}}_i = a_i \hat{\mathbf{1}} + \hat{\mathbf{T}}_i$. The first term in Eq. (1.22) contains electron Zeeman and spin-orbit interactions and it has been discussed in the previous paragraphs, whereas the second term describes the hyperfine interaction of the unpaired electron with n magnetic nuclei. The hyperfine Hamiltonian for each of the n nuclei includes two terms; the first term is the Fermi contact one which is isotropic, whereas the second term is due to the dipolar interaction between electron and nuclear magnetic moments and is anisotropic [12-15]. In the simple case in which the unpaired electron interacts with a single magnetic nucleus, Eq. (1.22) simplifies as follows

$$\hat{\mathcal{H}}_{spin} = \mu_B \mathbf{H}^T \cdot \hat{\mathbf{g}} \cdot \mathbf{S} + \mathbf{S}^T \cdot \hat{\mathbf{A}} \cdot \mathbf{I} \qquad (1.23)$$

From a general point of view, the problem to obtain eigenstates and eigenvalues of the paramagnetic system described by the spin Hamiltonian of Eq. (1.23) can be solved in different ways, depending on the relative strength of the various interaction terms involved [12-15]. In one of the most common situations, which is of particular interest here, the hyperfine terms are small compared to the Zeeman electronic one and the principal axes of $\hat{\mathbf{g}}$ and $\hat{\mathbf{A}}$ point in the same direction. In this special case, the eigenstates and the eigenvalues of the system can be obtained from Eq. (1.23) by a sequence of first-order perturbative treatments which first involve the Zeeman plus spin-orbit terms and then the hyperfine ones [12]. With this hypothesis and referring to an axial symmetric paramagnetic center, for simplicity, the eigenvalues of the spin Hamiltonian of Eq. (1.23) can be written as follows [12]

$$\varepsilon_{m'_s, m'_I} = \mu_B g_{eff} H m'_s + A_{eff}\, m'_s m'_I \qquad (1.24)$$

where $m'_I$ is the eigenvalue of the component of the operator **I** along the direction of quantization, $g_{eff} = \left(g_{||}^2 \cos^2\theta + g_\perp^2 \sin^2\theta\right)^{½}$ and $A_{eff} = g_{eff}^{-1}\left(A_{||}^2 g_{||}^2 \cos^2\theta + A_\perp^2 g_\perp^2 \sin^2\theta\right)^{½}$, in which $A_{||}$ and $A_\perp$ are the principal values of the matrix $\hat{\mathbf{A}}$ and $\theta$ is the angle between **H** and the symmetry axis of the paramagnetic center [12]. From Eq. (1.24) it follows that the energy levels separated by the Zeeman and spin-orbit interactions, are successively splitted by an amount $A_{eff} \, m'_s \, m'_I$ by the action of the hyperfine interaction. The strongly allowed transitions of the system are obtained from the selection rules $\Delta m'_s = 1$ and $\Delta m'_I = 0$ and give rise to (2I+1) equally spaced lines whose center of gravity falls at $H_r = \dfrac{h\nu}{g_{eff} \, \mu_B}$ [12]. It is worth to note that, due to the anisotropic term contained in the matrix $\hat{\mathbf{A}}$ [Eq. (1.23)], the position in the spectrum of each component line of the hyperfine multiplet depends on the relative orientation of the external magnetic field with respect to the sample under study [12-15] and the value of the constants $g_{||}$, $g_{\perp}$, $A_{||}$ and $A_{\perp}$ can be experimentally estimated by measuring the positions of the hyperfine lines for different orientations of the external magnetic field relative to the sample axes [15]. Furthermore, when the paramagnetic sample under study consists of powdered crystal or it is in amorphous form, analogous considerations to those discussed in Paragraphs 1.1.2.1 and 1.1.2.2 apply for the hyperfine lines too [15].

The occurrence of the hyperfine interaction in a system furnishes one of the most useful tools to obtain many important chemical and structural information on the paramagnetic centers involved [2, 16-18]. In the following two paragraphs, some of the methods used to pick up these information are described, limiting ourself to consider specific physical systems which are of particular interest in the present Thesis.

### 1.1.3.1 *EPR intensity ratio between the hyperfine structure and the main resonance line*

The first step towards the identification of the relevant properties of a paramagnetic center consists in the identification of the chemical species involved. First we consider the simplest case in which the unpaired electron is localized on a single chemical species which, however, possesses $\varkappa$ isotopes with abundance $z_i$ (which can be the natural occurring ones or those imposed by a specific production process of the sample), nuclear magnetic moments $\mu_i$, and nuclear spin $I_i$, with i=1, …, $\varkappa$. In this case, Eq. (1.22) splits in a set of $\varkappa$ equations, each one describing the hyperfine interaction occurring in the paramagnetic centers involving one of the possible isotopes

$$\left(\hat{\mathcal{H}}_{\mathbf{spin}}\right)_i = \mu_B \, \mathbf{H}^T \cdot \hat{\mathbf{g}} \cdot \mathbf{S} + \mathbf{S}^T \cdot \hat{\mathbf{A}}_i \cdot \mathbf{I}_i \qquad (1.25)$$

where i=1, …, $\varkappa$ and the nuclear magnetic moments $\mu_i$ are contained in $\hat{\mathbf{A}}_i$. For such a system, in general, the EPR spectrum consists of $\varkappa$ multiplets. In particular, the mutiplet arising from the interaction of the unpaired electron with the isotope i comprises $(2I_i+1)$ lines whose specific spectroscopic features are determined by $\hat{\mathbf{A}}_i$. Furthermore, no matter how complex the system is, the overall intensity of the multiplet i relative to that of the multiplet j is simply given by the quantity $z_i/z_j$, which is the abundances ratio of the two isotopes involved. These simple considerations can be extended to comprise the situations in which the electronic magnetic moment experiences hyperfine interaction with more than a single nucleus. In the rest of the present paragraph we will focus on a specific example of this class of systems, which is interesting in connection with the arguments of the present Thesis.

In the previous discussion it has been assumed that the component lines belonging to an hyperfine multiplet can be isolated experimentally from those belonging to the other multiplets. However, in many physical systems of interest, this assumption does not hold. For example, such a problem occurs when an electronic magnetic moment undergoes an hyperfine interaction with more then one *equivalent* nuclei. In this case, in fact, the hyperfine lines of many distinct multiplets superimpose in the EPR spectrum and, consequently, a more appropriated statistical approach has to be considered in order to describe the relative overall intensity of each multiplet. To treat these type of systems it is simpler to consider a specific example. We assume that the paramagnetic center consists in an electronic magnetic moment undergoing hyperfine interaction with n equivalent nuclei of the same chemical species, which possesses only two isotopes: the first one with nuclear spin I=0 and abundance $z_0$ and the second one with nuclear spin I=½, nuclear magnetic moment $\mu_1$ and abundance $z_1=1-z_0$. In such a system (n+1) physical situations can be statistically found, depending on the number k of magnetic nuclei involved in the paramagnetic center: k=0, 1, 2, …, n. The EPR spectrum arising from each one of these statistically possible cases consists of a multiplet of (k + 1) lines. Of course, in the experimental spectrum all the multiplets corresponding to different value of k are simultaneously present, even though with different relative intensities. The overall intensity of a generic hyperfine multiplet is proportional to the probability, $P(k;n,z_1)$, that k magnetic nuclei occur in the same paramagnetic center, which, in the simple case we have considered, is simply given by the Binomial statistical distribution:

$$P(k;n,z_1) = \frac{n!}{k!(n-k)!}(z_1)^k(1-z_1)^{n-k} \tag{1.26}$$

In Eq. (1.26), the binomial coefficient $\frac{n!}{k!(n-k)!}$ takes into account the number of ways in which k magnetic nuclei can be distributed among the n possible equivalent nuclear positions. The relative overall intensity of two distinct hyperfine multiplets is simply given by the ratio of the probabilities associated to them, which are obtained from Eq. (1.26).

## 1.1.3.2 *Estimation of the percentage of atomic s and p orbital components of an unpaired electron wave function*

The ability of the EPR spectroscopy to determine important information on the microscopic structure of a paramagnetic center can be recognized if one considers the case in which the unpaired electron wave function consists prevalently of a mixture of s and p orbitals of an atom with non-zero nuclear magnetic moment [2, 16-18]. Note that the broken tetrahedron structure discussed in Paragraph 1.1.2 is an example of such a system, provided that the atom A (see Figure 1.1) has a non-zero nuclear spin. The wave function of the unpaired electron involved in the system can be written as

$$|G\rangle = c_s |ns\rangle + c_p |np\rangle + \sum_j c_j |k_j\rangle \tag{1.27}$$

where $|ns\rangle$ and $|np\rangle$ are the ns-state and np-state atomic orbitals of the central atom (atom A in Figure 1.1), $|k_j\rangle$ are the orbitals of the ligand atoms (atoms X in Figure 1.1), whereas $c_s$, $c_p$ and $c_j$ are the coefficients of the linear combination, with $|c_j| \ll |c_s|$ and $|c_j| \ll |c_p|$. If one assumes that [16]: i) both charge and spin polarization effects can be neglected, ii) only the orbitals of the central atom (atom A in Figure 1.1) contribute significantly to the hyperfine interaction, iii) the use of the *isolated* atomic orbitals is legitimate in the integral calculations, iv) the $\hat{\mathbf{g}}$ tensor anisotropy can be neglected in the calculation of the hyperfine isotropic constant a and of the anisotropic matrix $\hat{\mathbf{T}}$, allowing one to consider simply an average value $g_{\text{eff}}^{\text{mean}}$, then it can be shown that the hyperfine operator $\hat{\mathbf{A}}$ of Eq. (1.23), expressed with respect to its principal-axes system, reduces to

$$\hat{\mathbf{A}} = a\,\hat{\mathbf{1}} + \hat{\mathbf{T}} = a \begin{Vmatrix} 1 & 0 & 0 \\ 0 & 1 & 0 \\ 0 & 0 & 1 \end{Vmatrix} + t \begin{Vmatrix} -1 & 0 & 0 \\ 0 & -1 & 0 \\ 0 & 0 & 2 \end{Vmatrix} = \begin{Vmatrix} a-t & 0 & 0 \\ 0 & a-t & 0 \\ 0 & 0 & a+2t \end{Vmatrix} \tag{1.28}$$

where

$$a = \frac{2}{3} \mu_0 \, g_{\text{eff}}^{\text{mean}} \, \mu_B \, g_N \, \mu_N \, c_s^2 \, |\varphi(0)|^2 \tag{1.29}$$

$$t = \frac{\mu_0}{4\pi} \frac{g_{\text{eff}}^{\text{mean}} \mu_B \, g_N \, \mu_N}{\langle np|r^3|np\rangle} c_p^2 \tag{1.30}$$

in which r is the electronic radial coordinate relative to the position of the nucleus, $\mu_0$ is the magnetic susceptivity in vacuum, $g_N$ and $\mu_N$ are the spectroscopic splitting factor and the magnetic moment of the nucleus, respectively, whereas $|\varphi(0)|$ is the electron spin density at the position of the nucleus. Comparing Eq. (1.28) with the general form of an axial symmetric $\hat{\mathbf{A}}$ matrix

$$\hat{\mathbf{A}} = \begin{Vmatrix} A_\perp & 0 & 0 \\ 0 & A_\perp & 0 \\ 0 & 0 & A_{||} \end{Vmatrix} \qquad (1.31)$$

one simply recognizes that

$$A_\perp = a - t \qquad (1.32)$$

$$A_{||} = a + 2t \qquad (1.33)$$

The quantities a and t of Eqs. (1.29) and (1.30) coincide with the Fermi contact and the dipolar interactions of the corresponding free atom $a_{free}$ and $t_{free}$, respectively, except for the presence of the expansion constants $c_s^2$ and $c_p^2$ in the formers. Consequently, these latter constants can be easily obtained from the following ratios

$$c_s^2 = \frac{a}{a_{free}} \qquad (1.34)$$

$$c_p^2 = \frac{t}{t_{free}} \qquad (1.35)$$

Finally, from Eqs. (1.32) – (1.35) one obtains

$$c_s^2 = \frac{A_{||} + 2A_\perp}{3a_{free}} \qquad (1.36)$$

$$c_p^2 = \frac{A_{||} - A_\perp}{3t_{free}} \qquad (1.37)$$

The latter equations permit to obtain experimental estimations of $c_s^2$ and $c_p^2$, after the two principal values $A_{||}$ and $A_\perp$ of the matrix $\hat{\mathbf{A}}$ are estimated from the experimental spectra, and making use of the known values of the constants $a_{free}$ and $t_{free}$ for the free atom [2, 16, 17].

Concerning the broken tetrahedron structure (see Figure 1.1), another important structural information that can be obtained consists in the angle ρ between the dangling orbital and the three A-X basal bonds [2, 16, 17]. It has been shown, in fact, that an estimation of this angle can be obtained from the following relation

$$\tan\rho = -\left\{2\left[1+\left(\frac{c_p^2}{c_s^2}\right)\right]\right\}^{1/2} = -\left\{2\left[1+\left(\frac{t}{t_{free}}\frac{a_{free}}{a}\right)\right]\right\}^{1/2} \qquad (1.38)$$

However, one has to take care in using Eq. (1.38) to obtain $\rho$ from the hyperfine data estimated experimentally. In fact, by comparing the dependence of the angle $\rho$ predicted from Eq. (1.38) with those obtained using various self-consistent calculations, a quite large disagreement has been found for three different types of paramagnetic centers (with the broken tetrahedron structure), among which the E'$_1$ center of α-quartz [26]. One of the most relevant factors determining the failure of Eq. (1.38) is connected with the partial ionicity of the A-X bonds (Figure 1.1), not accounted for in the treatment leading to Eq. (1.38) [26].

## 1.1.4 Spin Hamiltonian for a paramagnetic center with S=1

In this paragraph we consider a paramagnetic system consisting of two electrons located at distances lower than ~5 Å. Due to the low distance between the electrons, two other important interactions are effective and have to be considered in the spin Hamiltonian: the *electron-electron dipole* and the *electron-exchange* interactions [12, 15]. As can be easily shown [12, 15], the effect of the latter is to couple the electron spins $S_1=½$ and $S_2=½$ to give a diamagnetic singlet state with $S_{tot}=0$ and a paramagnetic triplet state with $S_{tot}=1$. If, for simplicity, we neglect the hyperfine interaction and the anisotropy of $\hat{g}$ ($g \cong 2$) and if we suppose that the wave functions of the electrons are in the form of a product of the orbital and of the spin components, then the spin Hamiltonian of two interacting electrons in the spin triplet state can be written as follows [15]:

$$\hat{\mathcal{H}}_{spin} = \mu_B\, g\, \mathbf{H}^T \cdot \mathbf{S}_{tot} + \mathbf{S}_{tot}^T \cdot \hat{\mathbf{D}}_{dip} \cdot \mathbf{S}_{tot} + J_0\left(½\, \mathbf{S}_{tot}^2 - ¾\, \mathbf{1}\right) \qquad (1.39)$$

with

$$\hat{\mathbf{D}}_{dip} = \frac{\mu_0}{8\pi}(g\mu_B)^2 \left\| \begin{array}{ccc} \left\langle\frac{r^2-3x^2}{r^5}\right\rangle & \left\langle\frac{-3xy}{r^5}\right\rangle & \left\langle\frac{-3xz}{r^5}\right\rangle \\ \left\langle\frac{-3xy}{r^5}\right\rangle & \left\langle\frac{r^2-3y^2}{r^5}\right\rangle & \left\langle\frac{-3yz}{r^5}\right\rangle \\ \left\langle\frac{-3xz}{r^5}\right\rangle & \left\langle\frac{-3yz}{r^5}\right\rangle & \left\langle\frac{r^2-3z^2}{r^5}\right\rangle \end{array} \right\| \qquad (1.40)$$

and

$$J_0 = -2\left\langle\varphi_a(1)\varphi_b(2)\left|\frac{e^2}{4\pi\varepsilon_0 r}\right|\varphi_a(2)\varphi_b(1)\right\rangle \qquad (1.41)$$

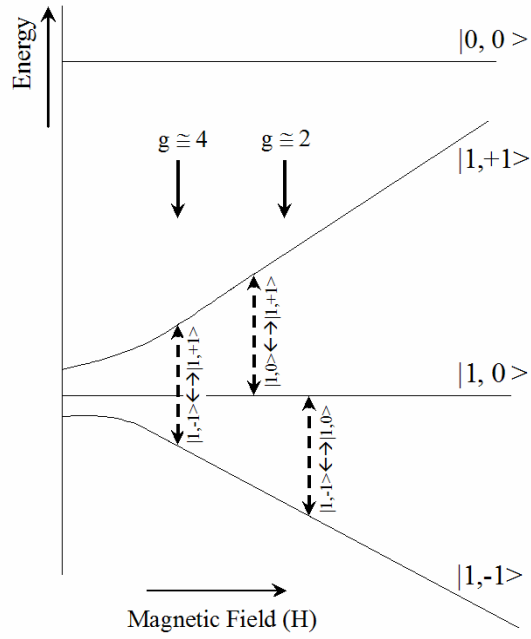

**Figure 1.4** Schematic representation of the energy levels of a pair of interacting electrons as a function of the modulus of the magnetic field **H**. The transitions observable in an EPR experiments are indicated by broken arrow. In this figure it has been supposed $J_0<0$.

where **S**$_{tot}$=**S**$_1$+**S**$_2$ is the total spin operator obtained summing over the spins angular momenta, **S**$_1$ and **S**$_2$, of the two electrons, whereas $\varphi_a$ and $\varphi_b$ are the spatial parts of the wave functions of the two electrons. In Eq. (1.39) the term containing the operator $\hat{\mathbf{D}}_{dip}$ describes the *electron-electron dipole interaction*, while the term with $J_0$ is the *electron-exchange interaction*. The constant $J_0$ is known as isotropic electron-exchange coupling constant [12, 15]. Choosing the eigenstates of the operator **S**$_{tot}^2$ and of the projection of **S**$_{tot}$ on the direction of **H** as basis set, indicated as $|S_{tot},M\rangle$, the energies of the system of coupled electrons are [15]:

$$E(S_{tot}=0) = -\tfrac{3}{4} J_0 \qquad (1.42)$$

$$E_{X,Y}(S_{tot}=1) = \tfrac{1}{4} J_0 + \tfrac{1}{2} \{D_Z \pm [4 g^2 \mu_B^2 H^2 + (D_X-D_Y)^2]^{1/2}\} \qquad (1.43)$$

$$E_Z(S_{tot}=1) = \tfrac{1}{4} J_0 - D_Z \qquad (1.44)$$

where X, Y and Z are the principal axes of the projection of the operator $\hat{\mathbf{D}}_{dip}$ in the subspace of states with $S_{tot}=1$ and $D_X$, $D_Y$ and $D_Z$ are its diagonal values, whereas H is the modulus of the magnetic field **H** supposed directed along Z. From Eq. (1.39) and Eqs. (1.42)-(1.44) the following properties of the system of two interacting electrons can be outlined (see Figure 1.4) [15]:

i) The exchange interaction separates the energies of the singlet ($S_{tot}=0$) with respect to those of the triplet states ($S_{tot}=1$). Furthermore, if $J_0<0$ and $|J_0|>>k_bT$, where $k_b$ is the Boltzmann constant and T is the temperature, only the paramagnetic triplet state is populated. Conversely, if $J_0>>k_bT$ only the diamagnetic singlet state is populated.

ii) The states $|1,-1\rangle$, $|1,0\rangle$ and $|1,+1\rangle$ are eigenstates of the Hamiltonian only for a magnetic field H large enough that the Zeeman interaction dominates on the dipolar term.

iii) Since, in general, the three triplet eigenstates of the system are linear combinations of pure $|1,-1\rangle$, $|1,0\rangle$ and $|1,+1\rangle$ states, the selection rule $\Delta M = \pm 1$ does not apply. Consequently, together with the allowed transitions $|1,-1\rangle \leftrightarrow |1,0\rangle$ and $|1,0\rangle \leftrightarrow |1,+1\rangle$ giving rise to a pair of lines with center of gravity at $g \cong 2$, the transition between the lowest and the highest energy levels of the triplet can be observed, giving an EPR line at $g \cong 4$.

iv) In general, due to the dipolar interaction, the energies of the levels are not linear functions of the amplitude of the field **H**.

v) The three states of the system are not degenerate for H=0. Furthermore, for H≠0, the energy splitting among the states depends on the relative orientations of the field **H** with respect to the principal-axes system of the operator $\hat{\mathbf{D}}_{dip}$.

## 1.2 Optical spectroscopy

Usually, in the study of a paramagnetic center, the first step consists in the investigation of its structural properties by EPR spectroscopy. Subsequently, other complementary information are obtained by exploring its energetic levels by optical spectroscopy, provided that the same paramagnetic center possesses also optical properties. Since EPR and optical properties are intimately related, as was put forward explicitly for some special systems in the previous Paragraph 1.1.2 [see Eqs. (1.15) and (1.17)], the use of both EPR and optical techniques could drive to a deeper comprehension of the structure and electronic configuration of the paramagnetic center. Finally, it is worth to note that in the study of *diamagnetic* centers the optical spectroscopy usually represents one of the most powerful methods, among the various spectroscopic techniques, to put forward relevant information on the system under study, as the EPR spectroscopy cannot be applied to them.

In the present Thesis, point defects are investigated by both EPR and optical absorption spectroscopy in order to look into the existing correlations between EPR and optical properties of the same center. Furthermore, we pay attention to the correlations existing between distinct defects, where each one can be diamagnetic or paramagnetic, as they represent determinant information to support structural models of defects and to get hints on their possible generation mechanisms.

### 1.2.1 Absorption

The presence of a point defect in a material introduces new electronic energy levels that usually belong neither to the valence band nor to the conduction band, but they are localized in the energy gap. Consequently, electronic excitation between them can be induced by electromagnetic radiation with energy lower than the band gap. Such transitions give rise to absorption bands that in principle should possess Lorentzian line shape. However, in the case of electronic systems embedded in solids the interactions with the atomic vibrations and the inhomogeneities of the local structure, especially in an amorphous matrix, lead to absorption bands that can be approximated by Gaussian shapes.

In a typical optical absorption experiment the incident electromagnetic radiation, with energy varied continuously in a chosen interval, traverses the sample and then is revealed. In accordance to the Bouguer-Lambert-Beer's empiric law the intensity $I(E)$ (energy per unit of area and unit of time) of radiation emerging from the sample is related to the incident intensity $I_0(E)$ by the macroscopic formula [27, 28]:

$$I(E) = I_0(E) e^{-\alpha(E)\,d} \qquad (1.45)$$

where $\alpha(E)$ is the *absorption coefficient* at the photon energy E and d is the sample thickness (crossed by the radiation). The absorption coefficient expresses the inverse of the length traversed by the light before it is attenuated by a factor 1/e, and is usually measured in cm$^{-1}$. However, in many cases, the adimensional quantity $A(E)=\text{Log}_{10}(I_0(E)/I(E))$ is reported. This latter equation defines the *absorbance*, which represents an absolute measure of the absorption and it is related to the absorption coefficient by

$$\alpha(E)=(2.303)\frac{A(E)}{d} \qquad (1.46)$$

To relate the experimental quantities to the theory we consider two non-degenerate electronic states $\varphi_i$ and $\varphi_f$ of the defects with energies $E_i < E_f$. In the radiation-matter interaction, an atomic system can be considered as an electric dipole with moment $\mathbf{M}=\sum_\nu e_\nu \mathbf{r}_\nu$, where $e_\nu$ are the electron charges and $\mathbf{r}_\nu$ their position vectors. Then, according to the quantum mechanics, the probability that the electromagnetic radiation induces an electronic transition between the levels $E_i$ and $E_f$ is proportional to the Einstein coefficient for absorption [29, 30]

$$B_{if}=\frac{2\pi}{3\hbar^2 c}|\mathbf{M}^{fi}|^2 \qquad (1.47)$$

where c is the speed of light and $\mathbf{M}^{fi}=\int \varphi_f^* \mathbf{M} \varphi_i \, d\tau$ is the electric dipole matrix element, in which $\varphi_f^*$ is the complex conjugate of the wave function $\varphi_f$ and $d\tau$ is a volume element. The integrated absorption coefficient is related to the Einstein coefficient $B_{if}$ and to $\mathbf{M}^{fi}$ by

$$\int\alpha(E)dE=\frac{NB_{if}\hbar\omega_{fi}}{c}=\frac{2\pi\omega_{fi}}{3\hbar c^2}N|\mathbf{M}^{fi}|^2 \qquad (1.48)$$

where N is the number of atoms per unit volume in the state $\varphi_i$, and $\omega_{fi}=\frac{E_f-E_i}{\hbar}$ is the Bohr frequency of the transition. The important result of Eq. (1.48) is that the absorption intensity experimentally measured can give useful information on $|\mathbf{M}^{fi}|$ and, possibly, on the wave functions of the states involved in the electronic transition.

A relevant quantity commonly used to compare the intensities of the absorption bands is the *oscillator strength*, *f*, of the electronic transition. This is a dimensionless quantity defined as the ratio between the integrated absorption over an experimental band peaked at $\omega_{fi}$ and the theoretical absorption calculated approximating the atomic system as a charged harmonic oscillator with frequency $\omega_{fi}$ [28, 29]. It can be shown that the oscillator strength for the transition $\varphi_i \to \varphi_f$ is given by [28-30]

$$f=\frac{2m\omega_{fi}}{3\hbar^2 e^2}|\mathbf{M}^{fi}| \qquad (1.49)$$

where m is the electron mass and in this equation $\mathbf{M}^{fi}$ is the dipole matrix element of the real transition. The oscillator strength is a parameter related to the selection rules of the electronic transition through $\mathbf{M}^{fi}$ and its determination can give information on the electronic orbitals involved. In particular, the influence of the symmetry properties or the spin multiplicity of the states involved in the transition establishes the value of $\mathbf{M}^{fi}$ and, as a consequence, if a given transition is possible or not. For example, $f \cong 1$ pertains to a strongly allowed electric dipole transition, whereas $f \ll 1$ characterizes a forbidden transition [28, 29]. We note that for the majority of the electronic transitions considered in the present thesis $f \cong 0.1$, which corresponds to a dipole allowed transition.

A quantitative relation between $f$ and the experimental quantity $\alpha_{max}$ is given by the Smakula's equation [31]:

$$Nf = (8.7 \times 10^{16}) \frac{n \alpha_{max} \Delta}{(n^2 + 2)^2} \quad (eV^{-1} cm^{-2}) \qquad (1.50)$$

where $N$ is expressed in cm$^{-3}$, n is the glass refractive index, $\alpha_{max}$ (cm$^{-1}$) is the absorption band height and $\Delta$ (eV) the Full-Width at Half-Maximum (FWHM) of the absorption band. This equation has been derived under the assumption of Gaussian absorption bands as expected in glasses due to the intrinsic inhomogeneities. It enables to derive the concentration of absorbing centers provided that the oscillator strength is known or, conversely, the oscillator strength from the concentration.

Another used quantity for the absorption effect is the cross section $\sigma(E)$. This is the total probability of absorption for a given absorbing species, it is measured in cm$^2$ and is related to the absorption coefficient by the equation

$$\alpha(E) = N \sigma(E) \qquad (1.51)$$

The cross section can be related to the oscillator strength by means of Eq. (1.50), and in this respect both quantities may be interchanged to characterize an optically active center when its absorption coefficient has been measured.

# Chapter 2

# *Crystalline and amorphous SiO$_2$: structures and point defects*

## 2.1 Structure of crystalline and amorphous SiO$_2$

Many forms of crystalline SiO$_2$ are present in nature and all of them, apart from the Stishovite (produced at very high pressure), consist in SiO$_4$ tetrahedra linked to each other by the terminal O atoms (see Figure 2.1) [1]. The most common form of crystalline SiO$_2$ is the α-quartz in which each O atom is bonded with two Si atoms and each Si atom is bonded with four O atoms. The bond length between an O atom and the two Si atoms are 0.1608 nm and 0.1611 nm, named short- and long-bond, respectively, and the bond angles are O-$\hat{\text{Si}}$-O ≅ 109.5° and Si-$\hat{\text{O}}$-Si = 143.6° [1, 32]. Furthermore, α-quartz is slightly birefringent and exhibits rotary dispersion of light rays transmitted along the crystal axis (c-axis), both right-hand and left-hand forms being known [1]. The crystalline forms of SiO$_2$ have been extensively studied by neutron and X-ray diffraction techniques and their structures are known to a very high degree of precision [1]. In contrast, a very poor knowledge of the amorphous structure of the SiO$_2$ has been reached up until now [1]. The most successful theory of the a-SiO$_2$ structure is due to Zachariasen [33]. In this theory it is assumed that the SiO$_4$ tetrahedra of the amorphous state are virtually identical with respect to those of the crystalline state of SiO$_2$. However, it is supposed that in a-SiO$_2$ the SiO$_4$ tetrahedra are linked together randomly to give a structure that lacks periodicity, symmetry and long range order [1, 33]. Mozzi and Warren [34] have shown that, in the context of the Zachariasen theory and supposing an uniform distribution of the dihedral angles (relative rotation angle of adjoining neighbouring SiO$_4$ tetrahedra) between 0° and 360° [34], the X-rays diffraction spectra of a-SiO$_2$ can be properly fitted assuming the statistical distribution of the Si-$\hat{O}$-Si bond angle shown in the inset of Figure 2.1. Consequently, it has been suggested that in a-SiO$_2$ the Si-$\hat{O}$-Si angle is not fixed to the value of 143.6°, characteristic of α-quartz, but it can take all the values from 120° to 180°, with different probabilities. However, it is worth to note that the bond angle distribution reported by Mozzi and Warren was obtained under opportune assumptions on the interconnection between SiO$_4$ tetrahedra and on the dihedral angles distributions, whose validity can not be independently proved. In addition, in a successive diffraction study in which neutron and X-rays data were combined, Neuefeind and Liss [35] obtained a statistical distribution of the Si-$\hat{O}$-Si angle that is nearly half the width found by Mozzi and Warren. This latter result was obtained without the assumption of uniformly distributed dihedral angles. For these reasons, the clarification of the structural properties of the a-SiO$_2$ still remains an open question [36, 37].

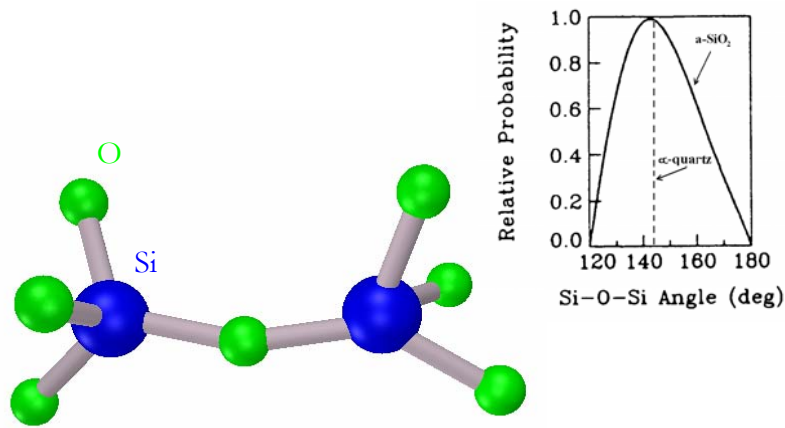

**Figure 2.1** Fragment of SiO$_2$ presenting the linking between two tethraedra. Inset: Statistical distribution of the Si-$\hat{\text{O}}$-Si angle in a-SiO$_2$, as determined from X-rays diffraction data by Mozzi and Warren [34].

When an SiO$_2$ sample, in crystalline or in amorphous forms, is subjected to irradiation with heavy particles (neutrons, protons, …) or ionizing radiation (UV photons, X-rays, β-rays, γ-rays) a large number of point defects is usually induced [1, 2]. Many types of paramagnetic defects, whose microscopic structures have been studied prevalently by EPR spectroscopy, have been reported since 1950 [1, 2].

The following two paragraphs are devoted to review the literature on the most relevant paramagnetic point defects induced by irradiation in crystalline and amorphous SiO$_2$. In particular, their microscopic structures and the current understanding of their physical properties are discussed.

## 2.2　Point defects in crystalline SiO$_2$

### 2.2.1　E'$_1$ center

The E'$_1$ center is the most important and fundamental point defect induced by neutron, electron and γ-ray irradiation in α-quartz [1]. It was first observed by Weeks [38] and the angular dependence of its EPR spectrum was subsequently studied by Silsbee [19] and Jani *et al.* [39]. As shown in Figure 2.2, the EPR spectrum of the E'$_1$ center, for a magnetic field parallel to the c-axis, consists of a single main resonance line and of four hyperfine doublets with splitting of ~40 mT (strong hyperfine), ~0.8 mT and ~0.9 mT (weak hyperfine) and ~0.05 mT (very weak hyperfine) [1, 2, 19, 39]. Each doublet has an EPR intensity relative to the main resonance line of ~5 %, in agreement with the 4.7 % natural abundance of the $^{29}$Si nuclei [1, 2, 19, 39]. The principal values of the $\hat{g}$ matrix of the E'$_1$ center, obtained by EPR measurements, are $g_1$=2.0018, $g_2$=2.0005 and $g_3$=2.0003 [1, 19, 39].

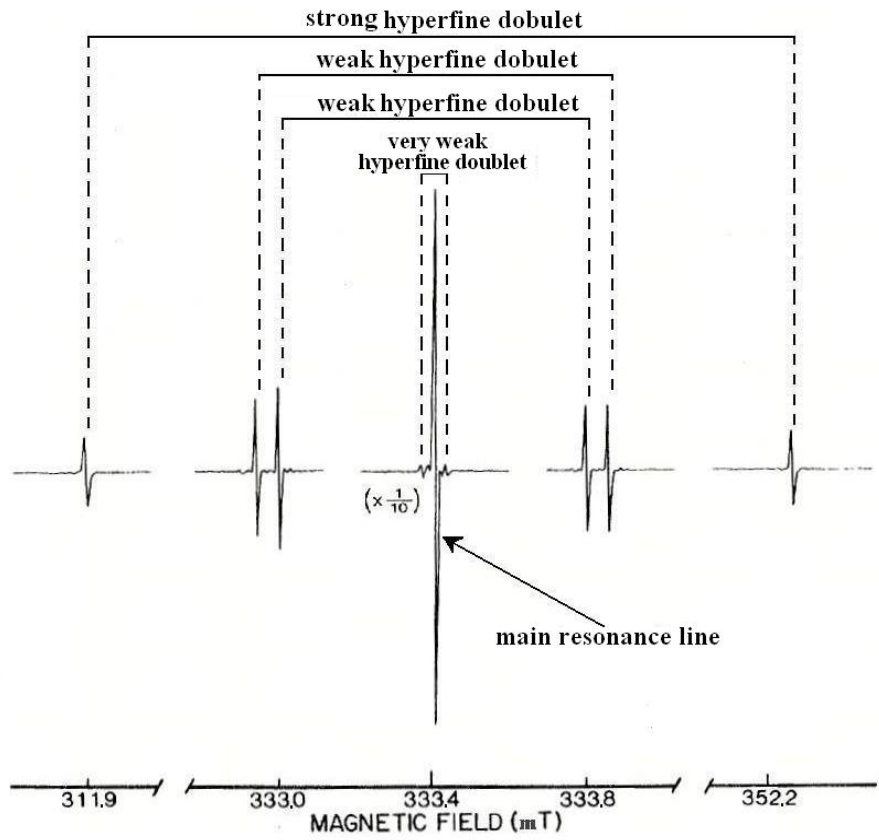

**Figure 2.2** EPR spectrum of the E'$_1$ center obtained in an electron irradiated and thermally treated α-quartz sample. The measurement has been performed at room temperature and with the magnetic field **H** parallel to the c-axis. The central region of the spectrum, containing the main resonance line and the very weak hyperfine doublet, has been acquired with a spectrometer gain reduced by a factor 10. Adapted from Ref. 39.

Silsbee [19], on the basis of the analysis of the strong hyperfine structure, deduced that the unpaired electron involved in the E'$_1$ center is localized in a sp$^3$ hybrid orbital of a single Si atom. The subsequent important step in the understanding of the microscopic structure of the E'$_1$ center was represented by the work of Feigl *et al.* [40], in which this defect was associated to an asymmetrically relaxed positively charged oxygen vacancy. In this model it was supposed [40] that the Si atom in the long-bond side with respect to the missing O atom moves backward in a planar configuration, whereas the unpaired electron localizes in an sp$^3$ hybrid orbital of the short-bond Si atom [40, 41]. This microscopic model was improved by successive theoretical works [42-46], in which it was suggested that the relaxation of the long-bond Si is stabilized by the formation of a threefold coordinated O atom, giving rise to a puckered configuration, as schematically show in Figure 2.3. In this model the strong hyperfine structure originates from the hyperfine interaction of the unpaired electron with a $^{29}$Si nucleus located in the Si(0) position, the two weak hyperfine doublets are due to $^{29}$Si atoms in the Si(2) and Si(3) positions, whereas the very weak hyperfine structure is due to a $^{29}$Si atom in the Si(4) position (see Figure 2.3) [42-46]. The inequivalence of the very weak hyperfine structure with respect to the weak ones is due to the inequivalence of the Si(4) atom position with respect to those of Si(2) and Si(3). In fact, it was shown by theoretical methods [42-46] that the unpaired electron wave function density at the Si(4) position is negligible, as the bonds Si(0)-Ô(4)-Si(4) fall near to the equator of a sphere centered on Si(0) and with its polar axis pointing along the direction of the dangling orbital [2, 47].

An OA band peaked at ~6.2 eV was initially attributed to the E'$_1$ center by Nelson and Weeks [48]. However, in a successive experimental work [49], this attribution was questioned, suggesting that the actual OA band of this center is peaked at ~5.9 eV with a full width at half maximum (FWHM) of ~1 eV. A general consensus on the OA properties of the E'$_1$ center has not been reached yet and, consequently, new experimental investigations are needed to clarify this aspect [49, 50].

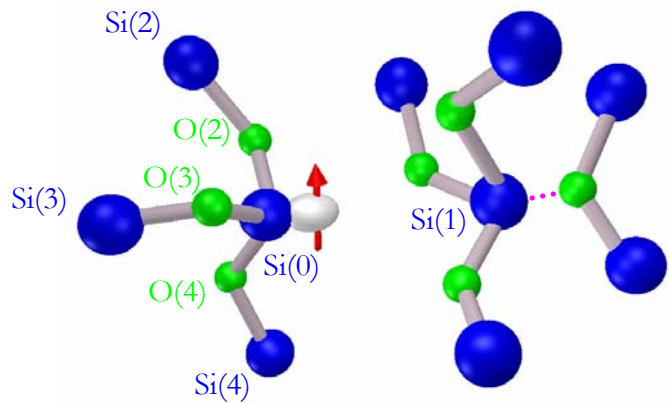

**Figure 2.3** Puckered positively charged oxygen vacancy structure proposed as a model for the E'$_1$ center in α-quartz [42-46].

## 2.2.2 E'$_4$ and E'$_2$ centers

The E'$_4$ center is the best known of the point defects in irradiated α-quartz and its main EPR spectrum, for a magnetic field parallel to the c-axis, consists of four primary lines, which at room temperature are nearly equally spaced and have nearly equal intensity (Figure 2.4). This center was first observed by Weeks and Nelson [51], while its $^{29}$Si hyperfine spectrum was first reported by Solntsev *et al.* [52] and subsequently studied by Isoya *et al.* [53]. A relevant variation with temperature of the spin Hamiltonian parameters of this defect was observed, suggesting a distortion of the center and related electronic structure changes [53]. The principal g values, for example, were found to change from $g_1$=2.00163, $g_2$=2.00073 and $g_3$=2.00053 at 40 K, to $g_1$=2.00154, $g_2$=2.00065 and $g_3$=2.00060 at 300 K [53]. The knowledge of the structure of the E'$_4$ center is due mostly to the EPR study and to the related theoretical analysis with the Hartree-Fock method carried out by Isoya *et al.* [53]. The authors have shown that the defect consists in a

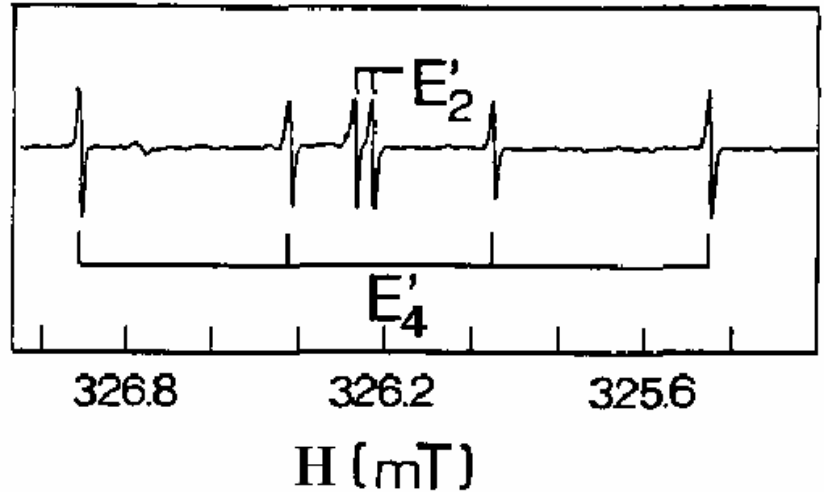

**Figure 2.4** EPR spectrum of the E'$_4$ and the E'$_2$ centers obtained in an X-ray irradiated α-quartz sample. The measurement has been performed at room temperature and with the magnetic field **H** parallel to the c-axis. Adapted from Ref. 49.

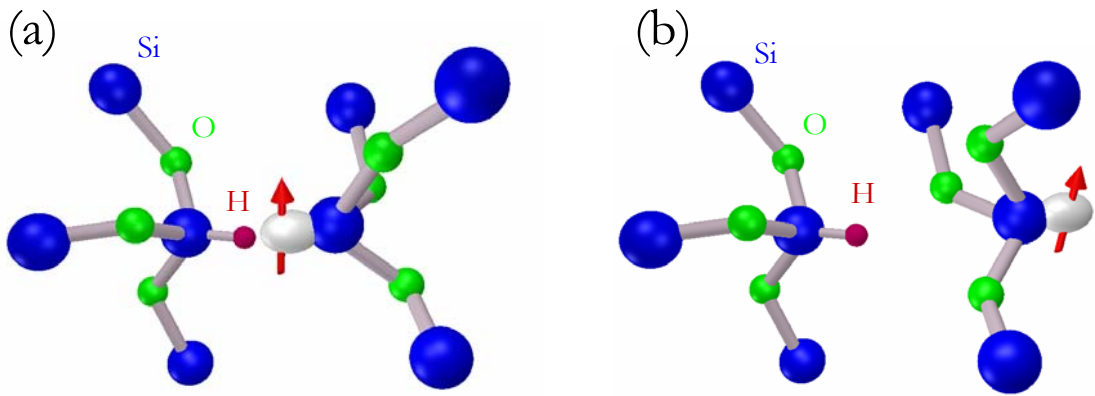

**Figure 2.5** (a) Microscopic structure of the E'$_4$ center at low temperature proposed by Isoya *et al.* [53]. (b) Microscopic structure of the E'$_2$ center proposed by Rudra *et al.* [58].

hydrogen atom trapped in a positively charged oxygen vacancy [53]. At low temperature the hydrogen atom bonds more strongly with the Si atom in the short-bond side with respect to the missing oxygen, and the unpaired electron is located mostly on the other Si atom of the vacancy [Figure 2.5 (a)] [53]. At higher temperature, the probability for the hydrogen atom to bond with the long-bond Si atom increases and consequently the EPR signal due to the unpaired electron wave function located on the short-bond Si atom of the vacancy becomes evident [53]. In this framework, the four primary EPR lines observed at room temperature arise from the superposition of the two physical situations corresponding to the unpaired electron located on the short-bond Si and undergoing hyperfine interaction with the nucleus of the H atom bonded to the long-bond Si and that in which the unpaired electron is located on the long-bond Si and interacts with the H atom bonded to the short-bond Si. The microscopic picture of the E'$_4$ center depicted by Isoya *et al.* [53] was confirmed by successive theoretical studies performed with various theoretical methods [46, 54-56].

The E'$_2$ center main EPR spectrum consists of two equally intense lines (Figure 2.4) [48, 51, 57]. Rudra *et al.* [58] have proposed that this defect possesses a microscopic structure similar to that of the E'$_4$ center, but for the fact that in the former the hydrogen atom is bonded to the short-bond Si atom and the unpaired electron localizes stably in the long-bond Si atom, which relaxes through the plane of its basal O atoms in a back-projected configuration, as schematically shown in Figure 2.5 (b). This configuration was found to be consistent with experimental data and it was predicted to be slightly lower in energy than that of the E'$_4$ center by ~0.09 eV with an energy barrier between the two configurations of about 0.6 eV [58]. In this framework, the two primary EPR lines of the E'$_2$ center arise from the hyperfine interaction the unpaired electron located on the lond-bond Si with the nucleus of the H atom bonded to the short-bond Si. The principal g values of the E'$_2$ center, estimated by EPR measurements, are $g_{||}$ = 2.0022 and $g_{\perp}$ = 2.0006 [48, 51, 57]

The OA properties of the E'$_4$ center are unknown, whereas those of the E'$_2$ center were investigated by Nelson and Weeks [48, 51]. Their studies have suggested that the E'$_2$ center possesses an OA band peaked at ~5.4 eV. However, since no successive works were undertaken confirming this attribution, it should be considered as tentative [50].

### 2.2.3 Triplet state centers

In irradiated α-quartz many distinguishable triplet state centers (pair of coupled electrons with total spin S=1) have been identified. The first ones to observe these centers in γ-ray irradiated quartz were Weeks and Abraham [59]. Subsequently, other three triplet state centers, named E"$_1$, E"$_2$ and E"$_3$, were reported and characterized [60-63].

The properties common to all these centers are [59-62]: i) the EPR spectrum includes a doublet of main resonance lines whose center of gravity falls at about g ≅ 2, and a single weak line with g ≅ 4, ii) the EPR lines are narrow (~0.005 mT), iii) the spin-lattice relaxation time is very long and iv) the principal g values are very little negatively shifted with respect to the free electron g value. On the basis of these properties and in direct analogy to the E' centers, the triplet state centers are believed to consist in a pair of nearby E' centers [59-63], as generically represented in Figure 2.6. In this scheme, the main difference between the various triplet state centers should consist in the distance between the two interacting spins [60-63]. However, the reason why four different distances are possible and the details on the relative orientations of the two E' centers in each of the four types of triplet centers are still open questions.

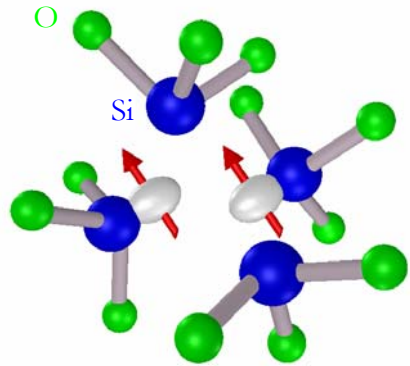

**Figure 2.6** Generic microscopic model of the triplet state centers in α-quartz.

## 2.2.4 The $[AlO_4]^0$ center

The $[AlO_4]^0$ center in irradiated α-quartz has been extensively studied in many experimental [64-69] and theoretical [70-76] works. Although it possesses a complex temperature-dependent EPR spectrum, the microscopic structure of the $[AlO_4]^0$ center has been well established and consists in a Al atom substituting for a four-coordinated Si in the lattice with a hole that, for temperature lower than ~30 K, is stably trapped in a nonbonding 2p orbital of a short-bond O atom adjacent to Al (Figure 2.7) [64-76]. On increasing the temperature, the hole first jumps in both the short-bond O atoms and then, at room temperature, it delocalizes over all the four tetrahedral O atoms [69].

Since the Al atom has three electrons in the outer shell, instead of the four of Si, in order to complete the four tetrahedral bonds it needs a surplus electron. For this reason, in the as grown crystal, the substitutional Al atom usually occurs with an alkali ion nearby, which has given the electron to the Al [69]. In this configuration the negatively charged Al atom has completed the tetrahedral bonds and attracts the alkali positive ion by electrostatic interaction. When an alkali ion is present nearby to the substitutional Al atom the defect is signed by $[AlO_4^-/M^+]^0$, where $AlO_4^-$ indicates that the Al atom has trapped an electron, $M^+$ indicates which alkali ion is involved in the defect (usually $M^+=Li^+$, $Na^+$ or $H^+$), and the superscript 0 indicates that the overall defect complex is neutral [69]. This structure is considered to be the precursor site of the $[AlO_4]^0$ center, as discussed in the next paragraph.

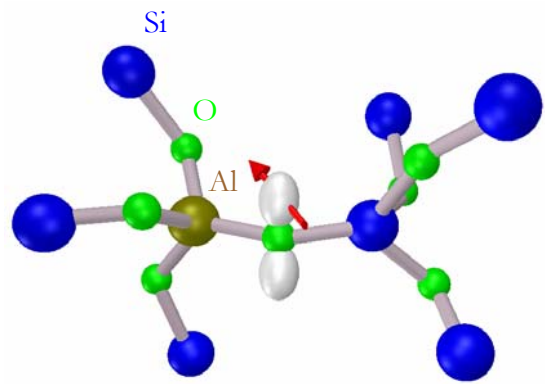

**Figure 2.7** Microscopic structure of the $[AlO_4]^0$ center in α-quartz [69].

## 2.2.5 Complex mechanisms of point defects generation

Although the microscopic structures of the relevant point defects of α-quartz have been established, their generation mechanisms have been found in many cases very complicated and deserve further investigation to be clarified [39, 48, 51, 60-62, 67, 69].

Irradiation experiments at variable temperatures have shown that the $[AlO_4]^0$ center is induced in α-quartz only if the irradiation is performed at temperature above ~100 K [69]. This experimental evidence has been attributed to the ability of alkali ion, $M^+$, to diffuse away from the Al site during irradiation [69]. In this frame it is believed that irradiation at temperature lower than ~100 K causes the creation of a hole on an O atom bonded to Al. At this temperature the $M^+$ ion does not have sufficient thermal energy to diffuse away, even though in $[AlO_4/M^+]^+$ complex it is no longer required as a charge compensator [69]. On increasing the temperature, or directly upon irradiation at room temperature, the $M^+$ ion diffuses away along the c-axis channels of the crystal becoming stably trapped at unknown sites and the EPR signal of the bare $[AlO_4]^0$ center is observed [69]. Finally, detailed studies by EPR and IR spectroscopies have established that, upon thermal treatments in the range from 500 K to 650 K of an irradiated sample, the as-grown configuration of the crystal is restored with the formation of the initial $[AlO_4^-/M^+]^0$ complexes [67].

The $E'_2$ and $E'_4$ centers are easily induced by room temperature neutrons, X- γ- and β-ray irradiation of α-quartz [48, 49, 51, 53, 57]. At variance, the $E'_1$ center generation mechanism has been found to be more complex [39, 61]. In fact, simple irradiation at room temperature does not produce an appreciable number of $E'_1$ centers, while upon successive thermal treatment above 500 K the concentration of $E'_1$ centers is observed to increase for more than one order of magnitude [39, 48, 51, 61]. A detailed EPR study by Jani *et al.* [39] has pointed out that this growth of concentration of $E'_1$ centers occurs in connection with the annealing of the $[AlO_4]^0$ center, as shown in Figure 2.8 [39]. Furthermore, nearly constant total concentration of $E'_1$ plus $[AlO_4]^0$ centers has been found for thermal treatments temperature up to 570 K [39]. On the basis of this experimental evidence the authors have proposed that holes are released from the $[AlO_4]^0$ centers upon thermal treatment in the range of temperatures from 500 K to 570 K and that each of these holes is trapped by the diamagnetic precursor site of the $E'_1$ center [39].

Complex generation mechanisms, as for the $E'_1$ center, have been also observed for the triplet state centers [60, 61]. In fact, it has been shown that an appreciable number of these centers cannot be induced by simple irradiation at 77 K. At variance, the same irradiation produces a large number of triplet state centers provided that a previous irradiation at room temperature has been performed [60, 61]. Although the exact mechanisms of triplet centers generation has not yet been established, the above discussed experimental evidences have suggested that they could be connected with the temperature-dependent processes of [AlO$_4$]$^0$ centers generation and of alkali ions diffusion [60, 61].

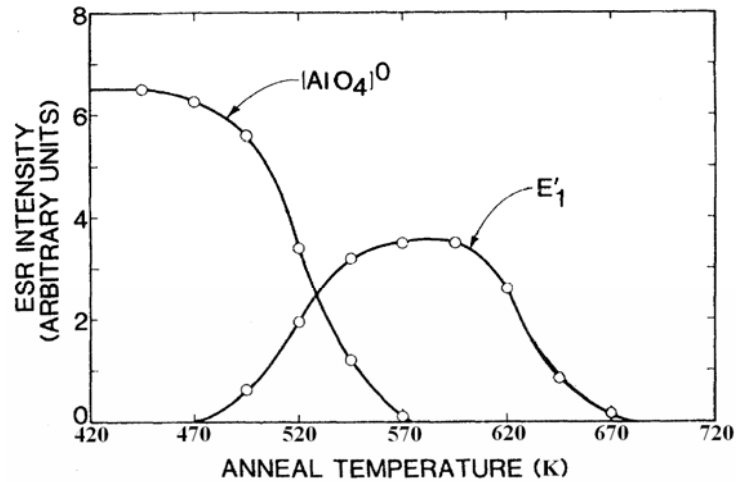

**Figure 2.8** Effects of an isochronal thermal treatment on the concentration of [AlO$_4$]$^0$ and E'$_1$ centers on a sample of α-quartz previously β-ray irradiated at 300 K and then at 77 K. The concentration of [AlO$_4$]$^0$ and E'$_1$ centers were measured by EPR measurements at 77 K and room temperature, respectively. Adapted from Ref. 39.

## 2.3   Point defects in amorphous SiO$_2$

### 2.3.1  E'$_\gamma$ center

The E'$_\gamma$ center is the most commonly observed point defect induced in a-SiO$_2$ by UV, X, γ and neutrons irradiation [1, 2]. Is was first observed by Weeks [38] and successively further characterized by Griscom [2, 47, 77-82]. The E'$_\gamma$ center exhibits an almost axially symmetric EPR line shape with principal g values $g_1=2.0018$, $g_2=2.0006$ and $g_3=2.0003$ [2]. Furthermore, it was shown that in order to obtain a good fit of its EPR line shape a distribution of the latter two principal g values has to be considered [79]. In Figure 2.9 (a) a comparison between the experimental EPR line shape of the E'$_\gamma$ center and the simulated one is reported [79]. The statistical distributions of the principal g values used in the simulation are shown in Figure 2.9 (b) [79].

To the E'$_\gamma$ center has been definitively attributed [2, 77, 78] a strong hyperfine structure consisting in a pair of lines split by ~42 mT, arising from the interaction of the unpaired electron

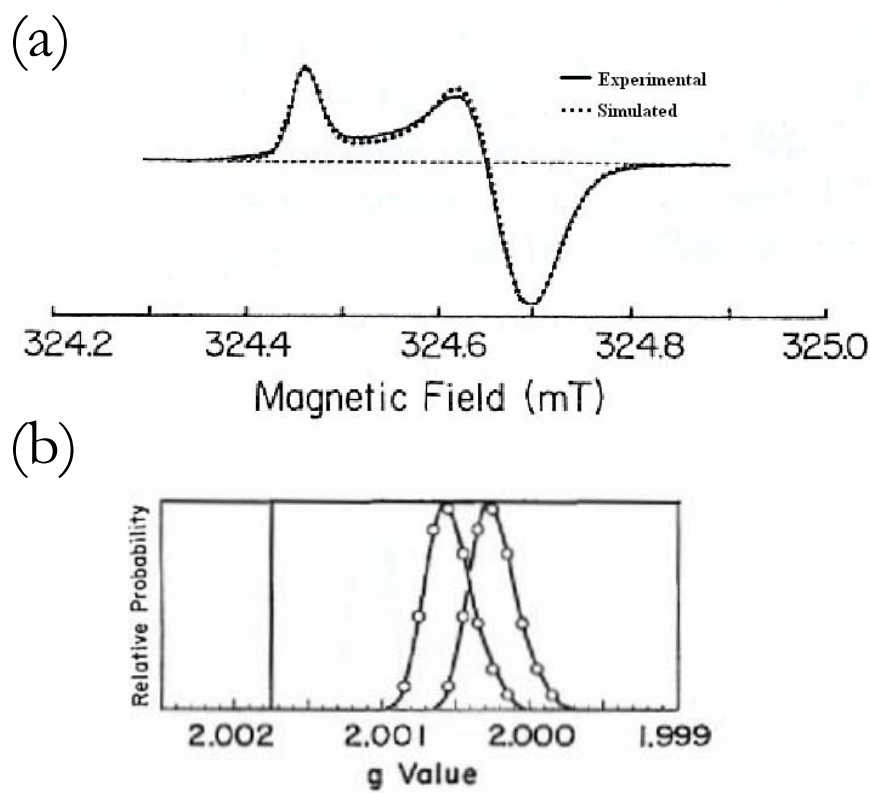

**Figure 2.9** (a) Comparison between the experimental EPR spectrum of the E'$_\gamma$ center and the line shape obtained by simulation. (b) Statistical distribution of the principal g values used in the simulated spectrum shown in (a). Adapted from Ref. 79.

with a $^{29}$Si nucleus. Due to the similarity of the principal g values and hyperfine structures of E'$_\gamma$ and E'$_1$ centers, it has been supposed that the two defects could share the same microscopic structure (Figure 2.3) and, consequently, the E'$_\gamma$ center has been suggested to be the defect equivalent of the E'$_1$ center of α-quartz in a-SiO$_2$ [2, 77, 78]. On the basis of this hypothesis, in analogy with E'$_1$ center, it is expected [2] that the E'$_\gamma$ center should possess a weak hyperfine doublet split by about 0.8 mT ÷ 0.9 mT. In agreement with this prediction, a doublet of lines with a peak-to-peak split of ~1.26 mT has been observed in irradiated a-SiO$_2$ [78, 81, 83-86] and, by computer line shape simulations, it was shown to correspond to an isotropic hyperfine splitting constant of ~0.8 mT [81]. However, in these experimental studies [78, 81, 83-86] the 1.26 mT doublet has been attributed to the hyperfine interaction of an unpaired electron with a proton, rather than with a $^{29}$Si nucleus. In contrast with this attribution, Boero *et al.* [45] suggested, on the basis of a first principles study, that the E'$_\gamma$ in a-SiO$_2$ possesses a weak hyperfine structure very similar to that of the E'$_1$ center in α-quartz, so supporting the attribution of the 1.26 mT doublet to the hyperfine interaction of the unpaired electron involved in the E'$_\gamma$ center with a second nearest $^{29}$Si atom. This conclusion was subsequently supported by an experimental work [87] in which the 1.26 mT doublet EPR signal was found to be strictly correlated with that of the E'$_\gamma$ center main EPR line in a wide γ- and β-ray irradiation dose range. The weak hyperfine structures of the E'$_\gamma$ center were also studied by Griscom and Cook [47] performing an EPR study of γ-ray irradiated $^{29}$Si enriched a-SiO$_2$ samples. Surprisingly, the authors have found that the hyperfine interaction strength of the unpaired electron of the E'$_\gamma$ centers with a second nearest $^{29}$Si atom agrees with that of the E'$_1$ center only in few cases, whereas in most cases it is negligible [2, 47]. In order to account for this experimental evidence, the authors proposed that in the majority of E'$_\gamma$ centers the Si atom on which the unpaired electron is localized relaxes through the plan of its basal O atoms in a back projected configuration [2, 47]. In fact, as the authors showed by simple tight-binding analysis [47], in this configuration the weak hyperfine interaction is significantly reduced. This conclusion was also supported by successive theoretical studies by Embedded-Cluster-Method focused on the stable configurations of the positively charged oxygen vacancy in a-SiO$_2$ [88, 89].

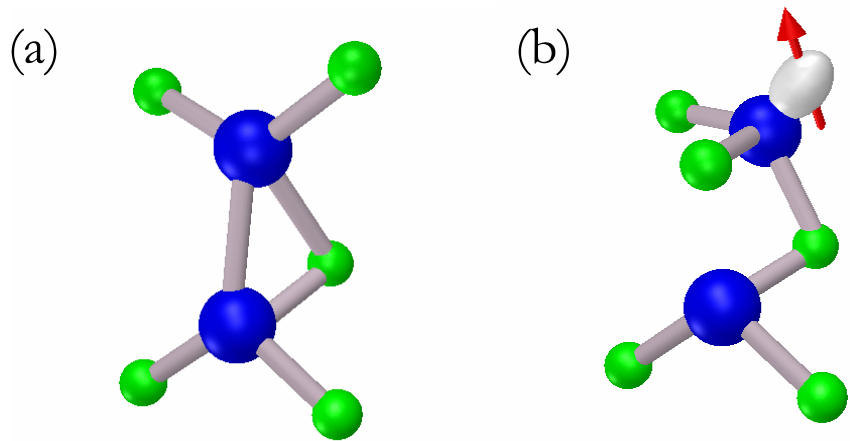

**Figure 2.10** Microscopic structures proposed by Uchino *et al.* [96] for (b) the E'$_\gamma$ center and (a) its precursor site.

The difficulties related to the lack of a certain identification of the weak hyperfine structure of the E'$_\gamma$ centers, as discussed above, rise the question if the microscopic structure of the E'$_\gamma$ is actually similar to that of the E'$_1$ center, as initially suggested [2, 77, 78]. Furthermore, the problem of the attribution of a specific microscopic structure to the E'$_\gamma$ center becomes more complicated in the light of some experimental evidences found in a-SiO$_2$ films on crystalline Si [90-95]. In fact, in these systems it has been shown, by capacitance-voltage measurements and charge injection experiments, that at least two types of E'$_\gamma$ centers exist [92, 94]. Although these two E'$_\gamma$ centers possess similar EPR features, they differ in the charge state which can be neutral or positive [90-94]. While the positively charged E'$_\gamma$ could be, again, considered the equivalent in a-SiO$_2$ of the E'$_1$ center, the neutral E'$_\gamma$ is not consistent with an oxygen vacancy model. It has been proposed that the latter defect could possess a microscopic structure similar to the former but for the lack of the positively charged group of atoms facing the unpaired electron [91, 94, 95].

Recent theoretical investigations have suggested that the problem concerning the microscopic structure of the E'$_\gamma$ center in a-SiO$_2$ is more complex. Uchino *et al.* [96-98], on the basis of quantum chemical calculation, have suggested that the E'$_\gamma$ could originate from an "edge-sharing" oxygen vacancy (triangular oxygen-deficient center), schematically represented in Figure 2.10 (a), by trapping a hole and relaxing in the structure of Figure 2.10 (b) (bridged hole-trapping

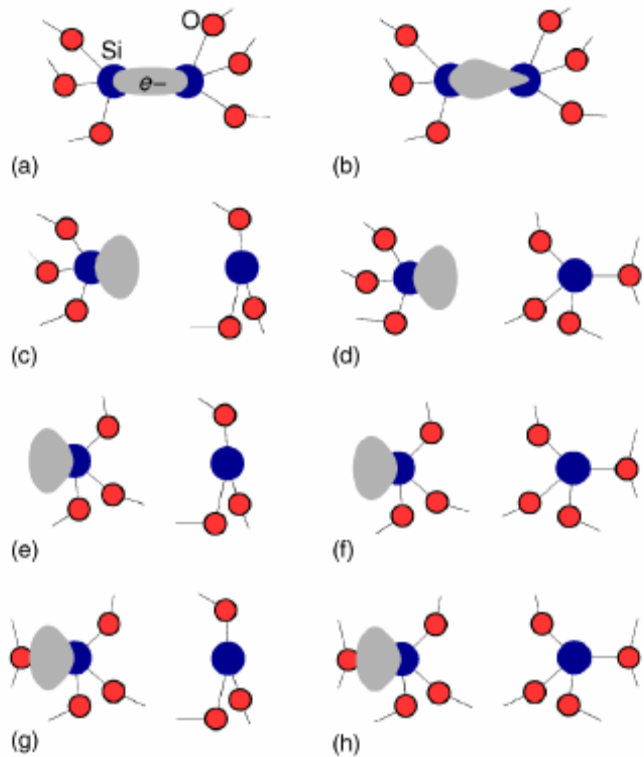

**Figure 2.11**  (a)-(h) Stable microscopic structures originating from ionization of a single oxygen vacancy in a-SiO$_2$. After Ref. 88.

oxygen-deficiency center). In this model, the unpaired electron is localized on one Si atom and the trapped hole on the other. The structural model of Figure 2.10 (b) succeeds in accounting for the experimental value, 42 mT, of the strong $^{29}$Si hyperfine structure of the E'$_\gamma$ center [96-98]. Successively, in a theoretical investigation by Lu *et al.* [99], the simple oxygen vacancy has been reconsidered as a precursor of the E'$_\gamma$ center. It has been suggested that in a-SiO$_2$, at variance with respect to α-quartz, two distinct E'$_\gamma$ centers could arise, differing in the positively charged puckered Si atom facing the unpaired electron, which can bond with one or two back O atoms. A more complex situation has been found by a theoretical study with the Embedded-Cluster-Method [88, 89], which has suggested that many microscopic structures can stabilize in a-SiO$_2$ after ionization of an oxygen vacancy, as summarized in Figure 2.11. The structures reported in Figures 2.11 (a) and (b) were suggested by the authors [88, 89] to pertain to the E'$_\delta$ center and will be discussed in more details in Paragraph 2.3.4. The structure of Figure 2.11 (d) essentially corresponds to that of the E'$_1$ center in α-quartz (Figure 2.3), whereas the other structures are believed to be peculiar of the amorphous state of SiO$_2$ [88, 89]. In the structure of Figure 2.11 (c) an incomplete puckering of the positively charged Si atom of the vacancy occurs, and it remains in a planar configuration with respect to its basal O atoms. The structures of Figures 2.11 (e) and (f) are similar to those of Figures 2.11 (c) and (d), respectively, but for the fact that the Si atom on which the unpaired electron wave function is localized has relaxed in a back projected configuration. The structures of Figures 2.11 (g) and (h) are similar to those of Figures 2.11 (e) and (f), respectively, but in the former two structures an extra O atom of the a-SiO$_2$ is nearby and interacts with the unpaired electron. The hyperfine parameters corresponding to the microscopic structures of Figure 2.11 are reported in Table 2.1. As shown, the strong hyperfine isotropic constants of the structures of Figures 2.11 (c), (d) and (e) are consistent with that of the E'$_\gamma$ center, but that of Figure 2.11 (g) is somewhat larger. Furthermore, one can expect that to this latter structure, which will be discussed in more details in Paragraph 2.3.3 in connection with the E'$_\alpha$ center, should pertain a **ĝ** tensor with a symmetry lower than axial, due to the perturbation of the nearby O atom to the unpaired electron wave function.

**Table 2.1** Hyperfine constants (mT) for the microscopic structures reported in Figure 2.9. Si$_n$ and Si$_{nk}$ are the nearest and the second nearest neighbours of the oxygen vacancy, respectively. After Ref. 88.

| Type of E' defect | QM cluster | Figures | Si$_1$ hf | Si$_{1l}$ shf | | | Si$_2$ hf | Si$_{2l}$ shf | | |
|---|---|---|---|---|---|---|---|---|---|---|
| | | | | Si$_{11}$ | Si$_{12}$ | Si$_{13}$ | | Si$_{21}$ | Si$_{22}$ | Si$_{23}$ |
| Dimerlike | Si$_{13}$O$_{37}$Si*$_{22}$ | 2.11(a) | 10.9 | 0.08 | 0.17 | 0.04 | 12.1 | 0.05 | 0.35 | 0.35 |
| E'$_{DB}$, unpuckered | Si$_{10}$O$_{29}$Si*$_{18}$ | 2.11(c) | 44.3 | 1.33 | 0.50 | 0.07 | 0.0 | 0.03 | 0.00 | 0.00 |
| E'$_{DB}$, puckered with O$_B$ | Si$_{13}$O$_{37}$Si*$_{22}$ | 2.11(d) | 43.0 | 1.13 | 0.80 | 0.20 | 0.0 | 0.00 | 0.00 | 0.00 |
| E'$_{DB}$, back-projected | Si$_8$O$_{24}$Si*$_{16}$ | 2.11(e) | 43.1 | 0.26 | 0.23 | 0.29 | 0.0 | 0.00 | 0.00 | 0.00 |
| E'$_{DB}$, back-projected with O$_B$ | Si$_{13}$O$_{37}$Si*$_{22}$ | 2.11(g) | 48.9 | 0.45 | 0.02 | 0.25 | 0.0 | 0.00 | 0.00 | 0.00 |

Interestingly, the observation of two slightly different but distinguishable E'$_\gamma$ center main EPR line shapes [100], named L1 and L2, in γ-ray irradiated a-SiO$_2$ could also indicate the existence of at least two types of E'$_\gamma$ centers, in agreement with the experimental data on the charge state of the E'$_\gamma$ center and on the theoretical results discussed above. However, since the two E'$_\gamma$ center EPR line shapes L1 and L2 differ very little, their attribution to specific structural models is not a simple task. In future, an important step towards the solution of this problem could derive from the measurement of the charge state of the E'$_\gamma$ centers with EPR line shapes L1 and L2 by, for example, simultaneous EPR and charge state studies.

An OA band peaked at 5.8 with FWHM of ~0.8 eV and oscillator strength ~0.14 has been attributed to the E'$_\gamma$ center by Weeks and Sonders [101] by EPR and OA measurements in irradiated a-SiO$_2$. This attribution was subsequently confirmed by other experimental works [102, 103]. Although this attribution is considered certain [50], the outstanding question concerns the specific electronic levels involved in this OA band, for which two possible electronic processes have been proposed [50]. Griscom and Fowler [104] first suggested that this absorption involves a charge transfer from the unpaired electron level to that of the facing Si atom of the oxygen vacancy. This hypothesis was supported subsequently by theoretical investigations by Edwards [105] and Pacchioni and Ieranò [106]. Alternatively, it has been proposed [80, 89, 107, 108] that both the initial and the final states involved in the electron transition are confined in the O≡Si• group of the E'$_\gamma$ center. This latter attribution is strongly supported [80] by the invariance of the spectral properties of the 5.8 eV absorption band observed in various a-SiO$_2$ samples whether irradiated by UV laser [2, 50, 109-112], X-rays [113, 114], γ-rays [48, 51, 101], neutrons [115] and heavy ions [116, 117]. In fact, if the 5.8 eV absorption band is due to a charge transfer process, then its peak position should be extremely sensitive to the internuclear distance between the two facing Si atoms of the oxygen vacancy and a dependence of its spectroscopic properties on the specific material or on the method of irradiation is expected to be observable, in disagreement with the experimental data [50, 80].

In summary, although the attribution of an OA band peaked at 5.8 eV to the E'$_\gamma$ center is certain, up to now a general consensus on the energy levels involved in this electronic transition has not been reached and further experimental and theoretical studies are needed to clarify this point [50].

### 2.3.2 E'$_\beta$ center

The E'$_\beta$ center was first observed and characterized by Griscom [2, 79, 80, 82]. Its main EPR line shape is similar to that of the E'$_\gamma$ center, but it possesses a more pronounced axial symmetry [79]. Its principal g values are $g_\parallel$ = 2.0018 and $g_\perp$ = 2.0004, with the latter g value statistically distributed [2, 79]. In Figure 2.12 (a) a comparison between the experimental EPR line shape of the E'$_\beta$ center and the simulated one is reported [79]. The statistical distributions of the principal g values used in the simulation are shown in Figure 2.12 (b) [79].

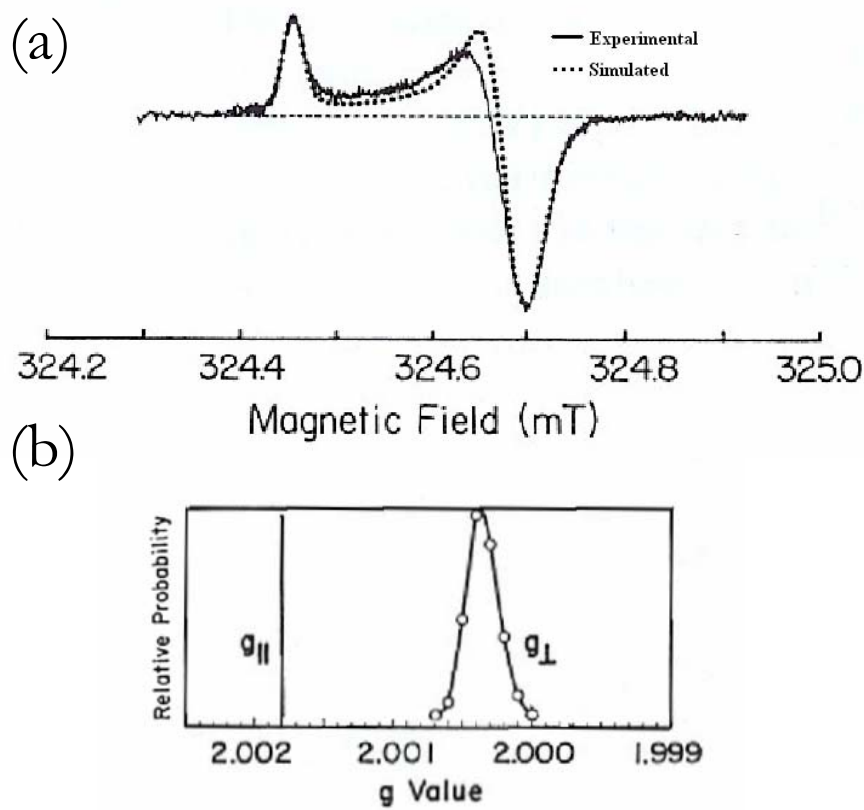

**Figure 2.12** (a) Comparison between the experimental EPR spectrum of the E'$_\beta$ center and the line shape obtained by simulation. (b) Statistical distribution of the principal g values used in the simulated line shown in (a). Adapted from Ref. 79.

An hyperfine structure consisting in a pair of lines split by 42 mT, indistinguishable with respect to that of the E'$_\gamma$ center, has been attributed to the E'$_\beta$ center [2, 79, 80, 82]. This experimental evidence has suggested that in the E'$_\beta$ center an O≡Si$^\bullet$ moiety similar to that of the E'$_\gamma$ center is involved [2, 79, 80, 82].

The E'$_\beta$ center has been observed in synthetic a-SiO$_2$ materials containing ~1200 ppm OH groups X-ray irradiated at 77 K [79]. Furthermore, it was shown that if the same material is subsequently subjected to an isochronal thermal treatment at higher temperatures, the post-irradiation concentration of E'$_\beta$ centers increases by a factor of about five [79]. In particular, two stages of E'$_\beta$ centers concentration increase have been observed, occurring in the temperature ranges 100 K ÷ 130 K and 180 K ÷ 270 K [79]. Since in these temperature ranges the thermal diffusion of H and H$_2$, respectively, occurs, it was supposed that the E'$_\beta$ centers could originate from the interaction of H atoms with some precursor site in a-SiO$_2$ [2, 79, 80]. At present, the most reasonable hypothesis on the microscopic structure of the E'$_\beta$ center is that it represents the equivalent defect in a-SiO$_2$ of the E'$_2$ center of α-quartz [Figure 2.5 (b)].

An OA band peaked at 5.4 eV, similar to that attributed to the E'$_2$ center in α-quartz, has been tentatively attributed to the E'$_\beta$ center in a-SiO$_2$, basing on the conjecture that these two point defects could share the same microscopic structure [50, 82, 118, 119].

### 2.3.3 E'$_\alpha$ center

The E'$_\alpha$ center[1] was first observed and characterized by Griscom [2, 79, 80, 120]. The principal g values of this defect are g$_1$=2.0018, g$_2$=2.0013 and g$_3$=1.9998, with the latter two g values statistically distributed [2, 79, 80]. In Figure 2.13 (a) a comparison between the experimental EPR line shape of the E'$_\alpha$ center and the simulated one is reported. The discrepancies between the experimental spectrum and the simulated line shape, evident in Figure 2.13 (a), are due to the presence in the former of the EPR signal of the E'$_\gamma$ center partially superimposed to that of the E'$_\alpha$ center, whereas the simulated line refers to the E'$_\alpha$ center alone [79]. The statistical distributions of the principal g values used in the simulation of the E'$_\alpha$ center EPR line shape are shown in Figure 2.13 (b) [79].

---

[1] In Ref. 79 Griscom defined two types of E'$_\alpha$ centers: E'$_{\alpha 1}$ and E'$_{\alpha 2}$. However, since the E'$_{\alpha 2}$ was found to possess EPR features indistinguishable with respect to those of the E'$_\gamma$, in successive papers [2, 80, 82, 120] Griscom have named E'$_\alpha$ the former, and E'$_\gamma$ the latter. Here the latter nomenclature is used.

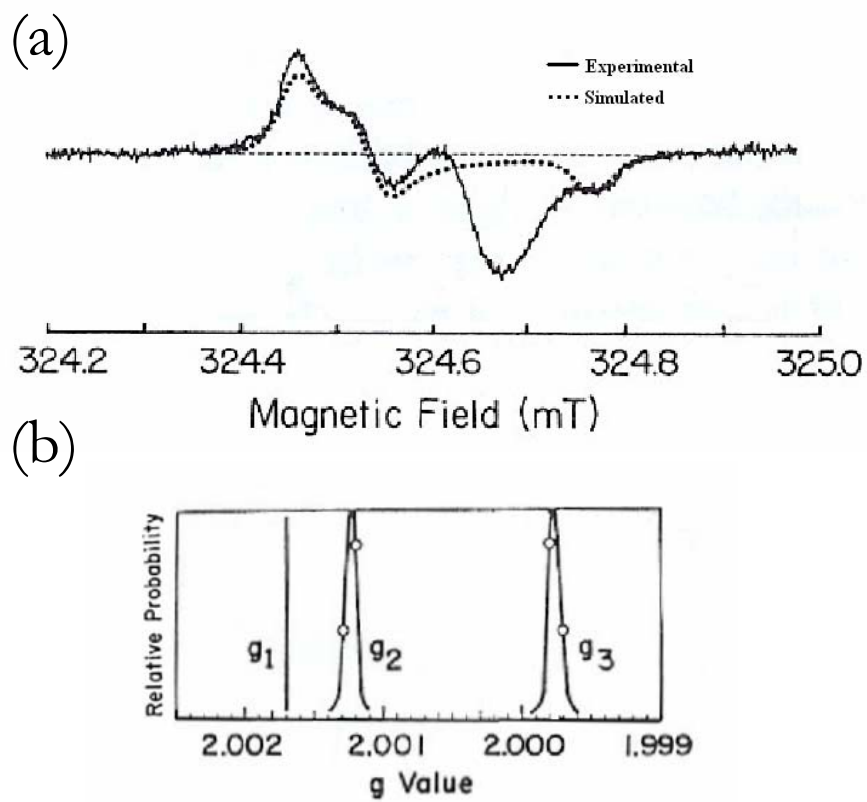

**Figure 2.13** (a) Comparison between the experimental EPR spectrum of the E'$_\alpha$ center and the line shape obtained by simulation. (b) Statistical distribution of the principal g values used in the simulated line shown in (a). Adapted from Ref. 79.

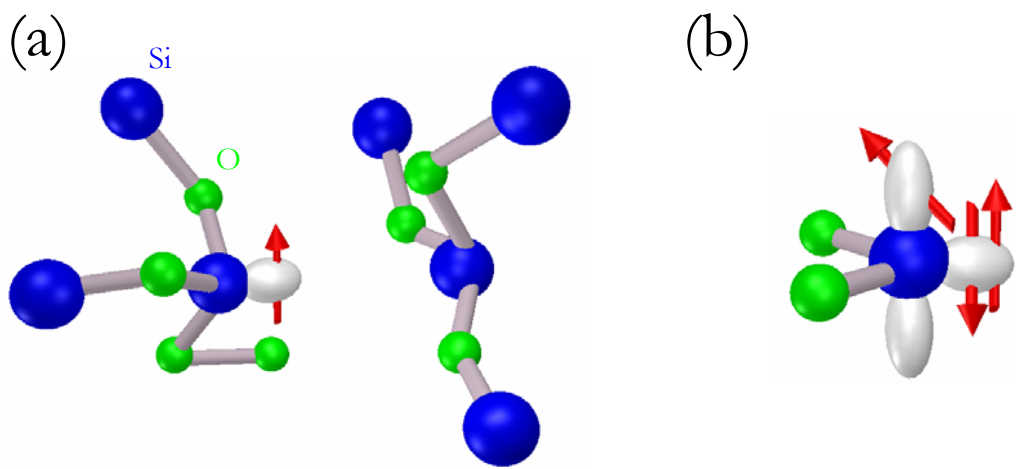

**Figure 2.14** Microscopic structures proposed for the E'$_\alpha$ centers by Griscom (a) in 1984 [79] and (b) in 2000 [2].

In a first work, Griscom observed this defect in high purity (stoichiometric) a-SiO$_2$ materials X-ray irradiated at 77 K [79]. In that work the author found that the E'$_\alpha$ is stable only below T≅200 K and that it converts to the E'$_\gamma$ upon exposure to room light [79]. In a successive experimental investigation, Griscom and Friebele [120] observed the E'$_\alpha$ defect in nonstoichiometric a-SiO$_2$ γ-ray irradiated at room temperature. However, at variance to the first experimental investigation, the E'$_\alpha$ defect was found to be thermally stable at room temperature and no light induced conversion to E'$_\gamma$ was observed [120].

In 1984 Griscom suggested [79] that the $^{29}$Si hyperfine structure of the E'$_\alpha$ center consists in a doublet split by ~42 mT, as for E'$_\gamma$ and E'$_\beta$ centers. On the basis of this attribution, he proposed that the E'$_\alpha$ center could originate by an irradiation induced displacement of an O atom and an electron from a regular site of the a-SiO$_2$ matrix [79]. This mechanism generates a positively charged oxygen vacancy similar to the one involved in the E'$_\gamma$ center, but for the displaced O atom forming a peroxy bridge with one of the basal O atoms of the O≡Si$^\bullet$ moiety [Figure 2.14 (a)] [79]. The perturbation induced by the knocked O atom on the unpaired electron wave function being responsible for the orthorhombic g matrix of the E'$_\alpha$ center [79]. In 2000, after a reexamination of the previously published data, Griscom proposed [2] that the $^{29}$Si hyperfine structure of the E'$_\alpha$ center should consist in a pair of lines split by ~14 mT. On the basis of this latter attribution, a model consisting in a twofold coordinated Si (O=Si:, where : represents a lone pair) having trapped an electron was put forward [Figure 2.14 (b)] [2].

Successively, Uchino *et al.* [97, 121] on the basis of quantum-chemical calculations suggested that the E'$_\alpha$ center could originate from an hole trapped in a twofold coordinated Si. The authors showed that upon hole capture the structure relaxes to a metastable state characterized by a sp$^3$-like unpaired electron wave function and with an expected value of the isotropic hyperfine coupling constant of ~44 mT [97, 121]. Furthermore, the system was found to easily relax to a more stable structure in which an O≡Si$^\bullet$ moiety is formed, giving a possible explanation of the experimentally observed room light induced conversion from E'$_\alpha$ to E'$_\gamma$ [79, 97, 121].

Finally, it is worth to note that the stable structures originating from the positively charged oxygen vacancy reported in Figures 2.11 (g) and (h) should also account for the spectroscopic features characterizing the E'$_\alpha$ center, although this possibility was not explicitly discussed by the authors [88]. In fact, as a consequence of the perturbation of the nearby O atom on the unpaired electron wave function, the expected $\hat{\mathbf{g}}$ matrix of these structures should possess low symmetry, as experimentally observed for the E'$_\alpha$ center [see Figure 2.13 (b)]. Consequently, the models of Figures 2.11 (g) and (h) could pertain to the E'$_\alpha$ center. In future, the validity of this attribution could be evaluated comparing the hyperfine doublet of the E'$_\alpha$ center with that of the structure of Figures 2.11 (g) and (h), for which an isotropic hyperfine constant of ~48.9 mT has been calculated (see Table 2.1).

Summarizing, a general consensus on the actual microscopic structure of the E'$_\alpha$ center has not yet been reached [2, 121]. However, as it comes from the above discussion, a way to establish a definitive structure for the E'$_\alpha$ center could result by the experimental identification of its $^{29}$Si hyperfine structure.

## 2.3.4 E'$_\delta$ center

The E'$_\delta$ center was first observed and characterized by Griscom [120]. It possesses an highly symmetric EPR line shape, as shown in Figure 2.15, with principal g values $g_{\parallel}$ = 2.0018 and $g_\perp$ = 2.0021 [120].

To the E'$_\delta$ center has been attributed an hyperfine structure consisting in a pair of lines split by ~10 mT, supposed to arise from the hyperfine interaction of the unpaired electron with a $^{29}$Si nucleus [120]. The E'$_\delta$ point defect has been observed in bulk a-SiO$_2$ [120, 122-124], in thermally grown a-SiO$_2$ films on Si [92, 125-136], and in buried oxide layer of separation by implantation of oxygen (SIMOX) systems [125, 131, 137-145]. These experimental works have pointed out that the E'$_\delta$ center can be induced in a-SiO$_2$ by X- and γ-ray irradiation, hole injection and by bombardment with Ar$^+$ ions. Although all these treatments are able to induce the E'$_\delta$ center, large differences in the generation efficiency have been found. Bombardment with Ar$^+$ ions, for example, was found to be at least three orders of magnitude more efficient in generating E'$_\delta$ center with respect to hole injection and X- or γ-ray irradiation [144]. Furthermore, it has been pointed out that holes injection is able to induce E'$_\delta$ centers in many a-SiO$_2$ on Si systems, whereas an equal number of injected electrons is not [92, 127, 141]. This experimental evidence has been considered as an indication of the hole trapped nature of the E'$_\delta$ center [92, 127, 141].

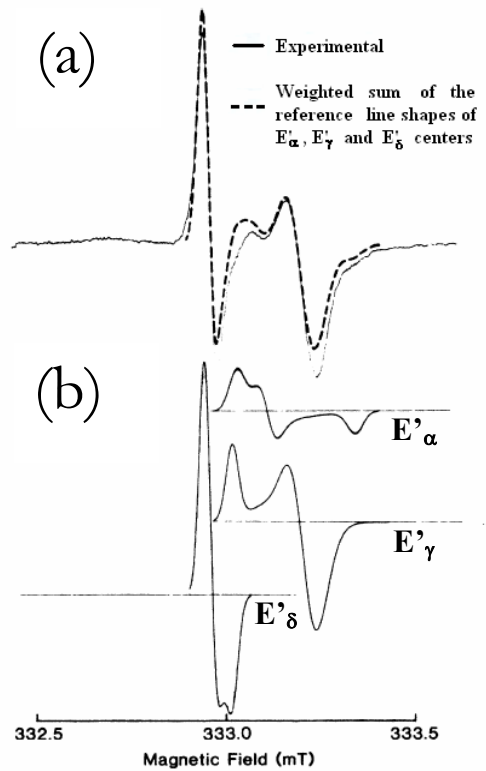

**Figure 2.15** (a) Experimental EPR spectrum compared to the line shape obtained as a weighted sum of the reference lines (b) for E'$_\alpha$, E'$_\gamma$ and E'$_\delta$ centers. Adapted from Ref. 120.

A key experimental evidence on the E'$_\delta$ center consists in the observation that a large number of its precursors are induced during high temperature (T>800 K) annealing in different atmospheres of buried [126, 131, 125, 134] and unburied [131, 131-136] a-SiO$_2$ films on Si. In the same works a similar generation process has been also observed for the precursors of the E'$_\gamma$ center. To explain these findings, it has been proposed that during thermal treatments O atoms could diffuse from the a-SiO$_2$ layer to the substrate and to the polysilicon overlayer, the driving force of this process being the different solubility limit of O in Si and SiO$_2$ [126, 132, 146]. In this scheme, the generation of precursors of E'$_\gamma$ and E'$_\delta$ centers in the a-SiO$_2$ layer should be due to the out-diffusion of O atoms from the oxide and the formation of oxygen vacancies [126, 132, 146]. Nevertheless, the observation of SiO gas production during low pressure oxidation of silicon [147], has inspired a different model in which it is supposed that during high temperature treatments the freeing of volatile SiO at the Si/SiO$_2$ interface, through the reduction reaction Si + SiO$_2$ → 2 SiO (volatile), could occur [131, 133, 135, 136, 147]. In this case, oxygen deficient defects could be induced by the rearrangement within the oxide network of volatile SiO, released from the interface and diffusing through the oxide [136]. Although the exact process responsible for the thermally induced degradation of MOS structures is not fully clarified, its occurrence clearly indicates the intrinsic and oxygen deficiency related nature of the E'$_\delta$ center.

The microscopic structure of the E'$_\delta$ center is not yet univocally determined. Since its first observation, many distinct microscopic models have been proposed. Griscom and Friebele [120] observed that: i) the $^{29}$Si hyperfine splitting of the E'$_\delta$ center (~10 mT) is ~4 times smaller than that of the E'$_\gamma$ center (~42 mT), ii) the g tensor of the E'$_\delta$ center is nearly isotropic. These features were tentatively explained supposing that the unpaired electron of the E'$_\delta$ center is delocalized over four symmetrically disposed Si sp$^3$ orbitals similar to the one involved in the E'$_\gamma$ center [120]. Furthermore, since the concentration of E'$_\delta$ center was found to correlate with the Cl content of the materials, a model consisting in an electron delocalized over four Si-sp$^3$ orbitals of an [SiO$_4$]$^{4+}$ vacancy decorated by three Cl$^-$ ions was proposed (4-Si Cl-containing model) [120]. However, as the same authors pointed out, the absence of the EPR lines due to the hyperfine interaction of the unpaired electron with the I=3/2 nuclei of $^{35}$Cl and $^{37}$Cl (with 75.4% and 24.6% natural abundance, respectively) represented a serious difficulty for the reliability of this model. The possibility that F atoms, together with Cl, could be involved in the microscopic structure of the E'$_\delta$ center has also been raised by Tohmon *et al.* [122] However, in successive works it has been reported that the E'$_\delta$ defect can be equivalently induced in Cl- and F-free samples, ruling out definitively the direct involvement of these impurities in the E'$_\delta$ center [92, 127, 142, 143].

Tohmon *et al.* [122] have pointed out that a necessary condition for the formation of the E'$_\delta$ center is the oxygen deficiency of the material, estimated by measuring the intensity of the OA band peaked at ~5.0 eV [50]. Furthermore, the authors have shown that as a consequence of a thermal treatment at 500 °C in atmosphere of H$_2$, the 5.0 eV OA band disappears together with the precursors of the E'$_\delta$ centers [122]. On the basis of these observations a microscopic model was proposed for the E'$_\delta$ center consisting in an ionized single oxygen vacancy with the unpaired electron nearly equally shared by the two Si atoms (2-Si model) [Figure 2.16 (a)] [122].

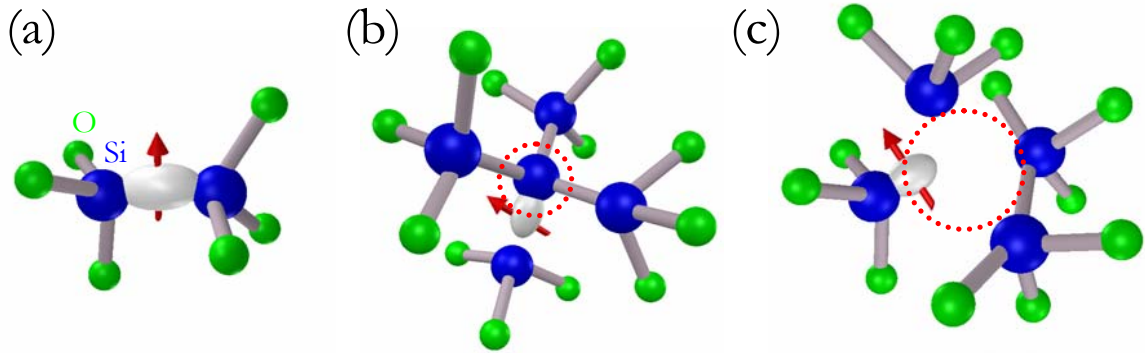

**Figure 2.16** Microscopic structures proposed for the E'$_\delta$ center by (a) Tohmon *et al.* [122], (b) Vanheusden and Stesmans [142] and (c) Zhang and Leisure [123].

Vanheusden and Stesmans [142, 143] reported that E'$_\gamma$ and E'$_\delta$ centers are induced in SIMOX samples. Furthermore, the authors have shown that E'$_\delta$ centers are prevalently induced in the region 200 Å÷700 Å away from the BOX/substrate interface, whereas E'$_\gamma$ centers are localized in a more extended region in the BOX. Since it was known that a large number of Si inclusions occur in the same region of the BOX in which the E'$_\delta$ centers are induced [148, 149], a new microscopic model was proposed in which the unpaired electron of the defect was supposed to be delocalized over the four sp$^3$ hybrid orbitals of an Si atom disposed at the center of a five Si cluster (5-Si model) [Figure 2.16 (b)] [142, 143].

Zhang and Leisure [123] have focussed on the experimental estimation of the EPR intensity ratio, ζ, between the 10 mT doublet and the E'$_\delta$ main line. However, due to the low concentration of defects, the authors [123] have detected the 10 mT doublet in the high-power second-harmonic mode (SH-EPR), which allows high sensitivity. Unfortunately, this detection scheme cannot give quantitative information on the number of defects responsible for the EPR signal and consequently the ratio ζ cannot be determined. Nevertheless, *postulating* a strict similarity between the properties of E'$_\gamma$ and E'$_\delta$ centers' SH-EPR signals, the authors could estimate ζ≅0.175, indicating a delocalization of the unpaired electron over four equivalent Si atoms [123]. On the basis of this estimation, it has been proposed a microscopic model for the E'$_\delta$ center consisting in a pairs of nearby oxygen vacancies, involving four Si neighboring atoms, with the unpaired electron delocalized over the four sp$^3$ hybrid orbitals of the Si atoms (4-Si model) [Figure 2.16 (c)] [123].

Conley and Lenahan [145] have studied the effects on E'$_\gamma$ and E'$_\delta$ centers of a room temperature treatment in hydrogen atmosphere (10% H$_2$ + 90% N$_2$). The authors have found that, as a consequence of the interaction with H$_2$, the EPR intensity of the E'$_\gamma$ center decreases in concomitance with the growth of a doublet split by 7.4 mT. Similarly, the EPR intensity of the E'$_\delta$ center decreased with a simultaneous growth of a doublet split by 7.8 mT. Furthermore, both these conversion processes were found to take place in few minutes and saturate within two hours. The two doublets split by 7.4 mT and 7.8 mT were associated to hydrogen complexed E'$_\gamma$ [81] and E'$_\delta$ centers, respectively. The authors [145], on the basis of the similarity in the doublets splitting and in the time scales of the processes of interaction with H$_2$, have proposed that the E'$_\delta$ center could possess a microscopic structure consisting in an unpaired electron strongly localized on a single Si atom, as for the E'$_\gamma$ center (1-Si model).

The atomic and electronic structures of the E'$_\delta$ center have been explored in many simulative calculations using Density Functional (DFT) [56, 96, 99, 150, 151, 152], Hartree-Fock (HF) [153-157], and Embedded Cluster [88, 89, 108, 158] methods. Chavez *et al.* [153] and Karna *et al.* [154] have studied the electronic structure of 2-Si, 4-Si and 5-Si models of the E'$_\delta$ center, and have shown that in all the cases considered the unpaired spin preferentially localizes on a single pair of Si atoms, so supporting the 2-Si model. However, in these calculations the atoms of the clusters were not allowed to relax after ionization. Consequently if, as suggested [120, 123, 142, 143], the delocalization of the unpaired electron results from a structural relaxation following the ionization of the precursor, then the conclusions outlined in these works could be questioned. Successive works focussed on the electronic properties of the ionized single oxygen vacancy (2-Si model) [56, 88, 89, 96, 99, 108, 150, 151, 155-158]. These works have pointed out that the ionized single oxygen vacancy in a-SiO$_2$, at variance to quartz, could admit a stable configuration in which the unpaired electron is nearly equally shared by the two Si atoms (2-Si model). This structure differs from that of E'$_\gamma$ because the puckering does not occur. To test if the 2-Si model could actually represent a realistic model for the E'$_\delta$ center, the hyperfine structure [56, 88, 89, 96, 108, 152, 153, 154] and the principal g values [89, 108] have been predicted. The hyperfine structure has been found to consist of a doublet of lines split by 8 mT - 13.5 mT, in good agreement with the experimental observations [120, 122, 123]. At variance, the calculated principal g values differ significantly from those obtained by EPR spectroscopy [92, 120, 127, 132, 141-144]. Furthermore, the calculated g values for the 2-Si model point out that this structure has a very low symmetry, whereas the EPR spectrum of the E'$_\delta$ center indicates an almost spherical symmetric unpaired electron wave function [120].

### 2.3.5 Triplet state center

Griscom [120] reported on the observation in X- and γ-ray irradiated Cl-doped a-SiO$_2$ of a pair of lines with peak-to-peak split of ~13.4 mT and center of gravity at g≈2 [Figure 2.17 (a)] correlated to a characteristic weak EPR line with g≈4 [Figure 2.17 (b)]. These EPR lines where attributed by the author to a point defect in a triplet state (pair of coupled electrons with total spin S=1) [120]. The observation of the same triplet state center has been also reported by other authors in X- and γ-ray irradiated a-SiO$_2$ at various Cl and F doping levels [122, 123]. Although at present an exact correspondence has not been established, it is believed that this center represents the equivalent in a-SiO$_2$ of one of the triplet centers observed in α-quartz and that the other types of triplet centers observed in α-quartz should also exists in a-SiO$_2$ [63].

The triplet state center has been observed in the same irradiated a-SiO$_2$ materials in which the E'$_\delta$ center was also induced [120, 122, 123]. Furthermore, a similar growth of concentration as a function of the X-ray irradiation dose has been reported for both these point defects [123].

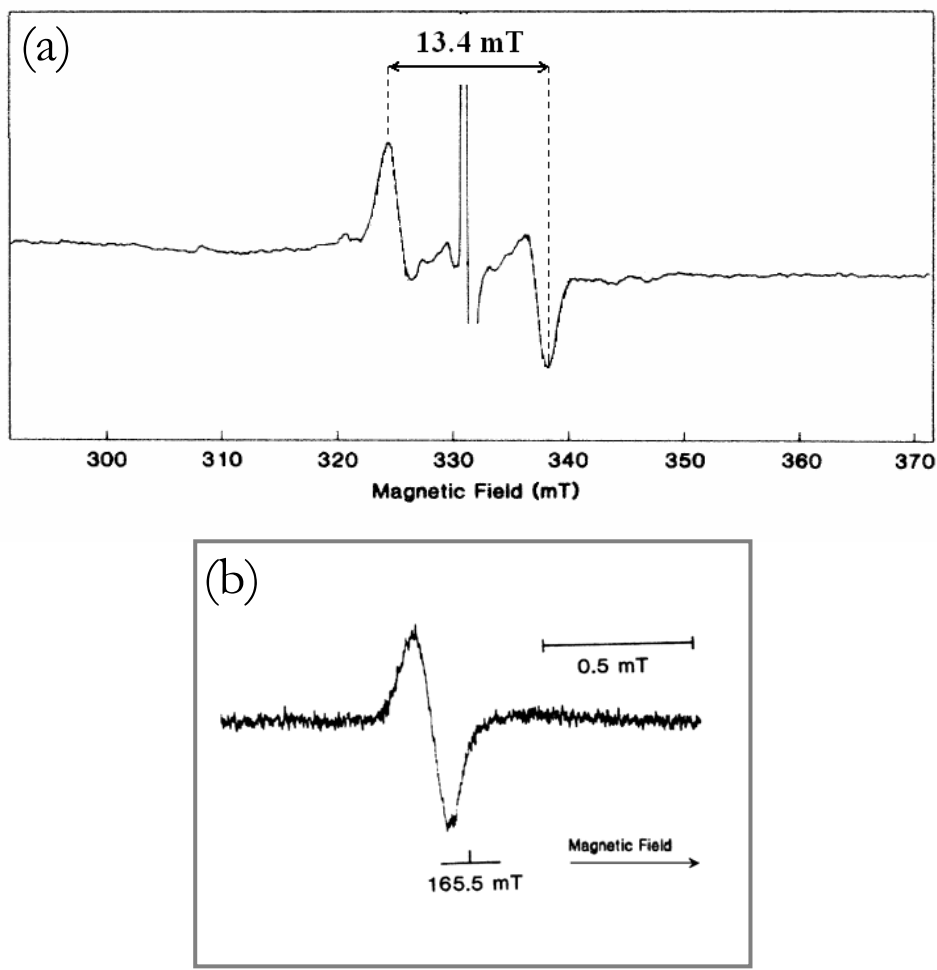

**Figure 2.17** EPR spectrum of the triplet state center acquired in correspondence of (a) g~2 and (b) g~4 in a sample of synthetic a-SiO$_2$ X-ray irradiated at 77 K. The measurements have been performed at 225 K. Adapted from Ref. 120.

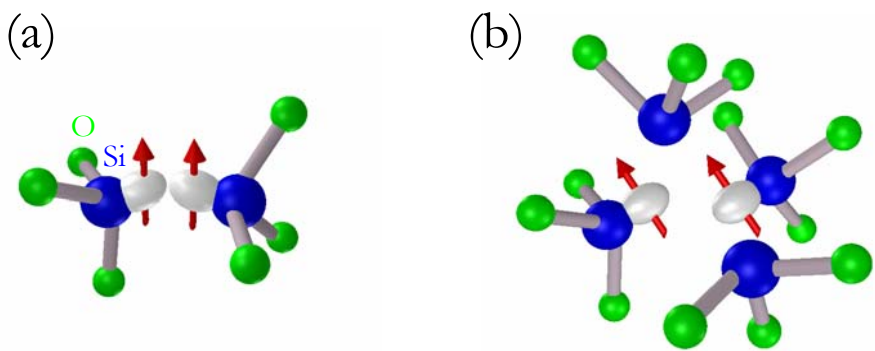

**Figure 2.18** Possible microscopic structures for the triplet state center in a-SiO$_2$.

On the basis of these experimental evidences, it has been proposed that they could share the same precursor site [120, 122, 123]. In this scheme, the proposed microscopic structures of the triplet state center corresponding to those reported in Figure 2.16 (a) and (c) for the E'$_\delta$ center are shown in Figure 2.18 (a) [122] and (b) [123], respectively. At variance, no specific microscopic structure has been proposed for the triplet state center originating from the 5-Si cluster.

## 2.3.6 The [AlO$_4$]$^0$ center

An Al-related EPR signal in irradiated a-SiO$_2$ was first reported by Fröman *et al.* [159], and subsequently observed by Lee and Bray [160] in irradiated Al$_2$O$_3$-SiO$_2$ glasses. However, the effective attribution of this resonance to the [AlO$_4$]$^0$ is due to Schnadt and Räuber [161]. The authors compared the EPR spectrum obtained in irradiated powdered Al-doped quartz with that of a glass prepared from the same material [161]. In Figure 2.19 (a)-(d) the EPR spectra recorded at 77 K and at 300 K for both these materials are reported [161]. As shown, a strict correspondence occurs between the EPR line shapes in the two materials, but for a discrepancy in correspondence to g ≅ 2.061. This difference has been attributed to a vitreous-state induced distribution in the g$_3$ values, smearing out this part of the a-SiO$_2$ spectrum (see Paragraph 1.1.2.2) [161]. The attribution of the EPR line shape observed in Al-doped a-SiO$_2$ to the [AlO$_4$]$^0$ center has been subsequently supported by computer line shape simulations [23].

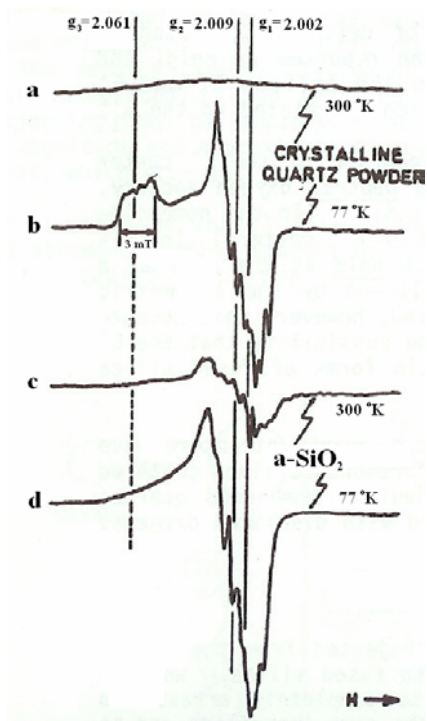

**Figure 2.19** (a)-(d) EPR spectra recorded at 300 K and 77 K in powdered Al-doped quartz and in a-SiO₂ samples obtained from the same material. Adapted from Ref. 161.

# Chapter 3

# *Experimental set-ups*

This chapter summarizes the experimental procedures and the properties of the instruments employed in the present Thesis. In particular, a detailed description of the EPR instrumentation is given, since it is the principal measurement technique employed.

## 3.1 The EPR spectrometer

### 3.1.1 Working principle and main components

In a typical EPR experiment the sample under study is subjected to two external magnetic fields pointing in orthogonal directions [11-15]. The first field, **H**, which can be considered static with respect to the intrinsic relaxation times of the paramagnetic centers, has the effect to spread the ground state energetic levels, as a consequence of the interaction between the magnetic moment of the system and **H** (see Chapter 1) [11-15]. The second external magnetic field, **H₁** $\left(|\mathbf{H}_1| \ll |\mathbf{H}|\right)$, with amplitude oscillating at a microwave frequency, is used to induce resonant transitions between pairs of states splitted by **H** [11-15]. The acquisition of an EPR spectrum consists in the measurement of the energy absorbed by the paramagnetic system as a function of the amplitude of **H**, at fixed amplitude and frequency of the magnetic field **H₁** [11-15].

In Figure 3.1 we report a scheme of the X-band continuous wave EPR spectrometer we have used, that is the Bruker EMX [162]. The microwave radiation, produced by a Gunn diode contained in the *source* group, travels through a rectangular waveguide and is directed to the resonant cavity containing the sample under study. The cavity is exposed to two magnetic fields: the static one, **H**, produced by the electromagnet, and a modulated magnetic field, **H_m**, parallel to **H** and produced by the *magnetic field modulation system* through the modulation coils. Once the microwave radiation has reached the cavity, it is partially absorbed by the sample and partially reflected back into the waveguide. The microwave radiation reflected back from the cavity is directed to the *detection system*, where it is converted by a Schottky diode in a current signal, and then phase detected and rectified. Finally, the current signal is digitalized in the *output circuit* and its value is displayed on the screen of the computer connected to the EPR spectrometer. The acquisition of an EPR spectrum consists in the measurement of the microwave power reflected by the cavity containing the paramagnetic system when the amplitude of the magnetic field **H** is varied linearly in time across the resonant value H$_r$ (defined in Chapter 1).

In the following Paragraphs 3.1.1.1 - 3.1.1.6 we discuss the functions of the individual EPR spectrometer components within each of the blocks indicated in Figure 3.1 [11, 15, 162].

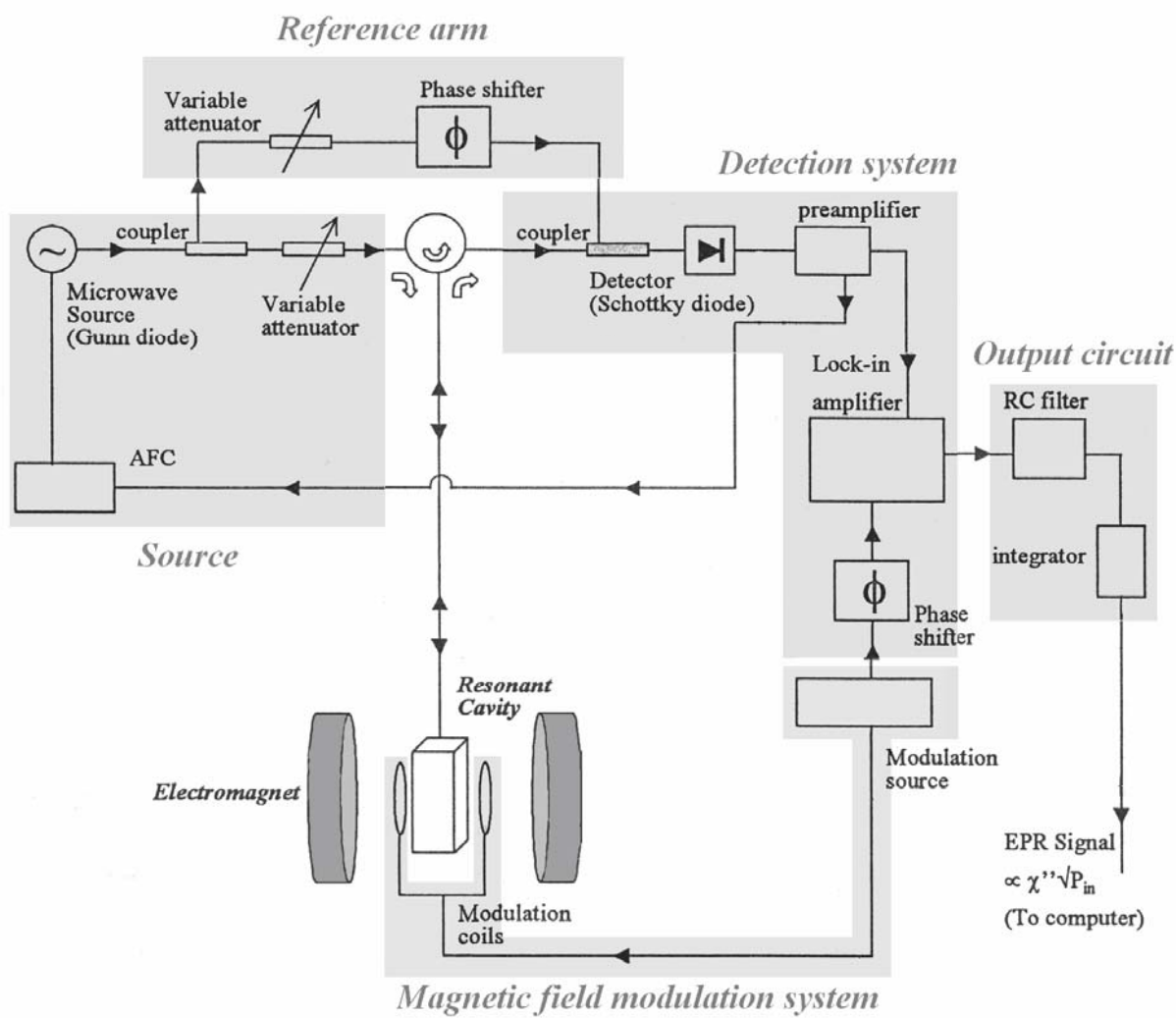

**Figure 3.1** Block scheme of the Bruker EMX EPR spectrometer [162].

### *3.1.1.1 Source*

The most commonly used source is the klystron, which is a vacuum tube that can produce microwaves centered in a small range of frequencies [11, 15]. At variance, in the EPR spectrometer we used, the microwave source is a Gunn diode. It consists in a GaAs junction which, as discovered by J. B. Gunn, if subjected to a d.c. voltage of about 10 volt, produces electromagnetic waves at microwave frequencies [163].

In order to obtain an undistorted EPR spectrum, the resonant frequency of the cavity containing the sample has to match that of the Gunn diode for all the duration of the acquisition, meanwhile the modulus of the external field **H** changes linearly in time. However, it can be shown that in correspondence to $H_r$ the frequency of the cavity undergoes a shift in connection with the change of the impedance of the cavity-plus-sample system [11, 12]. To compensate this effect, the EPR spectrometer is equipped with an automatic frequency control (AFC) system. In the EPR spectrometer we used, the AFC system introduces an amplitude modulation, at a frequency of 76.8 kHz, on the microwave radiation produced by the Gunn diode [162]. A second microwave signal, taken from the source and opportunely corrected in amplitude and phase, is superimposed to that reflected from the cavity so that, when the resonant frequency of the cavity-plus-sample system matches that of the Gunn diode, the total signal results to be not modulated at 76.8 kHz. Inversely, when the frequency of the cavity drifts off that of the Gunn diode, an error signal is produced and, by imposing a variation of the voltage applied to the Gunn diode, the matching is automatically reestablished.

The microwave power emitted by the Gunn diode is modified before reaching the cavity by a *variable attenuator* (indicated in Figure 3.1). The power transmitted to the cavity is measured in dB of attenuation with respect to its maximum value, given by

$$\text{Att.(dB)} = -10 \log_{10}\left(\frac{\text{power incident on the cavity}}{\text{power produced by the source}}\right) \quad (3.1)$$

For the Gunn diode of the EPR spectrometer we used the maximum obtainable microwave power is of 200 mW [162].

### *3.1.1.2 Resonant cavity*

The heart of an EPR spectrometer is the resonant cavity. It is constituted by a metal box characterized by high conductibility side walls and it is employed to store the microwave energy. It is placed at the center of the gap between the poles of the electromagnet and, in the spectrometer we used, have a rectangular shape [see Figure 3.2 (a)]. Each normal mode of the cavity is characterized by specific distributions of electric and magnetic fields lines. Those pertaining to the normal mode we considered in our experiments, which is named $TE_{102}$, are represented in Figures 3.2 (b) and (c), respectively [11, 15]. For the resonant cavity we used, a

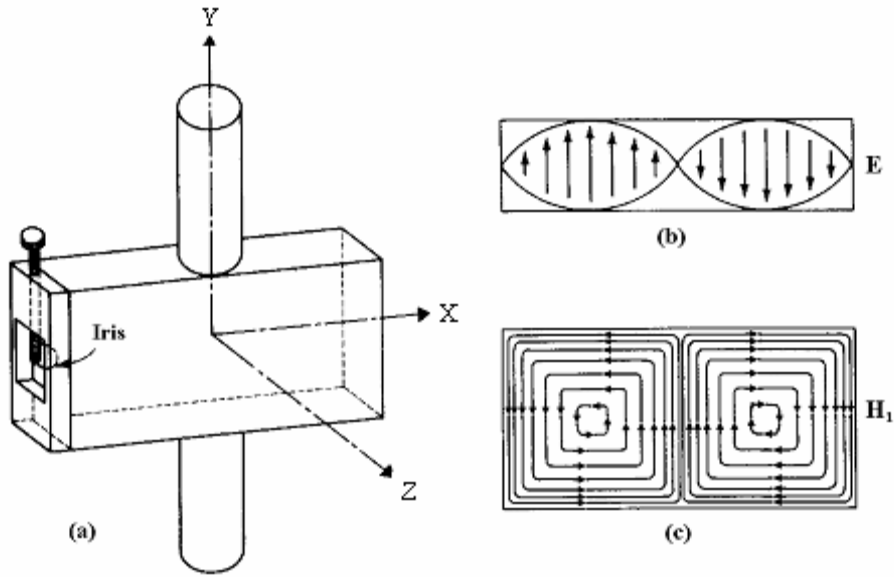

**Figure 3.2** (a) Resonant cavity of rectangular shape. The arrow indicates the iris, which connects the resonant cavity to the wave guide. Electric field lines in the XZ plan (b) and magnetic field lines in the XY plan (c) for the TE$_{102}$ mode. Adapted from Ref. 15.

microwave frequency of about 9.8 GHz corresponds to this mode, which falls into the range of frequencies known as EPR X-band [11]. The resonant cavity is coupled to the waveguide via a hole (iris), also indicated in Figure 3.2 (a). The size of the iris controls the amount of microwave power which enters the cavity. The iris accomplishes this task by carefully matching or transforming the impedances of the cavity and the waveguide. There is a screw in front of the iris which allows one to adjust this matching [11, 15].

The efficiency of a resonant cavity in accumulating microwave energy is measured by the quality factor defined as

$$Q_u = \frac{2\pi \ (\text{energy stored})}{(\text{energy dissipated per cicle})} \quad (3.2)$$

where the energy dissipated per cycle is the amount of energy lost during one microwave period. Energy can be lost in the side walls of the cavity because the microwaves generate electrical currents in them, which in turn generates heat. In order to limit this effect the side walls of the cavity are usually covered by a double layer of silver and gold. To take into account the energy absorbed by the sample due to dielectric losses and that dissipated in correspondence to the iris, it is possible to define two other contributions to the overall quality factor of the cavity as follows

$$Q_\varepsilon = \frac{2\pi \ (\text{energy stored})}{(\text{dielectric loss per cicle})} \quad (3.3)$$

$$Q_r = \frac{2\pi \ (\text{energy stored})}{(\text{iris loss per cicle})} \qquad (3.4)$$

and the total quality factor $Q_{tot}$ is given by

$$\frac{1}{Q_{tot}} = \frac{1}{Q_u} + \frac{1}{Q_e} + \frac{1}{Q_r} \qquad (3.5)$$

It is worth to note that, the higher is the quality factor of the cavity, the higher is the energy stored in the cavity and, consequently, the higher is the sensitivity of the EPR spectrometer [11, 15].

The sample, which in our case has the dimension of about 25 mm$^3$, is inserted into the resonant cavity from one of the cylindrical slots indicated in Figure 3.2 (a) and placed in the central position. This position, as indicated in Figures 3.2 (b) and (c), corresponds to the maximum amplitude of the magnetic field **H**$_1$ and to the minimum amplitude of the electric field **E**. In this way, the energy absorbed by the sample from the magnetic field **H**$_1$ is maximized, whereas the dielectric losses are minimized.

In the spectrometer we used the resonant cavity has been disposed in such a way that, referring to Figure 3.2, the static magnetic field **H** is directed along Z and the microwave field **H**$_1$ along Y.

### 3.1.1.3 Electromagnet

The purpose of the electromagnet shown in Figure 3.1 is to produce a magnetic field **H** as stable and homogenous as possible over the volume of the sample. The electromagnet employed in the EPR spectrometer we used can produce a maximum magnetic field amplitude of ~1 T, with an homogeneity of $10^{-3}$ mT on a volume $\Delta X \cdot \Delta Y \cdot \Delta Z = 5 \cdot 22 \cdot 10$ mm$^3$ [see Figure 3.2 (a)] located at the center of the gap between the poles of the electromagnet [162].

Absolute estimations of the amplitude of the magnetic field **H** were obtained by an Hall probe (located at a fixed position between the poles of the electromagnet) with a precision of $8 \times 10^{-2}$ mT [162]. As a consequence of this experimental uncertainty, the absolute position of a resonance in the spectrum is undetermined by the same amount. For this reason, to compare the line shapes of the same resonance in different spectra, in the present Thesis they will be horizontally shifted in order to superimpose a common EPR feature. In spite of the accuracy in determining absolute values of **H**, the estimation of the relative magnetic field distance of two features of an EPR spectrum is affected by a sensibly lower error, due to the presence of a system which imposes a low frequency modulation of the magnetic field **H** and guarantees a better stability. So, for distances of the features in the EPR spectrum of 1 mT, this error is ~$8 \times 10^{-4}$ mT [162].

### *3.1.1.4  Detector*

In the EPR spectrometer we used the key element of the detection system is constituted by a Schottky diode. The microwave signal reflected back from the cavity reaches the Schottky diode which converts it in a current signal, whose amplitude is proportional to the square root of the microwave power incident on the diode (linear region of work), provided that the incident microwave power is higher than ~1 mW. To assure that the detector operates at this level, the system we used employs a *reference arm* (see Figure 3.1) which supplies the detector with some extra microwave power. Some of the source power is tapped off into the reference arm, where a second attenuator controls the power level (and consequently the diode current) for optimal performance. There is also a phase shifter to assure that the reference arm microwaves are in phase with the signal reflected from the cavity when they combine at the detector diode.

### *3.1.1.5  Magnetic-field modulation system*

As anticipated above, in a typical EPR spectrometer a second magnetic field $\mathbf{H_m}$ ($|\mathbf{H_m}| \ll |\mathbf{H}|$), named modulation field, is superimposed to $\mathbf{H}$ [11, 15]. It is produced by a pair of Helmholtz coils placed on each side of the cavity along the axis of the static field $\mathbf{H}$ and its amplitude oscillates at a frequency that, in our EPR spectrometer, can be fixed at a value ranging from 5 kHz up to 100 kHz with step of 1 kHz. The effects of the modulation field is to induce a modulation of the microwave signal reflected from the cavity, which is converted by the Schottky diode in a periodic electric current F(t). During the modulation cycles the total amplitude of the magnetic field acting on the sample under study oscillates between the limit values $H-H_m/2$ and $H+H_m/2$ following a sinusoidal dependence. However, since in general the absorption profile of the paramagnetic system is not necessarily linear within these two magnetic field values, then the function F(t) cannot be described by a simple sinusoidal dependence. At variance, it can be expressed by a Fourier superposition of many sinusoidal components, oscillating at the fundamental modulation frequency $\omega_m$ and at a large number of harmonics with frequencies $n\omega_m$, with n = 2, 3, … [11]. The periodic current signal produced by the Schottky diode is directed to the subsequent stage of the EPR spectrometer, which consists in a lock-in amplifier. This device amplifies the signal and permits to select, by a band-pass filter, the harmonic and the phase of the component of F(t) to be acquired. The main advantage in using the lock-in amplifier is connected with its ability, by the band-pass filter, to cut off many noise components, so significantly enhancing the signal-to-noise ratio of the EPR spectra [11, 15].

In conclusion, the quantity measured by the lock-in amplifier is the amplitude of the component of the current signal produced by the Schottky diode which oscillates at the selected frequency and with the opportune phase. In the EPR spectrometer we used, the first- and the second-harmonic components can be acquired, whereas the phase shift can be fixed from 0° up to 360° with step of 1°.

### *3.1.1.6 Output circuit*

The final stage of our EPR spectrometer consists in two components: an RC filter and an integrator. The first one is a low-pass filter with a response time $\tau_{RC}$, which can be fixed by the operator, and it is used to cut off all the noise components with frequencies higher than $1/\tau_{RC}$. As a consequence of this filtering process the signal-to-noise of the EPR spectrum is enhanced by a factor $\sqrt{\tau_{RC}}$ [15]. The disadvantage of the use of the RC filter is connected with its finite response time which limits the maximum magnetic field sweep rate of the EPR spectrum. As a general rule, in order to avoid distortion of the spectra, we have combined the acquisition time and the $\tau_{RC}$ so that $t_{pp} > 10\,\tau_{RC}$, where $t_{pp}$ is the time occurring to scan the narrowest structure of interest in the spectrum.

The second component of the final stage integrates the EPR signal for a time duration $T_{conv}$, fixed by the operator. Finally, the digitalized EPR signal is shown on the screen of the computer controlling the spectrometer.

### 3.1.2 Different types of acquisition conditions

In order to present the two different acquisition schemes of the EPR spectra we considered in this Thesis, we introduce the concept of relaxation times of a paramagnetic center embedded in a solid matrix. A simple way to accomplish this objective is through the introduction of the *phenomenological Bloch equations* [164]. These equations, applicable to systems with $S=\frac{1}{2}$, describe the motion of the magnetic moments of the system subjected to external magnetic fields and permit to take into account the interactions among the magnetic moments and the interactions between them and the lattice [164]. These equations, including the modulation magnetic field, are [13, 164-166]

$$\frac{dM_X(t)}{dt} = \gamma \left[\mathbf{H_T} \wedge \mathbf{M}(t)\right]_X - \frac{M_X(t)}{T_2} \qquad (3.6)$$

$$\frac{dM_Y(t)}{dt} = \gamma \left[\mathbf{H_T} \wedge \mathbf{M}(t)\right]_Y - \frac{M_Y(t)}{T_2} \qquad (3.7)$$

$$\frac{dM_Z(t)}{dt} = \gamma \left[\mathbf{H_T} \wedge \mathbf{M}(t)\right]_Z - \frac{M_Z(t) - M_0}{T_1} \qquad (3.8)$$

were γ is the giromagnetic ratio, $\mathbf{M}=(\text{volume})^{-1}\sum_i \boldsymbol{\mu}_i$ is the sum of the microscopic magnetic moments $\boldsymbol{\mu}_i$ for an unitary volume, $M_X(t)$, $M_Y(t)$ and $M_Z(t)$ are the three components of $\mathbf{M}(t)$ with respect to the laboratory frame of reference, and $\mathbf{H_T} = H_1 \cos(\omega t)\,\hat{\mathbf{y}} + [H(t) + \frac{H_m}{2}\cos(\omega_m t)]\,\hat{\mathbf{z}}$, where $\hat{\mathbf{y}}$ and $\hat{\mathbf{z}}$ are the unitary vectors directed along the directions Y and Z, respectively, whereas ω is the microwave frequency. $M_0$ represents the modulus of the magnetization of the paramagnetic system obtained in stationary conditions, supposed to be directed along the direction of **H**, as $|\mathbf{H_1}| \ll |\mathbf{H}|$. The constants $T_2$ and $T_1$ define the *transversal* and *longitudinal* relaxation times and characterize the interaction among the paramagnetic centers and that of the paramagnetic centers with the lattice, respectively [164].

It is worth to note that, in Eqs. (3.6)-(3.8) an implicit time dependence is contained through the magnetic field $\mathbf{H_T}$. In fact, during the acquisition of an EPR spectrum the amplitude of **H** is varied linearly in time across the resonant value $H_r$, while the amplitudes of $\mathbf{H_m}$ and $\mathbf{H_1}$ oscillate with sinusoidal dependences. In definitive, the solution of the set of equations (3.6)-(3.8) strictly depends on the specific time dependence of these fields. Since an analytical solution of the Eqs. (3.6)-(3.8) for a general case cannot be obtained, some specific cases have to be considered. The most important of them occurs when

$$\frac{H_1}{d|\mathbf{H}+\mathbf{H_m}|/dt} \gg \sqrt{T_1 T_2} \tag{3.9}$$

which is known as *slow-passage* condition [167]. Roughly speaking, this condition states that the rates of change in time of **H** and $\mathbf{H_m}$ are slow with respect to the main relaxation rates of the paramagnetic centers [166, 167]. As a consequence, in slow-passage condition, $\mathbf{M}(t)$ assumes the stationary value pertaining to the total external magnetic field $\mathbf{H_T}$, at each time instant. At variance, in the opposite case in which

$$\frac{H_1}{d|\mathbf{H}+\mathbf{H_m}|/dt} \ll \sqrt{T_1 T_2} \tag{3.10}$$

the acquisition is referred to as obtained in *rapid-passage* condition [166, 167].

It is worth to note that, while an analytic solution of the Eqs. (3.6)-(3.8) can be easily obtained in *slow-passage* conditions, it is not the same for the *rapid-passage* case. Consequently, the latter case is usually studied by empirical experimental characterizations of the detected EPR signals or, more rigorously, by looking at the solutions of the Eqs. (3.6)-(3.8) for the specific case of interest by a computer numerical analysis. In the present Thesis the former approach has been used.

In the following paragraphs we describe the main properties of the EPR signals acquired under *slow-* and *rapid-passage* conditions. Furthermore, the main difficulties connected with the acquisition and the interpretation of the EPR spectra acquired in *rapid-passage* conditions are discussed.

### 3.1.2.1 EPR *measurements in slow-passage conditions*

Starting from the Eqs. (3.6)-(3.8), assuming the *slow-passage* condition valid [Eq. (3.10)] and neglecting the modulation magnetic field, for simplicity, by straightforward calculations it is possible to obtain that the energy for unit volume, f(H-$H_r$), absorbed by the paramagnetic system is given by [13]

$$f(H-H_r) = \frac{\omega}{2\pi} \int_0^{2\pi} H_1 \frac{dM}{dt} dt \propto \frac{NH_1^2}{1 + T_2^2 \gamma^2 (H - H_r)^2 + \gamma^2 H_1^2 T_1 T_2} \quad (3.11)$$

were N is the number of paramagnetic centers for unit volume. According to Eq. (3.11), for $\gamma^2 H_1^2 T_1 T_2 \ll 1$, the absorption is proportional to the square of the microwave field amplitude $H_1$. When this low-power condition is satisfied the EPR signal is said to be recorded in unsaturating conditions and a Lorentzian line shape is observed, which has the maximum in correspondence to H=$H_r$, full width at half maximum equal to $1/(\gamma T_2)$ and its area is proportional to the number of absorbing paramagnetic centers. At variance, when the condition $\gamma^2 H_1^2 T_1 T_2 \ll 1$ is not fulfilled, the line shape deviates from a Lorentzian and the absorption f(H-$H_r$) is no more proportional to $H_1^2$.

In the case of an inhomogeneous paramagnetic system, as for powdered crystals or for amorphous solids (see Paragraph 1.1.2.1), a distribution of the resonance fields $H_r$ occurs. Then the overall absorption of the paramagnetic centers is obtained by multiplying the single center line f(H-$H_r$), given by Eq. (3.11), by the inhomogeneous distribution function D($H_r$) and by integrating over the entire spectrum, as described in Paragraph 1.1.2.1. Typical dependences of the intensity of the EPR signal on increasing microwave power (which is proportional to $H_1^2$) will be reported in Figure 5.3, as obtained experimentally for the E'$_\gamma$ centers in a-SiO$_2$.

In conformity with previous experimental investigations [1, 2, 38, 77, 79, 120, 166], in the present Thesis the *slow-passage* EPR signal has been measured by revealing the component of the signal reflected from the resonant cavity oscillating at the same frequency and in phase with the modulation magnetic field. This detection scheme has been accomplished by opportune setting of the lock-in amplifier of the EPR spectrometer. Hereafter in the present Thesis, we will refer to the measurements performed in the conditions described above as *FH-EPR*. Furthermore, in order to obtain undistorted lines and to make quantitative estimation of the concentration of paramagnetic centers, we have acquired the FH-EPR spectra in the linear region of signals growth with microwave power, which corresponds to the range of microwave magnetic field values for which the condition $\gamma^2 H_1^2 T_1 T_2 \ll 1$ is satisfied.

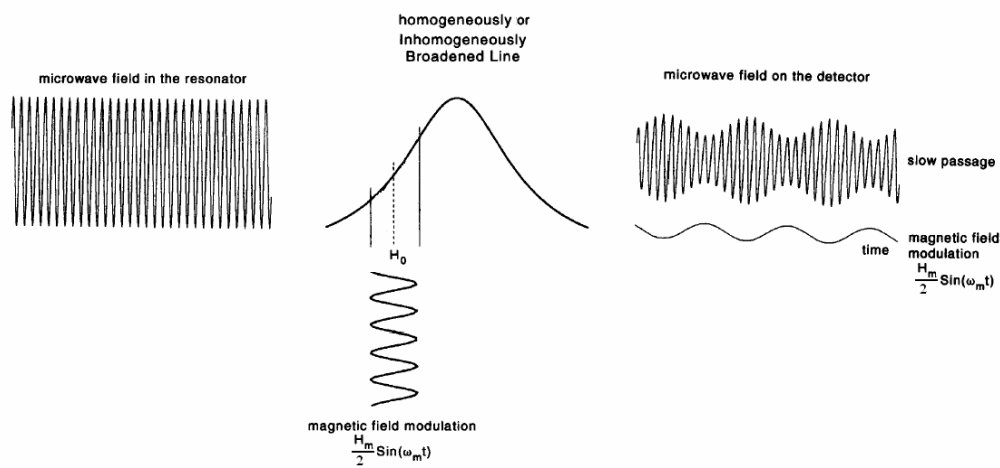

**Figure 3.3** Effects of the modulation magnetic field on the properties of the microwave signal reflected from the resonant cavity for the case of measurements performed in slow-passage conditions. Adapted from Ref. [168].

The effect of the modulation magnetic field, neglected in Eq. (3.11), on the properties of the microwave signal reflected from the resonant cavity is schematized in Figure 3.3. As shown, as a result of the magnetic field modulation, the position in the spectrum in correspondence to which the resonance condition is satisfied oscillates between the values $H_0-H_m/2$ and $H_0+H_m/2$, where $H_0$ is the value of the static magnetic field. As a consequence, the microwave signal reflected from the cavity, which is proportional to the value of the absorption in correspondence to the instantaneous value of the static-plus-modulation magnetic fields, becomes modulated. In particular, the amplitude of the microwave field oscillates with the same frequency and in phase with the modulation magnetic field, provided that the portion of the absorption profile spanned by the modulation magnetic field is small enough to be approximated as linear. Once this signal reaches the detector, it is converted in a current signal oscillating at the fundamental modulation frequency and with amplitude proportional to the difference between the values of the absorption in correspondence to $H_0+H_m/2$ and $H_0-H_m/2$. Finally, the amplitude of this oscillating current is measured by the lock-in amplifier and represents the detected EPR signal. From the above discussion, it follows that the FH-EPR signal reproduces the derivative of the absorption line, provided that the amplitude of the modulation magnetic field is much less than the width of the absorption profile [11].

In a similar way it can be shown that, for acquisitions in slow-passage conditions, the EPR signal oscillating at a double frequency and in phase relative to the modulation magnetic field is proportional to the second derivative of the absorption profile. This signal is revealed when the modulation amplitude is large enough that deviations from the linear dependence of the absorption profile are detectable between $H_0-H_m/2$ and $H_0+H_m/2$ [11].

We stress that, as it follows from the above discussion, when slow-passage conditions are satisfied, no EPR signals $\pi/2$-out-of-phase relative to the modulation magnetic field are detected, no matter the frequency component ($\omega_m$, $2\omega_m$, …) selected by the lock-in amplifier [11].

### *3.1.2.2 EPR measurements in rapid-passage conditions*

In many experimental works [2, 79, 81, 120, 166, 168-172] focused on paramagnetic centers with long relaxation times, it has been reported that, under high microwave power, an intense signal is detected when the lock-in is fixed to acquire the second-harmonic π/2-out-of-phase EPR signal. In particular, this EPR signal has been connected to the failure of the slow-passage condition of Eq. (3.9), due to the very long relaxation times of the paramagnetic centers, and to the consequent growth of the signal originating in rapid-passage conditions. In particular these effects should be induced by the rapid magnetic field sweep connected to the high modulation frequencies (typically 100 kHz) used in the experiments. Recently, this conjecture has been substantiated by looking at the solutions of the Bloch equations (3.6)-(3.8) by computer numerical methods [166]. This latter study has been focused on the EPR signal of the E'$_\gamma$ center in a-SiO$_2$ (see Paragraph 2.3.1), which has been taken as a reference system for the class of paramagnetic centers with long relaxation times.

In some experiments involved in the present Thesis, to enhance the detection sensitivity in revealing some EPR signals of interest, the rapid-passage EPR signal has been acquired which, in conformity with previous investigations, has been detected by revealing the second-harmonic π/2-out-of-phase EPR signal [2, 79, 81, 120, 166, 171]. Hereafter in the present Thesis, we will refer to these measurements as *SH-EPR*.

In all the experimental investigations in which the SH-EPR spectra of inhomogenously broadened EPR lines have been reported, it has been noted that they resemble the absorption profile [2, 79, 81, 120, 168-172]. Although a general analytical treatment explaining this effect has not been developed, it can be qualitatively understood in a rather simple way by considering the scheme described in Figure 3.4 [168], valid for inhomogenously broadened lines and for high microwave power. This scheme is based on the observation that during a magnetic field modulation cycle, the modulus of the sweep rate oscillates from 0 up to $\omega_m H_m/2$. As a consequence, two different types of passage effect could be induced in correspondence to the middle and to the turning points of the modulation cycles. In particular, near to the turning points of each modulation cycle, the modulation-induced field sweep rate tends to zero and, consequently, one should observe the slow-passage signal. At variance, in correspondence to the middle of each modulation cycle, the sweep rate is maximum and, if the modulation frequency is high enough, one should observe the rapid-passage signal. Furthermore, it is assumed that, since the resonance is dominated by inhomogenous broadening, the absorption profiles "as seen" in slow- and in rapid-passage conditions differ in intensity but not in shape (see Figure 3.4), as the details of the single center (homogenous) line shape pertaining to the two types of acquisitions become irrelevant. During a half modulation cycle the amplitude of the microwave signal reflected back from the resonant cavity will be proportional: to the slow-passage absorption profile at the low-field modulation turning point, to the rapid-passage absorption profile near $H_0$, to the slow-passage absorption profile at the high-field modulation turning point, and so on (see Figure 3.4). As a consequence, there will be a microwave signal reflected from the cavity that oscillates at a double frequency and $\pi/2$-out-of-phase with respect to the modulation magnetic field. Furthermore, the SH-EPR signal measured in these experiments is proportional to the difference between the rapid-passage and the slow-passage absorption profiles, and reproduces the *true* absorption profile, provided that inhomogenous broadening effects dominate.

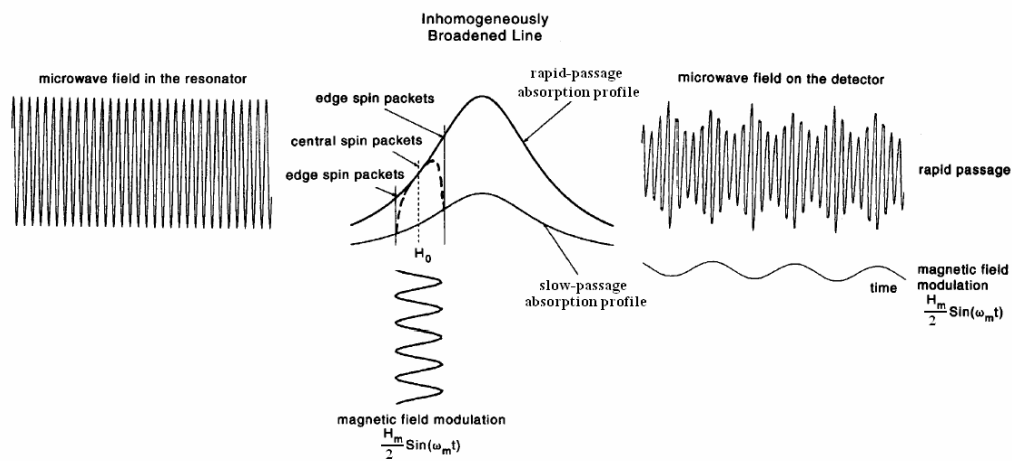

**Figure 3.4** Effects of the modulation magnetic field on the properties of the microwave signal reflected from the resonant cavity for the case of measurements performed in rapid-passage conditions. Adapted from Ref. [168].

As a final remark, it is worth to note that, at variance to the case of the FH-EPR signal, the SH-EPR signal growth with microwave power is neither linear nor it can be characterized by a simple power law of general validity [167].

In the present Thesis, SH-EPR measurements have been performed when a higher detection sensitivity was required. In the case when a resonance line of interest was detectable both with FH-EPR and with SH-EPR measurements, we have optimized the parameters of SH-EPR acquisition in order to maximize the sensitivity with minimum discrepancies between SH-EPR line shape and the integral of the corresponding FH-EPR one. At variance, when no FH-EPR signal was detectable, the optimal parameters of SH-EPR acquisition were chosen in analogy with those established for other fully characterized lines with similar spectral characteristics.

### 3.1.3  Spin concentration

In the present Thesis the concentrations of paramagnetic centers have been estimated by comparing the double integral (area) of the unsaturated FH-EPR spectra with that of a reference sample, consisting in a γ-ray irradiated a-SiO$_2$ with known number of E'$_\gamma$ centers (Paragraph 2.3.1). The spins concentration of the latter has been determined by the spin echo technique [173] with absolute accuracy of ~20 %. The concentration/area ratio has been estimated for reference. In particular, it was found that the area normalized by $\sqrt{P_{in}}$ (mW)$^{1/2}$, $H_m$ (mT), $T_{conv}$ (ms), sample weight (g), and receiver gain is related to the concentration of paramagnetic centers by the formula

$$\text{concentration} = (\text{normalized area}) \times (3 \pm 0.3) \times 10^{16} \text{ spins/cm}^3 \qquad (3.12)$$

The latter relation has been used for the determination of the other paramagnetic species concentration starting from their normalized area. From the reproducibility of the EPR measurements, a typical error of ± 10 % of the registered spectrum intensity has been estimated. As a consequence, we attribute this error to all the EPR intensity measurements reported in the present Thesis.

In one of the experiments reported in Chapter 5, we have estimated the concentration of E'$_\gamma$ point defects from SH-EPR measurements, since, due to the low concentration of defects, the FH-EPR signal was not detectable. These estimations were obtained by multiplying the integral of the SH-EPR spectrum by an empirical factor. This factor was obtained by the ratio between the concentration of E'$_\gamma$ centers, estimated from FH-EPR measurements, and the integral of the SH-EPR spectrum, as obtained in the same sample successively γ-ray irradiated at a higher dose so that both the FH- and SH-EPR signals were detectable.

## 3.2  The optical absorption spectrometer

The OA measurements were carried out in the UV range with a JASCO V560 spectrophotometer working in the NIR-VIS-UV. Its simplified block scheme is reported in Figure 3.5. The light source for UV range measurements is a deuterium discharge tube operating in the wavelength range 190-400 nm (3.1-6.5 eV). The wavelength of the impinging light is selected by a double monochromator employing a Czerny-Turner mount plane grating with 600 lines/mm. The light is then split into two paths, one passing through the sample and the other used as a reference. At the end of these paths the light is detected by a photomultiplier tube. The spectrophotometer measures the sample absorbance (Abs) from which the absorption coefficient can be determined, once the sample thickness is known [see Eq. (1.46)].

All the measurements reported in this Thesis were carried out in the range 200-400 nm (3.1-6.2 eV) by steps of 0.5 nm with a detection spectral bandwidth 2 nm, scan speed 40 nm/min and by averaging data value for ~1 sec at each data point.

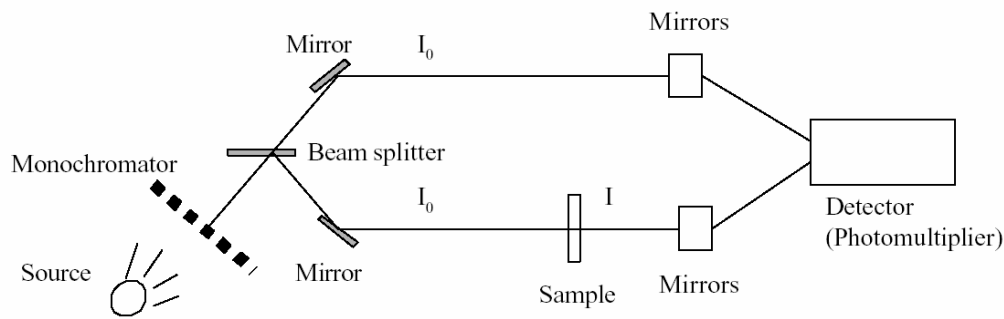

**Figure 3.5** Block scheme of the double-beam spectrophotometer employed for OA measurements.

# Chapter 4

# *Materials and treatments*

This chapter is devoted to describe the materials employed in the present Thesis and the treatments to which the samples were subjected. The principal details are reported to individuate those parameters relevant for the discussion of the experimental results.

## 4.1 Materials

The materials used in the present Thesis are of commercial origin. This choice is related to the rigidity of the industrial manufacturing and the tested high reproducibility of the material characteristics. One material, KUVI, is synthesized by the vapour-axial-deposition (VAD) technique [174, 175], whereas the other materials can be classified on the basis of the convention introduced in Refs. 176 and 177. In particular, four typologies of silica are distinguished on the basis of the main manufacturing features and the concentration of the most diffused impurity, the OH group, believed to be bonded to a Si atom, O≡Si-OH, normally bonded to three other oxygen atoms:

**Type I (natural dry)**: fusion of quartz powder by electric arc in a crucible in vacuum or inert gas atmosphere at low pressure. [OH] < 30 ppm[(1)]; other impurities, usually less than 10 ppm.

**Type II (natural wet)**: hydrogen/oxygen flame fusion of quartz powder. [OH]: 150 ÷ 400 ppm; other impurities less than the starting material because some of them are volatilized in the flame.

**Type III (Synthetic wet)**: hydrolysis of pure silicon compounds, usually $SiCl_4$, injected in gas-phase into a hydrogen/oxygen flame. Actually, the process is an oxidation, since the compound is transformed in the flame into fused drops of $SiO_2$. [OH] >> 100 ppm; other impurities content negligible since the starting material contains much less impurities than the natural quartz.

---

[1] The abbreviation ppm indicates parts per million by weight of $SiO_2$ groups. From this definition it follows that 1 ppm $\cong 7.7 \times 10^{16}$ cm$^{-3}$, valid for OH groups.

**Type IV (Synthetic dry)**: Reaction of O$_2$ with SiCl$_4$ in water-free-plasma. [OH] < 1 ppm; other relevant impurity: [Cl] $\cong$ 100 ppm.

A full list of the materials here used is reported in Table 4.1 specifying the type, the OH content and the manufacturer. In parenthesis is reported the nickname adopted hereafter. We have verified the OH content by IR absorption measurements by detecting the band at 2720 nm attributed to the OH stretching vibrational mode of Si-OH, and we have found fair agreement with the nominal values. Also reported in Table 4.1 are the concentrations of [AlO$_4$]$^0$ centers estimated by room temperature FH-EPR measurements in the various materials after γ-ray irradiation at a common dose of 10$^4$ kGy. The details of these measurements are reported in Paragraph 6.1.3.

In the present Thesis, the materials KI, KUVI, QC and P453, which after a γ-ray irradiation dose of 10$^4$ kGy exhibit an [AlO$_4$]$^0$ centers concentration higher than 2x10$^{17}$ spins/cm$^3$, will be referred to as Al-containing materials, in contrast with the other materials which exhibit lower [AlO$_4$]$^0$ centers concentrations.

The samples, received in the shape of slabs having sizes 50 x 5 x 1 mm$^3$, were cut in pieces of sizes 5 x 5 x 1 mm$^3$. Unless otherwise specified, all the results are reported for the latter size. In the following, to simplify reference to the various materials, the notation adopted is (nickname)/(dose), where the dose is reported in kGy.

We finally report that all the samples, also after the cut procedure, present no EPR signal before irradiation. Morever, all the native optical activities were found to be quite reproducible in samples of the same manufacturing but coming from different stocks.

## 4.2  Treatments

In this section we report the irradiation procedures and the thermal treatments employed in the present Thesis. We note that for each material all the irradiations were carried out in different pieces so, unless otherwise specified, series of irradiated samples were obtained. The results reported for each irradiation type refer to these series of samples.

### 4.2.1 γ-ray irradiation

Our study of point defects induced by ionizing radiation was performed prevalently by exposing the samples to γ-rays in a $^{60}$Co source (the irradiation was carried out in the irradiator IGS-3 of the Department of Nuclear Engineering, University of Palermo). γ-rays emitted by $^{60}$Co have energies 1.17 MeV and 1.33 MeV (mean energy 1.25 MeV). In the irradiator the dose rate for SiO$_2$ was ~3 kGy/h. All the irradiations were carried out at room temperature, in normal atmosphere, and in the dose range from 0.5 kGy up to 10$^4$ kGy.

**Table 4.1 List of the a-SiO$_2$ materials considered in the present Thesis**

| Material name (nickname) (supplier) | Type | [OH] ppm by weight; (cm$^{-3}$) | [AlO$_4$]$^0$ $^{(1)}$ spins/cm$^3$ |
|---|---|---|---|
| Infrasil 301 (I301) $^{(a)}$ | Natural dry (I) | $\leq 8$; ($\leq 6.2 \times 10^{17}$) | N. D. |
| Pursil 453 (P453) $^{(b)}$ | Natural dry (I) | < 5; (< $3.9 \times 10^{17}$) | ~$2.6 \times 10^{17}$ |
| Purosil A (QPA) $^{(b)}$ | Natural dry (I) | ~15; (~$1.2 \times 10^{18}$) | ~$6.5 \times 10^{16}$ |
| Silica EQ 906 (EQ906) $^{(b)}$ | Natural dry (I) | ~20; (~$1.5 \times 10^{18}$) | ~$7 \times 10^{16}$ |
| Silica EQ 912 (EQ912) $^{(b)}$ | Natural dry (I) | ~15; (~$1.2 \times 10^{18}$) | ~$6.7 \times 10^{16}$ |
| QC $^{(c)}$ | Natural dry (I) | < 1; (< $7.7 \times 10^{16}$) | ~$5.4 \times 10^{17}$ |
| KI $^{(d)}$ | Natural dry (I) | < 0.2; (<$1.5 \times 10^{16}$) | ~$8.6 \times 10^{17}$ |
| Vitreosil (VTS) $^{(e)}$ | Natural wet (II) | ~150; (~$1.2 \times 10^{19}$) | < $3 \times 10^{16}$ |
| Herasil 1 (H1) $^{(a)}$ | Natural wet (II) | ~150; (~$1.2 \times 10^{19}$) | ~$8.3 \times 10^{16}$ |
| Herasil 3 (H3) $^{(a)}$ | Natural wet (II) | ~150; (~$1.2 \times 10^{19}$) | ~$3.7 \times 10^{16}$ |
| Homosil (HM) $^{(a)}$ | Natural wet (II) | ~150; (~$1.2 \times 10^{19}$) | ~$9.4 \times 10^{16}$ |
| Suprasil 1 (S1) $^{(a)}$ | Synthetic wet (III) | ~1000; (~$7.7 \times 10^{19}$) | N. D. |
| Suprasil 311 (S311) $^{(a)}$ | Synthetic wet (III) | ~200; (~$1.5 \times 10^{19}$) | N. D. |
| Corning Substrate (CSB) $^{(f)}$ | Synthetic wet (III) | 800÷1000; ($6.2 \div 7.7 \times 10^{19}$) | N. D. |
| Corning 7940 (CNG5F) $^{(f)}$ | Synthetic wet (III) | 800÷1000; ($6.2 \div 7.7 \times 10^{19}$) | N. D. |
| Suprasil 300 (S300) $^{(a)}$ | Synthetic dry (IV) | < 1; (< $7.7 \times 10^{16}$) | N. D. |
| Suprasil F300 (F300) $^{(a)}$ | Synthetic dry (IV) | < 0.3; (< $2.3 \times 10^{16}$) | N. D. |
| KUVI $^{(d)}$ | VAD | < 0.2; (~$1.5 \times 10^{16}$) | ~$3.8 \times 10^{17}$ |

The materials reported in the Table were supplied by (a) Heraeus [178], (b) Quartz & Silice [179], (c) Starna Ltd (Romford, England), (d) Almaz Optics [180], (e) TSL [181], and (f) Corning [182], as indicated.

$^{(1)}$ Estimated by room-temperature FH-EPR measurements in the samples γ-ray irradiated at a dose of $10^4$ kGy. The details of these measurements are reported in Paragraph 6.1.3. "N. D." indicates that no EPR signal has been detected.

### 4.2.2 β-ray irradiation

Irradiation with electrons of 2.5 MeV and current 20 mA was carried out in a Van de Graaff accelerator (irradiation was performed at LSI laboratory, Palaiseau, France) [183]. The dose range investigated was $1.2 \times 10^3$ kGy ÷ $5 \times 10^6$ kGy, at the dose rate 20 kGy/sec (SiO$_2$) and at T = 330 K. Moreover, only for this irradiation the sample size was 5 x 5 x 0.5 mm$^3$, where the thickness was 0.5 mm for irradiation homogeneity reasons and to avoid charge trapping and consequent current leakage.

## 4.2.3 Thermal treatments of the irradiated samples

After the irradiation some samples were thermally treated to investigate the stability of the radiation-induced centers and, more in general, to study the thermally activated processes involving the defects. These treatments were performed in an electric furnace equipped with an internal thermometric sensor and a feedback electronic circuit to stabilize the temperature within ± 3 K.

In each thermal cycle, after the furnace had reached the pre set temperature, the sample was placed inside the furnace. It is worth to note that immediately after the sample insertion the temperature variation observed was not relevant for the overall treatment (maximum variation of –5 K balanced within 2-3 min). After the end of the thermal cycle the sample was removed from the furnace and was returned to room temperature at normal atmosphere. After thermalization to room temperature the sample was monitored by EPR and OA measurements.

Two thermal treatment sequences were executed. They can be distinguished according to usual nomenclature [184]. The *isochronal treatment* is the sequence of thermal treatments $q(T_i,t)$ each with constant time interval, t, but at different temperatures, $T_i$. The *isothermal treatment* is the sequence of treatments $q(T,t_i)$ with fixed temperature, T, but for a sequence of time intervals, $t_i$.

# Chapter 5

# *E'$_\gamma$ center in a-SiO$_2$ : variants and structural modifications*

In the present chapter we present a characterization of the E'$_\gamma$ center in a-SiO$_2$ by EPR and OA measurements. In particular, we focus our attention on the E'$_\gamma$ EPR main line as well as on the strong hyperfine structure, and on its OA band peaked at ~5.8 eV. In Paragraph 5.1 the effects on these features induced by γ-ray irradiation up to a maximum dose of ~10$^4$ kGy are reported and discussed together with the changes induced by thermal treatments of the irradiated samples, whereas the results obtained extending this study to β-ray irradiation up to a maximum dose of ~5x10$^6$ kGy are reported in Paragraph 5.2.

The materials considered here are of commercial origin, obtained by synthesis techniques or from fused quartz both of dry ([OH]<20 ppm) and wet ([OH]>150 ppm) types [see Table 4.1 (Chapter 4)]. In particular, the present chapter is devoted to the a-SiO$_2$ materials which, after γ-ray irradiation at 10$^4$ kGy, exhibit [AlO$_4$]$^0$ centers concentrations lower than ~10$^{17}$ spins/cm$^3$, whereas a further characterization of the E'$_\gamma$ center in the materials with higher [AlO$_4$]$^0$ centers concentration will be reported in the successive two chapters.

## 5.1  Effects of γ-ray irradiation and thermal treatment

### 5.1.1 Main resonance line

A typical FH-EPR spectrum obtained in materials γ-ray irradiated at high doses is shown in Figure 5.1, as measured in a sample S1/5x10$^3$ using P$_{in}$ = 8 x 10$^{-4}$ mW, ν$_m$ = ω$_m$/2π = 100 kHz and H$_m$ = 0.01 mT. In this figure, the upper scale relative to the g values has been obtained by fixing the first positive peak position of the E'$_\gamma$ center at g$_1$=2.00180 [2]. We estimated the g$_2$ and g$_3$ principal values directly from the magnetic field positions indicated by vertical lines in Figure 5.1 (see Paragraphs 1.1.2.1 and 1.1.2.2). The obtained values, g$_2$= 2.00063 ± 0.00002 and g$_3$=2.00036 ± 0.00002, are in quite good agreement with those corresponding to the maxima of the statistical distributions, g$_2$ = 2.0006 and g$_3$ = 2.0003, obtained by Griscom using a computer simulation of the experimental EPR spectrum [Figure 2.9 (b)].

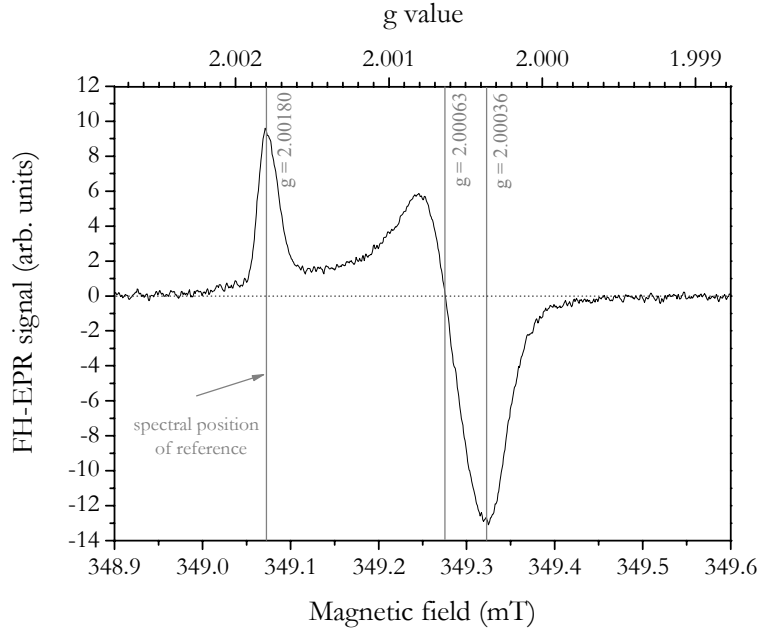

**Figure 5.1** EPR spectrum of the E'$_\gamma$ center as detected in the sample S1/5x10³. The scale of the g values is also reported, obtained by fixing the first positive peak position of the E'$_\gamma$ center at g=2.00180 [2].

In one of our previous works [100] we have pointed out that two EPR line shapes of the E'$_\gamma$ center, arising from two slightly different microscopic structures, can be distinguished in natural dry and wet and in synthetic dry a-SiO$_2$ materials. The main results found in that work are here summarized, for convenience. In Figure 5.2 (a) three normalized EPR spectra of the E'$_\gamma$ center are superimposed, measured in the natural dry I301 material γ-ray irradiated at three different doses: 0.5 kGy, 50 kGy and 5x10³ kGy. The curve of the E'$_\gamma$ centers concentration as a function of the irradiation dose is reported in the inset of the same figure. As shown, a modification of the E'$_\gamma$ center EPR line shape occurs on increasing the irradiation dose, even though no specific feature attributable to the occurrence of this line shape change can be recognized in the growth curve of the defects concentration. The difference $\Delta g_{1,2} = g_1 - g_2$ varies from 0.00124 at a dose of 0.5 kGy up to 0.00115 at 5x10³ kGy. The variation of $\Delta g_{1,3} = g_1 - g_3$ is less pronounced on increasing the dose, changing from 0.00147 at a dose of 0.5 kGy to 0.00142 at 5 x10³ kGy. The EPR line shapes reported for the sample irradiated at 0.5 kGy and 5x10³ kGy were observed for all the γ-ray doses lower than ~10 kGy (low-dose range) and higher then ~10³ kGy (high-dose range), respectively, whereas for doses between these two values an intermediate EPR line shape has been found. The low- and high-dose E'$_\gamma$ center FH-EPR signals were found to differ in the saturation properties with microwave power, as shown in Figure 5.3, presumably due to the different concentration of defects induced in the two dose ranges. For convenience, symbols L1 and L2 have been adopted for the low- and high-dose EPR line shapes, respectively. It has been found that the phenomenology reported for the I301 is a feature common to all the other natural dry and wet and synthetic dry materials. Indeed, in these materials it has been observed the line shape L1 after low γ-ray irradiation doses and the line shape L2 after high γ-ray doses, within an uncertainty in the principal g values less than 4x10⁻⁵.

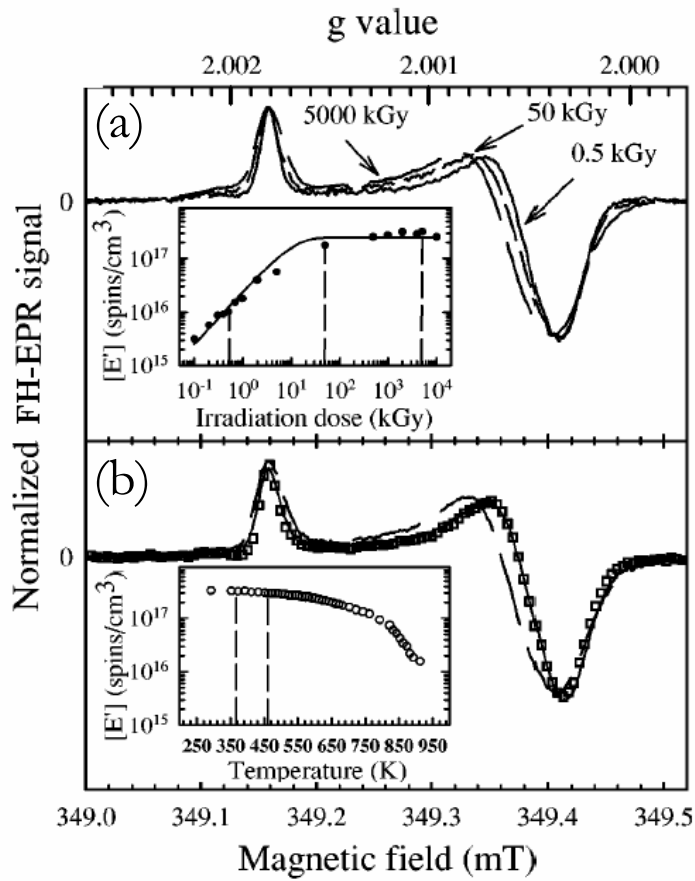

**Figure 5.2** FH-EPR spectra of the E'$_\gamma$ center normalized to peak-to-peak signal amplitude and horizontally shifted to overlap the first maximum. (a) I301 samples irradiated at γ-ray doses 0.5 kGy (solid line), 50 kGy (short-dashed line), and 5x10$^3$ kGy (long-dashed line); inset: E'$_\gamma$ center concentration as a function of the γ-ray dose (the solid line is a guide for the eye). (b) I301 sample after irradiation at 4x10$^3$ kGy (dashed line) and after isochronal thermal treatments up to T=460 K (solid line). The squares refer to the reference I301 irradiated at a dose of 0.5 kGy. Inset: E'$_\gamma$ center concentration as a function of the isochronal thermal treatment temperature. Adapted from Ref. 100.

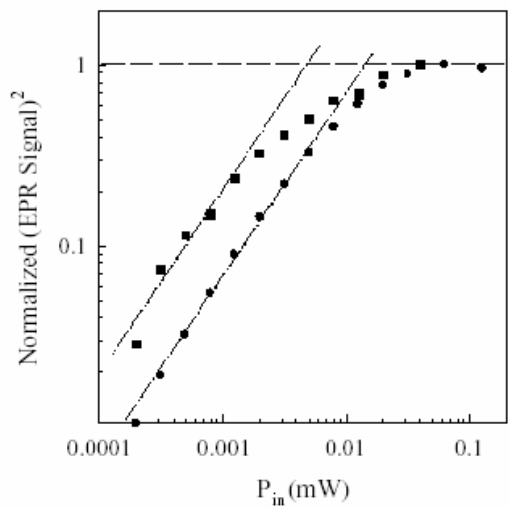

**Figure 5.3** Saturation with microwave power of the FH-EPR signal of the E'$_\gamma$ center in the samples I301/0.5 (squares) and I301/5x10$^3$ (circles). The linear dependence of the squared FH-EPR signal as a function of P$_{in}$ is evidenced by dash-dotted lines. After Ref. 185.

The effect of an isochronal thermal treatment on a sample I301/4x10$^3$ in which the E'$_\gamma$ center with the L2 line shape was preliminarily induced by γ-ray irradiation was studied in the same work [100]. In this experiment a gradual change from L2 toward L1 was found to occur [Figure 5.2 (b)], even if the concentration of E'$_\gamma$ centers remains almost constant [see inset of Figure 5.2 (b)]. This line shape change was found to occur in the temperature range from 370 K to 460 K, whereas for temperature T ≤ 370 K and T ≥ 460 K the line shapes L2 and L1, respectively, were observed. Results similar to those summarized in Figure 5.2 were also obtained in natural dry and wet and in synthetic dry a-SiO$_2$ materials [100].

In the synthetic wet materials, for γ-ray irradiation doses higher than ~200 kGy, the E'$_\gamma$ center EPR line shape of type L2 was observed. However, in this type of material, at variance with the other types, a complete study of the E'$_\gamma$ center EPR line shape in the low-dose limit was not possible, due to its high radiation resistance which prevents the observation of the FH-EPR signal of the E'$_\gamma$ center for γ-ray irradiation doses lower than ~200 kGy [100].

In the present Thesis, the above reported study has been extended. In order to investigate the E'$_\gamma$ center EPR line shape in synthetic wet materials for doses lower than ~200 kGy we have performed SH-EPR measurements, which allows to obtain an higher sensitivity in revealing the E'$_\gamma$ center. The SH-EPR spectra of the E'$_\gamma$ center were detected with $P_{in}$ = 5 mW, $\nu_m = \omega_m/2\pi$ = 100 kHz and $H_m$ = 0.01 mT. This experimental setting permits to distinguish between the two EPR line shapes L1 and L2, as shown in Figure 5.4 in which the normalized SH-EPR spectra of the E'$_\gamma$ center obtained for the same samples of Figure 5.2 (a) are superimposed. A similar study was performed in many natural dry and wet and synthetic dry samples and features similar to those obtained previously by using FH-EPR were derived [100].

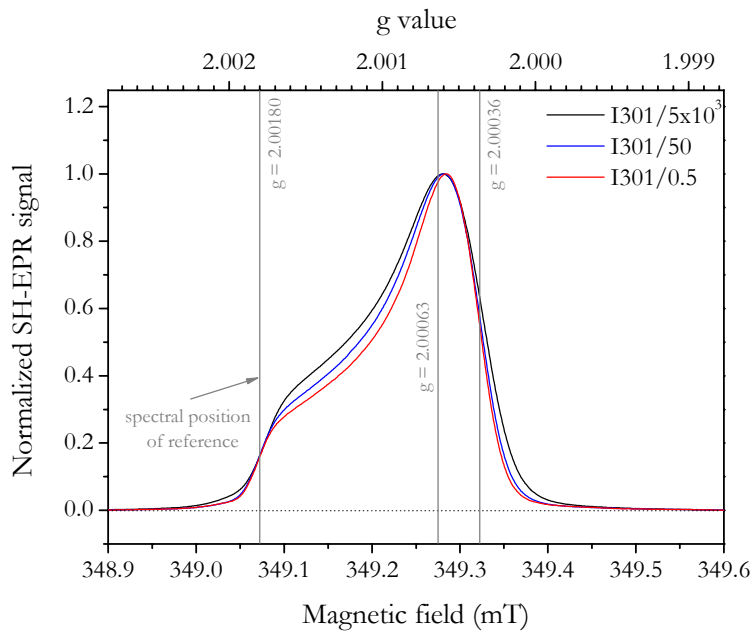

**Figure 5.4** SH-EPR spectra of the E'$_\gamma$ center in the material I301 irradiated at the γ-ray doses 0.5 kGy, 50 kGy and 5x10$^3$ kGy. The SH-EPR spectra have been normalized to the signal amplitude in correspondence to the maximum and have been horizontally shifted to overlap the first inflexion points.

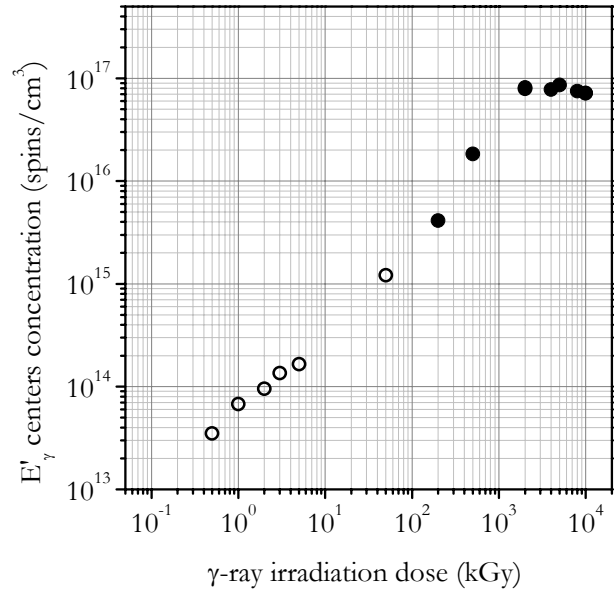

**Figure 5.5** Concentration of the E'$_\gamma$ centers in sample S1 as a function of the γ-ray irradiation dose as determined by FH-EPR (filled symbols) and SH-EPR (open symbols) measurements. The error of measurement is comparable with the size of the symbols.

Once this preliminary investigation was concluded, we have investigated the E'$_\gamma$ center EPR line shape in a S1 material γ-ray irradiated in the dose range from 0.5 kGy up to $10^4$ kGy, using both FH- and SH-EPR measurements. The growth of concentration with the irradiation dose obtained for this sample is shown in Figure 5.5. FH-EPR measurements (filled symbols) show that the concentration grows up to a dose of about $10^3$ kGy, where it becomes constant and unchanged up to the largest investigated doses. SH-EPR measurements (open symbols) were found to possess about two orders of magnitude higher sensitivity in revealing the E'$_\gamma$ center signal with respect to the FH-EPR ones, and they have permitted us to investigate the dose dependence of concentration of defects down to 0.5 kGy. To estimate the concentration of defects from SH-EPR measurements we have multiplied the integral of the SH-EPR spectrum by an empirical factor. This factor has been obtained in a sample S1/2x$10^3$ from the ratio between the concentration of E'$_\gamma$ centers, estimated from FH-EPR measurements, and the integral of the SH-EPR spectrum. Note that, in the range of doses from 2x$10^2$ kGy up to 1x$10^4$ kGy, in which both FH- and SH-EPR signals of the E'$_\gamma$ centers are detected, the concentrations of defects estimated with the two acquisition conditions coincide, within an experimental uncertainty of 10%.

Taking advantage of the high sensitivity of the SH-EPR measurements, we have investigated the E'$_\gamma$ center EPR line shape down to the lowest γ-ray doses considered. The normalized spectra recorded by SH-EPR in the S1 material after irradiation at the doses 2 kGy and $10^4$ kGy are superimposed in Figure 5.6. The two line shapes L1 and L2 are also shown in the figure, as a reference. From inspection of Figure 5.6, we conclude that the E'$_\gamma$ center EPR line shape in the synthetic wet material S1 is of type L2 in all the dose range investigated. It is worth to note that this experiment also permits to exclude that EPR line shape differences between L1 and L2 could be attributable to dipolar broadening effects. In fact, in the S1/2 sample the only detectable EPR signal is that of the E'$_\gamma$ center, whose concentration is ~$10^{14}$ spins/cm$^3$. At variance, in natural dry samples the line shape L1 is observed for concentration of E'$_\gamma$ center up to ~$10^{16}$ spins/cm$^3$, so excluding that the line shape L2, which has a little broader EPR line than L1, could result from dipolar broadening effects. In agreement with this conclusion, the expected single center dipolar line width contribution for a system of paramagnetic centers distributed homogenously in space and characterized by S=½, g $\cong$ 2 and having a concentration N $\cong$ $10^{14}$ spins/cm$^3$ is [11] $\Delta H_{dipolar} \cong 2.3$ g $\mu_B$ N $[S(S+1)]^{1/2}$ = $5 \times 10^{-7}$ mT. This quantity is too small to be responsible for the differences observed between L1 and L2 line shapes.

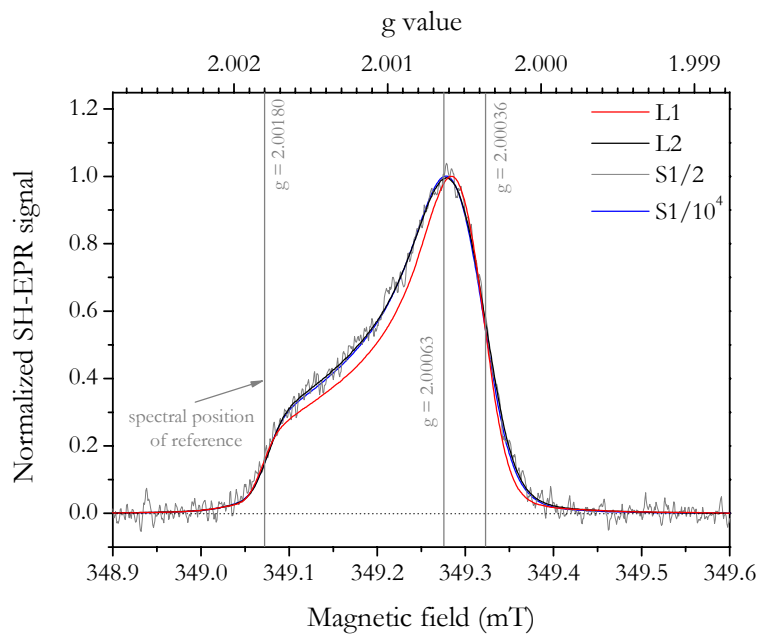

**Figure 5.6** SH-EPR spectra of the E'$_\gamma$ center in a sample of S1 after accumulation of γ-ray irradiation doses 2 kGy and 10⁴ kGy, compared to the reference line shapes L1 and L2. The SH-EPR spectra have been normalized to the signal amplitude in correspondence to the maximum and have been horizontally shifted to overlap the first inflexion points.

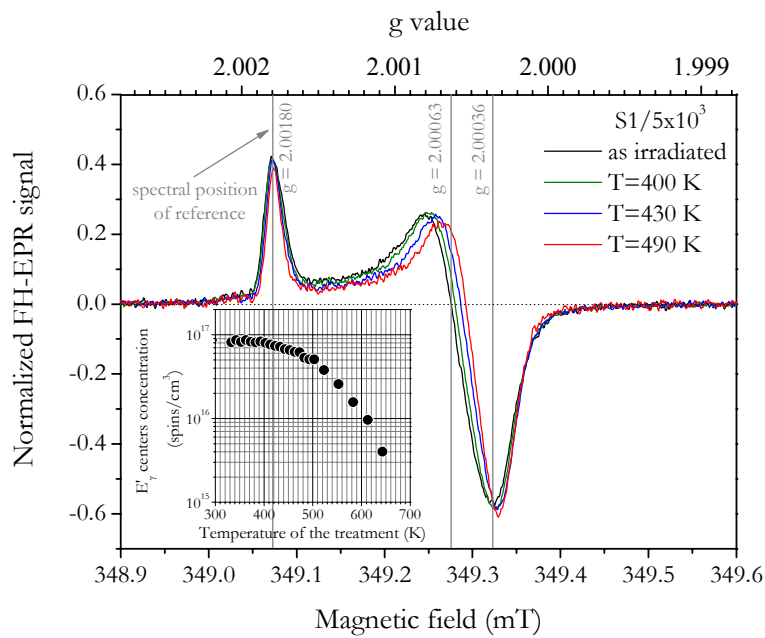

**Figure 5.7** FH-EPR spectrum of the E'$_\gamma$ center main line acquired in a sample of S1/5x10³ compared to those obtained for the same sample after isochronal thermal treatments up to three different temperatures. The FH-EPR spectra have been normalized to the peak-to-peak signal amplitude. Inset: E'$_\gamma$ center concentration as a function of the isochronal thermal treatment temperature.

Although the L1 line shape cannot be induced by irradiation in the S1 material, we have found that it is obtained by a successive thermal treatment of an irradiated sample. This experimental result is shown in Figure 5.7, in which the E'$_\gamma$ center EPR spectrum of a sample S1/5x10$^3$ is compared to those obtained after isochronal thermal treatments up to three different temperatures. As shown, thermal treatments of the sample induce a gradual EPR line shape conversion from L2 to L1. This variation has been found to occur in the temperature range from 370 K to 460 K, in quite good agreement with that at which the same process occurs in the other type of materials, even if for the synthetic wet a concentration reduction occurs, as evidenced in the inset.

Summarizing, upon γ-ray irradiation two types of E'$_\gamma$ centers can be induced in natural dry and wet and synthetic dry materials, distinguishable on the basis of their EPR line shape which is of type L1, in the low-doses range, and of type L2, in the high-doses range. At variance, in synthetic wet materials only the E'$_\gamma$ center with an EPR line shape L2 can be induced by γ-ray irradiation. Finally, in all the four types of materials the γ-ray induced L2 line shape converts to L1 by thermal treatments in the temperature range from 370 K to 460 K.

### 5.1.2 Strong hyperfine structure

As discussed in Paragraph 2.3.1, the strong hyperfine structure of the E'$_\gamma$ center consists in a pair of lines split by ~42 mT. The typical SH-EPR spectrum obtained in the materials γ-ray irradiated at high doses is shown in Figure 5.8, as measured in a sample S1/5x10$^3$ using P$_{in}$ = 50 mW, ν$_m$ = ω$_m$/2π = 100 kHz and H$_m$ = 0.3 mT. From inspection of the SH-EPR spectrum of Figure 5.8 we have estimated a full width at half maximum (FWHM) of ~4.4 mT and ~3.7 mT, for the low- and high-filed components of the E'$_\gamma$ center hyperfine doublet, respectively, and a splitting of ~41.8 mT, in quite good agreement with the values obtained in previous investigations [77, 79, 81].

The saturation properties with microwave power of the 42 mT doublet FH-EPR signal are reported in Figure 5.9 for a sample EQ906/10$^3$, as determined from the high-field component of the pair (shown in the inset). This curve pertains to the high-dose range and has been found to be characteristic of all the considered samples. At variance, a similar study in the low-dose range was prevented, as the concentration of defects has been found to be too low to detect the related hyperfine structure FH-EPR signal. From the comparison of Figure 5.9 with Figure 5.3 it is evident that the FH-EPR signal of the E'$_\gamma$ center hyperfine structure is less saturable with respect to that of the main resonance line, as already noted in previous investigations [77], probably due to the occurrence of relaxation mechanisms involving the nuclear magnetic moment of the $^{29}$Si [77].

We have verified the attribution of the 42 mT doublet to the strong hyperfine structure of the E'$_\gamma$ center by comparing the concentration of defects responsible for the doublet with that of the defects responsible for the main resonance line of the E'$_\gamma$ center. In fact, for a microscopic structure as that of the E'$_\gamma$ center, the value of this ratio is expected to be about equal to the natural abundance of $^{29}$Si nuclei (see Paragraph 1.1.3.1). This study has been performed considering natural, dry and wet, and synthetic, dry and wet, materials γ-ray irradiated in the dose range from 3x10² kGy up to 10⁴ kGy. The concentrations of defects were estimated by FH-EPR measurements and by using $P_{in}$ = 5 x 10⁻² mW, $\nu_m = \omega_m/2\pi$ = 100 kHz and $H_m$ = 0.8 mT, for the 42 mT doublet, and $P_{in}$ = 8 x 10⁻⁴ mW, $\nu_m = \omega_m/2\pi$ = 100 kHz and $H_m$ = 0.01 mT, for the main resonance line. These experimental parameters were adequately chosen in order to guarantee for all the investigated samples the acquisition of the FH-EPR signals in the region of linear growth with microwave power and, consequently, to avoid erroneous quantitative estimations due to the occurrence of saturation effects [77]. In Figure 5.10 the concentration of defects responsible for the 42 mT doublet is reported as a function of the concentration of defects responsible for the main resonance line of the E'$_\gamma$ center in many of the samples considered. As shown, the two concentrations are strictly correlated, and by a fit procedure we have estimated a concentration ratio of the hyperfine structure with respect to the E'$_\gamma$ center main line of 0.050 ± 0.005. This value is in quite good agreement with the ~4.7 % natural abundance of the $^{29}$Si nuclei and confirms the attribution of the 42 mT doublet to the strong hyperfine structure of the E'$_\gamma$ center.

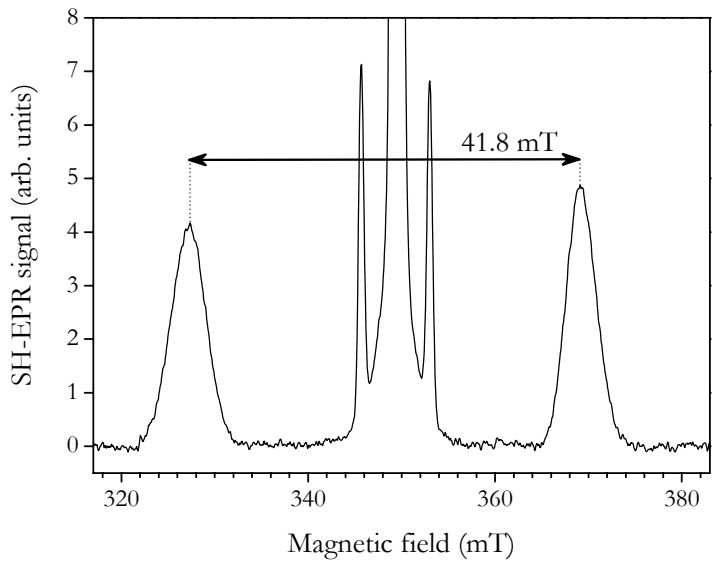

**Figure 5.8** SH-EPR spectrum of the ~42 mT hyperfine doublet of the E'$_\gamma$ center in a sample S1/5x10³.

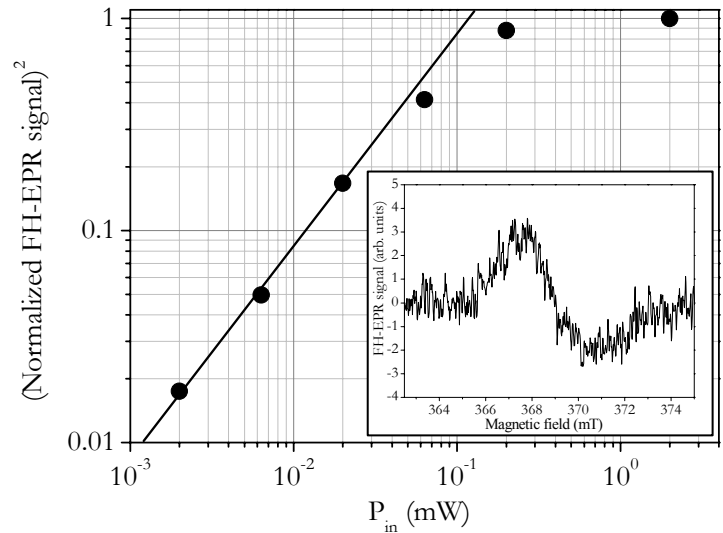

**Figure 5.9** Saturation with microwave power of the FH-EPR signal of the 42 mT doublet in the sample EQ906/10³, as determined by double integral of the high-field component of the doublet. The linear dependence of the squared FH-EPR signal as a function of $P_{in}$ is evidenced by the straight line. In the inset the FH-EPR spectrum obtained in the same sample for the high-field component of the 42 mT doublet is reported, measured using $P_{in}$ = 5 x 10⁻² mW, $\nu_m = \omega_m/2\pi$ = 100 kHz and $H_m$ = 0.8 mT.

As emerged in the discussion reported in the previous paragraph, two different E'$_\gamma$ centers can be distinguished in a-SiO$_2$ on the basis of their main EPR line shapes. Furthermore, it has been suggested that the differences in the two line shapes, L1 and L2, could have structural origin [100]. Under this hypothesis, it is expected that the strong hyperfine structures corresponding to E'$_\gamma$ centers characterized by L1 and L2 should also differ. In fact, as discussed in Paragraph 1.1.1.2, the isotropic and anisotropic hyperfine constants are strictly connected to the structural parameters of the defect. In order to verify this expectation, we have extended the study limited to the E'$_\gamma$ main resonance line previously discussed, carrying out a detailed investigation of the EPR line shape of the 42 mT hyperfine doublet. This study has been performed by using SH-EPR measurements, as they give the necessary sensitivity to perform line shape analysis. Furthermore, in the favourable cases in which the concentration of defects was high enough, the results obtained with SH-EPR measurements were also verified by the FH-EPR ones. In Figure 5.11 the SH-EPR hyperfine spectra obtained for the samples I301/0.5 and I301/5x10$^3$ are superimposed. As it can be observed, two distinguishable line shapes are found, as expected. These two characteristics EPR line shapes are observed for γ-ray irradiation doses lower than 10 kGy and higher than 10$^3$ kGy, whereas for doses between these two values the EPR line shape was found to be an intermediate one. From the spectra of Figure 5.11 we have estimated that in the dose range from 10 kGy to 10$^3$ kGy the FWHM of the low-field component of the doublet increases from ~3.7 mT to ~4.4 mT, whereas that of the high-field component increases from ~3.2 mT to ~3.7 mT, with an increase percentage in both cases of about 14 %. At variance, for all the doses investigated, no relevant change of the hyperfine splitting has been observed.

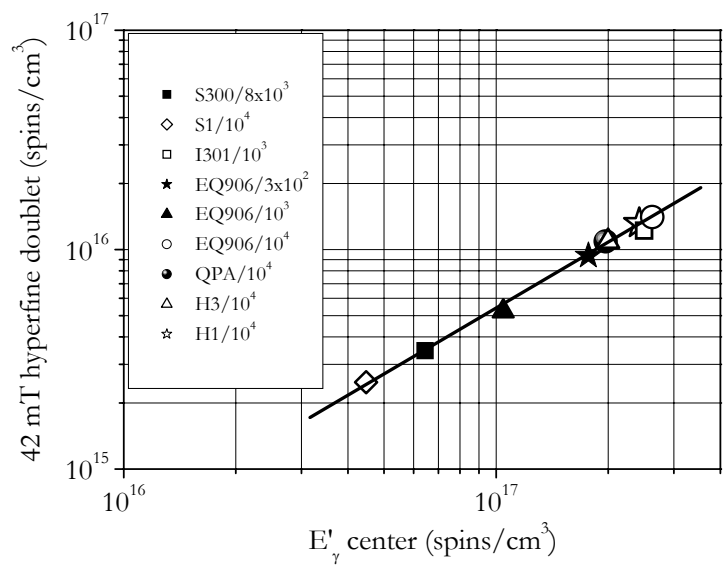

**Figure 5.10** Concentration of defects responsible for the 42 mT doublet as a function of the concentration of defects responsible for the main resonance line of the E'$_\gamma$ center in a-SiO$_2$ materials γ-ray irradiated in the dose range from 3x10$^2$ kGy to 10$^4$ kGy. The straight line, with slope 1, is superimposed to the data, for comparison.

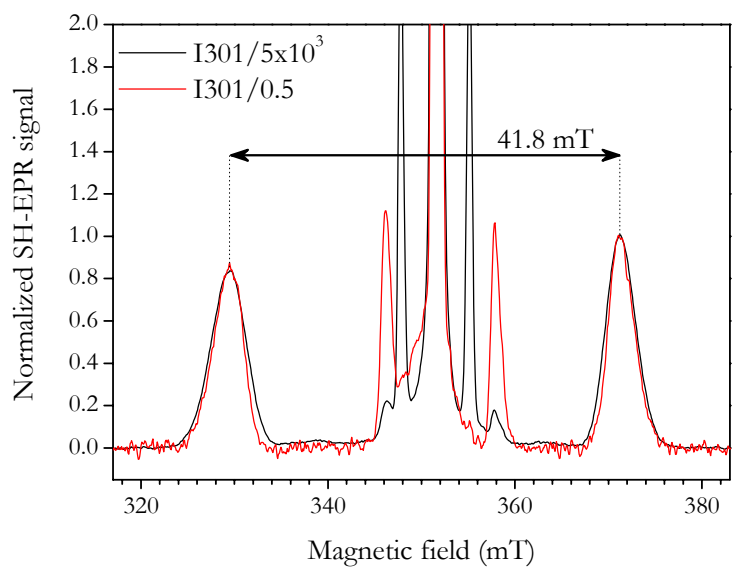

**Figure 5.11** SH-EPR spectra of the 42 mT doublet acquired in the I301/0.5 sample (red line) and 5x10$^3$ (black line). The SH-EPR spectra have been normalized to the signal amplitude in correspondence to the maximum of the high-field component of the 42 mT doublet.

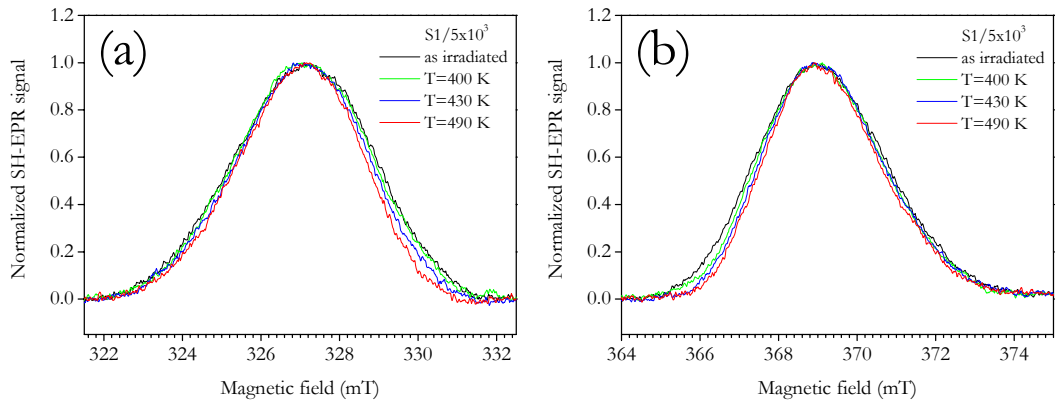

**Figure 5.12** SH-EPR spectra of the (a) low- and (b) high-field components of the 42 mT doublet acquired in a sample of S1 γ-ray irradiated at a dose of 5x10$^3$ kGy compared to the corresponding spectra obtained for the same material after isochronal thermal treatments at three different temperatures. The SH-EPR spectra have been normalized to the signal amplitude in correspondence to the maximum to the lines.

From the study over many irradiated and thermally treated materials, it emerged that a correspondence exists between the EPR line shape variation of the 42 doublet and that of the main line of the E'$_\gamma$ center. This property is further pointed out by Figure 5.12, in which the normalized SH-EPR spectra of the low- and high-field components of the 42 mT doublet, acquired in a sample of S1/5x10$^3$, are compared to those obtained for the same material after isochronal thermal treatments at three different temperatures. As reported, an inverse line shape variation occurs with respect to that observed by irradiation in the hyperfine structure. The analogous modifications observed in the main resonance line of the E'$_\gamma$ center in the same sample during the same thermal treatment experiment were discussed in the previous paragraph and are reported in Figure 5.7. The comparison of Figures 5.12 and 5.7 points out that the EPR line shapes changes observed both in the main line and in the hyperfine doublet are strictly correlated, suggesting that they could originate from the occurrence of a common relaxation process.

It is worth to note that a thermally induced reduction of the FWHM of the hyperfine lines similar to that reported in Figure 5.12 was also reported by Griscom [2, 79]. He showed that this feature is connected to a reduction of the width of the statistical distribution of the angle ϱ between the dangling bond orbital and the three basal oxygens [see Figure 1.1 (a)]. The relevance of this result consists in the fact that it associates a structural picture to the hyperfine line shape variation, even though the exact amount of the change occurring in the distribution of the angle ϱ is object of controversy [26].

### 5.1.3 Optical absorption band

As discussed in Paragraph 2.3.1 the E'$_\gamma$ center also possesses an OA band peaked at ~5.8 eV and, in fact, together to the EPR signal of the E'$_\gamma$ center, an OA band in the expected spectral position has been observed in all the irradiated samples we have considered. The typical OA spectrum in the UV energy region is shown in Figure 5.13, acquired for a sample S1/8x10$^3$. Also shown in Figure 5.13 are the four Gaussian profiles used to fit the spectrum with the aim to disentangle the contribution of the OA band associated to the E'$_\gamma$ center from those arising from other centers and contributing to the total absorption spectrum of the sample. In particular, to fit the experimental OA spectra discussed in the present paragraph we have used a band peaked at 4.67 eV with FWHM of 0.93 eV, a band peaked at 5.06 eV with FWHM = 0.46 eV, a band at 5.82 eV with FWHM = 0.78 eV, and a band at 6.34 eV with FWHM = 0.46 eV. The spectral features of the first three bands are in agreement with those attributed to the non-bridging oxygen hole center (NBOHC), the ODC(II) center and the E'$_\gamma$ center, respectively [50], whereas the latter band is introduced to take into account the absorption profile on the high energy shoulder of the E'$_\gamma$ centers OA band. It is worth to note that, we have verified that the introduction of this latter band in the fit of the OA spectra does not affect the results reported and discussed in the following.

By using the above described spectral decomposition and by fixing the spectral features of all the bands but those of the E'$_\gamma$ center OA band, we have studied many different samples, γ-ray irradiated in the dose range from 0.1 kGy up to 10$^4$ kGy. We have found that a shift towards lower energies of the peak position of the OA band associated to the E'$_\gamma$ centers occurs in concomitance to the change of the EPR line shape from L1 to L2. This effect is shown in Figure 5.14 (a), in which we report the OA spectra acquired in the sample EQ906 γ-ray irradiated at the

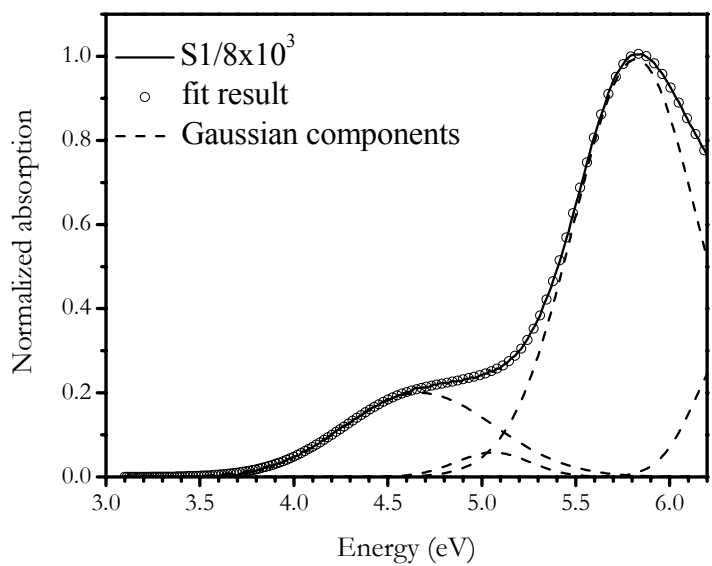

**Figure 5.13** OA spectrum (continuous line) normalized to the maximum absorption coefficient. The fit result is represented by circles and the components bands by dashed lines.

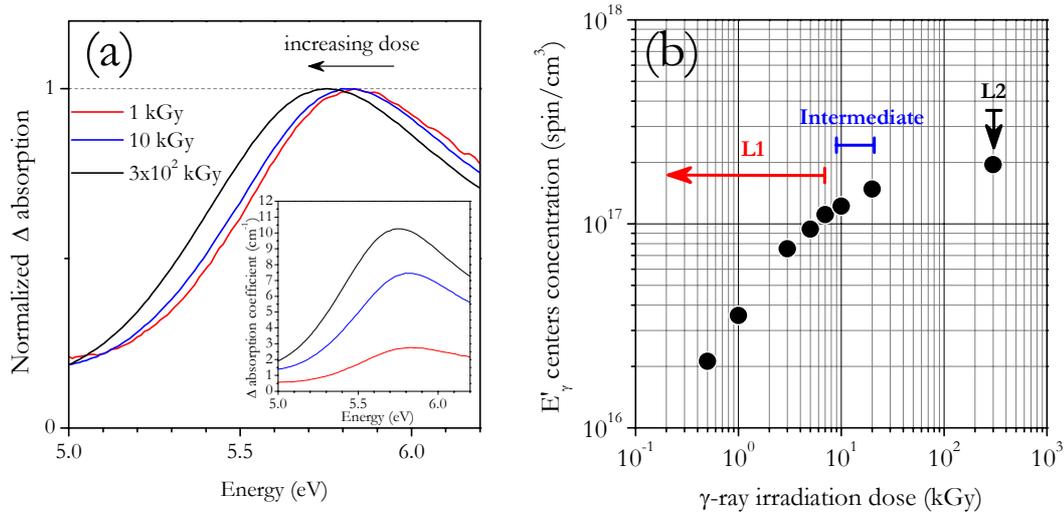

**Figure 5.14** (a) OA spectra acquired in the sample EQ906 γ-ray irradiated at the doses 1 kGy, 10 kGy and $3 \times 10^2$ kGy. The spectra have been obtained subtracting the spectrum of the not irradiated sample to those of the irradiated ones and normalizing to the maximum value of the Δ absorption at about 5.8 eV. The corresponding not normalized OA spectra are reported in the inset. (b) Concentration of the E'$_\gamma$ centers as a function of the γ-ray irradiation dose. In (b) it is indicated if the EPR line shape of the E'$_\gamma$ center is of type L1, L2 or intermediate, in correspondence to the various irradiation doses.

doses 1 kGy, 10 kGy and $3 \times 10^2$ kGy. The spectra of Figure 5.14 (a) have been obtained subtracting the spectrum of the not irradiated sample to those of the irradiated ones and normalizing to the maximum value of the Δ absorption at about 5.8 eV. The corresponding not normalized OA spectra are reported in the inset of the same figure, whereas the concentration growth of E'$_\gamma$ center with the irradiation dose is shown in Figure 5.14 (b). In this latter figure it is indicated if the EPR line shape of the E'$_\gamma$ center is of type L1, L2 or intermediate, in correspondence to the various irradiation doses considered. From the data reported in Figure 5.14 (a), by a fit procedure, we have estimated that the peak position of the E'$_\gamma$ canter OA band is centred at 5.83 ± 0.01 eV, when the defect exhibits the L1 EPR line shape, and at 5.77 ± 0.01 eV, when the defect exhibits the L2 line shape.

After irradiation at the maximum dose of $3 \times 10^2$ kGy, the same sample was isochronally thermally treated in the temperature range from 320 K up to 820 K in order to induce the E'$_\gamma$ EPR line shape conversion from L2 to L1. Also in this case we have monitored both the main resonance line of the E'$_\gamma$ center and its OA band. In Figure 5.15 (a) we report the OA spectra acquired in a sample EQ906/$3 \times 10^3$ before the thermal treatment and after the thermal treatment at two different temperatures of the isochronal treatment. The spectra of Figure 5.15 (a) have been obtained subtracting the spectrum of the not irradiated sample to those of the irradiated and thermally treated ones and normalizing to the maximum value of the Δ absorption at about 5.8 eV. The corresponding not normalized OA spectra are reported in the inset of the same figure, whereas the concentration of E'$_\gamma$ center during the thermal treatment experiment is shown in Figure 5.15 (b). As shown, upon thermal treatment, the OA band associated to the E'$_\gamma$ centers

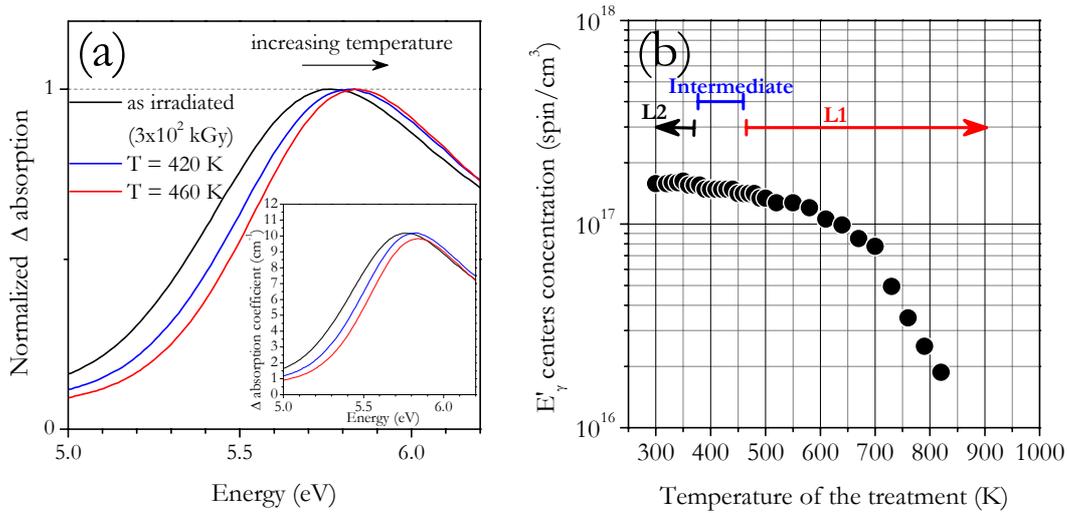

**Figure 5.15** (a) OA spectra acquired in the sample EQ906/3x10³ before thermal treatment and after thermal treatment at two different temperatures of an isochronal treatment. The spectra have been obtained subtracting the spectrum of the not irradiated sample to those of the irradiated and thermally treated ones and normalizing to the maximum value of the Δ absorption at about 5.8 eV. The corresponding not normalized OA spectra are reported in the inset. (b). Concentration of the E'$_\gamma$ centers as a function of the isochronal thermal treatment temperature. In (b) it is indicated if the EPR line shape of the E'$_\gamma$ center is of type L1, L2 or intermediate, during the thermal treatment at the various temperatures.

undergoes a peak shift from 5.77±0.01 eV to 5.86±0.01 eV. This blue shift is about of the same amount and in the opposite direction with respect to that observed under irradiation, in accordance with the observed thermally induced conversion of the E'$_\gamma$ EPR line shape from L2 to L1.

In order to study in more detail the correlation between the changes observed in the main EPR line shape and in the peak position of the 5.8 eV OA band we have considered four samples: EQ906 (natural dry), H1 (natural wet), S1 (synthetic wet) and an S300 (synthetic dry), γ-ray irradiated at the doses 1.8x10³ kGy, 3x10³ kGy, 8x10³ kGy and 8x10³ kGy, respectively. After γ-ray irradiation the samples have been subjected to the same sequence of isochronal thermal treatments in the temperature range from 330 K up to 950 K in order to induce the EPR line shape conversion of the E'$_\gamma$ centers from L2 to L1. Both the EPR and the OA spectra were acquired at each different temperature. In this study we have found that in all the investigated samples the area of the Gaussian OA band associated to the E'$_\gamma$ centers is correlated to their EPR signal, in agreement with the already known correlation of the EPR and OA spectral features associated to the E'$_\gamma$ centers [50]. In order to make a quantitative analysis of the correlation between the changes observed in the OA and in the EPR spectra of the E'$_\gamma$ centers, we have determined, for each sample and in correspondence to the various temperatures of the treatment, the shift of the peak position of the OA band and that of $\Delta g_{1,2}$ with respect to the values corresponding to the line shape of reference L2. In Figure 5.16, these shifts are reported for all the investigated samples and they show a pretty good correlation for all the thermal treatments suggesting a connection between the induced changes.

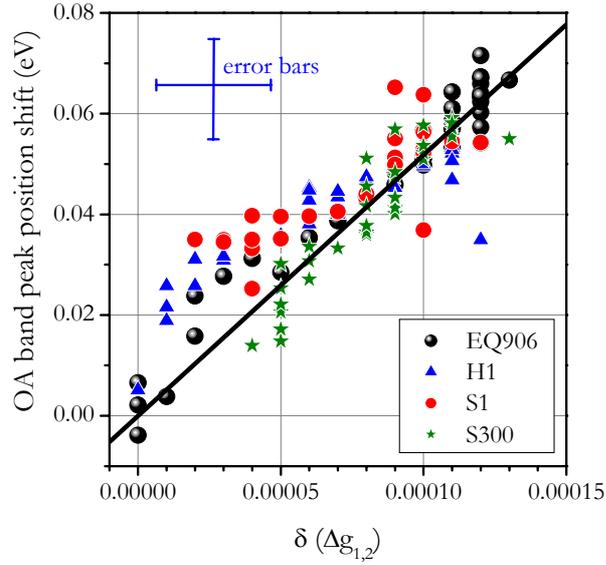

**Figure 5.16** Shift of the position of the Gaussian band at about 5.8 eV, derived from the fitting of the OA spectra, as a function of the variation of $\Delta g_{1,2} = (g_1 - g_2)$, derived from the FH-EPR spectra of the E'$_\gamma$ centers, measured during the thermal treatment experiments of the samples EQ906/1.8x10$^3$, H1/3x10$^3$, S1/8x10$^3$, S300/8x10$^3$. The differences are calculated with respect to the line shape of reference L2. The straight line has been obtained from Eq. (5.2) [see Paragraph 5.1.4]. The error bars for the quantities reported in abscissa and in ordinate axes are ±0.00002 and ±0.01 eV, respectively.

## 5.1.4 Discussion

As emerges from the data reported in the previous paragraphs, two structures of the E'$_\gamma$ center can be distinguished on the basis of their EPR and OA features. The main EPR and OA spectroscopic parameters characterizing these two types of E'$_\gamma$ centers are summarized in Table 5.1. In the following, for convenience, we will refer to the E'$_\gamma$ center with the spectral properties reported in the first and second rows of Table 5.1 as E'$_\gamma$ (1) and E'$_\gamma$ (2), respectively.

**Table 5.1 EPR parameters of the two type of E'$_\gamma$ centers**

|  | Type | Main EPR line $\Delta g_{1,2}$ | $\Delta g_{1,3}$ | Hyperfine splitting | FWHM of the hyperfine lines Low-field | High-field | Peak position of the OA band |
|---|---|---|---|---|---|---|---|
| **E'$_\gamma$ (1)** | L1 | (1.24 ±0.02)x10$^{-3}$ | (1.47 ±0.02)x10$^{-3}$ | ~ 41.8 mT | ~ 3.7 mT | ~ 3.2 mT | 5.83 eV ±0.01 |
| **E'$_\gamma$ (2)** | L2 | (1.15 ±0.02)x10$^{-3}$ | (1.42 ±0.02)x10$^{-3}$ | ~ 41.8 mT | ~ 4.4 mT | ~ 3.7 mT | 5.77 eV ±0.01 |

We have found that in the natural dry and wet and in synthetic dry materials for γ-ray irradiation doses lower than ~10 kGy and higher than ~10$^3$ kGy the E'$_\gamma$ (1) and E'$_\gamma$ (2) centers are observed, respectively, whereas in the synthetic wet materials only the E'$_\gamma$ (2) is detected in correspondence to all the irradiation doses considered. Furthermore, we have found that, upon thermal treatment in the temperature range from 370 K to 460 K, the E'$_\gamma$ (2) center undergoes to a structural conversion leading its spectral features to coincide with those of the E'$_\gamma$ (1) in all the four types of materials.

In principle, one can infer that the observed changes of the spectroscopic features from those of E'$_\gamma$ (1) center to those of the E'$_\gamma$ (2) on increasing the γ-ray dose can be an effect of the prolonged irradiation of the sample, which increases the structural disorder around the defect. In agreement with this hypothesis, the EPR line shapes of both the main resonance line and the hyperfine structure of the E'$_\gamma$ (2) center are wider than those of the E'$_\gamma$ (1). However, the results of the irradiation experiments on the synthetic wet materials disagree with this picture. In these materials, in fact, only the E'$_\gamma$ (2) center is observed, also for γ-ray irradiation doses as low as 0.5 kGy. Alternatively, the existence of two different precursors of the E'$_\gamma$ center can be supposed, which upon irradiation generate two spectroscopically distinguishable defects: the E'$_\gamma$ (1) and the E'$_\gamma$ (2) centers. Since the E'$_\gamma$ (1) center is observed in correspondence to lower γ-ray irradiation doses than the E'$_\gamma$ (2), then the precursor site of the latter should possess a radiation activation energy higher than that of the former. Furthermore, since the line shape L2 is observed in all the materials irradiated at doses higher than 10$^3$ kGy, it follows that the concentration of the precursors of the E'$_\gamma$ (2) center is higher than that of the E'$_\gamma$ (1) center in all the materials investigated. In this scheme, the observed change of the spectroscopic features in the dose range from 10 kGy to 10$^3$ kGy, observed in all the materials but the synthetic wet ones, can be attributed to a competitive growth of the E'$_\gamma$ (1) and the E'$_\gamma$ (2) centers, resulting in an overall line shape which is intermediate between L1 and L2. In contrast, in the synthetic wet materials the E'$_\gamma$ (1) is not induced by irradiation, indicating that its precursor site is ineffective (or absent at all) and consequently the materials exhibit higher radiation resistance and only the E'$_\gamma$ (2) center is observed.

It is worth to note that the existence of more than a single precursor site for the E'$_\gamma$ center in a-SiO$_2$ has been previously suggested on the basis of many experimental and theoretical works, as discussed in Paragraph 2.3.1. In this respect, the experimental characterization reported in the previous paragraphs could furnish a method to discern between E'$_\gamma$ centers belonging to two classes distinguished on the basis of the EPR and OA properties summarized in Table 5.1.

One of the most important results of our investigation consists in the observation of a thermally induced shift of the OA band of the E'$_\gamma$ (2) center strictly correlated to the variation observed in both the main EPR line [see Figure 5.16] and hyperfine doublet of the same center. As discussed in the following, this result permits to gain information on the nature of the electronic levels involved in the OA band of the E'$_\gamma$, which is one of the most relevant open questions concerning this point defect. In Figure 5.17 we report the oxygen vacancy structure of the E'$_\gamma$ center (top) and the corresponding energetic levels scheme (bottom). The schematic representation of the energy levels of the O≡Si$^\bullet$ moiety is that of the broken tetrahedron structure, already discussed in Paragraph 1.1.2 [see also Figure 1.1], whereas for the $^+$Si≡O structure facing the unpaired electron we consider only the energy level of its first unfilled state, which has been supposed by some authors to be involved in the transition responsible for the ~5.8 eV OA band of the E'$_\gamma$ centers [104, 106]. This latter transition is indicated by the arrow 3 in Figure 5.17, whereas the other two possible transitions, occurring between pairs of states both belonging to the O≡Si$^\bullet$ moiety, are indicated by arrows 1 and 2. As discussed in Paragraph 1.1.2, the $\Delta g_{\parallel,\perp} = g_\parallel - g_\perp$ for a paramagnetic center with the broken tetrahedron structure, as the O≡Si$^\bullet$ group in Figure 5.17 (top), is given by

$$\Delta g_{\parallel,\perp} = -|\lambda| C_{3p}^2 \left[ \frac{(1-\gamma)(1-\delta)}{E_b} - \frac{(1+\gamma)(1+\delta)}{E_a} \right] \quad (5.1)$$

where $C_{3p}^2$ is the 3p percentage of the atomic orbital of Si involved in the unpaired electron wave function, whereas $E_b$ and $E_a$ [shown in Figure 5.17] are the differences between the energy of the dangling bond state and those of valence (made by the Si-O bonding orbitals) and conduction (made by the Si-O antibonding orbital) states, respectively. The parameters $\gamma$ and $\delta$ are small corrections ($\gamma \ll 1$, $\delta \ll 1$) to the $g_\perp$ value introduced in order to make the hybrid orbitals of the outer shell of the Si atom orthogonal to the core states and to take into account a partial ionic character of the Si-O bonds, respectively [18]. The importance of Eq. (5.1) consists in the fact that it puts forward a connection between the quantity $\Delta g_{\parallel, \perp}$ and the energies $E_a$ and $E_b$. In particular, Eq. (5.1) predicts that a change of the energies $E_a$ and $E_b$ reflects in a change of $\Delta g_{\parallel, \perp}$, and vice versa. In order to compare the experimentally estimated values, shown in Figure 5.16, with those predicted from Eq. (5.1), we suppose that the ~5.8 eV OA band arises from the transition from the valence band states to the unpaired electron orbital (transition 1 in Figure 5.17) and that the sum $E_a + E_b$ does not change upon thermal treatment. Furthermore, we assume that $E_a + E_b$ is nearly equal to the gap energy of a-SiO₂, $E_g \cong 8.5$ eV [1]. In this scheme, the only energy level allowed to change is that of the unpaired electron state. In addition, we use the values $\gamma = -0.17$ [17], $|\lambda| = 0.02$ eV [17] and $C_{3p}^2 = 0.65$ [2]. Finally, we assume $\delta = 0$, which corresponds to neglect the partial ionic character of the Si-O bond [18]. Under these hypothesis, by differentiating Eq. (5.1) one obtains

$$\delta\left(\Delta g_{\parallel,\perp}\right) = |\lambda| C_{3p}^2 \left[ \frac{(1-\gamma)}{E_b^2} + \frac{(1+\gamma)}{(E_g - E_b)^2} \right] \delta E_b = (1.93 \times 10^{-3})\, \delta E_b \qquad (5.2)$$

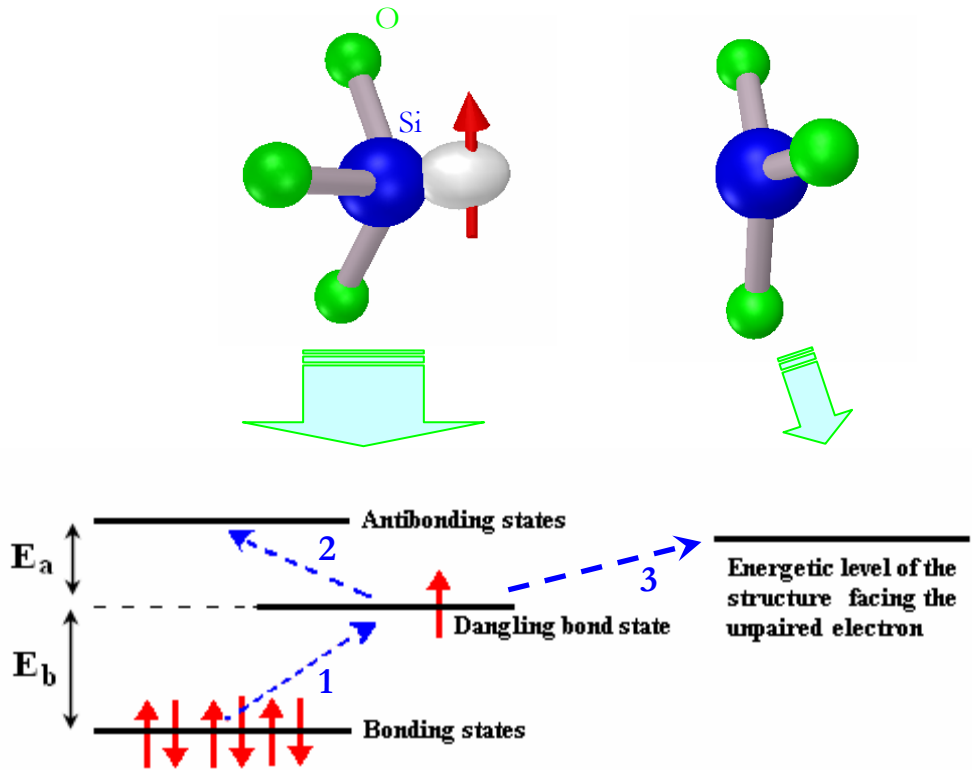

**Figure 5.17** Oxygen vacancy structure of the E'$_\gamma$ center (top) and the corresponding energetic levels scheme (bottom). The schematic representation of the energy levels of the O≡Si$^\bullet$ moiety (left) is that of the broken tetrahedron [17], discussed in Paragraph 1.1.2, whereas for the $^+$Si≡O structure facing the unpaired electron (right) only the energy level of its first unfilled state is indicated. Arrows 1 and 2 indicate the two transitions occurring among states of the O≡Si$^\bullet$ moiety, whereas arrow 3 indicates a charge transfer transition occurring from the O≡Si$^\bullet$ moiety to the $^+$Si≡O group.

The straight line defined by Eq. (5.2) is compared to the experimental data in Figure 5.16, assuming that for the E'$_\gamma$ center it is possible to approximate $\Delta g_{1,2} \cong \Delta g_{\parallel,\perp}$. As shown, an excellent agreement is found, strongly supporting the validity of the model and of the approximations used to obtain Eq. (5.2). Note that if one assumes that the OA band of the E'$_\gamma$ center is due to the transition indicated with the arrow 2 in Figure 5.17, to which corresponds an energy difference $E_a$, then the slope of the straight line indicated in Figure 5.16 changes sign, resulting in quite disagreement with the experimental data. Furthermore, it is worth to note that, if one assumes that the OA band of the E'$_\gamma$ center is associated to the transition indicated by arrow 3 in Figure 5.17, then no obvious correlation exists between the value of the peak position of the OA band and that of $\Delta g_{1,2}$, in disagreement with the experimental data reported in Figure 5.16. In Eqs. (5.1) and (5.2), in fact, no physical parameters depending on the $^+$Si≡O structure facing the unpaired electron are contained, as it has been supposed that only a negligible portion of the unpaired electron wave function is localized on it.

In view of the good agreement between the experimental data and the dependence predicted on the basis of Eq. (5.2), shown in Figure 5.16, one of our assumptions deserves to be discussed in more detail. As noted above, we have supposed that the only energy level which is allowed to shift during the structural change of the E'$_\gamma$ (2) upon thermal treatment, associated to the main line shape change from L2 to L1, is that of the occupied dangling bond state. It is worth to note that such a picture can be considered valid if the conversion from L2 to L1 is due to a change of the perturbative effect on the unpaired electron wave function arising from the structure facing the O≡Si$^\bullet$ moiety. In this case, in fact, it is reasonable that the energy of the occupied dangling bond state could change without a relevant variation, to a first approximation, of the energy levels of the three Si-O back bonds involved in the O≡Si$^\bullet$ molecule, as they are rigidly connected to the rest of the a-SiO$_2$ matrix. At variance, if the conversion from L2 to L1 is due to a structural change involving the O≡Si$^\bullet$ moiety, then all the energetic levels of this molecular group should change and the correlation described by the Eq. (5.2), in which E$_a$+E$_b$ has been considered a constant, should fail. Consequently, the quite good agreement shown on Figure 5.16 suggests that the thermally induced conversion from L2 to L1 should involve a change of the perturbative effect of the structure facing the O≡Si$^\bullet$ moiety, instead of a structural change of the O≡Si$^\bullet$ moiety itself. This conclusion also agrees with the fact that the main hyperfine splittings before and after thermal treatment are both approximately equal to 41.8 mT (see Figure 5.12), which indicates that the mean value of the angle between the unpaired electron wave function and the three Si-O back bonds is nearly the same in the two cases.

Summarizing, on the basis of the observed correlated variations of the EPR and OA features of the E'$_\gamma$ center upon thermal treatment, shown in Figure 5.16, we have found evidences suggesting that the OA band of the E'$_\gamma$ center is attributed to the O≡Si$^\bullet$ moiety and that it consists in a transition from the valence band states to that of the unpaired electron of the defect. Furthermore, evidences are found that the thermal induced change from L2 to the L1 arises from the weakening of the perturbative effect of the structure facing the O≡Si$^\bullet$ moiety of the defect on the unpaired electron wave function.

As a final remark, we note that the attribution of specific microscopic models to the E'$_\gamma$ (1) and the E'$_\gamma$ (2) centers is still an open question. Nevertheless, basing on the experimental data reported in the present Thesis, some suggestions on their probable structures will be discussed in Paragraph 8.2.

## 5.2 Effects of β-ray irradiation and thermal treatment

### 5.2.1 The main resonance line and the strong hyperfine structure

In the present paragraph the investigation of the irradiation effects on the E'$_\gamma$ center presented in the previous paragraph is extended to irradiation doses higher than $10^4$ kGy. Since the $^{60}$Co source we used has a dose rate of ~3 kGy/hr, too long irradiation time is needed in order to reach irradiation doses as high as those of interest here. For this reason, we have considered β-ray irradiation with a Van de Graff irradiator, which features a higher dose rate, of ~20 kGy/sec. In particular, the experimental data reported in the present paragraph have been obtained in the β-ray dose range between $1.2 \times 10^3$ kGy and $5 \times 10^6$ kGy.

In Figure 5.18 the growth curves of concentration of E'$_\gamma$ centers with γ- and β-ray dose are compared, as obtained in the I301 material. As shown, two stages of concentration growth are evident: the first occurs for doses up to $10^3$ kGy, and is covered by γ-ray irradiation, whereas the second stage of growth is observed for doses higher than $10^5$ kGy, covered by β-ray irradiation. At variance, in the dose range from $10^3$ kGy up to $10^4$ kGy, a nearly constant value of the concentration of defects is obtained, with both the two methods of irradiation. Growth curves of concentration of the E'$_\gamma$ center similar to that reported in Figure 5.18 have been also found in the other materials considered, even though the dose at which the second growth occurs has been found to depend on the sample, ranging from $10^5$ kGy to $10^6$ kGy.

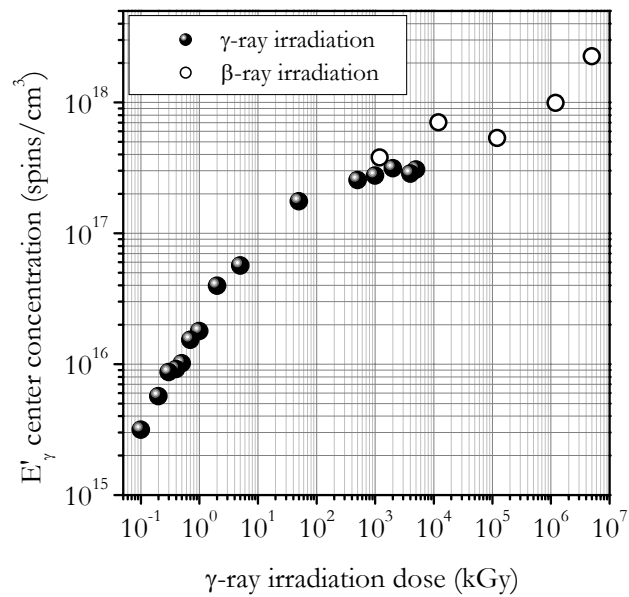

**Figure 5.18** Concentration of the E'$_\gamma$ centers in the material I301 as a function of the γ- and β-ray irradiation dose, as determined by FH-EPR measurements.

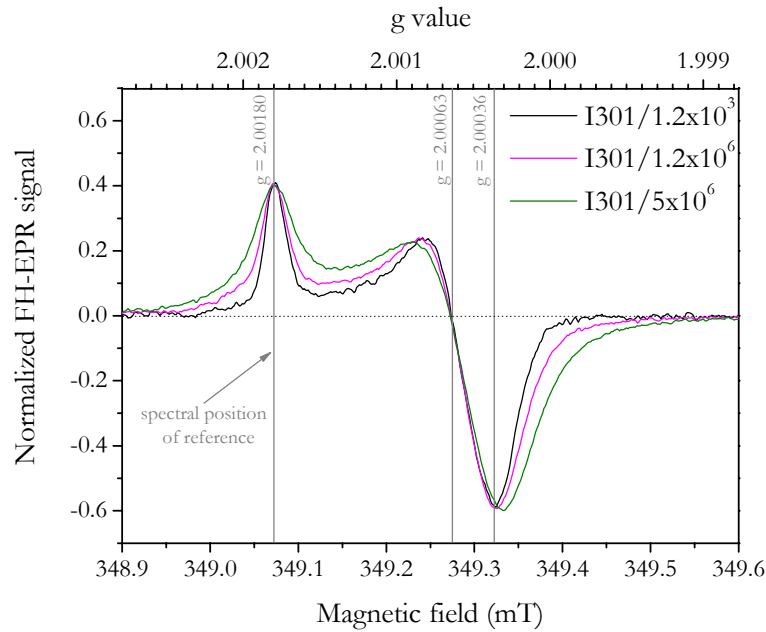

**Figure 5.19** FH-EPR spectra of the E'$_\gamma$ center observed in the material I301 β-ray irradiated at doses 1.2x10$^3$ kGy, 1.2x10$^6$ kGy and 5x10$^6$ kGy. The FH-EPR spectra have been normalized to the peak-to-peak signal amplitude. The scale of the g values is also reported, obtained fixing the first positive peak position of the E'$_\gamma$ center at g=2.00180 [2].

In principle, each concentration growth stage indicates that a generation process is activated, which could involve precursor sites of the E'$_\gamma$ center or, in the case of the second stage, the direct activation of regular sites of the a-SiO$_2$ matrix. In particular, the study performed in the previous paragraphs has concluded that in the first stage of growth two distinguishable precursor sites are activated. In order to investigate the generation process activated in the second stage of growth, occurring for doses higher than 10$^5$ kGy in Figure 5.18, and to put forward possible structural differences of the E'$_\gamma$ centers generated in this dose range, we have studied the E'$_\gamma$ center EPR main line shape. The normalized FH-EPR spectra of the main line of the E'$_\gamma$ centers induced in the material I301 β-ray irradiated at the doses 1.2x10$^3$ kGy, 1.2x10$^6$ kGy and 5x10$^6$ kGy are superimposed in Figure 5.19. We have verified that the EPR line shape in the sample I301/1.2x10$^3$ kGy is of type L2, in agreement with the data obtained by γ-ray irradiation in correspondence to the same doses. At variance, on increasing the β-ray irradiation dose, a gradual broadening of the EPR line occurs, in concomitance with a relevant increase of the concentration of defects, as it emerges from the inspection of both Figures 5.18 and 5.19. In principle, this broadening effect could be attributed to structural changes occurring in the defect or to dipolar interaction. In this respect, from the study of various samples we have found that a one to one correspondence exists between the overall concentration of paramagnetic defects, estimated by FH-EPR measurements, and the occurrence of the broadening effects on the EPR line shape of the E'$_\gamma$ center. Accordingly, it seems reasonable to assume that the change of shape shown in Figure 5.19 on increasing the β-ray irradiation dose is due to dipolar interaction between the paramagnetic centers induced by irradiation in the sample. This conclusion agrees with the results of previous investigations [186]. Note that, although the dipolar broadening influence the line shape, it does not affect the value of the double integral of the spectrum [11, 12], which can be still used to estimate the concentration of defects (Paragraph 3.1.3).

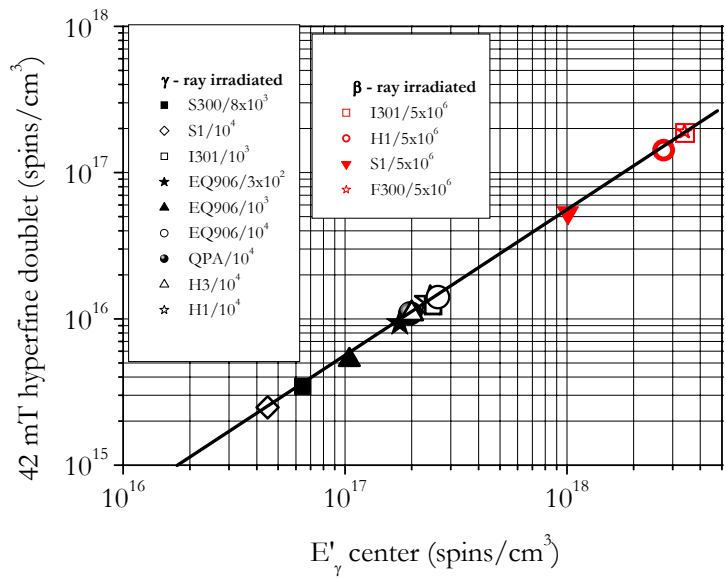

**Figure 5.20** Concentration of defects responsible for the 42 mT doublet as a function of the concentration of defects responsible for the main resonance line of the E'$_\gamma$ center in a-SiO$_2$ materials γ-ray irradiated in the dose range from 3x10² kGy to 10⁴ kGy (black points) and β-ray irradiated at the dose 5x10⁶ kGy (red points). The straight line, with slope 1, is superimposed to the data, for comparison.

The occurrence of the dipolar broadening of the main EPR line of the E'$_\gamma$ center invalidates the possibility to gain structural information on the point defects by line shape inspection. For this reason we have extended this study to the strong hyperfine structure of the E'$_\gamma$ center. As a first step we have verified that the correlation between the 42 mT hyperfine structure and the main resonance line of the E'$_\gamma$ center is maintained also in the dose range explored here. This study has been accomplished considering four different materials and choosing the same acquisition parameters indicated in Paragraph 5.1.2 for the FH-EPR measurements. The data obtained are reported in Figure 5.20 together with the data obtained by γ-ray irradiation (Figure 5.10) and discussed in Paragraph 5.1.2, for comparison. As shown, a quite good correlation is found for an overall variation of the concentration of defects of about two orders of magnitude.

In the successive step of this study we focused our attention on the hyperfine doublet EPR line shape and splitting. In Figure 5.21 we compare the SH-EPR spectra of the 42 mT doublet observed in the material I301 β-ray irradiated at the doses 1.2x10³ kGy, 1.2x10⁵ kGy and 5x10⁶ kGy. The SH-EPR spectra have been normalized and horizontally shifted in order to superimpose the high-field components of the doublet. From this figure, two relevant information can be obtained. First, it is evident that the EPR line shape of the high-field component of the doublet does not change in the range of doses considered, within the experimental error, even if in the samples I301/5x10⁶ the main line of the E'$_\gamma$ center is broadened with respect to that in the sample I301/1.2x10³. Similarly, the invariance of the EPR line shape of the low-field component of the doublet has been also verified. Furthermore, both line shapes of the low- and high-field components of the doublet are virtually identical to those of the E'$_\gamma$ (2) center induced by γ-ray irradiation in the range of doses from $10^3$ kGy up to $10^4$ kGy. This experimental evidence supports the above discussed occurrence of dipolar broadening effects on the E'$_\gamma$ center main EPR line. In fact, dipolar effects are expected not to affect appreciably the line shapes of the two components of the hyperfine doublet, as each of them is more than one order of magnitude wider than the main line, mainly due to inhomogenous distribution of the isotropic hyperfine constant. The second important information which can be obtained from the spectra of Figure 5.21 consists in the increase of the hyperfine splitting on increasing the β-ray irradiation dose, gradually changing from ~41.8 mT up to ~42.8 mT (see inset in Figure 5.21). In particular, the hyperfine splitting has been found to be affected appreciably for doses higher than ~$10^4$ kGy. This increase of the hyperfine splitting has been found to be general, as it has been observed in all the investigated samples in correspondence to the same irradiation doses. It is worth to note that, from the study of various materials it emerged that the increase of hyperfine splitting occurs, in some samples, without a relevant change of the concentration of E'$_\gamma$ centers, suggesting that the irradiation at doses higher than ~$10^4$ kGy affects the structure of already present centers instead to create new slightly structurally different ones.

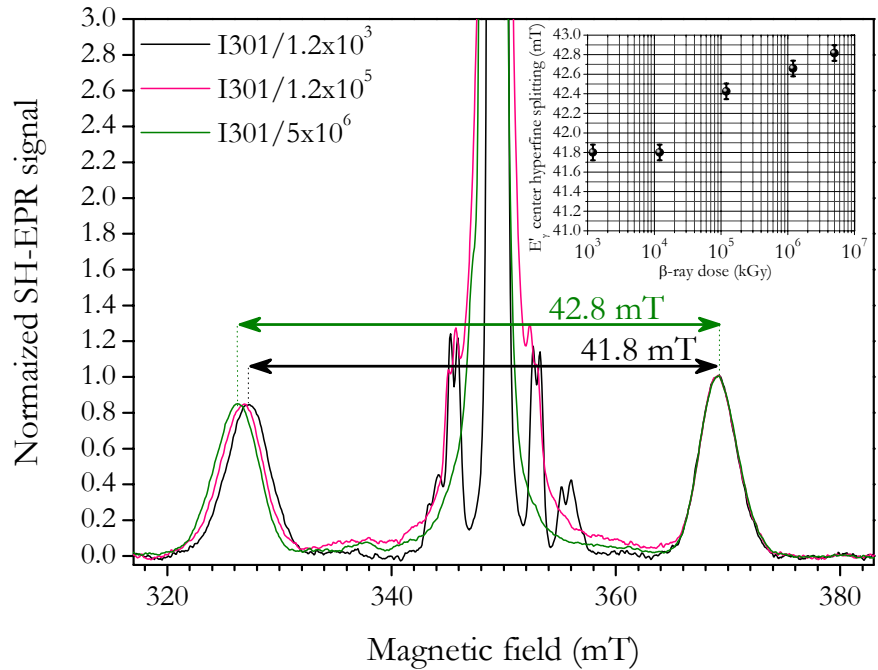

**Figure 5.21** SH-EPR spectra of the 42 mT doublet detected in the material I301 β-ray irradiated at the doses 1.2x10³ kGy, 1.2x10⁵ kGy and 5x10⁶ kGy. The SH-EPR spectra have been normalized and horizontally shifted in order to superimpose the high-field components of the doublet. Inset: Splitting of the hyperfine structure of the E'$_\gamma$ center in the material I301 as a function of the β-ray dose.

Finally, in order to investigate the thermally induced structural conversion of the E'$_\gamma$ center, we have studied the effects of a treatment in the range of temperature from 330 K up to

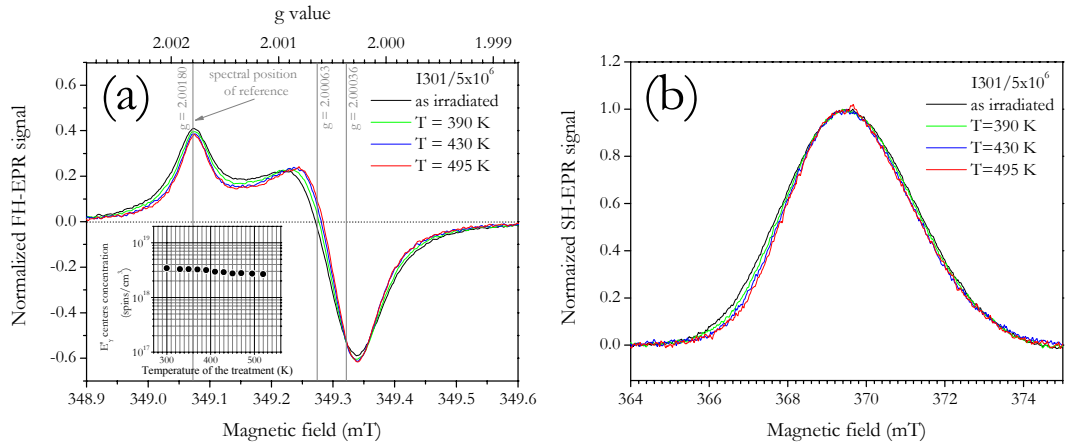

**Figure 5.22** (a) FH-EPR spectrum of the E'$_\gamma$ center main line and (b) SH-EPR spectrum of the high-field component of the 42 mT doublet acquired in a sample I301 irradiated at β-ray doses 5x10$^6$ kGy compared to the corresponding spectra obtained for the same sample after isochronal thermal treatment up to three different temperatures. Inset: Concentration of E'$_\gamma$ centers as a function of the isochronal thermal treatment temperature. The FH-EPR and SH-EPR spectra have been normalized to the peak-to-peak signal amplitude and to the signal amplitude in correspondence to the maximum of the line, respectively

520 K on both the main EPR line and hyperfine structure of the E'$_\gamma$ center. In this study we have considered four materials I301 (natural dry), H1 (natural wet), S1 (synthetic wet) and F300 (synthetic dry), previously β-ray irradiated at a dose of 5x10$^6$ kGy. The results obtained are summarized in Figure 5.22. As it emerges from the comparison of Figure 5.12 (b) with Figure 5.22 (b), a quite similar line shape variation of the high-field component of the 42 mT doublet occurs, indicating that a conversion of the E'$_\gamma$ (2) structure take place. Analogous comments apply to the low-field component of the doublet. In agreement with this conclusion, by comparing Figure 5.22 (a) with Figure 5.7, the correspondent variation of the main EPR line of the E'$_\gamma$ center from L2 to L1 is evident, even though it is somewhat masked in the thermally treated β-ray irradiated sample of Figure 5.22 (a) by the superimposed dipolar broadening effect. In this respect, it is worth to note, the line shape variations of Figure 5.22 (a) cannot be attributed to a reduction of the dipolar broadening arising from the partial decrease of the concentration of defects, of about the 20 % with respect to its value in the as irradiated sample [see inset of Figure 5.22 (a)]. From inspection of Figure 5.19, in fact, it is evident that the dipolar broadening does not affect the zero crossing g value of the E'$_\gamma$ EPR line in the range of concentration of defects of interest here. At variance, in Figure 5.22 (a), a relevant change of this spectral position is evident, which is compatible with that expected for a conversion from L2 to L1 [compare Figures 5.22 (a) and 5.7]. Finally, it is worth to note that the strong hyperfine splitting is not affected by these thermal treatments.

## 5.2.2 Discussion

From the data reported in the previous paragraph it can be concluded that, although the growth of concentration of Figure 5.18 suggests that β-ray irradiation above $10^5$ kGy activates a second generation process of E'$_\gamma$ centers, the spectral properties of the generated defects do not differ with respect to those of the E'$_\gamma$ (2) center, observed for γ-ray doses in the range from $10^3$ kGy up to $10^4$ kGy. Further support to this conclusion comes from the data of Figure 5.22, which point out that the thermally induced conversion process of the E'$_\gamma$ center is virtually identical to that induced in γ-ray irradiated samples.

Another important result regards the observed increase of the hyperfine splitting of the 42 mT doublet for β-ray irradiation dose higher than ~$10^4$ kGy. A similar increase of the hyperfine splitting has been reported by Devine and Arndt [187] in γ-ray irradiated pressure-densified a-SiO$_2$ on increasing the degree of densification of the material. In that work the increase of the hyperfine splitting has been attributed to a densification-induced increase of the mean bond angle ϱ between the dangling bond orbital and the three Si-O back bonds of the O≡Si$^\bullet$ moiety [see Figure 1.1 (a)] [187]. In particular, the authors have suggested that this effect reflects an overall structural change occurring in the a-SiO$_2$ matrix upon mechanical densification, involving a reduction of the O-Si-O and of the Si-O-Si bond angles and an increases of the Si-O bond length [187]. The amount of changes of the main values of these structural parameters was found to depend on the degree of densification of the material. It is worth to note that the materials we have considered here are not densified before irradiation and, in fact, an hyperfine doublet split by 41.8 mT is observed after γ- and β-ray irradiation up to a dose of $10^4$ kGy. At variance, for higher β-ray irradiation doses a gradual increase of the hyperfine splitting occurs. Furthermore, as noted in the previous paragraph, in this dose range the β-ray irradiation affects the microscopic structure of the already present defects, instead of inducing new structurally different ones, as the increase of the hyperfine splitting is observed also at nearly constant concentration of E'$_\gamma$ centers. These experimental evidences speak for a radiation-induced densification of the material which should occur for β-ray doses higher than ~$10^4$ kGy, in agreement with previous studies performed by Raman [188] and IR absorption [189] spectroscopies. If one assumes that a one to one correspondence exists between the effects of the mechanical densification studied by Devine and Arndt [187] and those induced by β-ray irradiation in the materials we have considered, then from the observed variation of about 2.4 % of the hyperfine splitting shown in Figure 5.21 we can estimate a radiation-induced densification of about 3 %.

We note that, in the light of the possibility, raised above, that relevant structural changes could occur in the O≡Si$^\bullet$ moiety upon β-ray irradiation at doses higher than ~$10^4$ kGy, the result, already mentioned, that the thermally induced structural variation of the E'$_\gamma$ (2) center occurs in quite similar way in samples γ-ray irradiated at 5x$10^3$ kGy (Figures 5.7 and 5.12) and in those β-ray irradiated at 5x$10^6$ kGy (Figure 5.22) becomes more surprising. A way to overcome this difficulty is to suppose that β-ray irradiation doses higher than ~$10^4$ kGy mainly affect the structure of the O≡Si$^\bullet$ moiety, as suggested by Devine and Arndt [187], whereas the thermally induced structural change of the E'$_\gamma$ (2) involves a group of atoms which is nearby but spatially separated with respect to the O≡Si$^\bullet$ molecule, as for the $^+$Si≡O group facing the unpaired electron in the positively charged oxygen vacancy model of the E'$_\gamma$ (2) center. This picture is in line and gives further support to the conclusions drawn in the discussion at the end of the first part of the present chapter.

# Chapter 6

## *Intrinsic point defects induced in Al-containing oxygen-deficient a-SiO$_2$*

In the present chapter we report a study by EPR spectroscopy on the intrinsic E'$_\gamma$, E'$_\delta$, E'$_\alpha$ and triplet state centers observed in Al-containing oxygen-deficient a-SiO$_2$ materials γ-ray irradiated in the dose range from 5 kGy to $10^4$ kGy and subsequently thermally treated at temperatures as high as 1000 K. The materials considered, after γ-ray irradiation at $10^4$ kGy, exhibit [AlO$_4$]$^0$ centers concentrations higher than ~2x10$^{17}$ spins/cm$^3$. This class of materials is treated separately from that possessing sensibly lower concentration of Al-related hole centers, studied in the previous chapter, as they share common properties upon thermal treatment on previously irradiated sample.

Four distinct Al-containing oxygen-deficient a-SiO$_2$ materials: KI, KUVI, QC and P453 [see Table 4.1 (Chapter 4)] are considered. This enables to disentangle the features connected with the intrinsic oxygen deficiency of the material from those connected with the presence of the Al-related hole centers. These materials, after γ-ray irradiation at $10^4$ kGy, exhibit a concentration of [AlO$_4$]$^0$ centers comprised from ~2.6 x10$^{17}$ spins/cm$^3$ to ~8.6 x10$^{17}$ spins/cm$^3$, whereas their intrinsic oxygen deficiency differs by more than two orders of magnitude. This latter property was proven performing OA measurements on the 7.6 eV band, attributed to the oxygen vacancies [50], and luminescence (PL) measurements on the 4.4 eV band excited at 5 eV, attributed to the twofold coordinated Si defect [50], as described in the following.

The OA spectra obtained for the materials KI and KUVI are reported in Figure 6.1. As shown, the 7.6 eV band is evident in the OA spectrum of the KUVI, whereas in that of the KI material a band peaked at 7.6 eV cannot be recognized, even though the OA signal at 7.6 eV is not zero. From the spectra reported in Figure 6.1, by applying a fit procedure, the contribution of the 7.6 eV band to the full OA spectra has been obtained. From this estimations and assuming an absorption cross section of 7.5x10$^{-17}$ cm$^{-2}$ [50] for the 7.6 eV OA band we have evaluated that the concentration of oxygen vacancies in the as-grown KI and KUVI are <10$^{16}$ cm$^{-3}$ and ~5x10$^{17}$ cm$^{-3}$, respectively. Note that, since the peak of the 7.6 eV band is not clearly observed in the OA spectrum of the KI, only an upper limit of the concentration of oxygen vacancies present in the material can be determined. OA spectra were also obtained for the QC and P453 as-grown materials. However, in these cases, due to the huge OA observed in correspondence to 7.6 eV, an exact quantitative analysis was prevented. Nevertheless, a rough estimation of the 7.6 OA band in these latter materials indicates that the concentration of oxygen vacancies is higher than ~10$^{18}$ cm$^{-3}$.

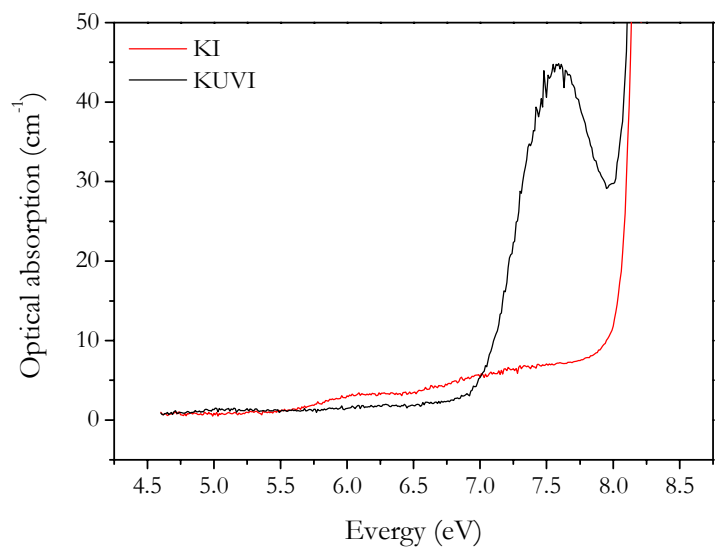

**Figure 6.1** OA spectra obtained for the as-grown KI and KUVI materials.

In principle, in order to estimate the concentration of twofold coordinated Si defects in a given material, the amplitude of the 5.0 eV band has to be determined. However, this OA band was difficult to detect in the spectra of the KI and KUVI (due to the low concentration of defects), whereas it superimposes to other OA bands in the spectra of QC and P453, so preventing a reliable estimation of its amplitude. For these reasons, we focused our study on the PL band peaked at 4.4 eV and excited at 5 eV. Once the amplitude of this band was determined in the KI, KUVI, QC and P453 materials, an empirical factor was used to obtain the amplitude of the related OA band peaked at 5 eV. This empirical factor was derived with reference to an as-grown sample of F300, in which both the OA band at 5 eV and the PL band at 4.4 eV were determined with high precision [190]. Finally, from the estimated amplitude of the 5 eV OA band and by using an oscillator strength for this band of 0.15 [50] we have obtained the concentration of twofold coordinated Si in the materials of interest. In particular, we have obtained that this concentration in the KUVI, P453 and QC as-grown materials is of $\sim 4 \times 10^{14}$ cm$^{-3}$, $2 \times 10^{15}$ cm$^{-3}$ and $5 \times 10^{15}$ cm$^{-3}$, respectively, whereas in the KI no 4.4 eV PL band has been detected, indicating that the concentration of twofold coordinated Si is lower than $\sim 10^{14}$ cm$^{-3}$.

Here we summarize the content of the present chapter. In Paragraph 6.1 we present a characterization of the paramagnetic point defects induced by γ-ray irradiation in the materials KI, KUVI, QC and P453 and we report their growth curves with irradiation dose from 5 kGy up to $10^4$ kGy. In Paragraph 6.2, we focus our attention on the effects induced by thermal treatment of previously irradiated materials. In Paragraphs 6.3 the effects of a γ-ray reirradiation on the materials already irradiated and thermally treated are reported. Finally, the experimental data are discussed in Paragraph 6.4.

## 6.1 γ-ray radiation induced point defects

### 6.1.1 E'$_\gamma$ and E'$_\delta$ centers

In all the samples considered no EPR signal was detected before irradiation. At variance, after irradiation many distinct signals are induced. In Figure 6.2 (a) the FH-EPR spectrum acquired in correspondence to g ≅ 2 for a sample P453/10³ (continuous line) is reported, obtained using $P_{in}$ = 8 x 10$^{-4}$ mW, $\nu_m = \omega_m/2\pi$ = 100 kHz and $H_m$ = 0.01 mT. This EPR signal arises from the partial superposition of two distinct resonance lines ascribed to E'$_\gamma$ and E'$_\delta$ centers [compare Figures 6.2 and 2.15]. These two contributions were determined by fitting the spectrum with a weighted sum of an experimental line shape of type L2 for the E'$_\gamma$ center (see previous chapter), and a simulated line shape for the E'$_\delta$ center, reported in Figure 6.2 (c) and (b),

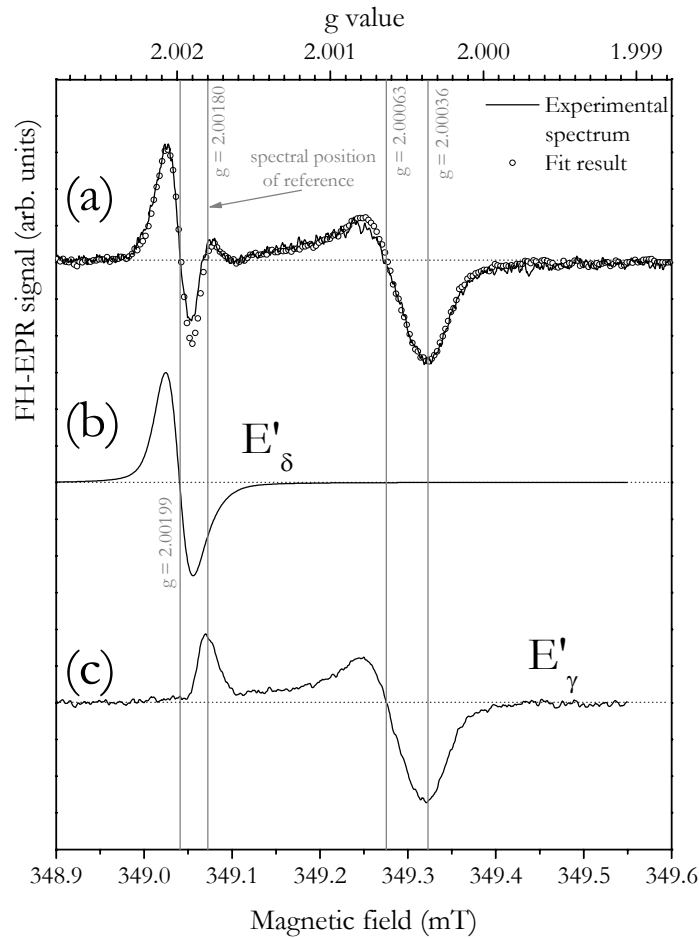

**Figure 6.2** (a) FH-EPR spectrum centred in correspondence to g ≅ 2 as detected in the sample P453/10³ (continuous line) compared to the line obtained as a weighted sum (circles) of the reference lines for (b) E'$_\delta$ and (c) E'$_\gamma$ centers. For the E'$_\gamma$ centers the L2 line shape is considered. The scale of the g values is also reported, obtained by fixing the first positive peak position of the E'$_\gamma$ centers at g=2.00180 [2].

respectively. The latter was obtained by the Bruker SIMFONIA software. The result of this procedure is reported in Figure 6.2 (a), where the weighted sum (circles) of the reference line shapes for E'$_\gamma$ and E'$_\delta$ is superimposed to the experimental spectrum (continuous line). By repeating the analysis reported in Figure 6.2 for many different samples and by fixing $g_{||}$=2.00180 for the E'$_\gamma$ [2], we have obtained an estimation of the E'$_\delta$ center zero crossing g value of 2.0020 ± 0.0001, in good agreement with other experimental evaluations [92, 120, 127, 132, 141-144].

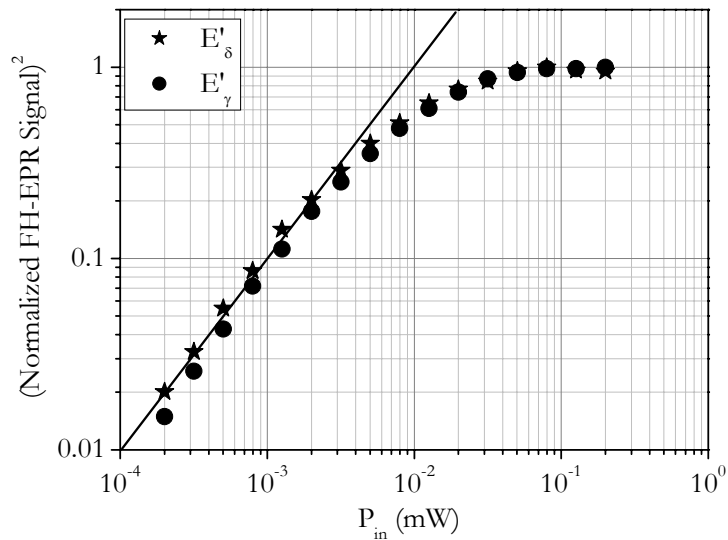

**Figure 6.3** Saturation with microwave power of the FH-EPR signal of E'$_\gamma$ (circles) and E'$_\delta$ (stars) centers in a sample P453/10⁴. The linear dependence of the squared FH-EPR signal as a function of $P_{in}$ is evidenced by the straight line.

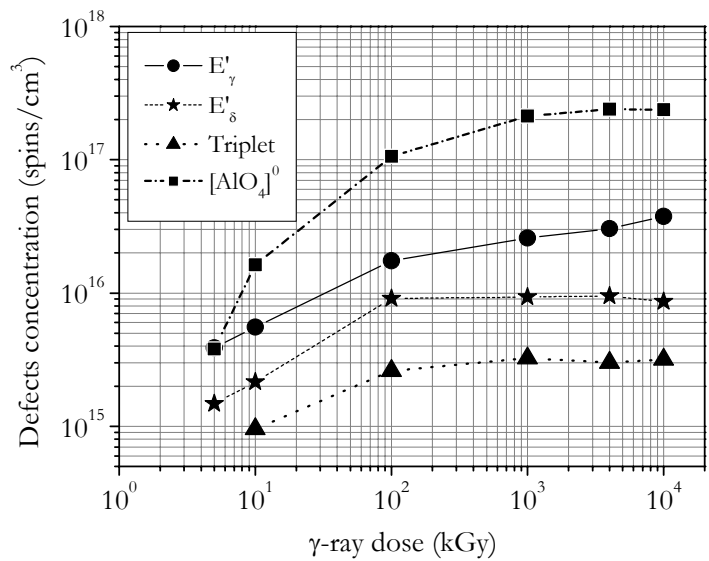

**Figure 6.4** Concentration of the E'$_\gamma$, E'$_\delta$, Triplet and [AlO$_4$]⁰ centers in the material P453 as a function of the γ-ray irradiation dose as determined by FH-EPR. The error of measurement is comparable with the size of the symbols.

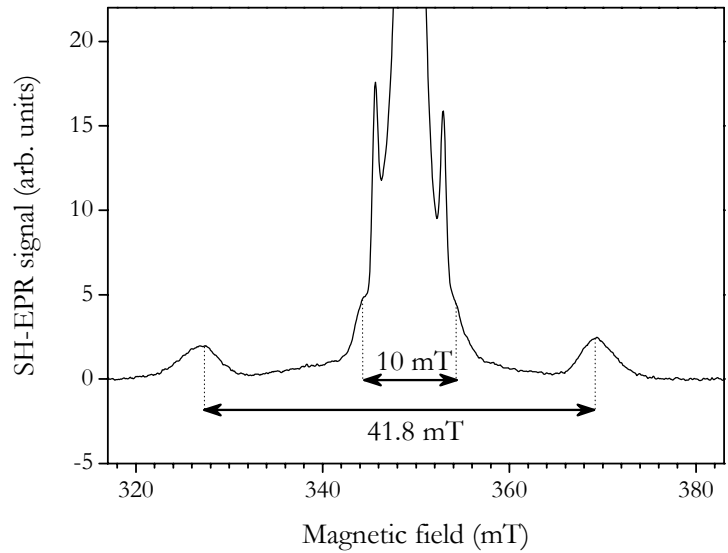

**Figure 6.5** SH-EPR spectrum on a wide field scan obtained for a sample P453/10$^3$. The arrows indicate the doublets split by 10 mT and 41.8 mT.

To better characterize E'$_\gamma$ and E'$_\delta$ defects, the room temperature saturation properties of their FH-EPR signals with microwave power have been studied. These data are reported in Figure 6.3 for a sample P453/10$^4$ and point out that the E'$_\gamma$ and E'$_\delta$ centers have virtually identical saturation properties. Moreover, these saturation curves reproduce those observed in all the other Type I-IV commercial a-SiO$_2$ for doses of 10$^3$ kGy or higher (see Figure 5.3) [91].

The line shapes reported in Figure 6.2 (b) and (c) were used to estimate the concentrations of E'$_\gamma$ and E'$_\delta$ induced in samples of P453 irradiated at different gamma ray doses in the range from 5 kGy up to 10$^4$ kGy. The growth curves of defect concentrations as a function of the γ-ray dose are reported in Figure 6.4. The concentration of E'$_\delta$ centers was found to increase with irradiation dose up to ~10$^2$ kGy. For higher doses a maximum concentration of ~10$^{16}$ spins/cm$^3$ is maintained, compatible with a generation process from precursor defects. At variance, the concentration of E'$_\gamma$ centers increases up to the highest dose considered, indicating a more complex generation process that could involve a direct activation of normal matrix sites or a not complete exhaustion of precursor defects. Similar concentration growths were previously reported for an X-ray irradiated synthetic a-SiO$_2$ material [123].

In the irradiated samples we have also looked for the EPR signals of the strong hyperfine structures of the E'$_\gamma$ and E'$_\delta$ centers. In Figure 6.5, the SH-EPR spectrum centered on g ≅ 2 and acquired over an extended range is reported. As shown, the hyperfine structure of the E'$_\gamma$ center is easily recognized and is indicated by a pair of lines split by 41.8 mT. At variance, the presence of the hyperfine structure of the E'$_\delta$ center, which as been proposed to consist in a pair of lines split by 10 mT (see Paragraph 2.3.4), cannot be unambiguously established. In fact, although a pair of lines approximately split by 10 mT can be recognized in the spectrum, as indicated in Figure 6.5, a reliable identification cannot be accomplished, due to the presence in the central region of the spectrum of a superimposing intense signal originating from other paramagnetic centers. FH-EPR measurements have been also performed to detect the hyperfine structures of the E'$_\gamma$ and E'$_\delta$ centers. However, due to the low concentration of defects, these FH-EPR signals were not observed.

Results similar to those above reported for the P453 have been obtained for the KUVI and QC materials, even though the maximum value of defects concentration was found to depend on the specific material. In particular, we have found that in correspondence to a dose of $10^2$ kGy in KUVI and QC materials the concentrations of the E'$_\gamma$ centers are $1\times10^{16}$ spins/cm$^3$ and $2\times10^{16}$ spins/cm$^3$, respectively. In correspondence to the same dose the concentration of E'$_\delta$ centers in KUVI and QC materials are $6\times10^{14}$ spins/cm$^3$ and $3\times10^{15}$ spins/cm$^3$, respectively. At variance, in the KI no EPR signal of this point defect has been detected, even if a concentration of E'$_\gamma$ centers $6\times10^{15}$ spins/cm$^3$ is observed.

### 6.1.2 The triplet state center

In the irradiated materials we looked for the $g \cong 4$ resonance of the triplet state center (see Paragraph 2.3.5). To this aim we have performed measurements setting the magnetic field at approximately half of the resonance field of the E' centers. As reported in Figure 6.6 for a sample P453/10$^4$, a FH-EPR signal was detected with line shape and resonance magnetic field compatible with those ascribed to the triplet center [120, 122]. The FH-EPR line of Figure 6.6

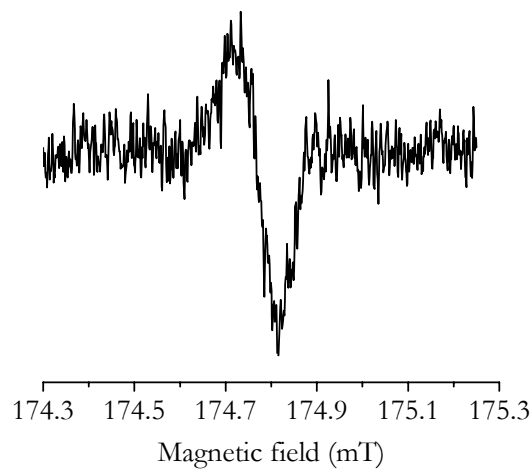

**Figure 6.6** FH-EPR spectrum acquired in correspondence to $g \cong 4$ for a sample P453/10$^4$.

has been acquired using $P_{in}$ = 50 mW, $\nu_m = \omega_m/2\pi$ = 100 kHz and $H_m$ = 0.01 mT.

To obtain an estimation of the triplet centers concentration, the intensity of the FH-EPR lines split by ~13 mT (see Paragraph 2.3.5), due to the allowed transitions between the states $|m_s=-1\rangle \leftrightarrow |m_s=0\rangle$ and $|m_s=0\rangle \leftrightarrow |m_s=+1\rangle$, has to be determined [120]. In our samples, due to the presence of the intense EPR signal of the $[AlO_4]^0$ centers (discussed in the successive paragraph), we were not able to isolate these lines. However, since it was reported that the ~13 mT pair is ~2500 times more intense than the $g \cong 4$ resonance [120], we have roughly estimated the concentration of triplet centers multiplying by a factor 2500 the double integral of the $g \cong 4$ FH-EPR signal. The values obtained for the material P453 in correspondence to various irradiation doses in the range from 5 kGy up to $10^4$ kGy are reported in Figure 6.4. As shown, the defects concentration grows up to ~$10^2$ kGy, and then it maintains a constant value of ~$3 \times 10^{15}$ spins/cm$^3$. Furthermore, from the comparison of the growth characteristics of E'$_\delta$ and triplet centers reported in Figure 6.4, we note that they reach the maximum value of concentration nearly in correspondence to the same dose.

The dependence of the FH-EPR signal on the microwave power for the $g \cong 4$ resonance was also studied and is reported in Figure 6.7. The square of the FH-EPR signal was found to grow *linearly* with the microwave power up to ~50 mW, whereas a deviation from the linear dependence was observed for higher power due to the occurrence of the saturation effect. These data point out that the saturation of the $g \cong 4$ line occurs at higher microwave power with respect to E'$_\gamma$ and E'$_\delta$, indicating that the triplet center possesses more effective relaxation channels with respect to the E' centers.

A triplet state center concentration growth with γ-ray irradiation dose virtually identical to that of Figure 6.4, obtained for P453, was also obtained for QC, whereas no FH-EPR signal arising from the triplet state centers was detected in KI and KUVI materials.

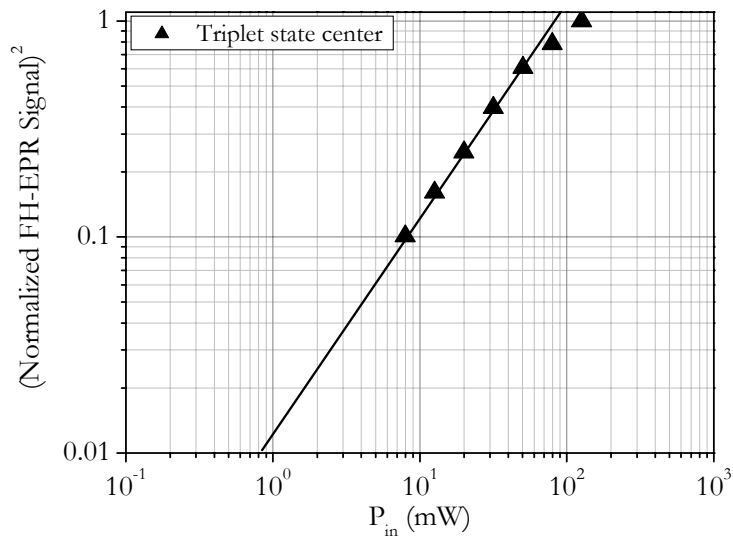

**Figure 6.7** Saturation with microwave power of the FH-EPR signal of the $g \cong 4$ line in a sample P453/10$^4$. The linear dependence of the squared FH-EPR signal as a function of $P_{in}$ is evidenced by the straight line.

## 6.1.3 The [AlO$_4$]$^0$ center

As reported in Table 4.1 (Chapter 4), all the materials considered in the present chapter, after γ-ray irradiation at $10^4$ kGy, exhibit a concentration of [AlO$_4$]$^0$ centers higher than $2 \times 10^{17}$ spins/cm$^3$, as estimated from the FH-EPR measurements discussed in the present paragraph. The typical spectrum we attributed to the [AlO$_4$]$^0$ centers is reported in Figure 6.8, as measured in a sample KI/80, using $P_{in}$ = 50 mW, $\nu_m = \omega_m/2\pi$ = 100 kHz and $H_m$ = 0.03 mT. This spectrum shows a structured signal mainly arising from the [AlO$_4$]$^0$ but for a feature imputable to the E'$_\gamma$ centers evidenced by the arrow in Figure 6.8. The g scale indicated in this figure has been obtained by fixing g = 2.0006 in correspondence to the E' centers peak in the EPR spectrum, corresponding to the zero crossing g value of the E'$_\gamma$ center. The positions in the spectrum corresponding to the three principal g values attributed to the [AlO$_4$]$^0$ center (see Figure 2.19) are also indicated in Figure 6.8. Our attribution of the FH-EPR signal shown in Figure 6.8 to the [AlO$_4$]$^0$ center is supported by the strict similarity found between this line shape and those obtained in previous investigations focused on Al-containing a-SiO$_2$ and powdered quartz materials (Figure 2.19) [161].

As for the other paramagnetic defects reported in previous paragraphs, the dependence of the FH-EPR signal on the microwave power has been studied for the [AlO$_4$]$^0$ center. However, at variance to the other defects, the [AlO$_4$]$^0$ centers squared FH-EPR signal was found to grow linearly up to the maximum microwave power obtainable with the EPR spectrometer we used.

In Figure 6.4 the growth of concentration of the [AlO$_4$]$^0$ centers on increasing the irradiation dose in the material P453 is reported, as estimated by FH-EPR measurements. As shown, the [AlO$_4$]$^0$ centers concentration was found to increase up to ~10$^3$ kGy, whereas a limit value of ~2 x 10$^{17}$ spins/cm$^3$ is maintained for higher doses. The reaching of a limit value should indicate the exhaustion of the precursor sites of the defect which, in turn, suggests that each AlO$_4$ site in the a-SiO$_2$ matrix has trapped a hole and contributes to the observed [AlO$_4$]$^0$ EPR signal.

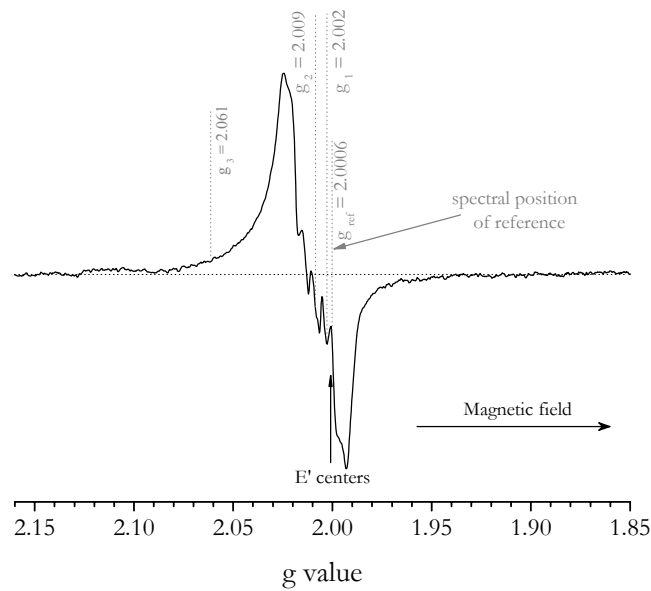

**Figure 6.8** FH-EPR spectrum of the [AlO$_4$]$^0$ center in a sample KI/80. The g scale has been obtained fixing the position of the peak in the spectrum due to the E' centers saturated signal at $g_{ref}$ = 2.0006. The positions in the spectrum of the three principal g values attributed to the [AlO$_4$]$^0$ are also shown [161].

## 6.2 Effects of thermal treatments on γ-ray irradiated materials

In order to further investigate the properties of point defects induced in the KI, KUVI, QC and P453 materials, we have performed thermal treatment experiments of the irradiated samples. The key experiment consisted in isochronal thermal treatments from 333 K up to 1023 K with a temperature step of 10 K on a sample P453/10$^3$. These treatments allowed us to observe the occurrence of thermally activated processes and to evaluate the temperatures at which they take place. Once these temperatures were established, isothermal treatments were performed in correspondence to T=580 K and T=630 K, as they permit to focus the attention on the specific thermally activated process of interest.

In Paragraph 6.2.1, we present a spectroscopic characterization of a point defect, the E'$_\alpha$ center, referring to the FH-EPR spectra acquired during the isochronal thermal treatment experiment on the sample P453/10$^3$. Once the FH-EPR line shape of the E'$_\alpha$ center is identified, in Paragraph 6.2.2 we discuss in detail the paramagnetic defects annealing curves obtained during isochronal and isothermal treatments experiments on irradiated materials.

### 6.2.1 Observation of the E'$_\alpha$ center

Upon thermal treatment of the γ-ray irradiated materials, a new contribution to the EPR spectrum of the E' centers main lines becomes evident, which is distinguishable from those of the E'$_\gamma$ and E'$_\delta$ centers. In Figure 6.9 (a) the spectrum obtained after isochronal thermal treatments of the sample P453/10$^3$ up to 750 K (continuous line) is reported, in which this new EPR line appears superimposed to that of the E'$_\gamma$ center. These two contributions were separated by fitting the spectrum with a weighted sum of an experimental line shape of type L1 for the E'$_\gamma$ center, characterized in the previous chapter, and a simulated line shape for the new paramagnetic center, reported in Figure 6.9 (c) and (b), respectively. The latter was obtained by the Bruker SIMFONIA software. The result of this procedure is reported in Figure 6.9 (a), where the weighted sum (circles) of the reference line shapes for E'$_\gamma$ and that of the new center is compared to the experimental spectrum (continuous line). By repeating the analysis reported in Figure 6.9 for many different samples thermally treated at different temperatures and by fixing $g_{||}$ =2.00180 for E'$_\gamma$ [2] we have obtained an estimation of the principal g values of the new center, which are $g_1$=2.0018 ± 0.0001, $g_2$=2.0009 ± 0.0001 and $g_3$=1.9997 ± 0.0001. The $g_1$ and $g_3$

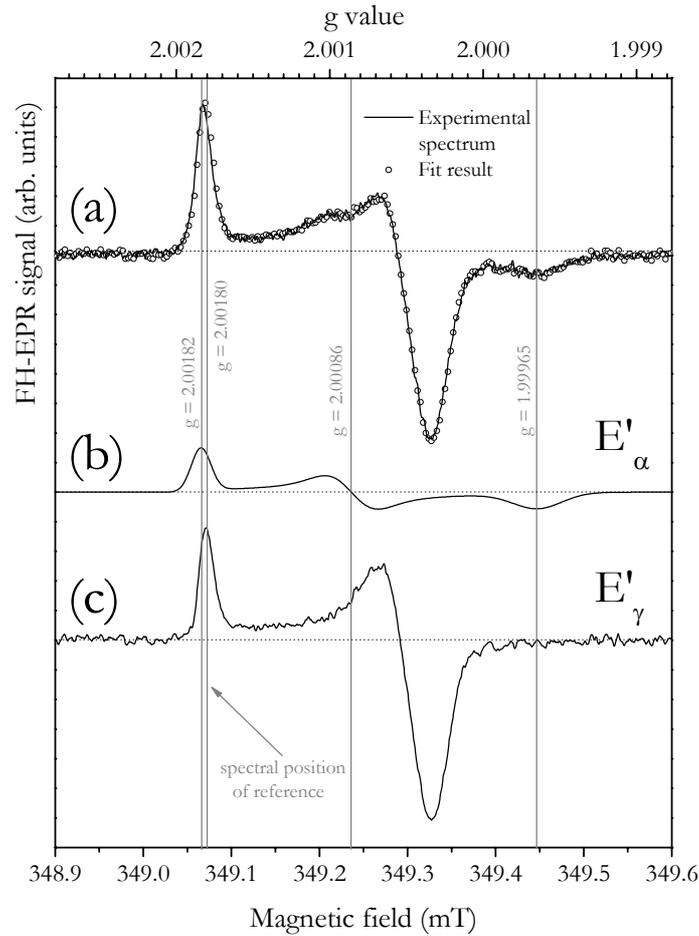

**Figure 6.9** (a) FH-EPR spectrum of the E' centers main lines acquired in the sample of P453/10$^3$ isochronally thermally treated up to 750 K (continuous line) compared to the line obtained as a weighted sum (circles) of the reference lines for (b) E'$_\alpha$ and (c) E'$_\gamma$ centers. For the E'$_\gamma$ centers the L1 line shape is considered. The scale of the g values is also reported, as obtained fixing the first positive peak position of the E'$_\gamma$ centers at g=2.0018 [2].

principal values obtained for this center agree with those attributed by Griscom [2, 79, 120] to the E'$_\alpha$ point defect (Paragraph 2.3.3), $g_1$=2.0018 and $g_3$=1.9998, whereas a worse agreement is found for the $g_2$ value, which for the E'$_\alpha$ is $g_2$=2.0013. In principle, this discrepancy could suggest that the orthorhombic line shape observed here is due to a variant of the E'$_\alpha$ center distinguishable from that previously reported. However, as discussed in detail in the previous chapter, the $g_2$ principal value of the E'$_\gamma$ center can shift by a maximum amount of about 0.0001 depending on the thermal history of the sample. Furthermore, in previous experimental investigations, also the zero crossing g value of another E' center, the E'$_\delta$ center, has been found to change by about 0.00025 depending on the method of defect generation [144]. Analogously, the discrepancy observed in the $g_2$ value of the E'$_\alpha$ center estimated in the present work with respect to that reported by Griscom [79] could simply reflect an intrinsic structural relaxation freedom of a-SiO$_2$. So, in the present Thesis we refer to the defect with the FH-EPR line shape shown in Figure 6.9 (b) as the E'$_\alpha$ center, without distinguishing it from that reported by Griscom [2, 79, 120].

In the most general case, upon thermal treatment of the sample P453 we have observed EPR spectra arising from the partial superposition of three contributions, due to E'$_\gamma$, E'$_\delta$ and E'$_\alpha$ centers. The result of the fit for this case is reported in Figure 6.10, in which the experimental spectrum obtained for the sample P453/10$^3$ isochronally thermally treated up to 580 K (continuous line) is compared to the weighted sum (circles) of the FH-EPR line shapes of reference for the three E' centers. By using this fit procedure we have separated the contributions of the three E' centers FH-EPR signals and we have studied their saturation properties with microwave power. The obtained data are reported in Figure 6.11 and point out that the three paramagnetic centers share virtually identical saturation properties with microwave power, getting stronger support to the attribution of the orthorhombic line shape to an E'-type center.

As a final remark we note that after the FH-EPR line shape of the E'$_\alpha$ center was determined, we have evaluated the possibility that this paramagnetic point defect was already

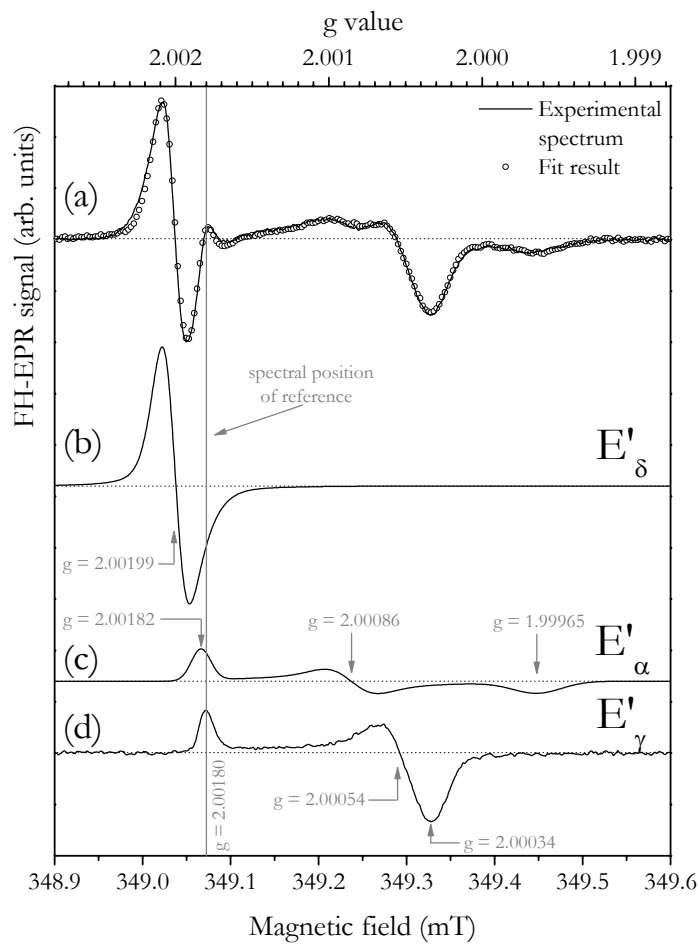

**Figure 6.10** (a) FH-EPR spectrum of the E' centers main lines acquired for a sample P453/10$^3$ isochronally thermally treated up to 580 K (continuous line) compared to the line obtained as a weighted sum (circles) of the reference lines for (b) E'$_\delta$, (c) E'$_\alpha$ and (d) E'$_\gamma$ centers. For the E'$_\gamma$ centers the L1 line shape is considered. The scale of the g values is also reported, as obtained fixing the first positive peak position of the E'$_\gamma$ centers at g=2.00180 [2].

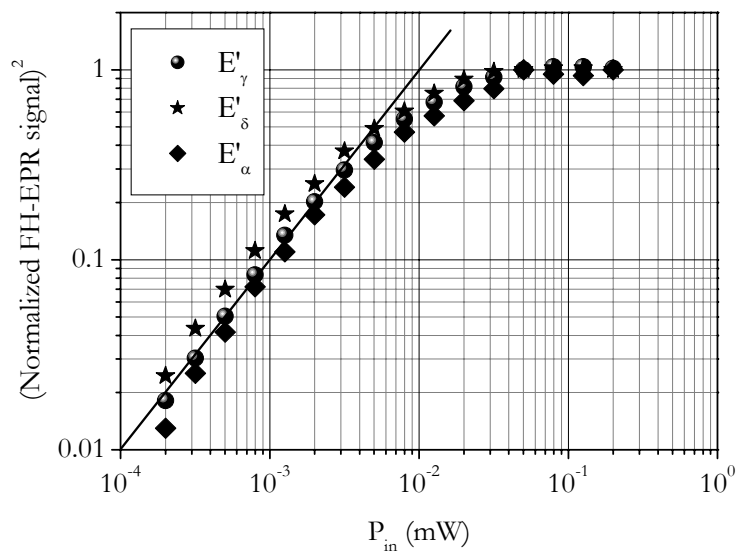

**Figure 6.11** Saturation with microwave power of the FH-EPR signal of E'$_\gamma$, E'$_\delta$ and E'$_\alpha$ centers in a sample P453/10$^4$ isochronally thermally treated up to 580 K. The linear dependence of the squared FH-EPR signal as a function of P$_{in}$ is evidenced by the straight line.

present in the as-irradiated samples. In fact, as emerges from inspection of the spectrum in Figure 6.2, a weak contribution to the overall spectrum arising from the FH-EPR signal of the E'$_\alpha$ center could be recognized. However, a reliable quantitative estimation of this contribution was found to be very difficult to obtain, as the FH-EPR signal of the E'$_\alpha$ center superimposes to that, more intense, of the E'$_\gamma$ center. Nevertheless, a rough estimation of the E'$_\alpha$ centers concentration in the P453 samples γ-ray irradiated at different doses has permitted us to obtain a value of about 10$^{15}$ spins/cm$^3$ in the dose range from 10$^2$ kGy up to 10$^4$ kGy, whereas no estimation was possible for lower doses. Results analogous to those obtained for P453 were found for irradiated QC, whereas no EPR signal of the E'$_\alpha$ center was detected in the KUVI and KI irradiated materials up to the maximum dose considered of 10$^4$ kGy.

### 6.2.2 Annealing curves of the paramagnetic centers

The concentration curves of the paramagnetic centers observed in the sample P453/10$^3$ after isochronal thermal treatments are reported in Figure 6.12. These data show that E'$_\gamma$, E'$_\delta$ and triplet state centers start to anneal at T~400 K.[1] However, while at higher temperature the triplet center anneals out definitively, the E'$_\gamma$ E'$_\delta$ and E'$_\alpha$ centers concentrations begin to increase for T

---

[1] In Figure 6.12 the concentration curve of the E'$_\alpha$ centers is reported for temperatures higher than 463 K, as at lower temperatures a reliable concentration estimation was prevented by the low concentration of defects.

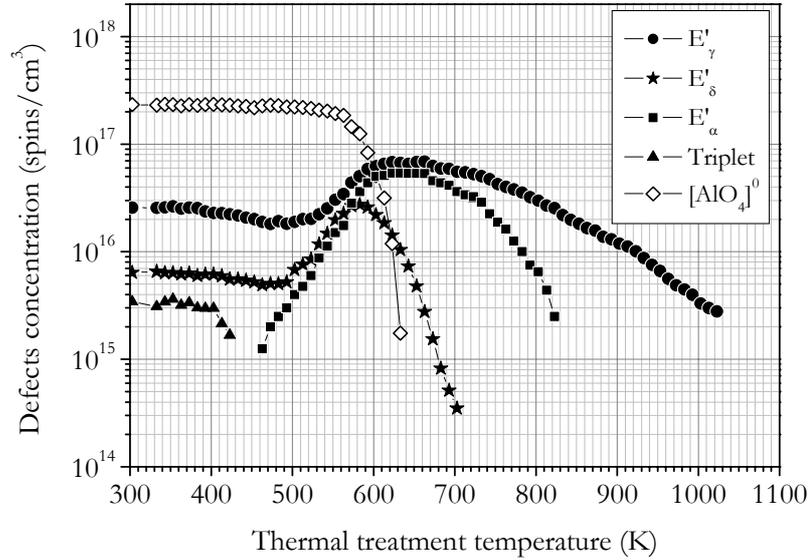

**Figure 6.12** Concentration of E'$_\gamma$, E'$_\delta$, E'$_\alpha$, Triplet and [AlO$_4$]$^0$ centers estimated by FH-EPR measurements in a sample P453/10$^4$ as a function of the isochronal thermal treatment temperature.

~ 500 K, indicating that a production mechanism is activated. Maximum concentrations of $7 \times 10^{16}$ spins/cm$^3$, $3 \times 10^{16}$ spins/cm$^3$ and $6 \times 10^{16}$ spins/cm$^3$ are obtained for E'$_\gamma$ E'$_\delta$ and E'$_\alpha$ centers, respectively. The FH-EPR spectrum obtained after thermal treatment at 580 K was reported in Figure 6.10.

Once the production mechanism of E'$_\gamma$, E'$_\delta$ and E'$_\alpha$ centers is exhausted, these point defects anneal out definitively. The annealing processes of the three E' centers take place with different rates and point out relevant differences in their overall thermal stability, indicating that different mechanisms are involved. In particular, the most efficient annealing process pertain to the E'$_\delta$, an intermediate one to the E'$_\alpha$, whereas the less efficient one is responsible for the disappearance of the E'$_\gamma$ center, whose FH-EPR signal was observed even at the highest temperature considered of 1023 K.

Quite different annealing features were found for the [AlO$_4$]$^0$ centers. As shown in Figure 6.12, thermal treatments up to T~500 K do not significantly change the concentration of these defects, while at higher temperatures the number of defects decreases, undergoing a more rapid annealing with respect to that of E' centers. In particular, [AlO$_4$]$^0$ centers anneal out in the same temperature range in which the growth of E'$_\gamma$, E'$_\delta$ and E'$_\alpha$ centers occurs and, after each thermal treatment, the total number of the generated E' centers is less than that of annealed [AlO$_4$]$^0$ centers.

Once the isochronal annealing curves of the paramagnetic centers were determined, as shown in Figure 6.12, we have studied the growth process of the E' centers concentrations in the KUVI, KI, and QC materials. These studies were undertaken by performing isothermal treatments at T=580 K and T=630 K on previously γ-ray irradiated samples. In correspondence to these temperatures, in fact, the growth process of the E' centers is efficient, whereas the annealing effect, which reduces the concentration of defects, is limited. A typical concentration curves of the E' centers obtained during isothermal treatments experiments is reported in Figure 6.13, as obtained for a sample KUVI/10² thermally treated at T = 630 K. As shown, the concentration of E'$_\delta$ centers grows up to ~4x10² s, reaching the highest values of 6.6x10¹⁵ spins/cm³, after that it gradually decreases. At variance, the concentrations of E'$_\gamma$ and E'$_\alpha$ centers grow up to ~3x10³ s, reaching the maximum values of 2x10¹⁶ spins/cm³ and 3x10¹⁶ spins/cm³, respectively, and then no other changes occur up to the maximum duration investigated of 7x10⁴ s. Similar results were also obtained upon isothermal treatment at T=580 K.

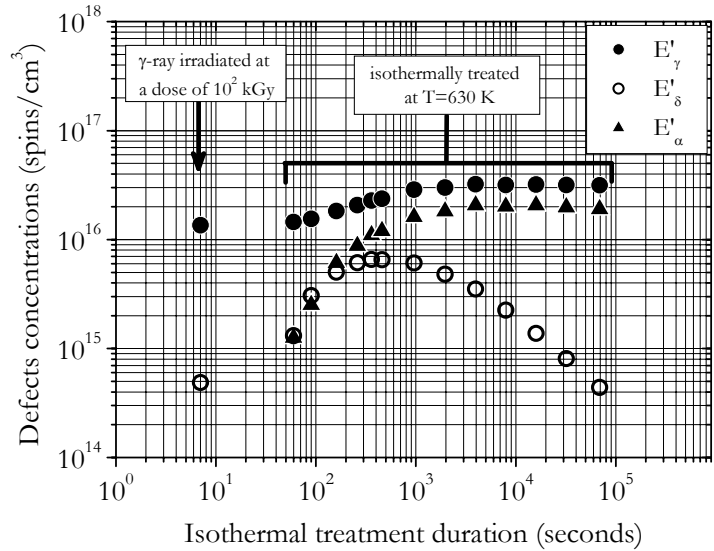

**Figure 6.13** Concentration of E'$_\gamma$, E'$_\delta$, E'$_\alpha$ centers estimated in a sample KUVI/10² after γ-ray irradiation and in correspondence to various durations of the subsequent isothermal treatment at T = 630 K.

Isothermal treatments experiments performed in previously irradiated samples of KI, KUVI, QC and P453 has permitted us to point out that the growth process of the E' centers occurs in all of them, even though the maximum concentration of defects obtained upon thermal treatment depends on the specific material. These data will be reported and discussed in Paragraph 6.4. Note that, the occurrence of the E' centers generation process in all the Al-containing materials (KI, KUVI, P453 and QC) and its absence in materials with sensibly lower Al content (see Chapter 5), suggests that the Al impurities are involved in this process.

As a final remark, we note that thermal treatment experiments were also performed in the as-grown materials, before any γ-ray irradiation, and no growth of E' centers concentration has been observed. This result indicates that the preliminary irradiation of the material is a necessary condition for the occurrence of the process of thermally induced defects generation.

## 6.3 Effects of γ-ray irradiation on materials previously irradiated and thermally treated

As emerges from the data reported in the previous paragraphs, upon γ-ray irradiation of the materials the concentration of induced E'$_\delta$ centers reaches a maximum value for doses of ~10$^2$ kGy (see Figure 6.4), suggesting that the sites precursors of the defect are exhausted. However, in contrast with this conjecture, if a sample γ-ray irradiated at a dose higher than ~10$^2$ kGy is thermally treated at temperatures above ~500 K, then the concentration of E'$_\delta$ centers starts again to increase (see Figures 6.12 and 6.13), indicating that other precursor sites of the defects are available in the material. In a similar way, we have found that upon γ-ray irradiation a weak EPR signal ascribed to the E'$_\alpha$ center is observed in the considered materials, whereas upon thermal treatment successive to the irradiation the concentration of these defects grows up and their EPR signal becomes easily detectable.

In order to further investigate these apparently contradictory results we have studied the effects of γ-ray irradiation on materials previously irradiated and thermally treated. The effects of the reirradiation on the E' centers main lines are reported in Figure 6.14, in which we compare the SH-EPR spectrum obtained for a sample P453/10$^2$ isothermally treated at T=630 K for 10$^3$ s with that obtained for the same sample after a subsequent γ-ray irradiation again at a dose of 10$^2$ kGy. Upon reirradiation three main effects are observed, as indicated by arrows I-III in figure: the SH-EPR signal amplitude of (I) the E'$_\delta$ and (II) the E'$_\alpha$ centers decrease, whereas that of the

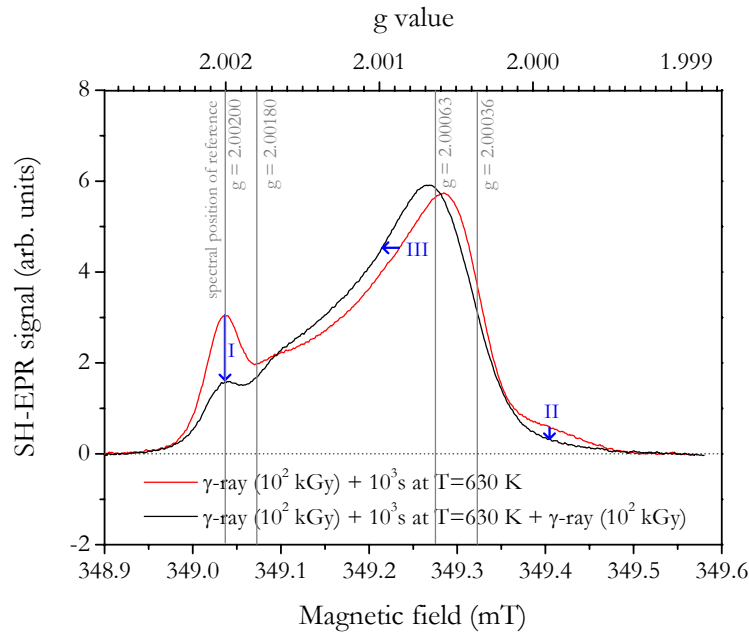

**Figure 6.14** (a) SH-EPR spectrum of the E' centers main lines acquired in a sample KUVI/10$^2$ isothermally treated at T=630 K for 10$^3$ s compared with the line obtained for the same sample after the subsequent γ-ray irradiation at a dose of 10$^2$ kGy. The three arrows indicate the main effects of the reirradiation. The scale of the g values is also reported, obtained fixing the signal peak position of the E'$_\delta$ centers at g=2.00200.

E'$_\gamma$ centers remains unchanged, within the experimental error, even though it changes shape from L1 to L2 (III). In addition, we have verified that the concentrations of E'$_\delta$ and E'$_\alpha$ centers obtained after the second γ-ray irradiation are virtually identical to those observed just after the first irradiation of the sample, namely before the thermal treatment.

The similarity in the effects observed for the E'$_\delta$ and E'$_\alpha$ centers upon reirradiation could rise the possibility that the disappearance of these two defects could be someway correlated. In order to investigate this point we have considered a sample KUVI/10² isothermally treated at T=630 K. The annealing curves of the E' centers obtained in correspondence to different duration of the thermal treatment are reported in Figure 6.13. As shown, the concentration of E'$_\delta$ centers in the as-irradiated sample is $5 \times 10^{14}$ spins/cm³. During the isothermal treatment it grows up to ~$5 \times 10^2$ s reaching a maximum values of $6.6 \times 10^{15}$ spins/cm³, after that it gradually decreases and, for a total duration of the thermal treatment of $7 \times 10^4$ s, its value is $4.5 \times 10^{14}$ spins/cm³, which is nearly the same as before the thermal treatment. At variance, the EPR signal of the E'$_\alpha$ centers increases monotonically up to $4 \times 10^4$ s and then it maintains a constant value of $2 \times 10^{16}$ spins/cm³. After thermal treatment up to $7 \times 10^4$ s, the same sample has been γ-ray irradiated at a dose of $10^2$ kGy. The FH-EPR spectra obtained before and after the reirradiation are compared in Figure 6.15. As shown, although the reirradiation process does not change the amplitude of the FH-EPR signal of the E'$_\delta$ centers, it causes the E'$_\alpha$ centers to disappear, indicating that the reirradiation-induced effects on these two defects are not correlated. Concerning the E'$_\gamma$ center,

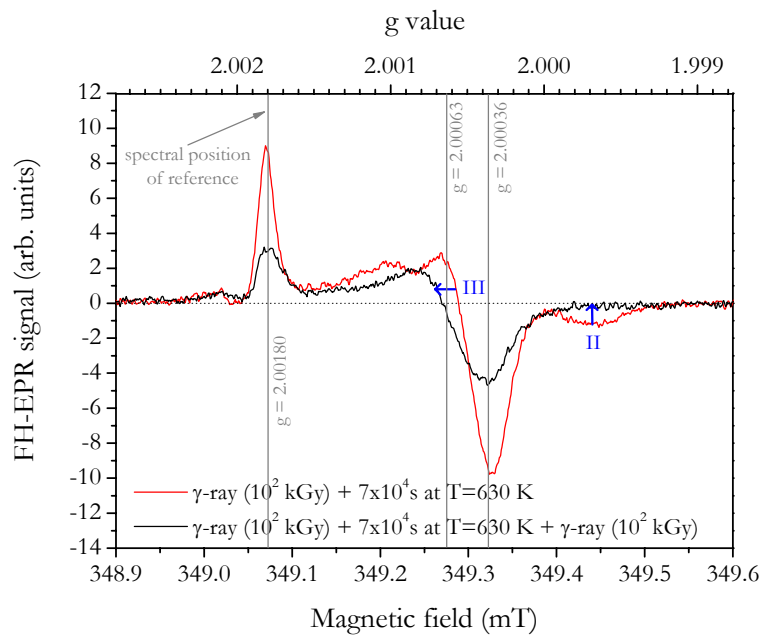

**Figure 6.15** (a) FH-EPR spectrum of the E' centers main lines acquired in a sample KUVI/10² isothermally treated up at T=630 K for $5 \times 10^4$ s compared to the line obtained after γ-ray reirradiation at the same dose of 10² kGy. The scale of the g values is also reported, obtained by fixing the first positive peak position of the E'$_\gamma$ centers at g=2.00180 [2].

Figure 6.15 shows that its EPR line shape converts from L1 to L2, in agreement with the data reported in Figure 6.14. Furthermore, a quantitative analysis of the FH-EPR spectra reported in Figure 6.15 has permitted us to verify that the concentration of E'$_\gamma$ center remains unchanged upon reirradiation, within an experimental uncertainty of 10%.

## 6.4 Discussion

In order to compare the properties of the point defects induced by γ-ray irradiation and thermal treatment in the various materials studied in the present chapter we have considered one sample of each type of them which was preliminarily irradiated at 80 kGy and then isothermally treated at T=630 K for 3x10$^2$ s. In Table 6.1 the concentration of E'$_\gamma$, E'$_\delta$ and E'$_\alpha$ centers are reported, as estimated in the as-irradiated samples and after the subsequent thermal treatment. The concentration of oxygen vacancies and twofold coordinated Si defects estimated in the as-grown samples are also reported, as they quantify the degree of oxygen deficiency of the various materials. Furthermore, we note that in the as-irradiated P453, QC, KUVI and KI samples concentrations of [AlO$_4$]$^0$ centers of ~9x10$^{16}$ spins/cm$^3$, ~2x10$^{17}$ spins/cm$^3$, ~1x10$^{17}$ spins/cm$^3$ and ~3x10$^{17}$ spins/cm$^3$, respectively, were estimated (not reported in Table 6.1).

From the data reported in Table 6.1, it is evident that the concentrations of E'$_\gamma$ and E'$_\delta$ centers obtained by irradiation are higher in the most oxygen-deficient materials. However, while the concentration of E'$_\delta$ centers changes in about the same proportion as the oxygen deficiency of the materials, this dependence is not respected in the case of the E'$_\gamma$ centers. This latter evidence agrees with the conjecture, already raised in literature (see Paragraph 2.3.1), that some precursor of this latter defect which is not oxygen-deficient could exist. At variance, for the E'$_\delta$ centers a direct dependence on the oxygen deficiency has been put forward [122]. It is worth to note that the concentration of the radiation induced E'$_\delta$ centers is uncorrelated with that of the [AlO$_4$]$^0$ centers in the four materials, indicating that this impurity center is not directly involved in

Table 6.1 Concentration of point defects in Al-containing oxygen-deficient a-SiO$_2$ materials

| Material nickname | Oxygen vacancies (cm$^{-3}$) as grown | Twofold coordinated Si (cm$^{-3}$) as grown | E'$_\alpha$ center (spins/cm$^3$) γ-ray | E'$_\alpha$ center γ-ray +TT | E'$_\gamma$ center (spins/cm$^3$) γ-ray | E'$_\gamma$ center γ-ray +TT | E'$_\delta$ center (spins/cm$^3$) γ-ray | E'$_\delta$ center γ-ray +TT | Triplet center (spins/cm$^3$) γ-ray | Triplet center γ-ray +TT |
|---|---|---|---|---|---|---|---|---|---|---|
| P453 | >10$^{18}$ | 2x10$^{15}$ | ~10$^{15}$ | 2x10$^{16}$ | 5x10$^{16}$ | 7x10$^{16}$ | 7x10$^{15}$ | 2x10$^{16}$ | ~2x10$^{15}$ | N. D. |
| QC | >10$^{18}$ | 5x10$^{15}$ | ~10$^{15}$ | 1x10$^{16}$ | 2x10$^{16}$ | 3x10$^{16}$ | 3x10$^{15}$ | 6x10$^{15}$ | ~1x10$^{15}$ | N. D. |
| KUVI | ~5x10$^{17}$ | ~4x10$^{14}$ | N. D. | 1x10$^{16}$ | 1x10$^{16}$ | 2x10$^{16}$ | 6x10$^{14}$ | 6x10$^{15}$ | N. D. | N. D. |
| KI | <10$^{16}$ | <10$^{14}$ | N. D. | ~10$^{15}$ | 6x10$^{15}$ | 6x10$^{15}$ | N.D. | 8x10$^{14}$ | N. D. | N. D. |

The γ-ray irradiation was performed at a dose of ~80 kGy, and the subsequent thermal treatments, indicated by TT in the table, at 630 K for 3x10$^2$ s.
N. D. indicates that no EPR signal has been detected.

the E'$_\delta$ centers generation process by γ-ray irradiation. So, the observation of an intense E'$_\delta$ centers EPR signal in the spectrum of the P453 and QC samples (see Figure 6.2 for a P453/10$^3$) is connected with the very large intrinsic oxygen deficiency of these materials rather than to the large concentration of [AlO$_4$]$^0$ centers induced by irradiation. Furthermore, in a study not reported in the present Thesis, we have verified that a concentration of E'$_\delta$ centers ≤ 6x10$^{14}$ spins/cm$^3$ is induced by γ-ray irradiation at a dose of 10$^4$ kGy in the materials reported in Table 4.1 (Chapter 4), but for P453 and QC, in agreements with the fact that in these materials the concentration of oxygen vacancies is ≤ 5x10$^{17}$ cm$^{-3}$ [191].

Another information coming from the irradiation data reported in Table 6.1 concerns the triplet state center. In fact this point defect is observed in the same materials in which the E'$_\delta$ centers is also induced. Furthermore, as reported in Paragraph 6.1.2, these two defects manifest similar growth curve with irradiation dose. These properties agree with the results of previous experimental investigations, as discussed in Paragraph 2.3.5, and suggest the existence of some correlation between these two centers. In particular, these features are compatible with the possibility, already raised in literature [120, 122, 123], that these two centers could originate from the same precursor site.

At variance to the irradiation data above discussed, which seem to be not affected by the [AlO$_4$]$^0$ centers induced in the material, the thermal treatment experiments have permitted us to point out the relevant role played by these defects. In fact, as reported in Paragraph 6.12, a generation process is activated for temperatures higher than ~500 K which involves E'$_\gamma$, E'$_\delta$ and E'$_\alpha$ centers, meanwhile the [AlO$_4$]$^0$ concentration decreases. It is worth to note that a growth process of defects concentration similar to the one reported in Figure 6.12 for the E'$_\gamma$ centers is usually observed in irradiated quartz, as discussed in Paragraph 2.2.5. In fact, in that case it has been found that for T~500 K the E'$_1$ center concentration grows in correspondence to the annealing of [AlO$_4$]$^0$ hole centers. By a detailed EPR analysis, a hole transfer process from [AlO$_4$]$^0$ to the sites precursors of E'$_1$ center was supposed to be responsible for these characteristic features [39]. The typical concentration curves of [AlO$_4$]$^0$ and E'$_1$ during an isochronal thermal treatment of an irradiated quartz sample are reported in Figure 2.8. A strict correspondence is evident from the comparison between these curves and those obtained for the [AlO$_4$]$^0$ and E'$_\gamma$ centers in a-SiO$_2$ reported in Figure 6.12. These strict analogies indicate that a similar process occurs in both systems and, by extension, they suggest that the growths of concentration of defects observed in a-SiO$_2$ materials are due to a hole transfer process from the [AlO$_4$]$^0$ to the sites precursors of E'$_\gamma$, E'$_\delta$ and E'$_\alpha$ centers.

We note that the occurrence of this hole transfer process in the a-SiO$_2$ samples considered in the present chapter, indicates that the thermally induced E'$_\gamma$, E'$_\delta$ and E'$_\alpha$ centers are positively charged. This feature agrees with the results of previous investigations by capacitance-voltage measurements and charge injection on the E'$_\delta$ and E'$_\gamma$ centers induced in a-SiO$_2$ films on crystalline Si (see Paragraphs 2.3.1 and 2.3.4) and indicates that the latter center consists in a positively charged oxygen vacancy (see Paragraphs 2.3.1). In addition, this experiment represents the first reported evidence of the positive charge state of the E'$_\alpha$ center.

As it is evident from the data reported in Table 6.1, the E'$_\alpha$ center is induced upon irradiation only in QC and P453 whereas, upon subsequent thermal treatment, it is observed in all the materials considered even if with different concentrations. In particular, a correlation exists between the concentration of E'$_\alpha$ centers and the oxygen deficiency of the materials (see Table 6.1), suggesting that the site precursor of the E'$_\alpha$ center is oxygen-deficient. Furthermore, as it follows from inspection of the data reported in Table 6.1, the concentration of thermally induced E'$_\alpha$ centers is significantly higher than the concentration of twofold coordinated Si defects in all the materials considered.[2] Consequently, it can be concluded that the E'$_\alpha$ centers does not originate by electron capture nor by hole capture on a twofold coordinated Si, as suggested by Griscom [2] and Uchino [97, 121], respectively (see Paragraph 2.3.3). In addition, the former model disagrees also with the experimental evidences, discussed above, that the E'$_\alpha$ centers is positively charged. It is worth to note that, at variance to the twofold coordinated Si, the concentration of oxygen vacancies is higher than the concentration of thermally induced E'$_\alpha$ centers in all the materials considered (see Table 6.1). Consequently, it can be supposed that the E'$_\alpha$ center could arise by hole capture by an oxygen vacancy. Further evidences on this hypothesis are discussed in the successive chapter in connection with the results we found on the strong hyperfine structure of this defect.

An intriguing point concerns the curves of defects concentrations as a function of the γ-ray irradiation dose reported in Figure 6.4. As shown, the concentration of E'$_\delta$ centers is found to reach a constant value for doses higher than $10^2$ kGy. Furthermore, a rough analysis on the irradiation data on the E'$_\alpha$ centers has conducted to a similar result. In principle, the fact that for irradiation doses higher than $10^2$ kGy the concentrations of E'$_\delta$ and E'$_\alpha$ centers remain constant could indicate that their precursors are exhausted and no other defects can be induced by further irradiation. However, in contrast with this picture, if a sample irradiated at a dose higher than $10^2$ kGy is subjected to a thermal treatment at temperatures above T$\cong$500 K, the concentrations of E'$_\delta$ and E'$_\alpha$ centers start again to increase, indicating that the precursor sites of the defects were not exhausted. This apparent contradiction is clarified by the results of the reirradiation experiments. In fact, as reported in Paragraph 6.3, if an irradiated and thermally treated sample is subjected to reirradiation, then the concentrations of E'$_\delta$ and E'$_\alpha$ centers are reduced, returning to the values they had just after the first irradiation. This experimental evidence suggests that the limit values of concentrations of E'$_\delta$ and E'$_\alpha$ centers obtained for γ-ray irradiation doses higher than ~$10^2$ kGy is the result of an equilibrium between the irradiation process, which generates these defects, and some other process, which induces their disappearance. In this scheme, the increase of E'$_\delta$ and E'$_\alpha$ centers concentrations upon subsequent thermal treatment should be the consequence of the breaking of this equilibrium connected with the change of the process of defects generation. In particular, upon thermal treatment the E'$_\delta$ and E'$_\alpha$ centers concentrations increase, indicating that the process which induces the defects to disappear is weakened (or it is absent at all) with respect to the irradiation. In this context, it is worth to note that one of the most evident differences between the irradiation and the thermally induced hole transfer processes consists in the fact that while in the former electron-hole pairs are generated, in the latter a selective hole migration should occur. In this regard, it is interesting to mention that the E'$_\delta$ center has been found to be induced in larger concentration by selective hole injection than by X-ray irradiation in a-SiO$_2$ films on crystalline Si [126]. Furthermore, in the same type of systems the electron capture cross section of the E'$_\delta$ center has been determined, and it was found to be about one order of magnitude larger than that of the E'$_\gamma$ center [92, 127]. Consequently, the different concentrations of E'$_\delta$ centers induced by irradiation and thermal treatment in the samples we considered in the present chapter could be attributed to the different efficiency of the E'$_\delta$ electron capture process. In fact, it is expected to be efficient during irradiation, as a large number of electrons are available, whereas it should be substantially absent in the process of defects generation by thermally induced selective hole transfer from the [AlO$_4$]$^0$ to the sites precursors of the E'$_\delta$ centers. This effect is negligible for the E'$_\gamma$ center, as it possesses a sensibly lower electron capture cross section than the E'$_\delta$. In this scheme, although no experimental estimation of the electron capture cross section of the E'$_\alpha$ centers has never been reported in literature, the quite similar results obtained in the reirradiation experiments for this defect and the E'$_\delta$ center suggest that analogous comments apply to both of them.

---

[2] It is worth to note that, whereas the concentration of E'$_\alpha$ centers were estimated in samples γ-ray irradiated at a dose of ~$10^2$ kGy, that of the twofold coordinated Si reported in Table 6.1 were obtained in the as-grown materials. However, by performing PL measurements on the 4.4 eV band in the irradiated samples, we verified that the irradiation at $10^2$ kGy does not affect appreciably the concentration of twofold coordinated Si.

Summarizing, the cycle consisting in irradiation → thermal treatment → reirradiation can be described as follows. Upon $\gamma$-ray irradiation the E'$_\delta$ and E'$_\alpha$ centers concentrations grow up to a dose of $\sim 10^2$ kGy. For higher doses a constant concentration value is maintained, as a consequence of the establishment of an equilibrium between the radiation induced ionization of the sites precursors of the E'$_\delta$ and E'$_\alpha$ centers and the inverse process in which these defects, once formed, capture an electron and return to their pristine structures. Upon subsequent thermal treatment, a selective hole-migration process is activated and the concentrations of E'$_\delta$ and E'$_\alpha$ centers increase, as well as that of the E'$_\gamma$, as no electron is available and consequently no electron capture process could occur. Finally, upon subsequent reirradiation of the sample, electrons are again available and the E'$_\delta$ and E'$_\alpha$ centers concentrations return to the values they had after the first irradiation, as imposed by the establishment of the equilibrium between the radiation-induced ionization of the sites precursors and the electron capture process by the generated defects.

# Chapter 7

## *Investigation on the strong hyperfine structures of E'$_\delta$ and E'$_\alpha$ centers*

In the present chapter we report on the strong hyperfine structures of the E'$_\delta$ and E'$_\alpha$ centers. As discussed in the previous chapter, these defects are induced by γ-ray irradiation and by a successive thermal treatment in Al-containing oxygen-deficient materials. We consider the sample P453/10$^3$ thermally treated by isochronal steps in the range of temperatures from 333 K up to 1023 K, and many samples of KI, KUVI, QC and P453 materials γ-ray irradiated at a dose of 10$^2$ kGy and then isothermally treated at T=580 K and T=630 K. The effects of these thermal treatments on the paramagnetic point defects were reported and discussed in the previous chapter. In the present chapter the role of the thermal treatments is to enhance the concentration of E'$_\delta$ and E'$_\alpha$ centers in order to make the usually weak EPR signal of their hyperfine structures more easily detectable. Furthermore, another advantage of making use of thermal treatments is that, whereas the EPR signals of the E' centers and of their hyperfine structures increase, those arising from the [AlO$_4$]$^0$ and from other not concerned paramagnetic centers decrease, so making the identification and the analysis of the hyperfine structures of the E'$_\delta$ and E'$_\alpha$ centers simpler.

### 7.1  The 10 mT doublet

#### 7.1.1 Correlation between the 10 mT doublet and the E'$_\delta$ center

In the isochronal thermal treatment experiment of the sample P453/10$^3$ [see Figure 6.12 (Chapter 6)], the 10 mT doublet was isolated in the SH-EPR spectra for temperature from 450 K to 650 K, as shown in Figure 7.1 (b) for T=580 K. In the same spectrum the 7.4 mT doublet is also evident, attributed to a point defect with a structure similar to that of the E'$_\gamma$ but involving an H atom [81]. Due to the superposition of many EPR lines, a fit procedure was necessary to estimate the SH-EPR signal of the 10 mT pair, in which the right and the left components of the 10 mT doublet were analyzed separately. The procedure used to fit the hyperfine spectra is illustrated in Figure 7.1 (c) for the right part of the spectrum. We have found that the experimental line can be properly fitted by a superposition of three Gaussian profiles: one describes the tail on the left of the spectra, while the other two Gaussians take into account the right components of the 7.4 mT and of the 10 mT doublets. Finally, the SH-EPR intensity of the right component of the 10 mT doublet was obtained by simple integration of the Gaussian profile peaked at ~354 mT. With a similar procedure the SH-EPR signal intensity of the left component was also estimated, and the total intensity was obtained by summing the contributions of both components. In addition, this fit procedure has permitted us to estimate that the low- and the high-field components of the 10 mT doublet have a full width at half maximum (FWHM) of ~1.7 mT and ~1.66 mT, respectively. In Figure 7.1 (a) the dependence of the SH-EPR intensity of the 10 mT doublet on the isochronal thermal treatment temperature is compared to the concentration of the E'_δ centers, estimated from FH-EPR measurements by using a fit procedure similar to that reported in Figure 6.10 (Chapter 6). As shown, a quite good correlation is found, suggesting that the two EPR signals are correlated.

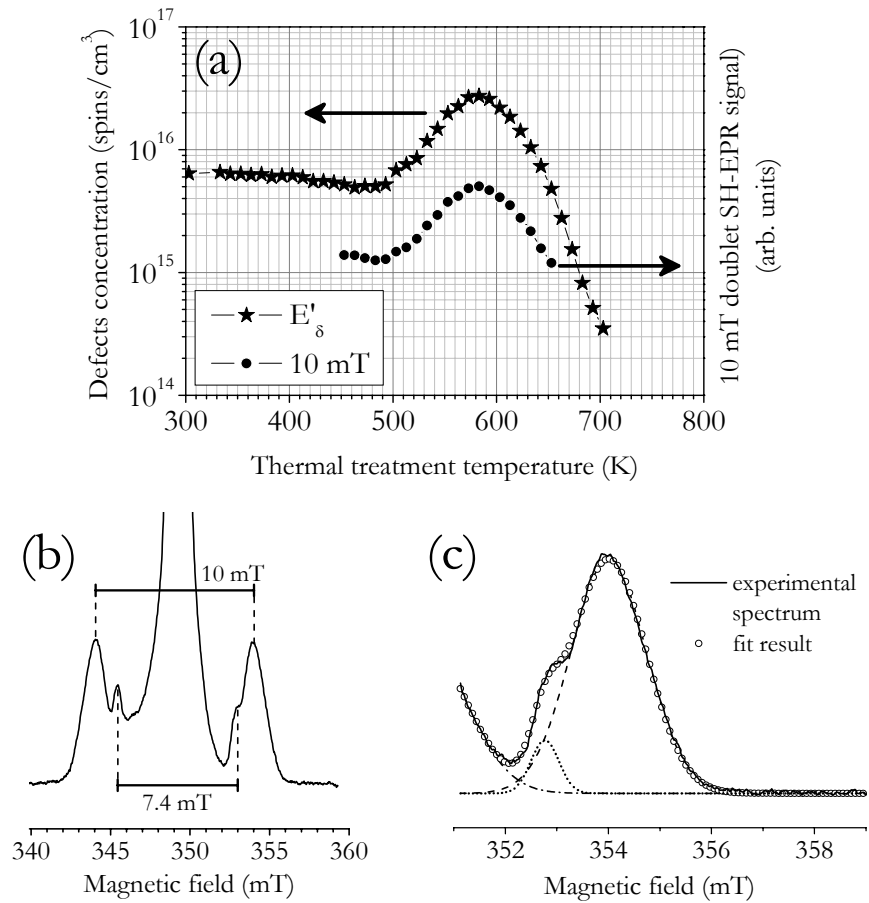

**Figure 7.1** P453/10³: (a) E'_δ center concentration (left vertical scale) and 10 mT doublet SH-EPR signal intensity (right vertical scale) as a function of the isochronal thermal treatment temperature. (b) SH-EPR spectrum of the 10 mT doublet acquired after thermal treatment at T=580 K. (c) Right side of the spectrum reported in (b) (continuous line) compared to the line obtained as a weighted sum (circles) of three Gaussian profiles (broken lines).

The study of the correlation between the 10 mT doublet and the E'_δ center has been extended to many samples of KUVI, QC and P453 materials γ-ray irradiated at ~$10^2$ kGy and isothermally treated at T=580 K and T=630 K. Similar studies were also attempted in the KI materials. However, a quantitative analysis was prevented due to the weakness of the 10 mT doublet SH-EPR signal observed in this latter material. The results obtained in the study of the 10 mT observed in the isothermally treated KUVI, QC and P453 samples are summarized in Figure 7.2, in which the SH-EPR intensity of the 10 mT doublet is reported as a function of the E'$_\delta$ center concentration. As shown, a quite good correlation is found in three distinct materials and for an overall variation of the defects concentration of about one order of magnitude, definitively supporting the attribution of the 10 mT doublet to the strong hyperfine structure of the E'$_\delta$ center. It is worth to note that in these experiments, in order to avoid a possible contribution under the 10 mT doublet arising from the transitions $|m_s=-1\rangle \leftrightarrow |m_s=0\rangle$ and $|m_s=0\rangle \leftrightarrow |m_s=+1\rangle$ of the triplet state center, before acquiring the 10 mT doublet spectra we have verified that the g $\cong$ 4 resonance was absent, assuring that the thermal treatments had annealed the triplet centers.

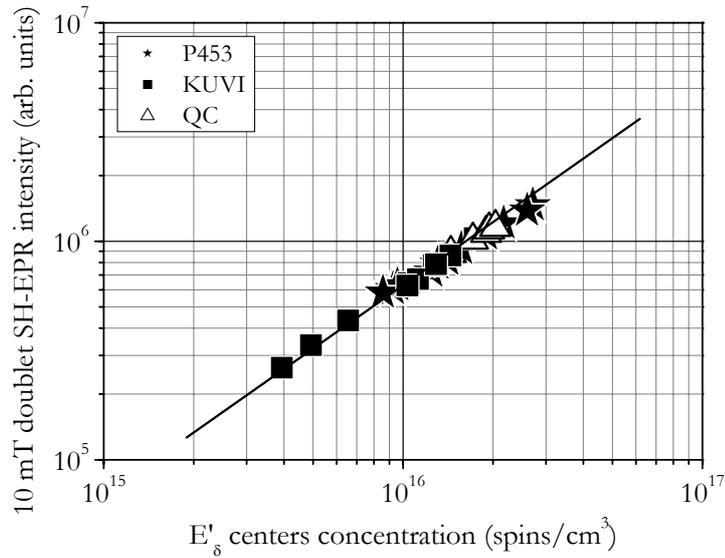

**Figure 7.2** SH-EPR signal of the 10 mT doublet as a function of the concentration of defects responsible for the main resonance line of the E'$_\delta$ center in P453, KUVI and QC samples γ-ray irradiated at ~$10^2$ kGy and subjected to isothermal treatments at T=580 K and T= 630 K. The dimensions of the symbols are comparable with the error on the measurements. The straight line, with slope 1, is superimposed to the data, for comparison.

## 7.1.2 Estimation of the ratio between the concentrations of defects responsible for the 10 mT doublet and the E'$_\delta$ main line

As discussed in Paragraph 2.3.4, the E'$_\delta$ center microscopic structure is still not univocally determined. In particular, it has been proposed that the unpaired electron involved in this defect could be localized on a single Si atom, or delocalized over two or four Si atoms. In this context, a direct experimental estimation of the number of Si atoms over which the unpaired electron wave function is delocalized could be determinant to identify the actual microscopic structure of this point defect of a-SiO$_2$. This information can be obtained from the comparison between the concentration of defects responsible for the 10 mT doublet and those responsible for the main resonance line of the E'$_\delta$ center. By using Eq. (1.26) it follows that if one supposes that the unpaired electron wave function of the E'$_\delta$ center is delocalized over n equivalent Si atoms, then the probability to observe the 10 mT doublet, which corresponds to observe a defect in which only one of the n involved Si atoms possesses a non zero nuclear spin, is given by (see Paragraph 1.1.3.1)

$$P(1;n,0.047) = n(0.047)(1-0.047)^{n-1} \qquad (7.1)$$

Similarly, the probability to observe the main resonance line of the E'$_\delta$ center, which corresponds to observe a defect in which all the n involved Si atoms possess zero nuclear spin, is given by

$$P(0;n,0.047) = (1-0.047)^n \qquad (7.2)$$

The ratio $\zeta$ (n) between these two probabilities is

$$\zeta(n) = \frac{P(1;n,0.047)}{P(0;n,0.047)} = n \frac{(0.047)}{(1-0.047)} \qquad (7.3)$$

Eq. (7.3) points out that the concentration of defects responsible for the 10 mT doublet is proportional to that of defects responsible for the main resonance line of the E'$_\delta$ center and that the coefficient of proportionality between these two concentrations depends on the number n of Si atoms over which the unpaired electron wave function is delocalized. Consequently, by determining experimentally the value of this proportionality coefficient, the number of Si atoms involved in the E'$_\delta$ center can be determined.

In order to obtain an experimental estimation of the concentration of defects responsible for the 10 mT doublet we have performed FH-EPR measurements in the γ-ray irradiated and thermally treated Al-containing samples. We have found that, upon thermal treatment, nearly in correspondence of the maximum concentration of E'$_\delta$ centers obtained, the FH-EPR signal of the 10 mT doublet is actually detectable. A typical FH-EPR spectrum for the right component of this doublet is reported in Figure 7.3, as acquired by using $P_{in} = 2.5 \times 10^{-2}$ mW, $\nu_m = \omega_m/2\pi = 100$ kHz and $H_m = 0.5$ mT. It has been obtained in a sample KUVI/10$^2$ isothermally treated at

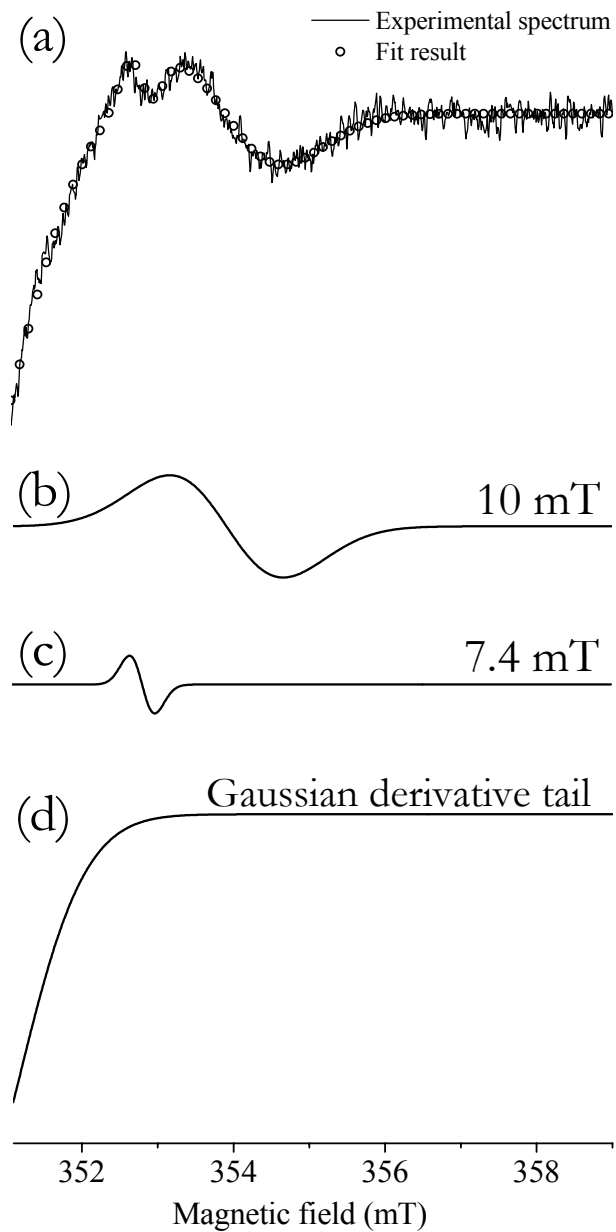

**Figure 7.3** (a) FH-EPR spectrum for a sample KUVI/10² isothermally treated at T=630 K for 510 s (continuous line) compared to the line obtained as a weighted sum (circles) of the reference lines (b)-(d).

T=630 K for 510 s (continuous line). To assure that the FH-EPR signal acquired at this microwave power level is not saturated we have studied the saturation properties of the FH-EPR signal of the 10 mT doublet. This study is reported in Figure 7.4, as obtained in the sample P453/10³ after isochronal thermal treatments up to 580 K. In this figure, the result of the same study performed for the 42 mT doublet in a sample EQ906/10³ [already presented in Figure 5.9 (Chapter 5)] is also reported, for comparison. As shown, the 10 mT and the 42 mT doublets share virtually identical properties with microwave power, as already pointed out for the main resonance lines of the E'$_\gamma$ and E'$_\delta$ centers [see Figures 6.3 and 6.11 (Chapter 6)].

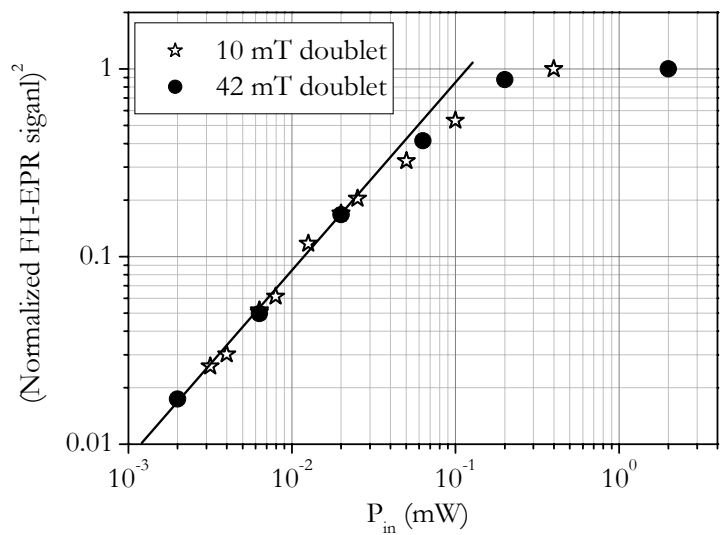

**Figure 7.4** Saturation with microwave power of the FH-EPR signal of the right component of the 10 mT doublet (stars) obtained in a sample P453/10³ isochronally thermally treated up to 580 K. The results obtained for the 42 mT doublet (circles) in a sample EQ906/10³ are also reported, for comparison. The linear dependence of the squared FH-EPR signal as a function of $P_{in}$ is evidenced by the straight line.

Once the experimental parameters for a reliable detection of the unsaturated FH-EPR signal of the 10 mT were established, they were used to estimate the concentration of defects responsible for this structure in many samples of KUVI, P453 and QC materials γ-ray irradiated at ~$10^2$ kGy and isothermally treated at T=580 K and T=630 K. Similar studies were also attempted in the KI material. However, the FH-EPR signal of the 10 mT doublet was not observed, due to the very low concentration of E'$_\delta$ defects induced in this material [see Table 6.1 (Chapter 6)]. The result obtained in isothermally treated samples of KUVI, P453 and QC are summarized in Figure 7.5, in which we report the concentration of defects responsible for the 10 mT doublet as a function of the concentration of defects responsible for the main resonance line of the E'$_\delta$ center, both estimated by FH-EPR measurements. Three straight lines are also shown in the figure, obtained from Eq. (7.3) assuming n=1, n=2 and n=4. These three n values are those pertaining to the various microscopic structures so far proposed for the E'$_\delta$ center (see Paragraph 2.3.4). As shown in Figure 7.5, the experimental data follow the dependence predicted on the basis of the Eq. (7.3) for n=4, indicating that the unpaired electron involved in the E'$_\delta$ center is actually delocalized over four nearly equivalent Si atoms. Performing a linear fit of the experimental data reported in Figure 7.5, we have obtained an estimation of the slope of the straight line, which represents the concentrations ratio between the detects responsible for the 10 mT doublet and the main resonance line of the E'$_\delta$ center. With this procedure we obtained $\zeta_{exp}$ = 0.18 ± 0.03, in agreement with the value $\zeta$ (n=4) = 0.197 expected for an unpaired electron delocalization over four Si atoms.

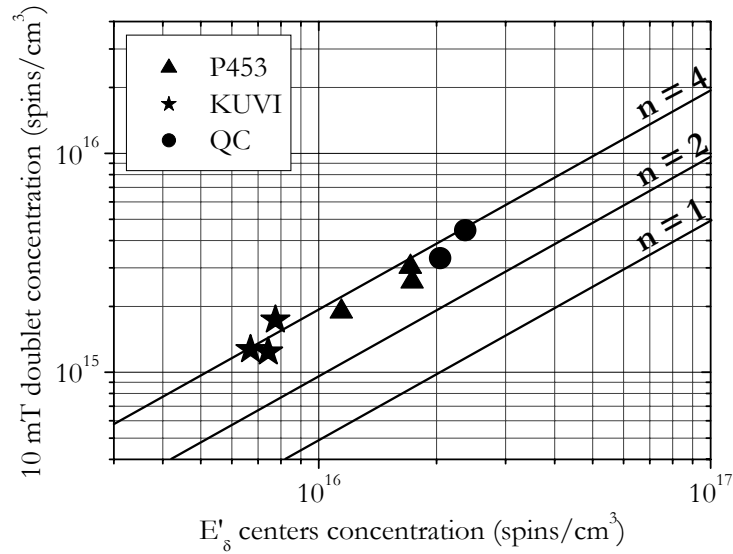

**Figure 7.5** Concentration of defects responsible for the 10 mT doublet as a function of the concentration of defects responsible for the main resonance line of the E'$_\delta$ center in P453, KUVI and QC samples γ-ray irradiated at ~$10^2$ kGy and subjected to isothermal treatments at T=580 K and T= 630 K. The dimensions of the symbols are comparable with the error on the measurements. The straight lines obtained from Eq. (7.3) for n=1, n=2 and n=4 are superimposed to the data, for comparison.

### 7.1.3 Discussion

The results reported in the previous paragraphs for the E'$_\delta$ center confirm that its strong hyperfine structure consists in a pair of lines split by 10 mT. Furthermore, we have determined that the ratio between the concentration of defects responsible for the 10 mT and that of defects responsible for the main resonance line of the E'$_\delta$ center is $\zeta_{exp} = 0.18 \pm 0.03$, indicating that the unpaired electron wave function involved in the defect is delocalized over four equivalent Si atoms, for which it is expected $\zeta$ (n=4) = 0.197. This result agrees with the model proposed by Zhang and Leisure (4-Si atoms) [123], in which the E'$_\delta$ center has been proposed to consist in an unpaired electron delocalized over the four sp$^3$ hybrid orbitals of the Si atoms involved in a pair of nearby oxygen vacancies.

At variance, the value we obtained for $\zeta_{exp}$ definitively rules out that the E'$_\delta$ center could consist in a ionized single oxygen vacancy (2-Si model). This structural model has been supported by several computational works giving as a result a $^{29}$Si hyperfine splitting compatible with that of the E'$_\delta$ center (see Paragraph 2.3.4). However, the expected value of $\zeta$ for this defect is 0.099, in disagreement, beyond any experimental uncertainty, with the value we have estimated. So, if this defect really exists in a-SiO$_2$, it should be well distinguishable from the E'$_\delta$ center. Furthermore, we note that, owing to its axial symmetry, the EPR signal of the ionized single oxygen vacancy should be different from that of E'$_\delta$ center and more similar to that of E'$_\gamma$.

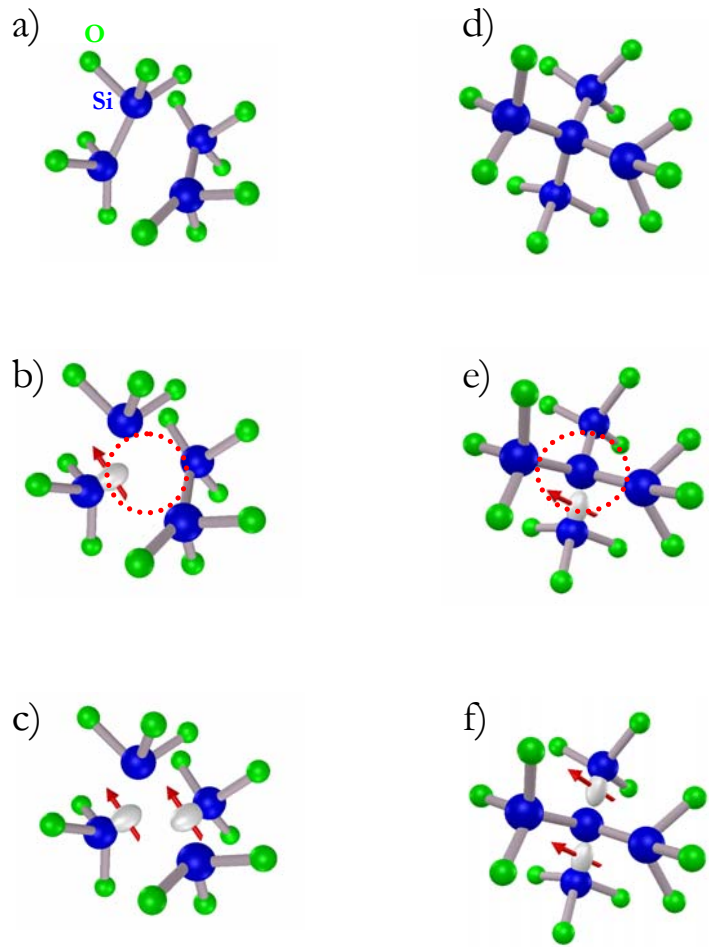

**Figure 7.6** 4-Si model for the site precursor of E'$_\delta$ and triplet centers (a), for the E'$_\delta$ center (b) and for the triplet center (c). 5-Si model for the site precursor of E'$_\delta$ and triplet centers (d), for the E'$_\delta$ center (e) and for the triplet center (f). Arrows represent unpaired electrons in Si-sp$^3$ orbitals.

Again on the basis of the value of $\zeta_{exp}$, a more strong disagreement is found for the model proposed by Conley and Lenahan [145], in which the E'$_\delta$ center has been supposed to consist in an unpaired electron localized on a single Si atom (1-Si model), similarly to the E'$_\gamma$ center. In this case, in fact, the expected value of $\zeta$ is 0.049, in disagreement with the value of $\zeta_{exp} = 0.18 \pm 0.03$ we obtained. Similar comments apply to the model proposed by Vanheusden and Stesmans [142, 143], in which it has been suggested that the E'$_\delta$ center could consist in an unpaired electron wave function resulting from the superposition of four sp$^3$ orbitals of the Si atom disposed at the center of a 5-Si cluster (5-Si model). However, in this case it is possible to propose a complementary view in which the unpaired electron is supposed to be delocalized over the four outermost Si atoms of the 5-Si cluster, with the overall unpaired electron wave function composed by the four sp$^3$ orbitals of these atoms. In this latter scheme, the expected value of $\zeta$ is 0.197, in agreement with our estimation.

Summarizing, our results suggest that the E'$_\delta$ center originates from ionization of a precursor site consisting in a pair of nearby oxygen vacancies [Figure 7.6 (a)] or in a 5-Si cluster [Figure 7.6 (d)]. In particular, after the ionization of the precursor site, a dynamical relaxation occurs and the remaining unpaired electron becomes delocalized over four symmetrically disposed Si-sp$^3$ orbitals of a pair of nearby oxygen vacancies [Figure 7.6 (b)] or of a 5-Si cluster [Figure 7.6 (e)]. We stress that our conclusions on the microscopic structure of the E'$_\delta$ center agree with the main experimental evidences of this defect, as described in the following. The g tensor is nearly isotropic, as expected for a delocalized highly symmetric electronic wave function. The hyperfine splitting of the E'$_\delta$ center is ~4 times smaller than that of E'$_\gamma$ center (10 mT ≈ ¼ · 42 mT), due to delocalization of the unpaired electron over four Si-sp$^3$ orbitals each one similar to that involved in the E'$_\gamma$ center. The ratio between the concentration of defects responsible for the 10 mT and that of defects responsible for the main resonance line of the E'$_\delta$ center is $\zeta_{exp}$ ~ 0.18. This is the consequence of the existence of four nearly equivalent sites of the defect in which the $^{29}$Si can be localized. Finally, the different depth profiles of E'$_\delta$ and E'$_\gamma$ centers observed in SIMOX samples (see Paragraph 2.3.4) are a direct consequence of the higher oxygen deficiency needed for the formation of the precursors of the E'$_\delta$ (two nearby oxygen vacancies or a small Si cluster) with respect to the E'$_\gamma$ centers (single oxygen vacancy).

As a final remark, we note that in the previous chapter evidences have been reported on the concomitant production of E'$_\delta$ and triplet centers and on their similar concentration growth as a function of the γ-ray irradiation dose, indicating that some correlation between these point defects could exist. In particular, on the basis of these experimental evidences, it can be supposed that the E'$_\delta$ and the triplet centers could share the same precursor site, as already proposed in previous works (see Paragraph 2.3.5). Under this hypothesis, our results suggest that the triplet state center could consist in two weakly interacting unpaired electrons localized in two different Si sp$^3$ orbitals of a pair of nearby oxygen vacancies [Figure 7.6 (c)] or of a 5-Si cluster [Figure 7.6 (f)]. In this scheme, a single and a double ionization of the same precursor site could be the processes responsible for the generation of the E'$_\delta$ and the triplet center, respectively.

## 7.2 The 49 mT doublet

### 7.2.1 Correlation between the 49 mT doublet and the E'$_\alpha$ center

In the γ-ray irradiated and thermally treated samples we have looked for the hyperfine structure of the E'$_\alpha$ center. We have found that in the samples in which the E'$_\gamma$ main resonance is detected and that of the E'$_\alpha$ is absent, a pair of lines split by 42 mT is observed, whose spectroscopic features coincide with those attributed to the hyperfine structure of the E'$_\gamma$ center. At variance, in the samples in which both E'$_\gamma$ and E'$_\alpha$ main EPR lines are detected, the hyperfine spectra show a composite nature. This experimental evidence is shown in Figure 7.7 in which the SH-EPR spectrum (continuous line) obtained for a sample KUVI/10$^2$ isothermally treated at T=630 K for ~7x10$^4$ seconds is reported. In order to account for these features the hyperfine spectra were fitted with a weighted sum (circles) of two pairs of lines split by 42 mT and 49 mT (broken lines), as shown in Figure 7.7. The 42 mT doublet considered in the fit was obtained experimentally in a sample EQ906/5 (in which E'$_\gamma$ but no E'$_\alpha$ centers are induced), whereas the pair split by 49 mT was obtained from the residual signal after the fitting of the experimental spectrum of Figure 7.7 with the 42 mT doublet alone. The spectral properties of the 42 mT doublet obtained in the EQ906/5 coincide with those reported in the first row of Table 5.1 for the E'$_\gamma$, whereas from the fit procedure reported in Figure 7.7 we have estimated that the low- and the high-field components of the 49 mT doublet have a FWHM of ~4.5 mT and ~4 mT, respectively. These values are about 25 % larger than those of the 42 mT doublet components (see first row in Table 5.1), suggesting a wider statistical distribution of the isotropic hyperfine constant. Once the line shapes of the 42 mT and 49 mT doublets were determined, they were used to fit the experimental hyperfine spectra obtained in all the other materials considered.

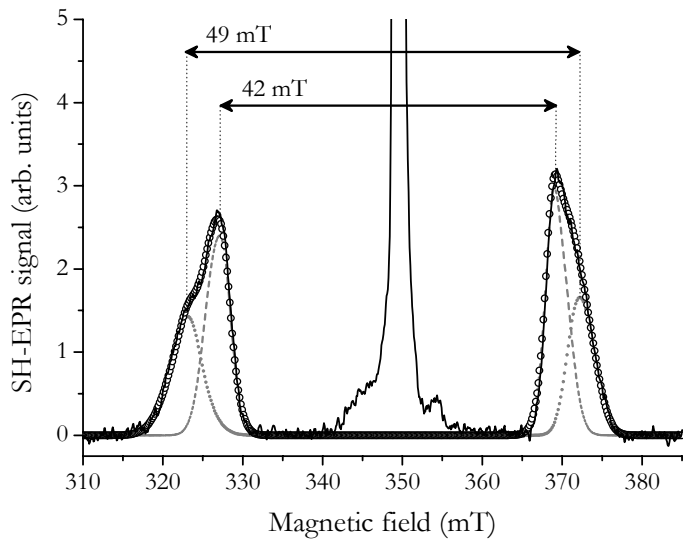

**Figure 7.7** SH-EPR spectrum on a wide field scan obtained for a sample KUVI/10$^2$ isothermally treated at T=630 K for ~7x10$^4$ s (continuous line) compared to the line obtained as a weighted sum (circles) of the reference lines for the 42 mT and 49 mT doublets (broken lines).

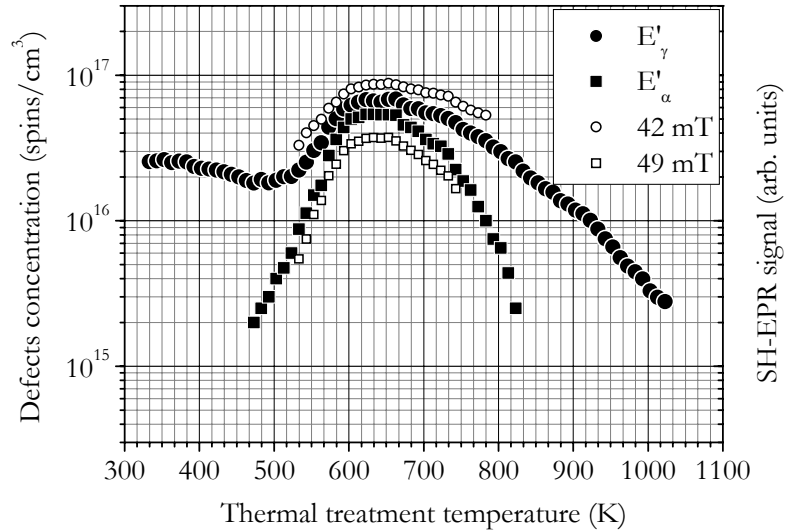

**Figure 7.8** Concentration of E'$_\gamma$ and E'$_\alpha$ centers in the sample P453/10$^3$ as a function of the isochronal thermal treatment temperature compared with the intensities of the 42 mT and 49 mT doublets SH-EPR signals. Filled symbols refer to the left scale, whereas open symbols refer to the right scale.

In Figure 7.8 we report the concentrations of E'$_\gamma$ and E'$_\alpha$ centers estimated in the sample P453/10$^3$ as a function of the isochronal thermal treatment temperature. In the same figure, the SH-EPR signal intensities of the 49 mT and 42 mT doublets are also shown and point out that the former correlates with the concentration of E'$_\alpha$ center, whereas, as expected, the latter correlates with the concentration of E'$_\gamma$ center. This study was extended to many samples of KI, KUVI, P453 and QC materials γ-ray irradiated at ~10$^2$ kGy and isothermally treated at T=580 K and T=630 K. The results obtained are summarized in Figure 7.9, in which we report the SH-EPR signal intensity of the 49 mT doublet as a function of the concentration of defects responsible for the main resonance line of the E'$_\alpha$ center in all the considered samples subjected to different thermal treatments. As shown, the SH-EPR intensity of the 49 mT doublet and the concentration of the E'$_\alpha$ center change in a strictly correlated way, supporting the attribution of the 49 mT pair to the hyperfine structure of the E'$_\alpha$ center, originating from the hyperfine interaction of the unpaired electron with a $^{29}$Si nucleus.

The attribution of the 49 mT doublet to the hyperfine structure of the E'$_\alpha$ center has been also investigated by measuring the relative concentration of defects responsible for the 49 doublet with respect to the main resonance line of the same center. However, a complete study over a large number of samples subjected to different treatments was not possible due to the low concentration of defects which prevents the FH-EPR signal of the 49 mT doublet to be detected. For these reasons we have considered a QC sample with about double volume with respect to the typical samples considered. This sample was γ-ray irradiated at a dose of ~10$^2$ kGy and was isothermally treated at T=630 K up to 5x10$^4$ s in order to obtain the maximum concentration of E'$_\alpha$ centers. In this sample a weak FH-EPR signal arising from the left-component of the 49 mT doublet has been observed by using the same acquisition parameters established for the 42 mT doublet. This choice followed the assumption that the 49 mT and 42 mT doublets FH-EPR signals dependencies on microwave power were similar. Finally, by applying a fit procedure to the obtained FH-EPR spectrum, we have verified that the relative concentration of defects responsible for the 49 mT doublet with respect to those responsible for the main resonance line of the E'$_\alpha$ center is ~5 %, in agreement with the ~4.7 % natural abundance of $^{29}$Si nuclei.

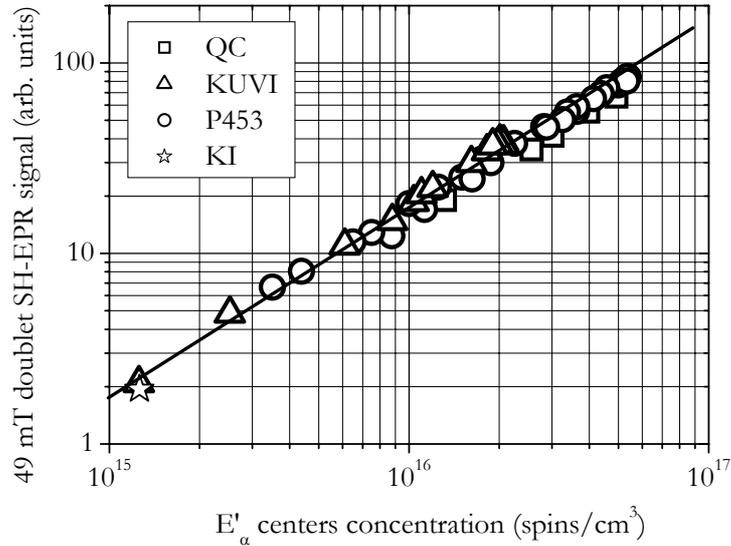

**Figure 7.9** SH-EPR signal of the 49 mT doublet as a function of the concentration of defects responsible for the main resonance line of the E'$_\alpha$ center in KI, KUVI, P453 and QC samples γ-ray irradiated at ~10$^2$ kGy and subjected to isothermal treatments at T=580 K and T= 630 K. The straight line, with slope 1, is superimposed to the data, for comparison.

### 7.2.2 Discussion

The results reported in the previous paragraph for the E'$_\alpha$ center have permitted us to point out that the strong hyperfine structure of this point defect consists in a pair of lines split by 49 mT. Furthermore, we have found indication that the ratio between the concentration of defects responsible for the 49 mT and that of defects responsible for the main resonance line of the E'$_\alpha$ center is ~ 0.05, suggesting that the unpaired electron wave function involved in the defect is localized on a single Si atom.

In the previous chapter we have found evidence that the E'$_\alpha$ center is a positively charged oxygen-deficient defect. Furthermore, our results do not support the E'$_\alpha$ structural models consisting in a twofold coordinated Si having trapped or lost an electron, as previously suggested by Griscom in 2000 [2] and by Uchino *et al.* [97, 121], respectively (see Paragraph 2.3.3). In fact,

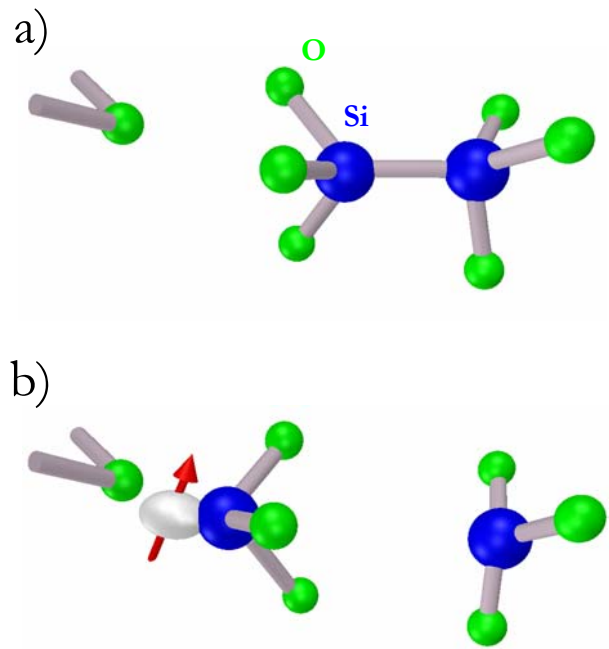

**Figure 7.10** (a) Oxygen vacancy with a nearby O atom. (b) Back projected dangling bond structure proposed to account for the E'$_\alpha$ center, supposed to originate from (a) by ionization of the Si-Si bond. Arrow represents an unpaired electron in an Si-sp$^3$ orbital.

the concentration of E'$_\alpha$ centers induced in all the materials considered is larger than the concentration of twofold coordinated Si estimated in the as-grown materials [see Table 6.1 (Chapter 6)].

Alternatively, it can be supposed that the E'$_\alpha$ center could originate from a hole trapped in a single oxygen vacancy. In this frame, since we exclude that, upon the isothermal treatment at T=630 K which generates the E'$_\alpha$ centers, oxygen atoms can be moved out from their regular site in the a-SiO$_2$ matrix, as suggested in the model proposed by Griscom in 1984 [79], we conclude that only oxygen vacancies already present in the material before the thermal treatment are involved in the generation process of the E'$_\alpha$ centers. This conclusion agrees with the fact, noted in the discussion at the end of the previous chapter, that the overall concentration of the induced E'$_\alpha$ centers has been found to be lower than the concentration of oxygen vacancies in all the as grown materials considered.

The strict similarity of the E'$_\gamma$ and E'$_\alpha$ centers hyperfine structures suggests that similar Si-sp$^3$ hybrid orbitals are involved in the two defects. On the other hand, the orthorhombic components of the E'$_\alpha$ center $\hat{g}$ tensor could indicate that a weak interaction of the unpaired electron with the atoms disposed close to the defect occurs. In order to take into account these two properties, we suggest that the microscopic structure of the E'$_\alpha$ center could consist in a hole trapped in an oxygen vacancy with the unpaired electron sp$^3$ hybrid orbital pointing away from the vacancy in a back-projected configuration and interacting with an extra oxygen atom of the a-SiO$_2$ matrix. In Figure 7.10 schematic representations of the E'$_\alpha$ center (b) and of its precursor site (a) are presented. The former is supposed to originate from the latter by ionization of the Si-Si bond. The existence in a-SiO$_2$ of a microscopic structure similar to the one we propose for the E'$_\alpha$ center has been previously suggested on the basis of experimental [47] and theoretical [88] studies. In particular, it is worth to note that in the latter study, on the basis of embedded cluster calculations, an hyperfine constant of ~48.9 mT [88] was calculated, in excellent agreement with our experimental observations.

# Chapter 8

# *Conclusions and suggestions for further investigations*

In the present Ph.D. Thesis we have reported an experimental investigation on the microscopic structures of the E'$_\gamma$, E'$_\delta$, E'$_\alpha$ and triplet state point defects in a-SiO$_2$. This study has been performed by investigating the effects of γ- and β-ray irradiation and of subsequent thermal treatment on many types of materials differing for the processing methods, OH- and Al-contents, and oxygen deficiencies. In the following we summarize the main experimental results reported in the present Thesis and we derive from them some suggestions for further investigations.

## 8.1 Hole transfer process between the [AlO$_4$]$^0$ centers to the sites precursors of the E' centers

Our investigation on the effects of thermal treatments on γ-ray irradiated Al-containing oxygen-deficient a-SiO$_2$ has permitted us to point out that a growth of concentration of E'$_\gamma$, E'$_\delta$, E'$_\alpha$ centers occurs for temperatures higher than ~500 K. The strict correspondence observed in the concentration growth curve of the E'$_\gamma$ centers in the materials we have considered and that typically observed for the E'$_1$ centers in α-quartz suggests that the same process occurs, which could consist in a thermally activated hole transfer process from the [AlO$_4$]$^0$ to the sites precursors of the E' centers. It is worth to note that, while in α-quartz this process involves only the E'$_1$ centers, in a-SiO$_2$ it involves the E'$_\gamma$, E'$_\alpha$ and E'$_\delta$ centers. In particular, while the E'$_\gamma$ center in a-SiO$_2$ should correspond to the E'$_1$ center of α-quartz, the latter two defects are peculiar to the amorphous structure and have no equivalent in α-quartz. The occurrence of the hole transfer process indicates that the thermally induced defects in a-SiO$_2$ are positively charged.

## 8.2   The E'$_\gamma$ center

The experimental investigation on γ-ray irradiated and thermally treated a-SiO$_2$ reported in Chapter 5, has permitted us to point out that two structures of the E'$_\gamma$ center, named E'$_\gamma$ (1) and E'$_\gamma$ (2) centers, can be distinguished on the basis of their EPR and OA features. We have found that in the natural dry and wet and in the synthetic dry materials for γ-ray irradiation doses lower than ~10 kGy and higher than ~10$^3$ kGy the E'$_\gamma$ (1) and E'$_\gamma$ (2) centers are observed, respectively, whereas in the synthetic wet materials only the E'$_\gamma$ (2) is detected for all the irradiation doses considered. On the basis of these data we have concluded that two distinct precursors of the E'$_\gamma$ center exist in a-SiO$_2$: the first one, with lower radiation activation energy, originates the E'$_\gamma$ (1), the second one, with higher radiation activation energy, originates the E'$_\gamma$ (2). Furthermore, we have found that in all the four types of materials the spectral features of the E'$_\gamma$ (2) convert to those of the E'$_\gamma$ (1) upon isochronal thermal treatments in the temperature range from 370 K to 460 K. During these thermal treatments a good correlation has been found between the variation of the $\Delta g_{1,2}$ in the EPR spectra and that of the peak position of the OA band of the E'$_\gamma$ center. On the basis of this experimental evidence and making use of the broken tetrahedron structural model, we have found indications that the OA transition of the E'$_\gamma$ center is localized in the O≡Si$^\bullet$ moiety and that it consists in an electron transition from a valence band state to the orbital of the unpaired electron of the defect. In addition, we have suggested that the different spectral features of the E'$_\gamma$ (2) center with respect to those of the E'$_\gamma$ (1) center could arise from the perturbative effect of the structure facing the O≡Si$^\bullet$ moiety of the defect on the unpaired electron wave function. In this scheme, a tentative model for the E'$_\gamma$ (2) center could be that of the positively charged oxygen vacancy, which comprises a structure facing the unpaired electron wave function constituted by the $^+$Si≡O group. In these hypothesis, the thermally induced change of the EPR and OA features of the E'$_\gamma$ (2) toward those of the E'$_\gamma$ (1) could be connected with the relaxation of the $^+$Si≡O group, which reduces (or cancels at all) its perturbative effect on the unpaired electron wave function. Since the EPR and OA features of the E'$_\gamma$ (1) center are virtually indistinguishable with respect to those reached by the E'$_\gamma$ (2) center after isochronal thermal treatment up to 460 K, it is natural to assume that this facing $^+$Si≡O structure is absent in the E'$_\gamma$ (1) center. In particular, it can be supposed that the E'$_\gamma$ (1) center originates from an O≡Si-H group embedded in the a-SiO$_2$ matrix, following the radiation-induced breaking of the Si-H bond and the subsequent diffusion of the H atom away from its initial site. The occurrence of this E'$_\gamma$ centers generation process in a-SiO$_2$ has been previously suggested on the basis of UV laser irradiation experiments [110, 111, 192-196]. It is worth to note that the attribution of the E'$_\gamma$ (2) center to the positively charged oxygen vacancy model is also supported by the data obtained upon thermal treatment of Al-containing oxygen-deficient a-SiO$_2$ materials, reported in Chapter 6. In that case, in fact, since the E'$_\gamma$ centers are induced by a hole transfer processes, their most probable structure is that of a positively charged oxygen vacancy. In agreement with this hypothesis, the reirradiation experiments have pointed out that a line shape change from L1 to L2 occurs, indicating that the E'$_\gamma$ centers induced by thermal treatments possess the two configuration minima corresponding to the observation of L1 or of L2 EPR line shapes, peculiar of the E'$_\gamma$ (2) center. However, at present, the specific structures above attributed to the E'$_\gamma$ (1) and E'$_\gamma$ (2) centers have to be considered only as tentative, waiting for new experiments giving further and more reliable validations of these attributions.

The experimental investigation on β-ray irradiated and thermally treated a-SiO$_2$ reported in Chapter 5 has pointed out that a gradual increase of the splitting of the E'$_\gamma$ strong hyperfine structure takes place on increasing the dose above $10^4$ kGy, indicating that an increase of the mean bond angle between the dangling bond orbital and the three Si-O back bonds of the O≡Si$^\bullet$ moiety occurs. These data agree with those reported by Devine and Arndt [187] for γ-ray irradiated pressure-densified a-SiO$_2$ materials, suggesting that the β-ray irradiation could induce a densification of the materials. Assuming that a one to one correspondence exists between the effects of the mechanical densification studied by Devine and Arndt [187] and those induced by β-ray irradiation in the materials we considered, a radiation-induced densification of about 3 % has been estimated to occur in the β-ray irradiation dose range from $10^4$ kGy up to $5\times10^6$ kGy. This result agrees with those previously obtained on the basis of Raman [188] and IR absorption [189] studies performed in the same materials. In the future, further support to these conclusions could be obtained by comparing the degree of densification estimated through the study of the splitting of the E'$_\gamma$ center strong hyperfine structure and by Raman and IR absorption spectroscopies with those obtained by direct density measurements. Another possible extension of the present study concerns the effects of β-ray irradiation in the dose range from $10^4$ kGy up to $5\times10^6$ kGy on the OA band of the E'$_\gamma$ center. In fact, the study of the modifications induced on the 5.8 eV band when the microscopic structure of the E'$_\gamma$ center is distorted by irradiation could give further insight on both the nature of the transition responsible for the OA band and the atomic structure of the defect.

## 8.3   The E'$_\delta$ center

In Chapter 6 we have reported evidences confirming that the E'$_\delta$ center is related to the oxygen deficiency of the material and that it is positively charged. In particular, this latter result extends to bulk systems a conclusion drawn in previous works focused on a-SiO$_2$ films on crystalline Si. In Chapter 7 we have confirmed that the strong hyperfine structure of the E'$_\delta$ center consists in a pair of lines split by 10 mT, and we have estimated that the ratio between the concentration of defects responsible for this doublet and that of defects responsible for the main resonance line of the E'$_\delta$ center is $\zeta_{exp} = 0.18 \pm 0.03$. This estimation has permitted us to rule out that this defect could consist in an unpaired electron localized on a single Si atom [145] or nearly equally shared between the two Si atoms of a single oxygen vacancy [56, 88, 89, 96, 108, 122, 153, 154]. At variance, the value we obtained for $\zeta_{exp}$ indicates that the E'$_\delta$ center consists in an unpaired electron delocalized over four symmetrically disposed Si-sp$^3$ orbitals of a pair of nearby oxygen vacancies [123] or of a 5-Si cluster. In addition, since we have observed that E'$_\delta$ and triplet centers are induced in the same materials and since their growth curves as a function of the γ-ray irradiation dose reach a constant value in correspondence to the same dose, we have supposed that these two defects could share a common precursor site. Under this hypothesis we have suggested that the triplet state center could consist in two weakly interacting unpaired electrons localized in two different Si-sp$^3$ orbitals of a pair of nearby oxygen vacancies [123] or of a 5-Si cluster. So, a single and a double ionization of the same precursor site could be the processes responsible for the generation of the E'$_\delta$ and the triplet center, respectively.

Our study on the microscopic structure of the E'$_\delta$ center could be further developed by investigating its weak hyperfine lines, arising from the hyperfine interaction of the unpaired electron involved in the defect with the second nearest Si atoms. Until now, the observation of these structures has never been reported. However, their identification and the estimation of their hyperfine splittings could give a relevant improvement on the knowledge of the structural properties of the E'$_\delta$ center.

Another point not investigated in the present Thesis which deserves attention in future works concerns the OA and PL properties of the E'$_\delta$ center. Although in a previous experimental work [124] it has been suggested that the E'$_\delta$ center could be responsible for a PL band peaked at 2.2 eV excited at 3.8 eV, no further confirmation of this result has been reported. Consequently, the OA and PL properties of the E'$_\delta$ center are at the present not definitively established. In the future, the investigation of these properties could provide relevant information on the electronic energy levels of the defect, helping to obtain a definitive picture of the E'$_\delta$ center microscopic structure. Similar comments apply to the triplet state center, for which until now no OA and PL data have been reported.

As a final remark, we note that a further extension of the study reported in the present Thesis for the E'$_\delta$ center could consist in the study of the effects of β-ray irradiation for doses higher than $10^4$ kGy on its strong hyperfine structure. In fact, in analogy to what observed for the E'$_\gamma$ center, an increase of the hyperfine splitting could occur, connected with the β-ray irradiation-induced densification of the material. As a successive step, by measuring the increase of the hyperfine splitting for the E'$_\delta$ center and by comparing it with that pertaining to the hyperfine structure of the E'$_\gamma$ center, further relevant structural information on the E'$_\delta$ center could be obtained.

## 8.4 The E'$_\alpha$ center

Concerning the E'$_\alpha$ center, in Chapter 6 we have pointed out that it is related to the oxygen-deficiency of the material and that it is positively charged. Furthermore, in Chapter 7 we have established that it possesses a strong hyperfine structure consisting in a pair of lines split by 49 mT. In addition, we have found indication that the ratio between the concentration of defects responsible for the 49 mT doublet and that of defects responsible for the main resonance line of the E'$_\alpha$ center is about 0.05, suggesting that the unpaired electron wave function involved in the defect is localized on a single Si atom. Our data have excluded that the E'$_\alpha$ center could consist in a twofold coordinated Si having trapped or lost an electron, as previously suggested by Griscom in 2000 [2] and by Uchino *et al.* [97, 121], respectively, as the concentration of E'$_\alpha$ centers induced in all the materials considered has been found to be larger than the concentration of twofold coordinated Si estimated in the as-grown materials. In contrast, our results have suggested that the E'$_\alpha$ center could originate from an oxygen vacancy. Furthermore, the similarities in the strong hyperfine structures of the E'$_\alpha$ and E'$_\gamma$ centers together with the orthorhombic symmetry of the E'$_\alpha$ center $\hat{g}$ matrix have permitted us to hypothesize that this latter defect could consist in a hole trapped in an oxygen vacancy with the unpaired electron Si-sp$^3$ orbital pointing away from the vacancy in a back-projected configuration and interacting with an extra oxygen atom of the a-SiO$_2$ matrix. This model is consistent with a previous work, based on embedded cluster calculations [88], where an hyperfine constant of ~48.9 mT has been obtained for such a structure.

Although the observation of the strong hyperfine structure of the E'$_\alpha$ center has permitted us to obtain one of the most relevant information on its microscopic structure, i.e., that the unpaired electron wave function consists of an Si-sp$^3$ hybrid orbital similar to that involved in the E'$_\gamma$ center, many other relevant properties remain unexplored. One of these is the weak hyperfine structure of the E'$_\alpha$ center. This point is particularly relevant in the light of the microscopic model we proposed for the E'$_\alpha$ center. In fact, it has been shown that, as a consequence of the back projected configuration of the O≡Si$^\bullet$ moiety involved in the E'$_\alpha$ center, the weak hyperfine lines are expected to exhibit a very small splitting [47]. In particular, this splitting should be significantly lower than ~0.8 mT, which is the value characterizing the weak hyperfine doublets of the E'$_1$ center in $\alpha$-quartz. Consequently, the observation of the weak hyperfine lines of the E'$_\alpha$ center and the estimation of their splitting could constitute a straightforward way to establish if the E'$_\alpha$ center actually consists in a dangling bond structure in a back projected configuration.

Another important point which deserves to be investigated in future works regards the OA properties of the E'$_\alpha$ center, which are at present unknown. On the basis of the quite similar structures of the E'$_\gamma$ and E'$_\alpha$ centers and assuming, as concluded in Chapter 5, that the OA band of the former is localized on the O≡Si$^\bullet$ moiety, then it is expected that the E'$_\alpha$ center should exhibit an OA band very similar to that of the E'$_\gamma$ center. Furthermore, in analogy with the 5.8 eV band, no luminescence activities should be related to the OA band of the E'$_\alpha$ center.

Finally, in future works we plan to perform experimental investigations on the effects of $\beta$-ray irradiation for doses higher than 10$^4$ kGy on the strong hyperfine structure of the E'$_\alpha$ center, in order to study how the $\beta$-ray irradiation-induced densification of the material affects the E'$_\alpha$ center microscopic structure.

# List of related publications

1. S. Agnello, R. Boscaino, G. Buscarino and F.M. Gelardi
   *Experimental evidence for two different precursors of $E'_\gamma$ center in silica*
   **J. Non-Cryst. Solids** 345&346 (2004) 505-508.

2. S. Agnello, G. Buscarino and F.M. Gelardi
   *Growth of paramagnetic defects by gamma rays irradiation in oxygen-deficient silica*
   **J. Non-Cryst. Solids** 351 (2005) 1787-1790.

3. S. Agnello, R. Boscaino, G. Buscarino and F.M. Gelardi
   *Modification of optical absorption band of $E'_\gamma$ center in silica*
   **J. Non-Cryst. Solids** 351 (2005) 1801-1804.

4. G. Buscarino, S. Agnello and F.M. Gelardi
   *Delocalized nature of the $E'_\delta$ center in amorphous silicon dioxide*
   **Phys. Rev. Lett.** 94 (2005) 125501 1-4.

5. G. Buscarino, S. Agnello and F. M. Gelardi
   *Characterization of $E'_\delta$ and triplet point defects in oxygen-deficient amorphous silicon dioxide*
   **Phys. Rev. B** 73 (2006) 045208 1-8.

6. G. Buscarino, S. Agnello and F. M. Gelardi
   *Investigation on the microscopic structure of $E'_\delta$ center in amorphous silicon dioxide by electron paramagnetic resonance spectroscopy* **(invited review paper)**
   **Mod. Phys. Lett. B** 20 (2006) 451-474.

7. G. Buscarino, S. Agnello and F. M. Gelardi
   *Hyperfine structure of the $E'_\delta$ centre in amorphous silicon dioxide*
   **J. Phys.: Conden. Matter** 18 (2006) 5213-5219.

8. G. Buscarino, S. Agnello and F. M. Gelardi
   *$^{29}Si$ hyperfine structure of the $E'_\alpha$ center in amorphous silicon dioxide*
   **Phys. Rev. Lett.** 97 (2006) 135502 1-4.

9. S. Agnello, G. Buscarino, M. Cannas, F. Messina, S. Grandi and A. Magistris
   *Structural inhomogeneity of Ge-doped amorphous $SiO_2$ probed by photoluminescence lifetime measurements under synchrotron radiation*
   **Phys. Status Solidi (C)** 4 (2007) 934.

10. G. Buscarino, S. Agnello and A. Parlato
   *Electron paramagnetic resonance line shape investigation of the $^{29}$Si hyperfine doublet of the E'$_\gamma$ center in a-SiO$_2$*
   **Phys. Status Solidi (C)** 4 (2007) 1301.

11. G. Buscarino and S. Agnello
   *Experimental evidence of E'$_\gamma$ centers generation from oxygen vacancies in a-SiO$_2$*
   **J. Non-Cryst. Solids** 353 (2007) 577.

12. G. Buscarino, S. Agnello, F. M. Gelardi and A. Parlato
   *Electron paramagnetic resonance investigation on the hyperfine structure of the E'$_\delta$ center in amorphous silicon dioxide*
   **J. Non-Cryst. Solids** 353 (2007) 518.

# *References*

1. *Structure and Imperfections in Amorphous and Crystalline Silicon Dioxide*, edited by R. A. B. Devine, J. P. Duraud, and E. Dooryhée (Wiley, New York, 2000); and references therein.
2. D. L. Griscom in *Defects in SiO$_2$ and related dielectrics: Science and Technology*, edited by G. Pacchioni, L. Skuja, and D. L. Griscom (Kluwer Academic, Dordrecht, 2000); and references therein.
3. D. L. Griscom, D. B. Brown and N. S. Saks, in *The Physics and Chemistry of SiO$_2$ and Si-SiO$_2$ Interface,* edited by C. R. Helms and B. E. Deal (Plenum, NewYork, 1988).
4. F. B. McLean, H. E. Boesch, Jr., and T. R. Oldham, in *Ionizing Radiation Effects in MOS Devices and Circuits*, edited by T.P. Ma and P.V. Dressendorfer (Wiley, New York, 1989).
5. R. A. B. Devine, *IEEE Trans. Nucl. Sci.* **41** (1994) 452.
6. E. H. Poindexter and W. L. Warren, *J. Electrochem. Soc.* **142** (1995) 2508.
7. P. M. Lenahan and J. F. Conley, Jr, *J. Vac. Sci. Tech.* B **16** (1998) 2134.
8. D. L. Griscom, *Rev. Solid State Science* **4** (1990) 565.
9. D. L. Griscom, P. J. Bray and R. E. Griscom, *J. Chem, Phys.* **47** (1967) 2711.
10. D. L. Griscom, *Glass: Science and Technology, Vol.* 4b, edited by D. R. Uhlmann and N. J. Kreidl (Academic, Boston, 1990), pp 151-251.
11. C. P. Poole Jr., *Electron spin resonance* (Wiley, New York, 1967).
12. A. Abragam and B. Bleaney, *Electron paramagnetic resonance of transition ions* (Clarendon, Oxford, 1970).
13. G. E. Pake and T. L. Estle; *The physical principles of electron paramagnetic resonance (*W. A. Benjamin Inc., Massachussets, 1973).
14. C. P. Slichter; *Principles of magnetic resonance* (Sprinlger-Verlag, Berlin Heidelberg, 1978).
15. J. A. Weil, J. R. Bolton and J. E. Wertz, *Electron Paramagnetic Resonance* (Wiley, New York, 1994).
16. M. Cook and C. T. White, *Semicond. Sci. Technol.* **4** (1989) 1012.
17. G. D. Watkins and J. W. Corbett, *Phys. Rev. B* **134** (1964) A1359.
18. F. J. Feigl and J. H. Anderson, *J. Phys. Chem. Sol.* **31** (1970) 575.
19. R. H. Silsbee, *J. Appl. Phys.* **32** (1961) 1459.
20. F. K. Kneubuhl, *J. Chem. Phys.* **33** (1960) 1074.
21. R. H. Sands, *Phys. Rev.* **99** (1955) 1222.
22. Y. Siderer and Z. Luz, *J. Magn. Res.* **37** (1980) 449.
23. D. L. Griscom in *The physics of SiO$_2$ and its interfaces,* edited by S. T. Pantelides (Pergamon, New York, 1978), p. 232.
24. P. C. Taylor and P. J. Bray, *J. Magn. Res.* **2** (1970) 305.
25. G. Morin and D. Bonnin, *J. Magn. Res.* **136** (1999) 176
26. A. H. Edwards and W. B. Fowler, *Phys. Rev. B* **41** (1990) 10816.
27. H. H. Jaffé and M. Orchin, *Theory and applications of UV spectroscopy* (Wiley, New York, 1970).

28. G. M. Barrow; *Introduction to Molecular Spectroscopy* (McGraw-Hill, 1962).
29. G. Herzberg; *Molecular Spectra and Molecular Structure* (Van Nostrand, New York, 1950).
30. B. H. Bransden and C. J. Joachain; *Physics of Atoms and Molecules* (Longman Scientific & Technical, UK, 1983).
31. A. Smakula; Z. Phys. 59 (1930) 603.
32. R. Wyckoff, *Crystal Structures* (Interscience, New York, 1951).
33. W. H. Zachariasen, *J. Am. Chem. Soc.* **54** (1932) 3841.
34. R. L. Mozzi and B. E. Warren, *J. Appl. Crystallogr.* **2** (1969) 164.
35. J. Neuefeind and K. -D. Liss, *Ber. Bunsenges. Phys. Chem.* **100** (1996) 1341.
36. F. Mauri, A. Pasquarello, B. G. Pfrommer, Y, -G, Yoon and S. G. Louie, *Phys. Rev. B* **62** (2000) 4786.
37. T. M. Clark, P. J. Grandinetti, P. Florian and J. F. Stebbis, *Phys. Rev. B* **70** (2004) 064202; and references therein.
38. R. A. Weeks, *J. Appl. Phys.* **27** (1956) 1376.
39. M. G. Jani, R. B. Bossoli and L. E. Halliburton, *Phys. Rev. B* **27** (1983) 2285.
40. F. J. Feigl, W. B. Fowler and K. L. Yip, *Solid State Commun.* **14** (1974) 225.
41. K. L. Yip and W. B. Fowler, *Phys. Rev. B* **11** (1975) 2327.
42. J. K. Rudra and W. B. Fowler, *Phys. Rev.* B **35** (1987) 8223.
43. D. C. Allan and M. P. Tetter, *J. Am. Ceram. Soc.* **73** (1990) 3247.
44. K. C. Snyder and W. B. Fowler, *Phys. Rev. B* **48** (1993) 13238.
45. M. Boero, A. Pasquarello, J. Sarnthein and R. Car, *Phys. Rev. Lett* **78** (1997) 887.
46. A. S. Mysovsky, P. V. Sushko, S. Mukhopadhyay, A. H. Edwards and A. L. Shluger, *Phys. Rev. B* **69** (2004) 85202.
47. D. L. Griscom and M. Cook, *J. Non-Cryst. Solids* **182** (1995) 119.
48. C. M. Nelson and R. A. Weeks, *J. Am. Ceram. Soc.* **43** (1960) 396.
49. M. Guzzi, F. Pio, G. Spinolo, A. Vedda, C. B. Azzoni and A. Paleari, *J. Phys.: Condens. Matter* **4** (1992) 8635.
50. L. Skuja, *J. Non-Cryst. Solids* **239** (1998) 16.
51. R. A. Weeks and C. M. Nelson, *J. Am. Ceram. Soc.* **43** (1960) 399.
52. V. P. Solntsev, R. I. Mashkovtsev and M. Y. Scherbakova, *J. Struct. Chem.* **18** (1977) 578.
53. J. Isoya, J. A. Weil and L. E. Halliburton, *J. Chem. Phys.* **74** (1981) 5436.
54. A. H. Edwards and W. B. Fowler, *J. Phys. Chem. Solids* **46** (1985) 841.
55. F. Sim, C. R. A. Catlow, M. Dupuis and J. D. Watts, *J. Chem. Phys.* 95 (1991) 4215.
56. P. E. Blöchl, *Phys. Rev.* B **62** (2000) 6158.
57. R. A. Weeks, *Phys. Rev.* **130** (1963) 570.
58. J. K. Rudra, W. B. Fowler and F. J. Feigl, *Phys. Rev. Lett.* **55** (1985) 2614.
59. R. A. Weeks and M. Abraham, *Bull. Am. Phys. Soc.* **10** (1965) 374.
60. R. B. Bossoli, M. G. Jani and L. E. Halliburton, *Solid State Commun.* **44** (1982) 213.
61. L. E. Halliburton, M. G. Jani and R. B. Bossoli, *Nucl. Instrum. Methods Phys. Res. B* **1** (1984) 192.
62. M. G. Jani and L. E. Halliburton, *J. Appl. Phys.* **56** (1984) 942.
63. R. A. Weeks, *J. Non-Cryst. Solids* **179** (1994) 1.
64. J. H. E. Griffiths, J. Owen and I. M. Ward, *Nature* (London) **173** (1954) 439.

65. M. C. M. O'Brien and M. H. L. Pryce, in *Defects in Crystalline Solids (*The Physical Society, London, 1955), p. 88.
66. M. C. M. O'Brien, *Proc. R. Soc. London, Ser. A* **231** (1955) 404.
67. L. E. Halliburton, N. Koumvakalis, M. E. Markes and J. J. Martin, *J. Appl. Phys.* **52** (1980) 3565.
68. R. H. D. Nuttall and J. A. Weil, *Can. J. Phys.* **59** (1981) 1696.
69. J. A. Weil, *Phys. Chem. Minerals* **10** (1984) 149; and references therein.
70. M. J. Mombourquette, J. A. Weil and P. G. Mezey, *Can. J. Phys.* **62** (1984) 21.
71. M. J. Mombourquette and J. A. Weil, *Can. J. Phys.* **63** (1985) 1282.
72. P. S. Rao, R. J. Mc Eachern and J. A. Weil, *J. Comput. Chem.* **12** (1991) 254.
73. A. Continenza and A. Di Pomponio, *Phys. Rev. B* **54** (1996) 13 687.
74. M. Magagnini, P. Giannozzi and A. Dal Corso, *Phys. Rev. B* **61** (2000) 2621.
75. J. Laegsgaard and K. Stokbro, *Phys. Rev. B* **61** (2000) 12590.
76. G. Pacchioni, F. Frigoli, D. Ricci and J. A. Weil, *Phys. Rev. B* **63** (2000) 054102.
77. D. L. Griscom, E. J. Friebele and G. H. Sigel, *Solid State Commun.* **15** (1974) 479; D. L. Griscom, *Phys. Rev. B* **20** (1979) 1823.
78. D. L. Griscom, *Phys. Rev. B* **22** (1980) 4192.
79. D. L. Griscom, *Nucl. Instrum. Methods Phys. Res., Sect. B* **1** (1984) 481.
80. D. L. Griscom, *J. Non-Cryst. Solids* **73** (1985) 51.
81. T. -E Tsai and D. L. Griscom, *J. Non-Cryst. Solids* **91** (1987) 170.
82. D. L. Griscom, *J. Ceram. Soc. Jpn* **99** (1991) 899.
83. A. V. Shendrik and D. M. Yudin, *Phys. Status Solidi B* **85** (1978) 343.
84. J. F. Conley and P. M. Lenahan, *Appl. Phys. Lett.* **62** (1993) 40.
85. J. Li, S. Kannan, R. L. Lahman and G. H. Sigel, Jr., *Appl. Phys. Lett.* **64** (1994) 2090.
86. J. Li, S. Kannan, R. L. Lahman and G. H. Sigel, Jr., *Appl. Phys. Lett.* **66** (1995) 2816.
87. S. Agnello, R. Boscaino, F. M. Gelardi and B. Boizot, *J. Appl. Phys.* **89** (2001) 6002.
88. S. Mukhopadhyay, P. V. Sushko, A. M. Stoneham and A. L. Shluger, *Phys. Rev. B* **70** (2004) 195203.
89. P. V. Sushko, S. Mukhopadhyay, A. S. Mysovsky, V. B. Sulimov, A. Taga and A. L. Shluger, *J. Phys.: Condens. Matter* **17** (2005) S2115.
90. W. L. Warren, P. M. Lenahan, B. Robinson and J. H. Statis, *Appl. Phys. Lett.* **53** (1988) 482.
91. M. E. Zvanut, F. J. Feigl, W. B. Fowler, J. K. Rudra, P. J. Caplan, E. H. Poindexter and J. D. Zook, *Appl. Phys. Lett.* **54** (1989) 2118.
92. J. F. Conley, Jr., P. M. Lenahan, H. L. Evans, R. K. Lowry and T. J. Morthorst, *J. Appl. Phys.* **76** (1994) 2872.
93. V. V. Afanas'ev, J. M. M. de Nijs, P. Balk and A. Stesmans, *J. Appl. Phys.* **78** (1995) 6481.
94. V. V. Afanas'ev and A. Stesmans, *J. Phys: Condens. Matter* **12** (2000) 2285.
95. A. Kalnitsky, J. P. Ellul, E. H. Poindexter, P. J. Caplan, R. A. Lux and A. R. Boothroyd, *J. Appl. Phys.* **67** (1990) 7359.
96. T. Uchino, M. Takahashi and T. Yoko, *Phys. Rev. Lett.* **86** (2001) 5522.
97. T. Uchino, *Curr. Opin. Sol. State Mat. Sci.* **5** (2001) 517.
98. T. Uchino, *J. Ceram. Soc. Jpn* **113** (2005) 17.

99. Z. -Y. Lu, C. J. Nichlaw, D. M. Fleetwood, R. D. Schrimpf and S. T. Pantelides, *Phys. Rev. Lett.* **98** (2002) 285505.
100. S. Agnello, R. Boscaino, G. Buscarino, M. Cannas and F. M. Gelardi, *Phys. Rev. B* **66** (2002) 113201.
101. R. A. Weeks and E. Sonder; in *Paramagnetic Resonance II*, Ed. Low. W., Academic Press, New York (1963) p.869.
102. H. Nishikawa, E. Watanabe, D. Ito and Y. Ohki, *J. Non-Cryst. Solids* **179** (1994) 179.
103. R. Boscaino, M. Cannas, F. M. Gelardi and M. Leone, *Nucl. Instrum. Methods Phys. Res., Sect. B* **116** (1996) 373.
104. D. L. Griscom and W. B. Fowler, in: G. Lucovsky, S.T. Pantelides, F.L. Galeener (Eds.), *The Physics of MOS Insulators*, Pergamon, New York, 1980, p. 97.
105. A. H. Edwards, *Mat. Res. Soc. Symp. Proc.* **61** (1986) 3.
106. G. Pacchioni and G. Ieranò, *Phys. Rev. B* **57** (1998) 818.
107. A. A. Bobyshev and V. A. Radtsig, *Fiz. Khim. Stekla* **14** (1988) 501
108. S. Mukhopadhyay, P. V. Sushko, V. A. Mashkov and A. Shluger, *J. Phys: Condens. Matter* **17** (2005) 1311.
109. M. Cannas and F. Messina, *J. Non-Cryst. Solids* **351** (2005) 1780.
110. F. Messina and M. Cannas, *J. Phys: Condens. Matter* **17** (2005) 3837.
111. F. Messina and M. Cannas, *J. Phys: Condens. Matter* **18** (2006) 9967.
112. F. Messina and M. Cannas, *J. Non-Cryst. Solids* (2007), in press.
113. G. W. Arnold and W. D. Compton, *Phys. Rev.* **116** (1959) 802.
114. W. D. Compton and G. W. Arnold, *Disc. Faraday Soc.* **31** (1961) 130.
115. E. W. J. Mitchell and E. G. S. Paige, *Phil. Mag.* **1** (1956) 1085.
116. M. Antonini, P. Camagni, P. N. Gibson and A. Manara, *Radiat. Eff.* **65** (1982) 41.
117. G. W. Arnold, *IEEE Trans. Nucl. Sci.* **NS-20** (1973) 220.
118. Y. Kawaguchi and N. Kuzuu, *J. Appl. Phys.* **80** (1996) 5633.
119. N. Kuzuu and H. Horikoshi, *J. Appl. Phys.* **97** (2005) 93508.
120. D. L. Griscom and E. J. Friebele, *Phys. Rev. B* **34** (1986) 7524.
121. T. Uchino, M. Takahashi and T. Yoko, *Appl. Phys. Lett.* **78** (2001) 2730.
122. R. Tohmon, Y. Shimogaichi, Y. Tsuta, S. Munekuni, Y. Ohki, Y. Hama and K. Nagasawa, *Phys. Rev. B* **41** (1990) 7258.
123. L. Zhang and R. G. Leisure, *J. Appl. Phys.* **80** (1996) 3744.
124. H. Nishikawa, E. Watanable, D. Ito, Y. Sakurai, K. Nagasawa and Y. Ohki, *J. Appl. Phys.* **80** (1996) 3513.
125. M. E. Zvanut, R. E. Stahlbush and W. E. Carlos, *Appl. Phys. Lett.* **60** (1992) 2989.
126. R.A.B. Devine, D. Mathiot, W. L. Warren, D. M. Fleetwood and B. Aspar, *Appl. Phys. Lett.* **63** (1993) 2926.
127. J. F. Conley, P. M. Lenahan, H. L. Evans, R. K. Lowry and T. J. Morthorst, *Appl. Phys. Lett.* **65** (1994) 2281.
128. W. L. Warren, M. R. Shaneyfelt, D. M. Fleetwood, J. R. Schwank, P. S. Winokur and R. A. B. Devine, *IEEE Trans. Nucl. Sci.* **41** (1994) 1817.
129. W. L. Warren, D. M. Fleetwood, M. R. Shaneyfelt, P. S. Winokur and R. A. B. Devine, *Phys. Rev. B* **50** (1994) 14710.

130. W. L. Warren, D. M. Fleetwood, M. R. Shaneyfelt, J. R. Schwank, P. S. Winokur, R. A. B. Devine and D. Mathiot, *Appl. Phys. Lett.* **64** (1994) 3452.
131. M. E. Zwanut, T. L. Chen, R. E. Stahlbush, E. S. Steigerwalt and G. A. Brown, *J. Appl. Phys.* **77** (1995) 4329.
132. R. A. B. Devine, W. L. Warren, J. B. Xu, I. H. Wilson, P. Paillet and J. L. Leray, *J. Appl. Phys.* **77** (1995) 175.
133. A. Stesmans and V. V. Afanas'ev, *Appl. Phys. Lett.* **69** (1996) 2056.
134. M. E. Zvanut and T. Chen, *Appl. Phys. Lett.* **69** (1996) 28.
135. A. Stesmans, B. Nouwen, D. Pierreux and V. V. Afanas'ev, *Appl. Phys. Lett.* **80** (2002) 4753.
136. A. Stesmans, B. Nouwen and V. V. Afanas'ev, *Phys. Rev. B* **66** (2002) 045307.
137. A. Stesmans, R. Devine, A. G. Revesz and H. Hughes, *IEEE Trans. Nucl. Sci.* **37** (1990) 2008.
138. J. F. Conley Jr, P. M. Lenahan and P. Roitman, *Appl. Phys. Lett.* **60** (1992) 2889.
139. J. F. Conley Jr, P. M. Lenahan and P. Roitman, *IEEE Trans. Nucl. Sci.* **39** (1992) 2114.
140. W. L. Warren, M. R. Shaneyfelt, J. R. Schwank, D. M. Fleetwood, P. S. Winokur, R. A. B. Devine, W. P. Maszara and J. B. McKitterick, *IEEE Trans. Nucl. Sci.* **40** (1993) 1755.
141. W. L. Warren, D. M. Fleetwood, M. R. Shaneyfelt, J. R. Schwank, P. S. Winokur and R. A. B. Devine, *Appl. Phys. Lett.* **62** (1993) 3330.
142. K. Vanheusden and A. Stesmans, *J. Appl. Phys.* **74** (1993) 275.
143. K. Vanheusden and A. Stesmans, *Appl. Phys. Lett.* **62** (1993) 2405.
144. A. Stesmans and K. Vanheusden, *J. Appl. Phys.* **76** (1994) 1681.
145. J. F. Conley and P. M. Lenahan, *IEEE Trans. Nucl. Sci.* **42** (1995) 1740.
146. R. A. B. Devine, D. Mathiot, W. L. Warren and B. Aspar, *J. Appl. Phys.* **79** (1996) 2302.
147. R. E. Walkup and S. I. Raider, *Appl. Phys. Lett.* **53** (1988) 888.
148. G. K. Celler, P. L. F. Hemment, K. W. West and J. M. Gibson, *Appl. Phys. Lett.* **48** (1986) 532.
149. J. Stoemenos, L. Margail, C. Jaussaud, M. Dupuy and M. Bruel, *Appl. Phys. Lett.* **48** (1986) 1470.
150. N. Lopez, F. Illas and G. Pacchioni, *J. Phys. Chem. B* **104** (2000) 5471.
151. C. J. Nicklaw, Z. -Y. Lu, D. M. Fleetwood, R. D. Schrimpf and S. T. Pantelides, *IEEE Trans. Nucl. Sci.* **49** (2002) 2667.
152. M. M. G. Alemany and J. R. Chelikowsky, *Phys. Rev. B* **73** (2006) 235211.
153. J. R. Chavez, S. P. Karna, K. Vanheusden, C. P. Brothers, R. D. Pugh, B. K. Singaraju, W. L. Warren and R. A. B. Devine, *IEEE trans. nucl. sci.* **44** (1997) 1799.
154. S. P. Karna, A. C. Pineda, W. M. Shedd and B. K. Singaraju, *Electrochem. Soc. Proc.* **99-3** (1999) 161.
155. S. P. Karna, H. A. Kurtz, A. C. Pineda, W. M. Shedd and R. D. Pugh, in *Defects in SiO$_2$ and related dielectrics: Science and Technology*, edited by G. Pacchioni, L. Skuja, and D.L. Griscom (Kluwer Academic, Dordrecht, 2000) p. 599.
156. A. C. Pineda and S. P. Karna, *J. Phys. Chem. A* **104** (2000) 4699.
157. S. P. Karna, A. C. Pineda, R. D. Pugh, W. M. Shed and T. R. Oldham, *IEEE Trans. Nucl. Sci.* **47** (2000) 2316.

158. S. Mukhopadhyay, P. V. Sushko, A. H. Edwards and A. L. Shluger, *J. Non-Cryst. Solids* **345&346** (2004) 703.
159. P. O. Fröman, R. Petterson and R. Vänngard, *Ark. Fys.* **15** (1959) 559.
160. S. Lee and P. J. Bray, *Phys. Chem. Gasses* **3** (1962) 37.
161. R. Schnadt and A. Räuber, *Solids State Commun.* **9** (1971) 159.
162. Bruker EMX user manual.
163. S. M. Sze, *Physics of semiconductor device*, (Wiley, 1969).
164. F. Bloch, *Phys. Rev.* **70** (1946) 460.
165. J. S. Hyde, *Phys. Rev.* **119** (1960) 1483.
166. J. R. Harbridge, G. A. Rinard, R. W. Quine, S. S. Eaton and G. R. Eaton, *J. Magn. Res.* **156** (2002) 41.
167. M. Weger, *Bell. Syst. Tech. J.* **39** (1960) 1013.
168. V. E. Galtsev, E. V. Galtseva and Y. A. S. Lebedev, *Appl. Radiat. Isot.* **47** (1996) 1311.
169. V. E. Galtsev, O. Y. Grinberg, Y. S. Lebedev and E. V. Galtseva, *Appl. Magn. Res.* **4** (1993) 331.
170. V. E. Galtsev, E. V. Galtseva, O. Y. Grinberg and Y. S. Lebedev, *J. Radioanal. Nucl. Chem. Lett.* **186** (1994) 35.
171. M. Hiray and M. Ikeya, *Phys. Stat. Sol.* **209** (1998) 449.
172. B. Yan, N. A. Schultz, A. L. Efros and P. C. Taylor, *Phys. Rev. Lett.* **84** (2000) 4180.
173. S. Agnello, R. Boscaino, M. Cannas and F. M. Gelardi, *Phys. Rev. B* **64** (2001) 174423.
174. T. Izawa and S. Sudo; *Optical Fibers: Material and Fabrication* (ADOP) (KTK, Tokyo, 1987), p. 123.
175. T. Izawa, *IEEE journal on selected topics in quantum electronics* **6** (2000) 1220.
176. R. Brukner, *J. Non-Cryst. Solids* **5** (1970) 123.
177. G. Hetherington, K. H. Jack, M. W. Ramsay, *Physics and Chemistry of Glasses* **6** (1965) 6.
178. Heraeus Quartzglas, Hanau, Germany, catalogue POL-0/102/E.
179. Quartz and Silice, Nemours, France, catalogue OPT-91-3.
180. Almaz Optics online catalog.
181. TSL Group PLC, Wallsend, England, Optical Products Cataogue.
182. Corning Spa ISPD Europe, Tecnottica Consonni S.N.C., Calco (LC), Italy.
183. B. Boizot, G. Petite, D. Ghaleb, B. Reynard and G. Calas, *J. Non-Cryst. Solids* **243** (1999) 268.
184. D. L. Griscom in *Structure and Bonding in NonCrystalline Solids*, edited by G. E. Walrafen and A. G. Revesz (Plenum, 1986), pp. 369-384; C. Chabrerie, J. L. Autran, P. Paillet, O. Flament, J. L. Leray and J. C. Boudenot, *IEEE Trans. Nucl. Sci.* **44** (1997) 2007.
185. S. Agnello; *Gamma ray induced process of point defect conversion in silica,* PhD thesis, Department of Physical and Astronomical Sciences, Palermo (2000).
186. S. Agnello, F. M. Gelardi, R. Boscaino, M. Cannas, B. Boizot and G. Petite, *Nucl. Instrum. Methods Phys. Res., Sect. B* **191** (2002) 387.
187. R. A. B. Devine and J. Arndt, *Phys. Rev. B* **35** (1987) 9376.
188. B. Boizot, S. Agnello, B. Reynard, R. Boscaino and G. Petite, *J. Non-Cryst. Solids* **325** (2003) 22.
189. G. Navarra, R. Boscaino, M. Leone and B. Boizot, *J. Non-Cryst. Solids* **353** (2007) 555.
190. S. Agnello and L. Nuccio, *Phys. Rev. B* **73** (2006) 115203.

191. M. Cannas; *Point defects in amorphous SiO$_2$: optical activity in the visible, UV and Vacuum-UV spectral regions,* PhD thesis, Department of Physical and Astronomical Sciences, Palermo (1998); M. Cannas, F. M. Gelardi, F. Pullara, M. Barbera, A. Collura and S. Varisco, *J. Non-Cryst. Solids* **280** (2001) 188.
192. H. Nishikawa, R. Nakamura, R. Tohmon, Y. Ohki, Y. Sakurai, K. Nagasawa and Y. Hama, *Phys. Rev. B* **41** (1990) 7828.
193. H. Imai, K. Arai, H. Hosono, Y. Abe, T. Arai and H. Imagawa, *Phys. Rev. B* **44** (1991) 4812.
194. N. Leclerc, C. Pfleiderer, H. Hitzler, J. Wolfrum, K. O. Greulich, S. Thomas and W. Englisch, *J. Non-Cryst. Solids* **149** (1992) 115.
195. H. Nishikawa, R. Nakamura, Y. Ohki and Y. Hama, *Phys. Rev. B* **48** (1993) 2968.
196. F. Messina; *Role of hydrogen on the generation and decay of point defects in amorphous silica exposed to UV laser radiation,* PhD thesis, Department of Physical and Astronomical Sciences, Palermo (2006).